\documentclass [a4paper, 10pt]{article}
\pdfoutput=1
\usepackage{jheppub1, amsmath, amsfonts, amssymb, amssymb,amsxtra, mathrsfs, makeidx, amsthm, youngtab, stmaryrd, bm, coffee4}

\usepackage{bm}
\usepackage{fancyhdr, dsfont}
\usepackage[latin1]{inputenc}
\usepackage{graphicx}
\usepackage{float}
\usepackage{rotating}
\usepackage{longtable}

\usepackage{color}
\definecolor{dgreen}{rgb}{0,0.70,0.30}
\definecolor{gold}{rgb}{0.85,.66,0}
\definecolor{purple}{rgb}{1.0,0.3,0.6}

\usepackage{tikz}
\usetikzlibrary{calc} \usetikzlibrary{patterns} \usetikzlibrary{decorations.pathreplacing} \usetikzlibrary{decorations.markings} \usetikzlibrary{decorations.pathmorphing} \usetikzlibrary{positioning}

\restylefloat{figure}
\usepackage{epstopdf}

\newcommand{\nwc}{\newcommand}
\nwc{\ba}  {\begin{array}}
\nwc{\ea}  {\end{array}}
\nwc{\bdm} {\begin{displaymath}}
\nwc{\edm} {\end{displaymath}}

\nwc{\bda} {\bdm\ba{lcl}} 
\nwc{\eda} {\ea\edm}

\nwc{\bc}  {\begin{center}}
\nwc{\ec}  {\end{center}}

\nwc{\ds}  {\displaystyle}
\nwc{\bmat}{\left(\ba}
\nwc{\emat}{\ea\right)}
\nwc{\nn}  {\nonumber}
\nwc{\nnn} {\nonumber \vspace{.2cm} \\ }
\nwc{\ra}  {\rightarrow}
\nwc{\lra} {\longrightarrow}

\def\beq{\begin{equation}}
\def\eeq{\end{equation}}

\newcommand{\vecb}{\left(\begin{array}{c}}
\newcommand{\vece}{\end{array}\right)}
\newcommand{\ccb}{\left(\begin{array}{cc}}
\newcommand{\cce}{\end{array}\right)}
\newcommand{\cccb}{\left(\begin{array}{ccc}}
\newcommand{\ccce}{\end{array}\right)}
\newcommand{\ccccb}{\left(\begin{array}{cccc}}
\newcommand{\cccce}{\end{array}\right)}
\newcommand{\cccccb}{\left(\begin{array}{ccccc}}
\newcommand{\ccccce}{\end{array}\right)}
\newcommand{\pa}{\partial}

\newcommand{\la}{\lambda}

\newcommand{\te}{\textrm}
\newcommand{\eq}{ \ \ = \ \ }
\newcommand{\co}{\ , \ \ \ \ \ \ }

\newcommand{\ap}{\alpha'}

\newcommand{\NN}{\mathbb N}
\newcommand{\ZZ}{\mathbb Z}

\newcommand{\tspinb}{\Theta^{\vec{a}}_{\vec{b}} \left[ \begin{smallmatrix}}
\newcommand{\tspine}{\end{smallmatrix} \right]}

\def\bea#1\eea{\allowdisplaybreaks \begin{align}#1\end{align}}
\def\vec#1{\bm{#1}}
\newcommand{\ben}{\begin{enumerate}}
\newcommand{\een}{\end{enumerate}}
\newcommand{\eref}[1]{(\ref{#1})}

\newcommand{\tref}[1]{Table~\ref{#1}}
\newcommand{\fref}[1]{Figure~\ref{#1}}

\newcommand{\PE}{\mathop{\rm PE}}

\newcommand{\BZ}{\mathbb{Z}}

\newcommand{\comment}[1]{}
\newcommand{\CC}{{\cal C}}
\newcommand{\CF}{{\cal F}}

\newcommand{\CN}{{\cal N}}

\newcommand{\CI}{{\cal I}}

\newcommand{\frM}{\mathfrak{M}}

\newcommand{\ie}{{\it i.e.}}
\newcommand{\eg}{{\it e.g.}}
\newcommand{\sgn}{\mathrm{sgn}}
\newcommand{\ud}{\mathrm{d}}

\newcommand{\Sym}{\mathrm{Sym}}

\def\th{\vartheta}

\newcommand{\NS}{\mathrm{NS}}
\newcommand{\Ra}{\mathrm{R}}
\newcommand{\GSO}{\mathrm{GSO}}

\newcommand{\lb}{\llbracket}
\newcommand{\rb}{\rrbracket}

\title{Refined Partition Functions for Open Superstrings with $\mathbf{4}$, $\mathbf{8}$ and $\mathbf{16}$ Supercharges }
\author[a,b,c]{Dieter L\"ust,}
\author[b]{Noppadol Mekareeya,}
\author[d]{Oliver Schlotterer}
\author[e]{and Andrew Thomson}

\affiliation[a]{Arnold-Sommerfeld-Center f\"ur Theoretische Physik \\
Department f\"ur Physik, Ludwig-Maximilians-Universit\"at M\"unchen \\
Theresienstra\ss e 37, 80333 M\"unchen, Germany}
\affiliation[b]{Max-Planck-Institut f\"ur Physik (Werner-Heisenberg-Institut), \\
F\"ohringer Ring 6, 80805 M\"unchen, Germany}
\affiliation[c]{CERN, Theory Group, 1211 Geneva 23, Switzerland}
\affiliation[d]{Max--Planck--Institut f\"ur Gravitationsphysik, \\
Albert--Einstein--Institut, 14476 Golm, Germany}
\affiliation[e]{Theoretical Physics Group, The Blackett Laboratory \\
Imperial College London, SW7 2AZ, United Kingdom}
\emailAdd{dieter.luest@lmu.de}
\emailAdd{noppadol@mpp.mpg.de}
\emailAdd{oliver.schlotterer@aei.mpg.de}
\emailAdd{andrew.thomson09@imperial.ac.uk}
\abstract{We analyse the perturbative massive open string spectrum of even dimensional superstring compactifications with four, eight and sixteen supercharges. In each of such cases, we focus on universal states that exist independently on the internal geometry and other compatification details. We analytically compute refined partition functions that count these states at each mass level.  Such refined partition functions are written in a super-Poincar\'e covariant form, providing information on how supermultiplets transform under the little group and the $R$ symmetry.  Various asymptotic limits of the partition functions and their associated quantities, such as the leading and subleading Regge trajectories, are studied empirically and analytically. In the phenomenologically relevant case of four supercharges, the partition function can be cast into the most compact form and the asymptotic formula in the large spin limit is derived explicitly.}

\begin{document}

\setcounter{tocdepth}{2}
\maketitle

\section{Introduction}

The purpose of this article is to compute the super Poincar\'e covariant perturbative open superstring spectrum which is completely universal to all compactifications to 4, 6 or 8 spacetime dimensions with 4, 8 or 16 supercharges. The number of such models is, of course, enormous and generic representatives have their own characteristic spectrum. Nevertheless, for each given number of supersymmetries (SUSYs), one can identify a set of physical states that exist {\it independently} on the internal geometry and any other compatification details. In this sense, one of the main aims of this work is to focus on universal statements about the spectrum in scenarios with various (even) numbers of spacetime dimensions and supercharges. The basic quantity we compute in different contexts is the number of model independent super Poincar\'e multiplets of given Lorentz and $R$ symmetry quantum numbers on each mass level of the superstring.

\begin{figure}[h]
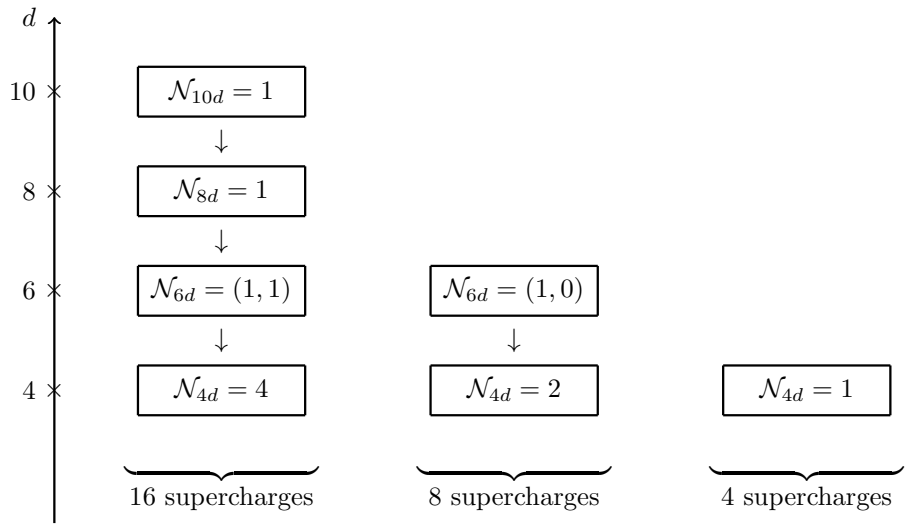

\centerline{
\tikzpicture [scale=1.1,line width=0.30mm]
\draw[line width=0.30mm,->] (0,-1.6) -- (0,4.5);
\draw (-0.1,4.5) node[left]{$d$};
%\draw[line width=0.50mm,->] (-0.7,-1.2) -- (10,-1.2) node[below]{$\#$ supercharges};
\draw (2,-1) node{$\underbrace{ \phantom{XXXXXXXX}}_{}$};
\draw (2,-1.3) node{16 supercharges};
\draw (5.5,-1) node{$\underbrace{ \phantom{XXXXXXXX}}_{}$};
\draw (5.5,-1.3) node{8 supercharges};
\draw (9,-1) node{$\underbrace{ \phantom{XXXXXXXX}}_{}$};
\draw (9,-1.3) node{4 supercharges};
\draw (2,0.6) node{$\downarrow$};
\draw (2,1.8) node{$\downarrow$};
\draw (2,3) node{$\downarrow$};
\draw (5.5,0.6) node{$\downarrow$};
\draw (0,0) node{$\times$}; \draw(-0.1,0)node[left]{$4$} ;
\draw (1,-0.3) -- (1,0.3);
\draw (3,-0.3) -- (3,0.3);
\draw (1,-0.3) -- (3,-0.3);
\draw (1,0.3) -- (3,0.3);
\draw (2,0) node{${\cal N}_{4d}=4$};
\draw (4.5,-0.3) -- (4.5,0.3);
\draw (6.5,-0.3) -- (6.5,0.3);
\draw (4.5,-0.3) -- (6.5,-0.3);
\draw (4.5,0.3) -- (6.5,0.3);
\draw (5.5,0) node{${\cal N}_{4d}=2$};
\draw (8,-0.3) -- (8,0.3);
\draw (10,-0.3) -- (10,0.3);
\draw (8,-0.3) -- (10,-0.3);
\draw (8,0.3) -- (10,0.3);
\draw (9,0) node{${\cal N}_{4d}=1$};
\draw (0,1.2) node{$\times$}; \draw(-0.1,1.2) node[left]{$6$}   ;
\draw (1,0.9) -- (1,1.5);
\draw (3,0.9) -- (3,1.5);
\draw (1,0.9) -- (3,0.9);
\draw (1,1.5) -- (3,1.5);
\draw (2,1.2) node{${\cal N}_{6d}=(1,1)$};
\draw (4.5,0.9) -- (4.5,1.5);
\draw (6.5,0.9) -- (6.5,1.5);
\draw (4.5,0.9) -- (6.5,0.9);
\draw (4.5,1.5) -- (6.5,1.5);
\draw (5.5,1.2) node{${\cal N}_{6d}=(1,0)$};
\draw  (0,2.4) node{$\times$}; \draw(-0.1,2.4) node[left]{$8$} ;
\draw (1,2.1) -- (1,2.7);
\draw (3,2.1) -- (3,2.7);
\draw (1,2.1) -- (3,2.1);
\draw (1,2.7) -- (3,2.7);
\draw (2,2.4) node{${\cal N}_{8d}=1$};
\draw (0,3.6) node{$\times$}; \draw(-0.1,3.6) node[left]{$10$};
\draw (1,3.3) -- (1,3.9);
\draw (3,3.3) -- (3,3.9);
\draw (1,3.3) -- (3,3.3);
\draw (1,3.9) -- (3,3.9);
\draw (2,3.6) node{${\cal N}_{10d}=1$};
\endtikzpicture
}
\caption{Classes of superstring compactifications for which we will discuss the universal particle content. The arrows within the columns represent dimensional reduction on a $T^2$ torus.}
\label{roadmap}
\end{figure}

The existence of 4, 8 and 16 supercharges is compatible with various spacetime dimensions. Theories with a fixed number of supercharges are related to each other through dimensional reduction. Note that the minimum number of supercharges existing in 4, 6 and 10 dimensions is 4, 8 and 16, labelled by ${\cal N}_{4d}=1$, ${\cal N}_{6d}=(1,0)$ and ${\cal N}_{10d}=1$ respectively. From each of such theories, one can therefore obtain theories with the same amount of SUSYs in lower dimensions via toroidal compactifications which preserve all the SUSYs \cite{Narain:1986am}. In this paper, Kaluza-Klein and winding modes are neglected, as these depend on compactification details.  Thus, determining the lower dimensional spectra becomes a group theoretical problem of branching the associated Lorentz and $R$ symmetry groups. The following \fref{roadmap} gives an overview of the supersymmetric theories for which we will work out the model independent subset of the open superstring spectrum.

In this paper we are providing  for the first time a complete investigation of the universal massive open string states of higher spin within
supersymmetric compactifications of the open Type I superstring. In particular, we will compute the partition functions of the universal open string spectra for all type I compactifications
with 4, 8 and 16 preserved supercharges.
Spectra of the associated Type IIA/B closed superstring theories with twice as many preserved supercharges can be easily inferred from our open string results through a double copy of the open string Hilbert space, that is why they will not be explicitly addressed in this paper.\footnote{For the case of heterotic string compactifications, the charged matter fields originate from closed strings, and hence a priori one expects a different pattern of massive string states. In order to match the heterotic-type I massive string spectrum via heterotic-type I string duality also non-perturbative states are needed.}
Four dimensional superstrings subject to ${\cal N}_{4d}=1$ super Poincar\'e invariance are especially worth to be studied, since  ${\cal N}_{4d}=1$ compactifications with 
broken supersymmetry are expected to provide phenomenologically interesting  string solutions at low energies with the spectrum of certain extensions of the supersymmetric Standard Model
(see \eg~\cite{Anchordoqui:2012fq} for a stable low energy open string vacuum, the Standard Model$^{++}$ with two Higgs fields).
In addition to the light states,
the knowledge of the universal massive string spectrum is also important in order to compute string scattering amplitudes of massive open string states in ${\cal N}_{4d}=1$ string 
compactifications \cite{Feng:2010yx}.
This task is particularly relevant, if the string scale is low compared to the Planck mass, as it is true in large volume compactifications.

\subsubsection*{Refined partition functions}

A convenient way to study the spectrum of string states is to compute a {\it partition function} that counts such states with respect to their mass levels.  Since the string states transform under representations of super-Poincar\'e algebra, such a counting can be done in a representation theoretic way, namely the partition function can be written in terms of an infinite power series such that each power keeps track of the mass level and the coefficient of each term in the series comprises irreducible characters of the super-Poincar\'e algebra. In this way, the symmetry of the problem is manifest in the partition function and the characters contain information on how a supermultiplet transform under the little group and the $R$ symmetry.  Moreover, knowing a partition function is equivalent to knowing how many times a given representation appears at each mass level -- also known as the multiplicity.  Hence, given a representation of super-Poincar\'e algebra, our aim is to compute its {\it multiplicity generating function}, a power series such that each power keeps track of the mass level and each coefficient are the multiplicity of this particular representation.

Such a way of counting of string states was already performed explicitly in \cite{Curtright:1986di, Hanany:2010da}
for the case of uncompactified (ten-dimensional) string theories. It has also been extensively applied to the study of moduli spaces of supersymmetric gauge theories \cite{Feng:2007ur, Benvenuti:2006qr, Forcella:2008bb, Gray:2008yu, Hanany:2008kn,Hanany:2008qc, Benvenuti:2010pq, Hanany:2010qu, Hanany:2012dm}; in such a context the partition function is also known as the Hilbert series. 

One can also view the partition function we are considering as a trace over the space of physical states. In the trace, we grade the states according to their mass levels and global charges, but {\it not} their spacetime fermion numbers.  The variables used in keeping track of these levels and charges are called fugacities.  The fugacities for the global charges are indeed the ones that appear in the character of a representation of the super-Poincar\'e algebra.  In general, the partition function is therefore a multivariate function.  We call the insertion of global fugacities into the trace so as to make the global symmetry manifest a {\it refinement}, and we refer to the corresponding partition function as a {\it refined partition function}.  On the other hand, in order to compute the total number of states at each mass level, one can set the fugacities in the characters to unity. This amounts to computing the dimension of the corresponding representation, and we call the resulting partition function an {\it unrefined} one.

The term `refinement' as for the insertion of the aforementioned types of fugacities has also been used recently in various papers on elliptic genera and loop amplitudes.  There are various `synonyms' that have been adopted in the literature, \eg~ McKay-Thompson series \cite{Cheng:2010pq}, twining characters \cite{Gaberdiel:2010ch, Gaberdiel:2010ca} and twisted elliptic genera \cite{Eguchi:2010fg, Govindarajan:2011em, Cheng:2012tq}.  We emphasise  that, on the contrary to elliptic genera or other types of characters that are used in loop amplitude computations, the states that we trace over are not graded with a minus sign for spacetime fermions\footnote{To illustrate this point, let us look at the first mass level for a 10d theory with 16 supercharges, there are 256 states in total (see \tref{numberofstates}).  This number comes from (a) 44 spin two degrees of freedom and 84 three-form degrees of freedom constituting the spacetime bosonic states, and (b) 128 spin 3/2 degrees of freedom constituting the spacetime fermonic states.  If we had included the grading with a minus sign for spacetime fermions into  the trace, we would have a zero here.}.  As a result, the partition functions we are considering in this paper do not exhibit a modular invariant property\footnote{To illustrate this point, we compare the unrefined partition functions presented in (5.3.37) of \cite{Green:1987sp} and  (9.1.14), (9.1.15) of \cite{Green:1987mn}. The former is the partition function we are interested in and it is clear that such a partition function does not possess a modular invariant property.  On the other hand, observe in the latter that if the grading with a minus sign for spacetime fermions is introduced in the trace, the contributions from the fermionic and bosonic excited states precisely cancel in the unrefined partition function, as exemplified in the preceding footnote.}.

%The open string states we are discussing here can also carry Chan-Paton factors. The massless states transform the adjoint representation of the Chan-Paton gauge group and their character can therefore be obtained by multiplying the character discussed here by the character of the adjoint representation.  The massive states, on the other hand, transform in various representations according to the gauge symmetry (see Page 294 of \cite{Green:1987sp} for further details), and the character can be computed by multiplying an appropriate character of the gauge group to our existing character at a given mass level.  This argument applies in general for all partition functions we computed in this paper; hence we shall not discuss Chan-Paton factors in the subsequent.

Open string states also carry Chan-Paton factors. The massless states and their massive excitations that arise from open strings with both endpoints attached to a stack of D branes transform in the adjoint representation of the Chan-Paton gauge group. Their character can therefore be obtained by multiplying the character discussed here by the character of the adjoint representation\footnote{Furthermore, compactifications with intersecting D branes give rise to model dependent excitations of open strings that end on different stacks of D-branes \cite{Berkooz:1996km}. These non-universal states beyond the scope of this work transform in the bifundamental representation.}.  The massive states corresponding to unoriented strings, on the other hand, transform in various representations according to the gauge symmetry (see e.g. Page 294 of \cite{Green:1987sp} for further details), and the character can be computed by multiplying an appropriate character of the gauge group to our existing character at a given mass level.  All partition functions computed in this paper allow for a straightforward inclusion of the Chan-Paton contributions; hence, we shall not discuss Chan-Paton factors in the subsequent.

%The universal states we are discussing stem from open strings with both endpoints attached to a stack of D branes and transform in the adjoint representation of the Chan-Paton gauge group. Their character can therefore be obtained by multiplying the character discussed here by the character of the adjoint representation. On top of that, compactifications with intersecting D branes give rise to model dependent excitations of open strings that end on different stacks of D-branes \cite{Berkooz:1996km}. These non-universal states beyond the scope of this work transform in bifundamental representations. All partition functions computed in this paper allow for a straightforward inclusion of the adjoint Chan-Paton degrees of freedom; hence, we shall not discuss Chan-Paton factors in the subsequent.

\subsubsection*{The number of universal open string states}

\begin{figure}[htbp]
\begin{center}
\includegraphics[scale=0.8]{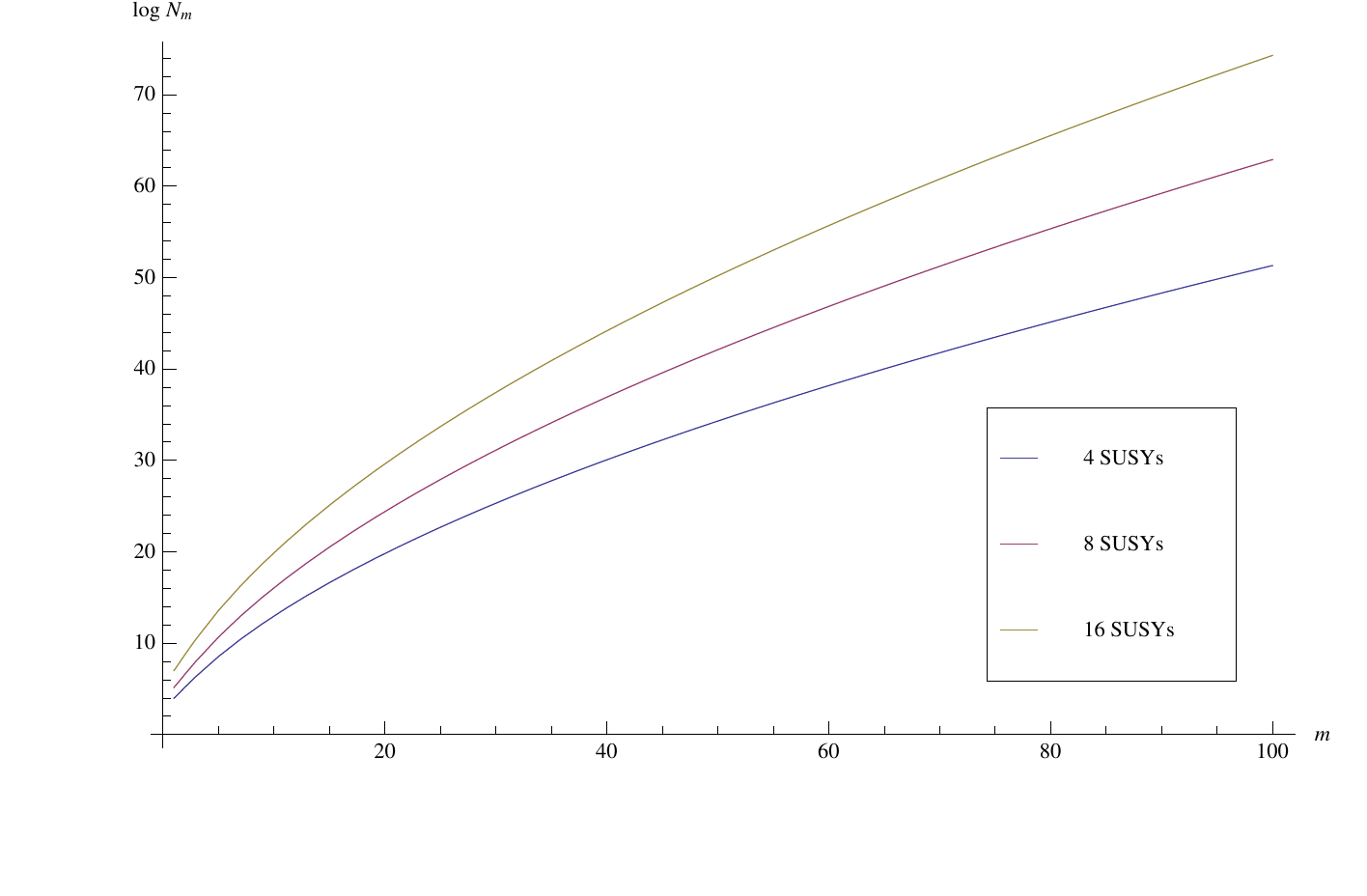}
\caption{The logarithmic plot of the number of states $N_m$ against the mass level $m$ for the case of 4 and 16 supercharges. The values of $N_m$ are taken from the asymptotic formulae \eref{asympNm4d}, \eref{asympNm6d} and \eref{asympNm10d}, which work well for large $m$.}
\vspace{-1.2cm}
\label{fig:asympstates}
\end{center}
\end{figure}

To give a first idea of the orders of magnitude governing the number of universal open string states at individual mass levels, the following \tref{numberofstates} summarizes their numbers at low levels $\leq 9$ in scenarios with 4, 8 and 16 supercharges, respectively. They are obtained by expanding the associated {\it unrefined partition functions}. For the cases of 4, 8 and 16 supercharges, the exact generating functions are respectively given by \eref{exactunref4d}, \eref{exactunref6d}, \eref{exactunref10d} and their asymptotics at large mass levels are respectively given by \eref{asympNm4d}, \eref{asympNm6d}, \eref{asympNm10d}. Roughly speaking, the number of states increases exponentially with respect to the square root of the mass level. The plot is depicted in \fref{fig:asympstates}.

\vspace{-0.3cm}
\noindent
\begin{table}[htdp]
\begin{tabular}{|c|c|c|c|}\hline  $\ap m^2$ &\# states for 4 supercharges&\# states for 8 supercharges &\# states for 16 supercharges \\ \hline \hline   
  0 & 4 &8 &16  \\\hline
  1 & 24 &80 &256  \\\hline
   2 & 104 &512 &2.304  \\\hline
   3 & 384 &2.576 &15.360  \\\hline
   4 & 1.240 &11.008 &84.224  \\\hline
   5 &  3.648 &41.792 &400.896  \\\hline
      6 &  9.992 &144.784 &1.711.104  \\\hline
         7 &  25.792 & 465.856 &6.690.816   \\\hline
            8 &  63.392 &1.409.792 &24.332.544  \\\hline
               9 &  149.464 &4.050.112 &83.219.712  \\\hline
\end{tabular}
\caption{The number of model independent open string states in compactifications with 4, 8 and 16 supercharges, respectively, up to mass level $\ap m^2=9$.}
\label{numberofstates}
\end{table}

\subsubsection*{Stable patterns and Regge trajectories}
For any number of dimensions and supercharges, we can examine the multiplicities of a supermultiplet transforming under a given super-Poincar\'e representation.  In four spacetime dimensions, such a representation contains an $SO(3)$ spin quantum number; this is {\it half} of the $SO(3)$ Dynkin label.  For dimensions $d > 4$, we refer to the first $SO(d-1)$ Dynkin label as `spin' in slight abuse of terminology. This allows for the generalised notion of spin in higher dimensions. It is interesting to study the multiplicities associated with large spin quantum numbers, \ie~ a large spin limit.   

There are certain crucial asymptotic patterns that universally appear for families of supermultiplets, regardless of the number of dimensions and supercharges. In particular, there are certain sets of numbers that repeatedly appear at various mass levels when spins are sufficiently large (and other quantum numbers are kept fixed).  As an example, it is convenient to consider table \ref{tab4d0} where such numbers are written in red.  Since this set of numbers stabilises in the large spin limit, we refer to it as a {\it stable pattern}.  In fact, such a pattern appears not only in superstring spectra we are considering, it also does so in spectra of the bosonic and various other types of string theories as pointed out in \cite{Curtright:1986di}; there, the stable pattern is referred to as the {\it leading Regge trajectory}.  We shall henceforth use these two terms interchangably.

Let us explore the stable pattern in more details. For a fixed sufficiently large mass level $M$, the stable pattern for a certain supermultiplet family starts appearing when the spin $j$ increases and reaches a certain value $j_{\te{min}}(M)$. It then extends up to some maximum value $j_{\te{max}}(M)$ where the multiplicity becomes zero for spins $j>j_{\te{max}}(M)$. As an empirical speculation, we observe that for a sufficiently large $M$, the stable pattern appearing in the spin range $j_{\te{min}}(M) \leq j \leq j_{\te{max}}(M)$ occupies approximately {\it half} of the spin range $0 \leq j \leq j_{\te{max}}(M)$ of all non-zero multiplicities. 
As an example, such a phenomenon is highlighted in red in each row of table \ref{tab4d0}.  

Stated differently, for a given super-Poincar\'e representation, the highest spin with nonzero multiplicity approximately scales linearly $j_{\te{max}}(M) \approx (M-M_0)$ for large $M$ where $M_0$ is the mass level at which the first non-zero red number appears. The onset of the stable pattern, on the other hand, roughly follows a linear scaling, $j_{\te{min}}(M) \approx \frac{1}{2}(M-M_0)$.  The region of validity for the stable pattern is therefore bounded by two straight lines whose slopes have the ratio $\frac{1}{2}$. In this sense, the stable pattern gives control over the essential part of the spectrum.

In addition to the stable pattern or the leading trajectory, there is also a notion of subleading trajectories bounded by linear spin-mass relations with approximate slopes $\frac{1}{3},\frac{1}{4},\ldots$.  We shall not go over any detail here and postpone the quantitative discussions to subsequent sections.

%This reflects the organization of the superstring spectrum in terms of Regge trajectories subject to a linear relation between mass $m$ and spin $j$. Roughly speaking, the numbers of multiplets with the highest spins $j \approx \ap m^2,\ap m^2-1,\ldots,\ap m^2/2$ are reproduced from mass level to mass level. More precisely, supermultiplets of spin $\ap m^2 - \De j$ appear with an $m^2$ independent multiplicity provided that their deviation $\De j$ from the maximum spin $\approx \ap m^2$ stays well below $\ap m^2/2$. We call this phenomenon an {\it asymptotic stable pattern} (or simply a {\it stable pattern} for brevity), it allows exact predictions of multiplicities for certain families of states to all mass levels\footnote{In table \ref{tab4d0} for instance, one can observe that the multiplicities of ${\cal N}_{4d}=1$ supermultiplets highlighted in red reappear at all subsequent mass levels: At level $k$, multiplets of highest spins $k+\frac{1}{2}, k-\frac{1}{2},\ldots,\lceil[ \frac{k}{2}\rceil]+\frac{1}{2}$ enjoy $k$ independent multiplicities (where $\lceil[ x \rceil]$ denotes the ceiling function that yields smallest integer $n \in \NN$ subject to $n\geq x$).}. For ${\cal N}_{4d}=1$ multiplets, we managed to compute these asymptotics analytically, and for all other scenarios, we can at least present empirical results due to the first 25 levels. Since the vast majority of string states are of high spin, this approach gives control over the essential part of the spectrum.
 
\subsubsection*{Outlines and key results}

This article can be roughly divided into two parts. The first part develops the SCFT foundations for refined superstring partition functions, using conventions from appendix \ref{sec:not}. Section \ref{sec:spacetime} introduces $SO(d-1)$ covariant characters for the degrees of freedom due to the superstring oscillators from the spacetime SCFT. In order to describe compactification scenarios, the spacetime sector has to be supplemented by SCFTs describing the internal dimensions. The SCFTs discussed in section \ref{sec:internal} capture the universal states present in {\it any} compactification that preserves four and eight supercharges, respectively. 

Starting from section \ref{sec:4d}, we proceed to the second part of this work where spacetime- and internal characters are combined to super Poincar\'e covariant partition functions. Universal states of four dimensional ${\cal N}_{4d}=1$ supersymmetric string compactifications are thoroughly investigated in section \ref{sec:4d}: We analytically derive the stable pattern for supermultiplet multiplicities, in manifest agreement with the tabulated particle content up to mass level 25. Similarly, section \ref{sec:6d} is devoted to scenarios with eight supercharges -- in both six and four spacetime dimensions. Finally, spectra of maximally supersymmetric open superstring theories are discussed in section \ref{sec:10d}, a chain of dimensional reductions encompasses $d=10,8,6$ and $d=4$ compactifications.

The analysis of universal ${\cal N}_{4d}=1$ supermultiplets in section \ref{sec:4d} enjoys the highest phenomenological relevance and provides the most compact results. Hence, the reader might want to skip subsections \ref{sec:highdim}, \ref{sec:highdimm} and \ref{sec:42} on higher dimensional generalizations upon the first reading. 

Let us summarise the key results in this paper below.
\begin{itemize}
\item The exact unrefined partition functions and asymptotic expressions for the number of states at each large mass level are given in \eref{exactunref4d}--\eref{asympNm4d}, \eref{exactunref6d}--\eref{asympNm6d} and \eref{exactunref10d}--\eref{asympNm10d} for theories with four, eight and sixteen supercharges respectively.  The graphs of these numbers versus the mass level are depicted in \fref{fig:asympstates}.
\item The exact multiplicity generating functions for theories of four, eight and sixteen supercharges are respectively given in  \eref{mulsumoddeven0}--\eref{mulsumoddeven}, \eref{end6dim} and \eref{end10dim}.
%\item The recurrence relations for multiplicity generating functions in the theory with four supercharges are given by \eref{recrel1}--\eref{recrel5}.
\item  The asymptotic expressions for the multiplicity generating functions for the theory with four supercharges are presented in \eref{asympmuloddeven} and \eref{mainferm}
\end{itemize}

Even though the tools for expanding the refined partition function to any mass level are presented for all the scenarios, exact formulae for multiplicities of particular multiplets generically involve nested infinite sums. In particular, the bookkeeping of $SO(d-1)$ quantum numbers becomes increasingly difficult in $d>4$ spacetime dimensions. That is why we elaborate the phenomenologically relevant and mathematically most accessible ${\cal N}_{4d}=1$ case in particular depth.

%%%%%%%%%%%%%%%%%%%%%%%%%%%%%%
%%%%%%%%%%%%%%%%%%%%%%%%%%%%%%
%%%%%%%%%%%%%%%%%%%%%%%%%%%%%%
%%%%%%%%%%%%%%%%%%%%%%%%%%%%%%
%%%%%%%%%%%%%%%%%%%%%%%%%%%%%%

\section{The spacetime CFT in various dimensions}
\label{sec:spacetime}

The aim of this section is to derive a refined partition function for the oscillator modes of the worldsheet fields $\pa X^\mu, \psi^{\mu}$ associated with the $d$ directions of Minkowski spacetime. They carry an $SO(1,d-1)$ vector index $\mu=0,1,\ldots,d-1$. In the framework of lightcone quantization the physical spectrum is obtained from transverse oscillators $\pa X^{i=2,3,\ldots,d-1}, \psi^{i=2,3,\ldots,d-1}$ which carry charges with respect to the $\frac{1}{2}(d-2)$ Cartan generators of $SO(1,d-1)$ outside the lightcone directions. We assign a separate fugacity $y_k$ to each pair of $\pa X^i,\psi^i$ components (say $(\pa X^{2k}, \pa X^{2k+1})$) such that the fugacity subscript lies in the range $1 \leq k \leq \frac{1}{2}(d-2)$. Since massive particles with $d$ dimensional timelike momentum form representations of the little group $SO(d-1)$, the dependence on Lorentz fugacities $y_k$ necessarily arranges into characters of the massive little group.

It is instructive to first of all study the simplest non-trivial example $d=4$ with one spacetime fugacity. The first three subsections are devoted to the $SO(3)$ covariant partition function of the four dimensional spacetime SCFT. As we will explain in later subsections, higher dimensional cases follow by combining several copies of $d=4$ building blocks.

\subsection[Bosonic partition function in $d=4$]{Bosonic partition function in $\bm{d=4}$} 

The contribution of the lightcone bosons to the refined partition function is 
\bea 
\chi^{SO(3)}_B (q, y) &= \PE \left[ ([2]_y -1) \, (q +q^2+q^3+q^4+\ldots) \right] \eq \PE \left[ ([2]_y -1) \frac{q}{1-q} \right]  \nn \\
&= \prod_{n=1}^{\infty} \frac{1}{(1-y^2 q^n) \,(1-y^{-2} q^n)} = \frac{1}{(qy^2; q)_\infty (qy^{-2}; q)_\infty} \label{bosonic} \\
&=  -i q^{\frac{1}{12}} (y-y^{-1}) \frac{\eta(q)}{\vartheta_1 (y^2, q)}~. 
\eea  
The representation $[2]_y -1=y^2+y^{-2}$ in the plethystic exponential corresponds to the two components $\partial X^+ ,\partial X^-$ perpendicular to the lightcone. The geometric series $\frac{q}{1-q}=q+q^2+q^3+\ldots$, on the other hand, represents the infinite tower of positive frequency modes of $\partial X^{\pm}$ which act as creation operators.

Explicitly, the first few terms in the power series of $\chi^{SO(3)}_B (q, y)$ can be written in terms of $SO(3)$ characters $[k]_y$ as
\bea
\chi^{SO(3)}_B (q, y)  &= 1+ q ([2]_y-1)+ q^2 [4]_y+q^3 ([2]_y+[6]_y)+q^4 ([0]_y+2 [4]_y+[8]_y) \nn \\
& \quad +q^5 (2 [2]_y+[4]_y+2 [6]_y+[10]_y)  +q^6 (2 [0]_y+[2]_y+3 [4]_y+2 [6]_y+2 [8]_y+[12]_y) \nn \\
& \quad +q^7 (4 [2]_y+3 [4]_y+4 [6]_y+2 [8]_y+2 [10]_y+[14]_y)+ \ldots~. \label{char2}
\eea

From such a power series, we are motivated to rewrite \eref{bosonic} as an infinite sum of the form
\bea
\chi^{SO(3)}_B (q,y) = \sum_{k=0}^\infty [k]_y f_k(q)~,
\eea
for some function $f_k(q)$ which depends only on $q$ and not on $y$.  The use of this form of the partition function will become clear later.  

In order to do so, we rewrite \eref{bosonic} using the $q$-binomial theorem\footnote{The version we use states that $\frac{1}{(z;q)_\infty}= \sum_{n=0}^\infty \frac{z^n}{(q;q)_\infty}$.} as
\bea
\chi^{SO(3)}_B (q, y) 
= \sum_{m=0}^\infty \sum_{n=0}^\infty \frac{y^{2(m-n)}}{(q;q)_m (q;q)_{n}} q^{m+n} =:  \sum_{k=0}^\infty [k]_y f_k(q)~. \label{Rogers}
\eea
Before proceeding further, let us state an identity that we are going to use many times later.  From \eref{char1} and the residue theorem, we find that
\bea
\int \ud \mu_{SO(3)} (y)~ y^m [n]_y 
= \begin{cases}
\delta_{0,n} &\quad \text{for $m = 0$}~, \\
\frac{1}{2}(\delta_{|m|,n} -\delta_{|m|,n+2}) &\quad \text{for $m \neq 0$}~, \label{idenortho}
\end{cases}
\eea
where the Haar measure of $SO(3)$ is given by \eref{HaarSO3}.
It is clear from the absence of odd $y$ powers in (\ref{Rogers}) that only integer spin representations of $SO(3)$ occur. We therefore have $f_{2k+1}(q) =0$ for all $k$, and the nontrivial coefficients to compute are\footnote{In intermediate steps, we are making use of identities like $\sum_{r=0}^\infty q^{r(1+p)} (q^{1+r};q)_\infty=(q;q)_\infty (q^{1+p};q)_\infty$.}
\bea
f_{2k} (q) &= \int \ud \mu_{SO(3)} (y)~\chi^{SO(3)}_B (q, y) [2k]_y  \nn \\
&= \sum_{m=0}^\infty \Bigg[\frac{q^{2m}}{(q;q)_m^2} \delta_{k,0} + \frac{1}{2} \sum_{n=0}^{m-1} \frac{\delta_{m-n,k}-\delta_{m-n,k+1}}{(q;q)_m (q;q)_n} q^{m+n} + \frac{1}{2} \sum_{n=m+1}^\infty \frac{\delta_{n-m,k}-\delta_{n-m,k+1}}{(q;q)_m (q;q)_n} q^{m+n} \Bigg] \nn \\
%&=\sum_{m=0}^\infty \frac{q^{2m}}{(q;q)_m^2} \delta_{k,0} + \frac{1}{2} \sum_{\Delta=0}^\infty \sum_{n=0}^\infty  \frac{\delta_{\Delta+1,k}-\delta_{\Delta+1,k+1}}{(q;q)_{n+\Delta+1} (q;q)_n} q^{2n+\Delta+1} + \frac{1}{2} \sum_{m=0}^\infty \sum_{n=0}^\infty \frac{\delta_{n,k-1}-\delta_{n,k}}{(q;q)_m (q;q)_{m+n+1}} q^{2m+n+1} \nn \\
%&= \sum_{n=0}^\infty \Bigg[ \frac{q^{2n+k}}{2(q;q)_{n+k} (q;q)_n} -\frac{q^{2n+k+1}}{2(q;q)_{n+k+1} (q;q)_n} + \frac{q^{2n+k}}{2(q;q)_{n} (q;q)_{n+k}}- \frac{q^{2n+k+1}}{2(q;q)_{n} (q;q)_{n+k+1}}   \Bigg] \nn \\
%&= \sum_{n=0}^\infty \Bigg[  \frac{q^{2n+k}}{(q;q)_{n+k} (q;q)_n} -\frac{q^{2n+k+1}}{(q;q)_{n+k+1} (q;q)_n}   \Bigg] \nn \\
%&= \sum_{n=0}^\infty \frac{q^{2n+k}}{(q;q)_n} \Bigg[ \frac{1}{(q;q)_{n+k}} - \frac{q}{(q;q)_{n+k+1}} \Bigg] \nn \\
&=  \sum_{n=0}^\infty \frac{q^{2n+k}}{(q;q)_n (q;q)_{n+k+1}}  \left( 1-q -q^{n+k+1} \right) \nn \\
%&=  (q;q)_\infty^{-2} \sum_{n=0}^\infty q^{2n+k} (q^{n+1}, q)_\infty (q^{n+k+2},q)_\infty (1-q-q^{n+k+1}) \nn \\
%&= (q;q)_\infty^{-2} \sum_{n=0}^\infty q^{2n+k} \Bigg[ \sum_{m=0}^\infty (-1)^m q^{m(n+1)+\frac{1}{2}m(m-1)} (q;q)_m^{-1} \Bigg] \times \nn \\
%& \qquad \Bigg[ \sum_{p=0}^\infty (-1)^p q^{p(n+k+2)+\frac{1}{2}p(p-1)} (q;q)_p^{-1} \Bigg] (1-q-q^{n+k+1}) \nn \\
%&= (q;q)_\infty^{-2} \sum_{m=0}^\infty \sum_{p=0}^\infty (-1)^m q^{\frac{1}{2}m(m-1)} (q;q)_m^{-1} (-1)^p q^{\frac{1}{2}p(p-1)} (q;q)_p^{-1} \times \nn \\
%& \qquad \Bigg[ \sum_{n=0}^\infty q^{2n+k+m(n+1)+p(n+k+2)} (1-q-q^{n+k+1}) \Bigg] \nn \\
%&= (q;q)_\infty^{-2} \sum_{m=0}^\infty \sum_{p=0}^\infty (-1)^m q^{\frac{1}{2}m(m-1)} (q;q)_m^{-1} (-1)^p q^{\frac{1}{2}p(p-1)} (q;q)_p^{-1} \times \nn \\
%& \qquad \Bigg[ q^{k+m+kp+2p} (1-q) \sum_{n=0}^\infty q^{n(2+m+p)} - q^{2k+m+kp+2p+1} \sum_{n=0}^\infty q^{n(3+m+p)} \Bigg] \nn \\
%&= (q;q)_\infty^{-2} \sum_{m=0}^\infty \sum_{p=0}^\infty (-1)^m q^{\frac{1}{2}m(m-1)} (q;q)_m^{-1} (-1)^p q^{\frac{1}{2}p(p-1)} (q;q)_p^{-1} \times \nn \\
%& \qquad \Bigg[ \frac{q^{(p+1)k+2p+m} (1-q)}{1-q^{2+m+p}} - \frac{q^{(p+2)k+2p+m+1}}{1-q^{3+m+p}} \Bigg] \nn \\
%&= (q;q)_\infty^{-2} \sum_{m=0}^\infty (-1)^m q^{\frac{1}{2}m(m-1)} (q;q)_m^{-1} \times \nn \\
%& \qquad \sum_{p=0}^\infty \left[ (-1)^p \frac{q^{(p+1)k+2p+m} (1-q) q^{\frac{1}{2}p(p-1)}}{1-q^{2+m+p}} (q;q)_p^{-1} \right. \nn \\
%& \qquad \left. + (-1)^{p+1} \frac{q^{((p+1)+1)k+2(p+1)+m-1} q^{\frac{1}{2}(p+1)((p+1)-1)-(p+1)+1}}{1-q^{2+m+(p+1)}} (1-q^{p+1}) (q;q)_{p+1}^{-1} \right] \nn \\
&= (q;q)_\infty^{-2} \sum_{p=0}^\infty (-1)^p q^{(p+1)k+\frac{1}{2}p(p+1)} (1-q^{p+1}) (q;q)_p^{-1} \sum_{m=0}^\infty (-1)^m \frac{q^{\frac{1}{2}m(m+1)}}{1-q^{2+m+p}} (q;q)_m^{-1} \nn \\
%&= (q;q)_\infty^{-2} \sum_{p=1}^\infty (-1)^{p-1} q^{pk+\frac{1}{2}p(p-1)} (1-q^p) (q;q)_{p-1}^{-1} \sum_{m=0}^\infty (-1)^m \frac{q^{\frac{1}{2}m(m+1)}}{1-q^{1+m+p}} (q;q)_m^{-1} \nn \\
%&= (q;q)_\infty^{-2} \sum_{p=1}^\infty (-1)^{p-1} q^{pk+\frac{1}{2}p(p-1)} (1-q^p) (q;q)_{p-1}^{-1} \sum_{r=0}^\infty q^{r(1+p)} \times \nn \\
%& \qquad \sum_{m=0}^\infty (-1)^m q^{\frac{1}{2}m(m-1)+m(1+r)} (q;q)_m^{-1} \nn \\
%&= (q;q)_\infty^{-2} \sum_{p=1}^\infty (-1)^{p-1} q^{pk+\frac{1}{2}p(p-1)} (1-q^p) (q;q)_{p-1}^{-1} \sum_{r=0}^\infty q^{r(1+p)} (q^{1+r};q)_\infty \nn \\
%%&= (q;q)_\infty^{-2} \sum_{p=1}^\infty (-1)^{p-1} q^{pk+\frac{1}{2}p(p-1)} (1-q^p) (q;q)_{p-1}^{-1} (q;q)_\infty \sum_{r=0}^\infty q^{r(1+p)} (q;q)_r^{-1} \nn \\
%&= (q;q)_\infty^{-2} \sum_{p=1}^\infty (-1)^{p-1} q^{pk+\frac{1}{2}p(p-1)} (1-q^p) (q;q)_{p-1}^{-1} (q;q)_\infty (q^{1+p};q)_\infty \nn \\
%&= (q;q)_\infty^{-2} \sum_{p=1}^\infty (-1)^{p-1} q^{pk+\frac{1}{2}p(p-1)} (1-q^p) (q;q)_{p-1}^{-1} (q;q)_p \nn \\
&= (q;q)_\infty^{-2}  \sum_{n=1}^\infty (-1)^{n-1} (1-q^{n})^2   q^{nk + \frac{1}{2}n(n-1)} ~.
\eea
We obtain an $SO(3)$ character expansion of the bosonic partition function:
\bea
\chi^{SO(3)}_B(q,y) = (q;q)_\infty^{-2}  \sum_{n=1}^\infty (-1)^{n-1} (1-q^{n})^2  \sum_{k=0}^\infty q^{nk + \frac{1}{2}n(n-1)} [2k]_y~. \label{chiBSO3}
\eea
Note that the pattern $\sum_{n=1}^\infty (-1)^{n-1}q^{nk} [2k]_y \ldots$ (where the $\ldots$ ellipsis does not depend on $y$ and $k$) is described in section 6 of \cite{Curtright:1986di} as an alternating sequence of additive and subtractive Regge trajectories of slope $\frac{1}{n}$. This is the source of stable patterns as described in the introduction in bosonic string theory. We will rediscover these patterns in the counting of SUSY multiplets later on.

\subsubsection{Multiplicities of representations $[2m]$ and their asymptotics}  
Let us determine the multiplicity of irreducible $SO(3)$ representations $[2m]$ at each mass level.  Recall the orthogonality of characters with respect to the Haar measure:
\bea
\int \ud \mu_{SO(3)}(y) [m]_y [n]_y = \delta_{mn}~.
\eea
From \eref{chiBSO3}, we find that the generating function of the multiplicity of $[2m]$ is
\bea
M(\chi^{SO(3)}_B, [2m]; q) &=  \int \ud \mu_{SO(3)}(y) ~[2m]_y \chi^{SO(3)}_B (q,y) \nn \\
%&= (q;q)_\infty^{-2}  \sum_{n=1}^\infty (-1)^{n-1} (1-q^{n})^2 q^{ \frac{1}{2}n(n+2m-1)}  \nn \\
&= (q;q)_\infty^{-2}  \sum_{n=1}^\infty (-1)^{n-1} (1-q^{n})^2 q^{ \frac{1}{2}n(n-1)} q^{ nm} ~.
\label{genfnmul}
\eea

\paragraph{Asymptotics as $m \rightarrow \infty$.} The expression \eref{genfnmul} found for multiplicity generating functions greatly simplifies in the limit $m \rightarrow \infty$ of large spin and mass level. In order to compute an asymptotic formula in this regime, we apply Laplace's method (see \eg~ section 6.7 of \cite{BenderOrszag}) to our question. Since $0<q<1$, the terms in the series peak sharply near the $n=1$ term as $m \rightarrow \infty$.  Therefore, it is expected that for any $\epsilon >0$
\bea
M(\chi^{SO(3)}_B, [2m]; q) & \sim (q;q)_\infty^{-2} \sum_{n=1}^{1+ \lfloor \epsilon \rfloor} (-1)^{n-1} (1-q^{n})^2 q^{ \frac{1}{2}n(n-1)} q^{nm}, \qquad m \rightarrow \infty \label{mulepsilon}~.
\eea
Now let us write $n = 1+ t$, where $t$ is small compared with $1$.  Note that
\bea
q^{\frac{1}{2} n(n-1)} = 1+\frac{1}{2} (\log q) t+O(t^2)~, \label{expqnchoose2}
\eea 
Substituting the leading term of this power series into the right hand side of \eref{mulepsilon} and extending the region of summation to $\infty$, we find that the leading behaviour of $M(\chi^{SO(3)}_B, [2m]; q)$ is given by
\bea
M(\chi^{SO(3)}_B, [2m]; q) &\sim (q;q)_\infty^{-2} \sum_{t=0}^\infty (-1)^{t} (1-q^{t+1})^2  q^{m(t+1)} \nn \\
&=  (q;q)_\infty^{-2}  \frac{q^{m} (1-q)^2  \left(1-q^{1+m}\right)}{\left(1+q^{m}\right) \left(1+q^{1+m}\right) \left(1+q^{2+m}\right) } \nn \\
&= (q^2;q)_\infty^{-2} ~  \frac{q^{m} \left(1-q^{m}\right)}{\left(1+q^{m}\right)^3 } ~, \qquad m \rightarrow \infty~. \label{asympmul}
\eea
The higher order corrections can be computed by taking into account the subleading terms of \eref{expqnchoose2}.  Note that the next to leading term of \eref{asympmul} is of order $O(q^{2m} \log q)$.  Thus, asymptotic formula \eref{asympmul} reproduces the exact result up to $O(q^{2m-1})$.

\paragraph{Interpretation and stable pattern.}  We can extract some information about bosonic string states from \eref{asympmul}.
\begin{itemize}
\item The representation $[2m]$ appears first time in the bosonic partition function $\chi^{SO(3)}_B (q,y)$ at mass level $q^m$.
\item  The multiplicities of $[2m]$ at levels $q^{m+\ell}$, for $0 \leq \ell \leq m-1$, are independent of $m$.  We refer to a set of these numbers as a {\it stable pattern} for bosonic string theory.  The generating function for such a stable pattern can be determined by taking a formal limit $m \rightarrow \infty$ in \eref{asympmul}:
\bea
& \lim_{m \rightarrow \infty}  q^{-m} M(\chi^{SO(3)}_B, [2m]; q)  \nn\\
& \qquad \quad    = (q^2; q)^{-2}_{\infty} = \prod_{k=2}^\infty (1-q^k)^{-2}   \\
& \qquad  \quad  = 1+2 q^2+2 q^3+5 q^4+6 q^5+13 q^6+16 q^7+30 q^8+40 q^9+66 q^{10}+90 q^{11} \nn \\
& \qquad  \quad \quad  +142 q^{12}+192 q^{13}+290 q^{14}+396 q^{15}+575 q^{16}+782 q^{17}+1112 q^{18} \nn \\
& \qquad  \quad  \quad +1500 q^{19}+2092 q^{20}+2808 q^{21}+3848 q^{22}+5132 q^{23}+6945 q^{24} \nn \\
& \qquad  \quad  \quad +9192 q^{25}+O(q^{26})~.
\eea
Note that terms with low orders in this power series are in agreement with the data presented in Table  6b of \cite{Curtright:1986di}.
\end{itemize}

\subsection[The NS sector in $d=4$]{The NS sector in $\bm{d=4}$} 
\label{sec:4dimNS}

Under NS boundary conditions, the worldsheet superpartners $\psi^i$ of the lightcone bosons contribute
\bea
f_{\NS}(q;y) &= {\PE}_F \left[ ([2]_y-1) \frac{q^{\frac{1}{2}}}{1-q} \right] \nn \\
&= \prod_{n=1}^{\infty} (1+y^2 q^{n-1/2})  (1+y^{-2} q^{n-1/2}) \label{deffNS} \\
&=q^{\frac{1}{24}} \frac{\vartheta_3( y^2, q)}{\eta(q)}~.
\eea
to the spacetime partition functions. We shall rewrite this function as an infinite sum by means of Jacobi's triple product identity (see, \eg, subsection 19.8 of \cite{HardyWright}) :
\bea
\prod_{n=1}^\infty (1-x^{2n}) (1+ x^{2n-1} z)( 1+x^{2n-1} z^{-1})  
= \sum_{m = -\infty}^\infty x^{m^2} z^{m}~. \label{Jacobitriple}
\eea
Applying identity \eref{Jacobitriple} with $x=q^{1/2}$ and $z = y^2$ to \eref{deffNS}, we obtain
\bea
f_{\NS}(q, y)  &= (q;q)_\infty^{-1} \sum_{m= - \infty}^{+\infty} y^{2m} q^{m^2/2}  \label{fNSinterm} \\
& = (q;q)_\infty^{-1} \sum_{m=0}^\infty q^{\frac{1}{2}m^2} (1-q^{m+\frac{1}{2}}) [2m]_{y}~,
\label{char3}
\eea
where \eref{char3} can be obtained by applying \eref{idenortho} and the orthogonality of the characters to \eref{fNSinterm} as follows:
\bea
\int \ud \mu_{SO(3)} (y) ~ f_{\NS} (q,y) [2k]_y &= (q;q)_\infty^{-1} \left[ \sum_{m=0}^\infty q^{m^2/2} \delta_{m,k} - \sum_{m=-\infty}^{-1} q^{m^2/2} \delta_{-m, k+1} \right]  \nn \\
&= (q;q)_\infty^{-1} \left( q^{\frac{1}{2}k^2} - q^{\frac{1}{2}(k+1)^2} \right) \nn \\
&= (q;q)_\infty^{-1} q^{\frac{1}{2}k^2} (1 - q^{k+\frac{1}{2}} )~.
\eea

Let us combine the bosonic partition function with the NS-sector contribution. Using \eref{bosonic}, \eref{char3} and the multiplication rule $[2m] \cdot [2k] = \sum_{l=|k-m|}^{k+m} [2l]$, we find that
\bea
\chi_{\te{NS}}^{SO(3)} (q,y) &:=  \chi^{SO(3)}_B (q,y ) f_{\te{NS}} (q,y)  = -i q^{1/8} (y-y^{-1}) \frac{\vartheta_3 (y^2,q)}{\vartheta_1 (y^2,q)} 
 \\
&=  \frac{-1}{(q;q)_\infty^{3}} \sum_{m=0}^\infty \sum_{n=1}^\infty (-1)^{n} (1-q^{m+\frac{1}{2}}) (1-q^n)^2 q^{\frac{1}{2}n(n-1) + \frac{1}{2}m^2}  \sum_{k=0}^\infty q^{nk}\! \! \sum_{\ell = |k-m|}^{k+m} [2\ell]  ~.
\eea
The expression in the curly brackets $\{\cdots\}$ can be rewritten as $\sum_{k=0}^\infty f_{kmn} (q) [2k]$, for some function $f_{kmn} (q)$. In order to determine this function, we use the orthogonality of characters:
\bea
f_{kmn} (q) &= \int \ud \mu_{SO(3)} (y)~ \sum_{k'=0}^\infty q^{nk'} \sum_{\ell = |k'-m|}^{k'+m} [2\ell]_y~ [2k]_y \nn \\
&=  \sum_{k'=0}^\infty q^{nk'} \sum_{\ell = |k'-m|}^{k'+m} \delta_{\ell, k} \nn \\
&= \sum_{\ell' = |k-m|}^{k+m} q^{n\ell'} = \frac{q^{n|k-m|} - q^{n(k+m+1)}}{1-q^n} ~.
\eea
Therefore, we obtain
\bea
\chi_{\te{NS}}^{SO(3)} (q,y) 
&= (q;q)_\infty^{-3}  \sum_{m=0}^\infty \sum_{n=1}^\infty (-1)^{n-1} (1-q^{m+\frac{1}{2}}) (1-q^n) q^{\frac{1}{2}[n(n-1) + m^2]} \times \nn \\
& \hspace{4cm}  \sum_{k=0}^\infty  \; ( q^{n|k-m|} - q^{n(k+m+1)} ) \; [2k]~. \label{exactchNSSO3}
\eea
We emphasise that the $SO(3)$ irreducible representations with odd Dynkin labels do not appear in the partition function $\chi_{\te{NS}}^{SO(3)} (q,y)$.

In terms of a power series in $q$, this can be written as
\bea
\chi_{\te{NS}}^{SO(3)} (q,y) &=  1  +  q^{1/2} ( [2] - 1)  +q [2]+q^{3/2} ([0]+[4])+q^2 ([0]+2 [4])+q^{5/2} (2 [2]+[4]+[6]) \nn \\
& \quad +q^3 (3 [2]+[4]+2 [6])+q^{7/2} (2 [0]+2 [2]+4 [4]+[6]+[8]) \nn \\
& \quad +q^4 (3 [0]+3 [2]+5 [4]+2 [6]+2 [8]) \nn \\
&\quad +q^{9/2} ([0]+7 [2]+4 [4]+6 [6]+[8]+[10]) \nn \\
&\quad +q^5 ([0]+9 [2]+7 [4]+7 [6]+2 [8]+2 [10]) + \ldots \label{char5}
%& \quad +q^{11/2} (6 [0]+8 [2]+13 [4]+7 [6]+6 [8]+[10]+[12]) \nn \\
%&\quad +q^6 (8 [0]+11 [2]+17 [4]+11 [6]+8 [8]+2 [10]+2 [12]) \nn \\
%&\quad +q^{13/2} (4 [0]+20 [2]+19 [4]+18 [6]+9 [8]+6 [10]+[12]+[14]) \nn \\
%&\quad +q^7 (6 [0]+26 [2]+25 [4]+25 [6]+13 [8]+8 [10]+2 [12]+2 [14]) + \ldots \label{char5}
\eea

Setting $y=1$, we obtain the unrefined partition function
\bea
\chi_{\te{NS}}^{SO(3)} (q,y=1) &=  \chi^{SO(3)}_B (q,y ) f_{\te{NS}} (q,y)  \nn \\
&= \prod_{n=1}^{\infty} \left( \frac{1+ q^{n-1/2}}{1-q^n} \right)^2 \nn \\
&=  (q;q)^{-3}_\infty \vartheta_3(1, q) = q^{-1/8}\frac{\vartheta_3(1, q)}{\eta(q)^3}~. \label{unrefchNSSO3}
\eea

 % We apply the following identies to our question:
%\bea
%\sum_{k=m}^\infty \left[q^{n (k-m)}-q^{n(k+m+1)}\right](2k+1) &= \frac{\left(1-q^{n(1+2 m)}\right) \left[1+q^n+2 m \left(1-q^n\right)\right]}{\left(1-q^n\right)^2}~,  \\
%\sum_{k=0}^{m-1} \left[q^{n (m-k)}-q^{n(k+m+1)}\right](2k+1) &= \frac{q^n \left[ 2 m \left(1-q^n\right) \left(1+q^{2 m n}\right) - \left(1+q^n\right) \left(1-q^{2 m n}\right)\right]}{\left(1-q^n\right)^2}~. \notag
%\eea
%From \eref{exactchNSSO3}, we therefore have
%\bea
%\chi_{\te{NS}}^{SO(3)} (q,y=1) &= (q;q)^{-3}_\infty  \left \{ \sum_{m=0}^\infty (1+2 m) \left(1-q^{\frac{1}{2}+m} \right) q^{\frac{1}{2} m^2}  \right \} \times \left \{ \sum_{n=0}^\infty  (-1)^{n-1}q^{\frac{1}{2} n(n-1)}\left(1+q^n\right)  \right\} \nn \\
%&= \frac{1}{2} (q;q)^{-3}_\infty \sum_{m = -\infty}^\infty (1+2 m) \left(1-q^{\frac{1}{2}+m} \right) q^{\frac{1}{2} m^2}  \nn \\
%&=  \frac{1}{2} (q;q)^{-3}_\infty \left[ \sum_{m=-\infty}^\infty q^{\frac{m^2}{2}} + 2 \sum_{m=-\infty}^\infty m  q^{\frac{m^2}{2}} -  \sum_{m=-\infty}^\infty q^{\frac{1}{2}(m+1)^2} -2 \sum_{m=-\infty}^\infty m q^{\frac{1}{2}(m+1)^2}  \right] \nn \\
%&= \frac{1}{2} (q;q)^{-3}_\infty \left[ \vartheta_3 (0, \tau) + 0 - \vartheta_3 (0, \tau) + 2 \vartheta_3 (0, \tau)  \right] \nn \\
%&= (q;q)^{-3}_\infty \vartheta_3 (0, \tau)~,
%\eea
%where the summation over $n$ in the second bracket of the first line is equal to unity.

\subsubsection{Multiplicities of representations $[2j]$ and their asymptotics}
Similarly to the bosonic partition function, we can read off the generating function for the multiplicities of the representations $[2j]$ at different mass levels of the NS superstring
\bea
M(\chi_{\te{NS}}^{SO(3)}, [2j], q) &= (q;q)_\infty^{-3}  \sum_{m=0}^\infty (1-q^{m+\frac{1}{2}}) q^{\frac{1}{2}m^2} \sum_{n=1}^\infty (-1)^{n-1}  (1-q^n) q^{\frac{1}{2}n(n-1)} \times \nn \\
&  \quad  ( q^{n|j-m|} - q^{n(j+m+1)} ) ~. 
\label{2jmultNS}
\eea

\paragraph{Asymptotics as $j \rightarrow \infty$.}  In this limit, we have $|j-m| \sim j-m$ for a finite $m$. Futhermore, the summand as a function of $n$ is sharply peaked near $n=1$, and so we can determine the leading behaviour of the sum over $n$ using Laplace's method as follows:
\bea
 & \sum_{n=1}^\infty (-1)^{n-1}  (1-q^n) q^{\frac{1}{2}n(n-1)} ( q^{n(j-m)} - q^{n(j+m+1)} ) \nn \\
 &\sim (1-q) \sum_{n=1}^{1+\lfloor \epsilon \rfloor} \left[ q^{n(j-m)} - q^{n(j+m+1)} \right] \qquad \text{for $\epsilon >0$} \nn \\
 &\sim (1-q)  \sum_{t=0}^\infty \left[ q^{(t+1)(j-m)} - q^{(t+1)(j+m+1)} \right] \nn \\
 &= (1-q)  \left[ \frac{q^{j-m}}{1-q^{j-m}} - \frac{q^{j+m+1}}{1-q^{j+m+1}}\right] \nn \\
 &= q^{j-m} (1-q) \frac{1-q^{2m+1}}{(1-q^{1+j+m})(1-q^{j-m})}~.
\eea
Therefore, we find that
\bea
& M(\chi_{\te{NS}}^{SO(3)}, [2j], q) \nn \\
&\qquad \sim (q;q)_\infty^{-3}  q^j (1-q) \Bigg[  \sum_{m=0}^\infty q^{-m+\frac{m^2}{2}}   \frac{ \left(1-q^{2m+1}\right) \left(1-q^{\frac{1}{2}+m}\right)}{\left(1+q^{1+j -m}\right) \left(1+q^{j-m}\right)} \Bigg] \nn \\
& \qquad \sim (q;q)_\infty^{-3}  q^j \frac{1-q}{(1-q^j)^2}  \Bigg[  \sum_{m=0}^\infty q^{\frac{1}{2}(m-1)^2 - \frac{1}{2}}   \left(1-q^{2m+1}\right) \left(1-q^{\frac{1}{2}+m}\right)\Bigg] \nn \\
%& \qquad \sim (q;q)_\infty^{-3}  \sum_{m=0}^\infty (1-q^{m+\frac{1}{2}}) q^{\frac{1}{2}m^2} \Bigg[ \sum_{n=1}^\infty (-1)^{n-1} (1-q^{n}) \left \{ q^{n (j-m)} - q^{n (j+m+1)} \right\} \Bigg] \nn \\
 %&\qquad = (q;q)_\infty^{-3}  q^j (1-q) \Bigg[  \sum_{m=0}^\infty \frac{ q^{-m+\frac{m^2}{2}} \left(1-q^{\frac{1}{2}+m}\right)}{\left(1+q^{1+j -m}\right) \left(1+q^{j-m}\right)}  - q \sum_{m=0}^\infty \frac{ q^{m+\frac{m^2}{2}} \left(1-q^{\frac{1}{2}+m}\right)}{\left(1+q^{1+j +m}\right) \left(1+q^{2+j+m}\right)} \Bigg] \nn \\
% & \qquad \sim  (q;q)_\infty^{-3}  \frac{q^j (1-q)}{\left(1+q^{j}\right)^2} \Bigg[\sum_{m=0}^\infty  q^{-m+\frac{m^2}{2}} \left(1-q^{\frac{1}{2}+m}\right)  -q \sum_{m=0}^\infty  q^{m+\frac{m^2}{2}} \left(1-q^{\frac{1}{2}+m}\right) \Bigg]  \nn \\
% &  \qquad =  (q;q)_\infty^{-3}  \frac{q^j (1-q)}{\left(1+q^{j}\right)^2} \Bigg[ \frac{(1+2 \sqrt{q} -q )+(1-q) \vartheta_3(0,\sqrt{q})}{2 \sqrt{q}} - \frac{(1+2 \sqrt{q} -q )-(1-q) \vartheta_3(0,\sqrt{q})}{2 \sqrt{q}} \Bigg]   \nn \\
 & \qquad= (q;q)_\infty^{-3} q^{j-\frac{1}{2}} \left( \frac{1-q}{1+q^j}\right)^2  \vartheta_3(1,q)~. \label{asympchiSO3NS}
 %+ O(q^{2j-1}) ~, 
\eea
Note that asymptotic formula \eref{asympchiSO3NS} reproduces the exact result up to the order $q^{2j-\frac{3}{2}}$.
We emphasise that the representation $[2j]$ appears first time at mass level $q^{j-\frac{1}{2}}$.

In \cite{Curtright:1986di}, the individual $n$ summands of (\ref{2jmultNS}) are interpreted as an alternating sequence of additive and subtractive Regge trajectories of slope $\frac{1}{n}$. In the notation of equation (6.2) of that reference, the $M(\chi_{\te{NS}}^{SO(3)}, [2j], q)$ are expanded as
\begin{align}
M(\chi_{\te{NS}}^{SO(3)}, [2j], q) \eq &q^{j} \, \tau_1^{\te{NS}}(q) \ - \ q^{2j} \, \tau_2^{\te{NS}}(q) \ + \ q^{3j} \, \tau_3^{\te{NS}}(q) \ - \ \ldots \notag \\
\eq &\sum_{\ell=1}^{\infty} (-1)^{\ell-1} \, q^{  \ell j} \, \tau_\ell^{\te{NS}}(q) \ .
\label{tausum}
\end{align}
Setting $|j-m|=j-m$ in (\ref{2jmultNS}) leads to the following asymptotic expressions for the $\tau_\ell^{\te{NS}}$:
\bea
\tau_\ell^{\te{NS}}(q) \eq (q;q)_\infty^{-3} \,q^{ - \frac{1}{2}\ell} \, (1-q^\ell)   \sum_{m=0}^\infty q^{\frac{1}{2}(m-\ell)^2} \,(1-q^{m+\frac{1}{2}}) \, ( 1 - q^{2m\ell + \ell} )
\eea
We will later on rediscover this trajectory structure in the counting of SUSY multiplets.

\paragraph{The stable pattern.} The generating function of the stable pattern can be determined by projecting the sum in (\ref{tausum}) to the first term (or, equivalently, by taking the limit $j \rightarrow \infty$):
\bea
& \lim_{j \rightarrow \infty}  q^{-j}  M(\chi_{\te{NS}}^{SO(3)}, [2j], q) \eq \tau_1^{\te{NS}}(q) \nn \\
& \qquad = (q;q)_\infty^{-3} q^{-1/2}  (1-q)^2 \vartheta_3(1, q) \\
%& \qquad = (q;q)_\infty^{-3} (1-q) \sum_{m=0}^\infty  q^{-m+\frac{m^2}{2}} (1-q^{2m+1}) \left( 1-q^{m+\frac{1}{2}} \right)  \nn \\
&  \qquad  =\Big( 2+2 q+8 q^2+14 q^3+34 q^4+58 q^5+120 q^6+204 q^7+378 q^8+632 q^9 \nn \\
& \quad \qquad +1096 q^{10}+1786 q^{11}+2968 q^{12}+4722 q^{13}+7578 q^{14}+11818 q^{15}+ \ldots \Big)+ \nn \\
& \quad \qquad + \Big( \frac{1}{\sqrt{q}}+\sqrt{q}+6 q^{3/2}+9 q^{5/2}+24 q^{7/2}+42 q^{9/2}+88 q^{11/2}+151 q^{13/2} \nn \\
& \quad \qquad +287 q^{15/2}+480 q^{17/2}+846 q^{19/2}+1388 q^{21/2}+2326 q^{23/2}+3724 q^{25/2} \nn \\
& \quad \qquad  +6025 q^{27/2}+9438 q^{29/2} + \ldots \Big) ~. \label{datachiSO3NS}
\eea
Note that terms with low orders in the power series \eref{datachiSO3NS} are in agreement with the data presented in Table  6c of \cite{Curtright:1986di}.

\subsection[The R sector in $d=4$]{The R sector in $\bm{d=4}$} 
\label{sec:4dimR}

The R sector of the worldsheet superpartners $\psi^i$ of the lightcone bosons contributes
\bea
f_{\Ra}(q, y)  &= (y+y^{-1}) {\PE}_F \left[ ([2]_y-1) \frac{q}{1-q} \right] \nn \\
&= (y+y^{-1}) \prod_{n=1}^{\infty} (1+y^2 q^{n})(1+y^{-2} q^{n}) \label{deffR} \\
&= q^{-\frac{1}{12}} \frac{\vartheta_2 (y^2, q)}{\eta(q)}
\eea
to the spacetime partition function. Again, it will turn out to be beneficial to rewrite this function as an infinite sum.  We proceed as follows.  Replacing $z$ by $xz$ in \eref{Jacobitriple}, we obtain
\bea
\prod_{n=1}^\infty (1-x^{2n}) (1+ x^{2n} z)( 1+x^{2n-2} z^{-1}) = \sum_{m = -\infty}^{+\infty} x^{m^2+m} z^{m}~.
\eea
Using the identity
\bea
\prod_{n=1}^\infty (1+ x^{2n-2} z^{-1}) = (1+z^{-1}) \prod_{n=1}^\infty (1+x^{2n} z^{-1})~,
\eea
we arrive at
\bea
(z^{\frac{1}{2}}+z^{-\frac{1}{2}}) \prod_{n=1}^\infty (1+ x^{2n} z)(1+x^{2n} z^{-1}) = \frac{ \sum_{m = -\infty}^{+\infty} x^{m^2+m} z^{m+1/2}}{\prod_{n=1}^\infty (1-x^{2n})}~.  \label{Jacobi2}
\eea
Applying identity \eref{Jacobi2} to \eref{deffR} with $x = q^{1/2}$ and $z=y^2$, we have
\bea
f_{\Ra}(q, y)  &= (q;q)_\infty^{-1} \sum_{m= - \infty}^{+\infty}  y^{2m+1}  q^{m(m+1)/2} \nn \\
&=  (q;q)_\infty^{-1} \sum_{m=0}^\infty q^{\frac{1}{2}m(m+1)} (1-q^{m+1}) [2m+1]_{y} \nn \\
&= q^{-1/8} (q;q)_\infty^{-1} \sum_{m \in \BZ_{\geq 0}+\frac{1}{2}} q^{\frac{1}{2}m^2} (1-q^{m+\frac{1}{2}}) [2m]_{y}~,
\eea
where the second equality follows from \eref{idenortho} and the orthogonality of the characters.

Let us combine the contribution from the R-sector with the bosonic part.  Using \eref{bosonic} and \eref{deffR}, we find that
\bea
\chi_{\te{R}}^{SO(3)} (q,y) &:=  \chi^{SO(3)}_B (q,y ) f_{\te{R}} (q,y)  = - i (y-y^{-1}) \frac{\vartheta_2(y^2, q)}{ \vartheta_1( y^2,q)} \\
&= q^{-\frac{1}{8}}  (q;q)_\infty^{-3}  \sum_{m \in \BZ_{\geq 0} + \frac{1}{2}}^\infty \sum_{n=1}^\infty (-1)^{n-1} (1-q^{m+\frac{1}{2}}) (1-q^n) q^{\frac{1}{2}[n(n-1) + m^2]}   \nn \\
& \hspace{4cm} \times \sum_{k =0 }^\infty \; ( q^{n|k-m|} - q^{n(k+m+2)} ) \; [2k+1] ~. \label{exactchRSO3}
\eea
This resembles (\ref{exactchNSSO3}) up to a shift in the summations over $m,k$ by $\pm \frac{1}{2}$. We emphasise that $SO(3)$ irreducible representation with even Dynkin labels do not appear in the R sector partition function $\chi_{\te{R}}^{SO(3)} (q,y)$.

In terms of a power series, this partition function can be written as
\begin{align}
\chi_{\te{R}}^{SO(3)} (q,y)  &= [1]  +  2 [3] q+2 ([1]+[3]+[5]) q^2+(4 [1]+4 [3]+4 [5]+2 [7]) q^3 \nn \\
& \quad +(6 [1]+10 [3]+8 [5]+4 [7]+2 [9]) q^4 \nn \\
& \quad +(12 [1]+18 [3]+16 [5]+10 [7]+4 [9]+2 [11]) q^5 \nn \\
& \quad +(22 [1]+32 [3]+30 [5]+22 [7]+10 [9]+4 [11]+2 [13]) q^6 \nn \\
& \quad +(36 [1]+58 [3]+56 [5]+40 [7]+24 [9]+10 [11]+4 [13]+2 [15]) q^7+  \ldots
\label{char6}
\end{align}

Setting $y=1$, we obtain the unrefined partition function
\bea
\chi_{\te{R}}^{SO(3)} (q,y=1) &= 2 \prod_{n=1}^{\infty} \left( \frac{1+ q^{n}}{1-q^n} \right)^2 =  q^{-\frac{1}{8}} (q;q)^{-3}_\infty \vartheta_2 (1, q) = \frac{\vartheta_2 (1, q)}{\eta(q)^3}~. \label{unrefchRSO3}
\eea 

% The calculation is very similar to that of the NS-sector; see the discussion around \eref{unrefchNSSO3}.
%\bea
%\chi_{\te{R}}^{SO(3)} (q,y=1) &= q^{-\frac{1}{8}} (q;q)^{-3}_\infty  \left \{ \sum_{m=0}^\infty 2 (1+ m) \left(1-q^{1+m} \right) q^{\frac{1}{2} (m+\frac{1}{2})^2}  \right \} \times \left \{ \sum_{n=0}^\infty  (-1)^{n-1}q^{\frac{1}{2} n(n-1)}\left(1+q^n\right)  \right\} \nn \\
%&= q^{-\frac{1}{8}} (q;q)^{-3}_\infty \sum_{m= -\infty}^\infty  (1+ m) \left(1-q^{1+m} \right) q^{\frac{1}{2} (m+\frac{1}{2})^2} \nn \\
%&= q^{-\frac{1}{8}} (q;q)^{-3}_\infty \vartheta_2 (0, \tau)~, \label{unrefchRSO3}
%\eea

\subsubsection{Multiplicities of representations $[2j+1]$ and their asymptotics}
The generating function for the multiplicities of the representations $[2j+1]$ at different mass levels are
\bea
M(\chi_{\te{R}}^{SO(3)}, [2j+1], q) &= q^{-\frac{1}{8}} (q;q)_\infty^{-3}  \sum_{m=0}^\infty (1-q^{m+1}) q^{\frac{1}{2}(m+\frac{1}{2})^2} \sum_{n=1}^\infty (-1)^{n-1}  (1-q^n) q^{\frac{1}{2}n(n-1)} \times \nn \\
&  \quad  ( q^{n|j-m|} - q^{n(j+m+2)} ) 
\eea
in close analogy to (\ref{2jmultNS}).  In fact, one can obtain the above formula by shifting $m \rightarrow m+\frac{1}{2}$ and $j \rightarrow j+\frac{1}{2}$ in (\ref{2jmultNS}) and multiply by an overall factor $q^{-\frac{1}{8}}$.

\paragraph{Asymptotics as $j \rightarrow \infty$.}  Similarly to the NS-sector, we find that the leading behaviour of $M(\chi_{\te{R}}^{SO(3)}, [2j+1], q)$ is
\bea
& M(\chi_{\te{R}}^{SO(3)}, [2j+1], q) \nn \\
& \qquad \sim q^{-\frac{1}{8}} (q;q)_\infty^{-3}  q^{j+\frac{1}{2}} \frac{1-q}{(1-q^j)^2}  \Bigg[  \sum_{m=0}^\infty q^{\frac{1}{2}(m-\frac{1}{2})^2 - \frac{1}{2}}   \left(1-q^{2m+2}\right) \left(1-q^{m+1}\right)\Bigg] \nn \\
%& \qquad \sim  (q;q)_\infty^{-3}  \frac{q^{j -\frac{1}{8}} (1-q)}{\left(1+q^{j}\right)^2} \Bigg[\sum_{m=0}^\infty  q^{-m+\frac{1}{2}(m+\frac{1}{2})^2} \left(1-q^{m+1}\right)  -q^2 \sum_{m=0}^\infty  q^{m+\frac{1}{2}(m+\frac{1}{2})^2} \left(1-q^{m+1}\right) \Bigg]  \nn \\
%& \qquad =  (q;q)_\infty^{-3}  \frac{q^{j -\frac{1}{8}} (1-q)}{\left(1+q^{j}\right)^2} \Bigg[ \left \{ q^{\frac{1}{8}} +\frac{1}{2} (1-q) \vartheta_2(0, \tau) \right \} -  \left \{ q^{\frac{1}{8}} - \frac{1}{2} (1-q) \vartheta_2(0, \tau) \right \} \Bigg] \nn \\
& \qquad =  (q;q)_\infty^{-3} q^{j -\frac{1}{8}}  \left( \frac{1-q}{1+q^{j}} \right)^2 \vartheta_2(1,q) ~.
%+ O(q^{2j})~.
\eea
Note that the representation $[2j+1]$ appears first time at mass level $q^{j}$ and the asymptotic formula reproduces the exact result up to the order $q^{2j-1}$.

Also the multiplicity generating functions of the Ramond sector are suitable for an expansion in terms of Regge trajectories:
\begin{align}
M(\chi_{\te{R}}^{SO(3)}, [2j+1], q) \eq &q^{j} \, \tau_1^{\te{R}}(q) \ - \ q^{2j} \, \tau_2^{\te{R}}(q) \ + \ q^{3j} \, \tau_3^{\te{R}}(q) \ - \ \ldots \notag \\
 \eq &\sum_{\ell=1}^{\infty} (-1)^{\ell-1} \, q^{  \ell j} \, \tau_\ell^{\te{R}}(q) \ .
\label{tausumR}
\end{align}
The $|j-m|=j-m$ asymptotics yield the following expressions for the $\ell$'th Ramond trajectory $\tau_\ell^{\te{R}}$:
\bea
\tau_\ell^{\te{R}}(q) \eq (q;q)_\infty^{-3} \,q^{ -\frac{1}{8}} \, (1-q^\ell)   \sum_{m=\frac{1}{2}}^\infty q^{\frac{1}{2}(m-\ell)^2} \,(1-q^{m+\frac{1}{2}}) \, ( 1 - q^{2m\ell + \ell} )
\eea

\paragraph{The stable pattern.} The generating function of the stable pattern can be determined by taking the limit $j \rightarrow \infty$:
\bea
& \lim_{j \rightarrow \infty}  q^{-j }  M(\chi_{\te{R}}^{SO(3)}, [2j+1], q) \eq \tau_1^{\te{R}}(q) \nn \\
& \qquad = (q;q)_\infty^{-3} q^{-1/8}  (1-q)^2 \vartheta_2(1, q)  \\
& \qquad = 2+4 q+10 q^2+24 q^3+48 q^4+96 q^5+184 q^6+336 q^7+600 q^8+1048 q^9+1784 q^{10}\nn \\
& \qquad \quad +2984 q^{11}+4912 q^{12}+7952 q^{13}+12704 q^{14}+20048 q^{15}+31256 q^{16}+48224 q^{17} \nn \\
& \qquad \quad +73680 q^{18}+111520 q^{19}+167368 q^{20} + O(q^{21})~. \label{datachiSO3R}
\eea
Note that terms with low orders in the power series \eref{datachiSO3R} are in agreement with the data presented in Table 6d of \cite{Curtright:1986di}.

\subsection[Bosonic partition function in $d>4$]{Bosonic partition function in $\bm{d>4}$}
\label{sec:highdim}

The bosonic partition function in $d=2n+2$ space-time dimensions can be written as
\bea
\chi^{SO(2n+1)}_{B}(q, \vec y) = \PE \left[ ([1,0, \ldots, 0]^{SO(2n+1)}_y -1) \frac{q}{1-q} \right]~,\label{PEvec}
\eea
where $\vec y = (y_1, \ldots, y_n)$ and the character of the vector representation $[1,0, \ldots, 0]$ of $SO(2n+1)$ is given by \eref{chaBn}. The $2n=d-2$ summands in $[1,0, \ldots, 0]^{SO(2n+1)}_y -1=\sum_{k=1}^n (y_k^2+y_k^{-2})$ reflect the $\partial X^i$ components outside the lightcone. Using \eref{vecBn}, we see that this choice of character allows us to write
\bea
\chi^{SO(2n+1)}_{B}(q, \vec y) &= \PE \left[  \frac{q}{1-q}  \sum_{k=1}^n ([2]_{y_k}-1)  \right] \nn \\
&= \prod_{A=1}^n \chi^{SO(3)}_B (y_A)~.
%&= \prod_{A=1}^n -i q^{\frac{1}{12}} (y_A-y_A^{-1}) \frac{\eta(\tau)}{\vartheta_1 (y_A, q^{\frac{1}{2}})}~.
\eea
Observe that the $(2n+2)$-dimensional partition function is simply a product of $n$ copies of the four dimensional partition function.
From \eref{chiBSO3}, we have
\bea
\chi^{SO(2n+1)}_{B}(q, \vec y) = (q;q)_\infty^{-2n} \sum_{\vec n \in \BZ_+^n} \sum_{\vec k \in \BZ_{\geq 0}^n}\prod_{A=1}^n (-1)^{n_A-1} (1-q^{n_A})^2    q^{n_A k_A + \frac{1}{2}n_A(n_A-1)} [2k_A]_{y_A}\label{chiBSO5}
\eea 
with $\BZ_+$ denoting the set of positive integers and $\BZ_{\geq 0} = \BZ_+ \cup \{ 0 \}$. For our purpose of resolving the $SO(2n+1)$ content of the partition function, the aim is to rewrite \eref{chiBSO5} in the form 
\bea
\chi^{SO(2n+1)}_{B}(q, \vec y) = \sum_{\lambda_1  \geq \cdots \geq \lambda_n  \geq 0} (\lambda_1, \ldots, \lambda_n)_{\vec y}  ~ G^{B,SO(2n+1)}_{\lambda_1, \ldots,  \lambda_n} (q)~,
\eea
where the summations run over highest weight vectors $\vec \lambda := (\lambda_1, \ldots, \lambda_n) \in \BZ^n$ subject to inequalities $\lambda_1 \geq \cdots \geq \lambda_n \geq 0$, see Appendix \ref{sec:not} for the conversion rule to Dynkin label notation $[a_1,\ldots,a_n]$.  
Since \eref{PEvec} involves only the vector representation and the plethystic exponential generates symmetrisations of the representation, there is no spinor representation of $SO(2n+1)$ appearing in $\chi^{SO(2n+1)}_{B}(q, \vec y)$; therefore,
\bea
G^{B,SO(2n+1)}_{\lambda_1+ \frac{1}{2}, \ldots,  \lambda_n+\frac{1}{2}} (q) =0 \co \lambda_k \in \BZ ~.
\eea
In general, $G^{B,SO(2n+1)}_{\lambda_1, \ldots, \lambda_n} (q)$ can be interpreted as a {\it generating function for the multiplicities} of the $SO(2n+1)$ representation $(\lambda_1, \ldots, \lambda_n)$ in the bosonic string partition function.

\subsubsection*{Some useful relations between $SO(2n+1)$ and $SO(3)$ representations}

In order to obtain compact formulae for the multiplicity generating functions $G^{B,SO(2n+1)}_{\lambda_1, \ldots,  \lambda_n} (q)$, we have to convert the $SO(3)$ character products in (\ref{chiBSO5}) into a basis of $(\la_1,\ldots,\la_n)_{\vec{y}}$, i.e. we have to find the $\Delta$ coefficients in the basis transformation
\bea
\prod_{A=1}^n [2k_A]_{y_A}= \sum_{\lambda_1 \geq \ldots \geq \lambda_n \geq 0} \Delta\left(\lambda_1,\ldots, \lambda_n;2k_1, \ldots, 2k_n  \right) (\lambda_1, \ldots, \lambda_n)_{\vec y} \ .
\eea
In general, according to (5.10) of \cite{Curtright:1986di}, it can be shown that the coefficients in this basis transformation are given by
\bea
\Delta(\lambda_1, \ldots, \lambda_n; 2k_1, \ldots, 2k_n) &:= 
 \int \ud \mu_{SO(2n+1)}(\vec y) ~ (\lambda_1, \ldots, \lambda_n)_{\vec y} \prod_{A=1}^n [2k_A]_{y_A} \nn \\
%& = \det \left( \theta_{|\lambda_i-i+j|}^{2n+\lambda_i-i-j} (2k_i) \right)_{i,j =1}^{n}  \times
&= \frac{1}{n!}  \sum_{\sigma_1, \sigma_2 \in S_n} \sgn(\sigma_2) \prod_{A=1}^n \theta_{|\lambda_A - A + \sigma_2(A)|}^{2n + \lambda_A - A - \sigma_2(A)} \left(k_{\sigma_1 (A)} \right) \nn \\
&= \frac{1}{n!} \sum_{\sigma \in S_n} \det \left(\theta_{|\lambda_A-A+B|}^{2n+\lambda_A-A-B} \left(k_{\sigma (A)} \right)  \right)_{A,B=1}^n  \label{CTorthoeven}
\eea
where 
the function $\theta_m^n (k)$ is defined as
\bea
\theta_m^n (k) = \begin{cases} 1 \quad \text{if ~ $m \leq k \leq n$}~, \\
0 \quad \text{otherwise}~. \end{cases}
\eea
Note that for spinorial representations $(\lambda_1 + \frac{1}{2}, \ldots, \lambda_n+ \frac{1}{2})$, \eref{CTorthoeven} vanishes identically:
\bea
\Delta \left(\lambda_1 + \frac{1}{2}, \ldots, \lambda_n+ \frac{1}{2}; 2k_1, \ldots, 2k_n \right)=0~,  \quad \forall~\vec \lambda \in \BZ^n ~\text{and}~\lambda_1 \geq \cdots \geq \lambda_n \geq 0~.
\eea
Thus, Eq. \eref{CTorthoeven} implies the following expansion rule for $SO(3)$ character products in terms of $SO(2n+1)$ characters in  Dynkin label notation:
%also implies that
\bea
\prod_{A=1}^n [2k_A]_{y_A}= \sum_{\vec \ell \in \BZ_{\geq 0}^n} [\ell_1,\ldots, \ell_{n-1}, 2\ell_n]_{\vec y} \ \Delta\left(2k_1, \ldots, 2k_n ; \ell_1+\ell_2 + \ldots +\ell_n, \ell_2 + \ldots +\ell_n, \ldots, \ell_n \right) \label{decompeven}
\eea
The inverse decomposition formula for an integer spin representation follows from the $SO(2n+1)$ Haar measure (\ref{HaarBn}):
\bea
[\ell_1,\ldots, \ell_{n-1},2\ell_n]_{\vec y} &=\frac{1}{\rho(\vec y)} \sum_{\vec k \in \BZ_{\geq 0}^n}  \prod_{A=1}^n [2k_A]_{y_A} \ \Delta\left(\ell_1+\ell_2 + \ldots +\ell_n, \ell_2 + \ldots +\ell_n, \ldots, \ell_n ; 2k_1, \ldots, 2k_n \right) ~, \label{decompnonspin}
\eea
where $\rho(\vec y)$ is defined as in \eref{defrho} and $\vec \ell = (\ell_1, \ldots, \ell_n) \in \BZ_{\geq 0}^n$.
\\~\\
\indent Similarly, one can convert spinorial $SO(3)$ character products to $SO(2n+1)$ characters via
\bea \label{CTorthoodd}
\Delta(\lambda_1, \ldots, \lambda_n; 2k_1+1, \ldots, 2k_n+1) &:=  \int \ud \mu_{SO(2n+1)}(\vec y) ~ (\lambda_1, \ldots, \lambda_n)_{\vec y} \prod_{A=1}^n [2k_A+1]_{y_A} \nn \\
&= \frac{1}{n!} \sum_{\sigma \in S_n} \det \left(\theta_{|\lambda_A-A+B|}^{2n+\lambda_A-A-B} \left(k_{\sigma (A)}+\frac{1}{2} \right)  \right)_{A,B=1}^n~. 
\eea
For integer spin representations of $SO(2n+1)$, \eref{CTorthoodd} vanishes identically:
\bea
\Delta \left(\lambda_1 \ldots, \lambda_n; 2k_1+1, \ldots, 2k_n+1 \right)=0~,  \quad \forall~\vec \lambda \in \BZ^n ~\text{and}~\lambda_1 \geq \cdots \geq \lambda_n \geq 0~.
\eea
We thus have the following decomposition for products of spinorial $SO(3)$ characters
\bea
\prod_{A=1}^n [2k_A&+1]_{y_A} = \sum_{\vec \ell \in \BZ_{\geq 0}^n}  [\ell_1,\ldots, \ell_{n-1}, 2 \ell_n+1]_{\vec y}  \nn \\ 
&\times \Delta\left(2k_1+1, \ldots, 2k_n+1 ; \ell_1+\ell_2 + \ldots +\frac{1}{2}\ell_n, \ell_2 + \ldots +\frac{1}{2}\ell_n, \ldots, \frac{1}{2}\ell_n \right)~, \label{decompodd}
\eea
with inverse
\bea
[\ell_1,&\ldots, , \ell_{n-1}, 2\ell_n+1]_{\vec y} =\frac{1}{\rho(\vec y)} \sum_{\vec k \in \BZ_{\geq 0}^n}  \prod_{A=1}^n [2k_A+1]_{y_A} \nn \\
&\times  \Delta\left(\ell_1+\ell_2 + \ldots +\ell_n+\frac{1}{2}, \ell_2 + \ldots +\ell_n+\frac{1}{2}, \ldots, \ell_n+\frac{1}{2}; 2k_1+1, \ldots, 2k_n+1 \right) ~. \label{decompspin}
\eea

\subsubsection*{Generating function for the multiplicities}
According to \eref{chiBSO5}, the bosonic spacetime partition function in $2n+2$ dimensions depends on Lorentz fugacities through the factor
\bea
& \sum_{k_1, \ldots, k_n \geq 0} \Delta( \lambda_1, \ldots, \lambda_n; 2k_1, \ldots, 2k_n)  q^{n_1 k_1+ \ldots +n_n k_n}  \nn \\
&= \sum_{k_1, \ldots, k_n \geq 0} \det (\theta_{|\lambda_A-A+B|}^{2n+\lambda_A-A-B} (k_A))_{A,B=1}^n  q^{n_1k_1 + \ldots+ n_n k_n} \nn \\
&= \det \left(\sum_{k_A \geq 0}  \theta_{|\lambda_A-A+B|}^{2n+\lambda_A-A-B} (k_A) q^{n_A k_A} \right)_{A,B=1}^n~. 
\eea

Let us apply this to \eref{chiBSO5} to compute $G^{B,SO(2n+1)}_{\lambda_1, \ldots,  \lambda_n} (q)$.
For $\lambda_1 \geq \cdots \geq \lambda_n \geq n-1$, the argument in the absolute value is non-negative and so
\bea
 & \sum_{k_1, \ldots, k_n \geq 0} \Delta( \lambda_1, \ldots, \lambda_n;2k_1, \ldots, 2k_n)  q^{n_1 k_1+ \ldots n_n k_n}  \nn \\
 &= \prod_{A=1}^n q^{n_A( \lambda_A-A+1)} \prod_{1 \leq B<C \leq n} (q^{n_C}-q^{n_B})(1-q^{n_C+n_B})\quad  \text{for $\lambda_1 \geq \cdots \geq \lambda_n \geq n-1$}~.
\eea
It is pointed out by \cite{Curtright:1986di} and can be checked directly that the contribution from $\lambda_n < n-1$ to the bosonic string partition function is zero.  Therefore, we have
\bea
G^{B,SO(2n+1)}_{\lambda_1, \ldots, \lambda_n} (q) &= (q;q)_\infty^{-2n} \sum_{\vec n \in \BZ_+^n}  \prod_{A=1}^n (-1)^{n_A-1} (1-q^{n_A})^2 q^{n_A( \lambda_A-A+1) +\frac{1}{2}n_A(n_A-1)}   \nn \\
&  \ \ \ \ \ \ \ \ \times \prod_{1 \leq B<C \leq n} (q^{n_C}-q^{n_B})(1-q^{n_C+n_B})~,
\label{GBSO(2n+1)}
\eea
for all $\lambda_1, \ldots, \lambda_n \in \BZ$ and $\lambda_1 \geq \ldots \geq  \lambda_n \geq 0$.

%\subsubsection{Decompositions of $SO(2n+1)$ representations} 
%Using the orthogonality of characters of irreducible representations, we find that \eref{CTorthoeven} 
%%also implies that
%%\bea
%%\prod_{A=1}^n [2k_n]_{y_n}= \sum_{\vec l \in \BZ_{\geq 0}^n} \Delta\left(2k_1, \ldots, 2k_n ; m_1+m_2 + \ldots +\frac{1}{2}m_n, m_2 + \ldots +\frac{1}{2}m_n, \ldots, \frac{1}{2}m_n \right) [m_1,\ldots, 2m_n]_{\vec y}~, \label{decomp1}
%%\eea
%and \eref{HaarSO5}, we obtain
%\bea
%[l_1,\ldots, l_n]_{\vec y} &=\frac{1}{\rho(\vec z)} \sum_{\vec k \in \BZ_{\geq 0}^n} \Delta\left(2k_1, \ldots, 2k_n ; l_1+l_2 + \ldots +\frac{1}{2}l_n, l_2 + \ldots +\frac{1}{2}l_n, \ldots, \frac{1}{2}l_n \right)  \prod_{A=1}^n [2k_A]_{y_A}~, \label{decomp2}
%\eea
%where $\rho(\vec z)$ is defined as in \eref{defrho}. Similarly, 
%\bea
%\prod_{A=1}^n [2k_n+1]_{y_n} &= \sum_{\vec l \in \BZ_{\geq 0}^n} \Delta\left(2k_1+1, \ldots, 2k_n +1; l_1+l_2 + \ldots +\frac{1}{2}l_n, l_2 + \ldots +\frac{1}{2}l_n, \ldots, \frac{1}{2}l_n \right) [l_1,\ldots, l_n]_{\vec y}~, \label{decomp3} \\ 
%[l_1,\ldots, l_n]_{\vec y} &=\frac{1}{\rho(\vec z)} \sum_{\vec k \in \BZ_{\geq 0}^n} \Delta\left(2k_1+1, \ldots, 2k_n+1 ; l_1+l_2 + \ldots +\frac{1}{2}l_n, l_2 + \ldots +\frac{1}{2}l_n, \ldots, \frac{1}{2}l_n \right)  \prod_{A=1}^n [2k_A+1]_{y_A}~. \label{decomp4}
%\eea
%These decompositions are useful in subsequent sections of the paper.

\subsection[The contributions from the NS and R sectors in $d>4$]{The contributions from the NS and R sectors in $\bm{d>4}$}
\label{sec:highdimm}

The contribution from the NS sector can be obtained by taking a product of $n$ copies of \eref{exactchNSSO3}: 
\bea \label{BnNS}
\chi^{SO(2n+1)}_{\NS} (q,\vec y) 
&= \prod_{A=1}^n \chi^{SO(3)}_{\NS} (q; y_A) \nn \\
&= (q;q)_\infty^{-3n} \sum_{\vec m  \in\BZ_{\geq 0}^n} \sum_{\vec n \in \BZ_+^n}  \prod_{A=1}^n (-1)^{n_A+1}\left(1-q^{m_A+\frac{1}{2}}\right)\left(1-q^{n_A}\right)q^{\frac{1}{2}[n_A(n_A-1)+m_A^2]} \times \nn \\
& \hspace{2.4cm}   \sum_{\vec k \in \BZ_{\geq 0}^n}^\infty  \prod_{A=1}^n ( q^{n_A|k_A-m_A|} - q^{n_A(k_A+m_A+1)} ) \; [2k_A]_{y_A}~.
\eea
Similarly for the contribution from the R sector, the product of $n$ copies of \eref{exactchRSO3}:
\bea \label{BnR}
\chi^{SO(2n+1)}_{\Ra} (q,\vec y) 
&= \prod_{A=1}^n \chi^{SO(3)}_{\Ra} (q; y_A) \nn \\
&= q^{-\frac{n}{8}}  (q;q)_\infty^{-3n}  \sum_{\vec m  \in\BZ_{\geq 0}^n} \sum_{\vec n \in \BZ_+^n}   \prod_{A=1}^n (-1)^{n_A+1} (1-q^{m_A+1}) (1-q^{n_A}) q^{\frac{1}{2}[n_A(n_A-1) + \left(m_A+\frac{1}{2} \right)^2]} \times \nn \\
& \hspace{2.4cm}  \sum_{\vec k \in \BZ_{\geq 0}^n}  \prod_{A=1}^n  ( q^{n_A|k_A-m_A|} - q^{n_A(k_A+m_A+2)} ) \; [2k_A+1]_{y_A} ~.
\eea
The unrefined partition functions can be written as
\bea
\chi^{SO(2n+1)}_{\NS} (q,\{ y_i =1 \}) &=  q^{-n/8}\frac{\vartheta_3(1, q)^n}{\eta(q)^{3n}}~,  \label{unrefBnNS} \\
\chi^{SO(2n+1)}_{\Ra} (q,\{ y_i =1 \}) &=   \frac{\vartheta_2 (1,q)^n}{\eta(q)^{3n}}~.\label{unrefBnR}
\eea

%%%%%%

\section{Internal SCFTs}
\label{sec:internal}

The SCFT description of four- and six dimensional string compactifications with ${\cal N}_{4d}=1, {\cal N}_{4d}=2$ or ${\cal N}_{6d}=(1,0)$ spacetime SUSY comprises universal sectors with enhanced ${\cal N}_{2d}=2,4$ worldsheet SUSY \cite{Banks:1987cy, Banks:1988yz, Ferrara:1989ud, Feng:2012bb}. The purpose of this section is to collect the associated charged characters, starting from the expressions given in \cite{Odake:1989dm, Eguchi:1988af} but adapting the dependence on fugacities $s,x$ and $z$ of the internal symmetries to the R symmetries of the spectrum.

%%%%%%%%%%%%%%%%%%%%%%
\subsection[${\cal N}_{2d}=2$ worldsheet superconformal algebra at $c=9$]{$\bm{{\cal N}_{2d}=2}$ worldsheet superconformal algebra at $\bm{c=9}$}
\label{sec:41}

The internal SCFT universal to any four dimensional string compactification with ${\cal N}_{4d}=1$ spacetime SUSY enjoys ${\cal N}_{2d}=2$ worldsheet SUSY.  The resulting model independent partition function receives contributions from characters of the ${\cal N}_{2d}=2$ superconformal algebra with central charge $c=9$.  Its representations are characterized by the conformal weight $h$ and the $U(1)$ charge $\ell$ of their highest weight state. The representations needed to describe ${\cal N}_{4d}=1$ compactifications have $(h,\ell) = (0,0)$ in the NS sector and $(h,\ell)= \left( \frac{3}{8},\frac{3}{2} \right)$ in the R sector.

%This enhanced SUSY algebra contains a $U(1)$ current ${\cal J}$ under which the two supercurrents $G_{\textrm{int}}^{\pm}$ have opposite charges. 

The $\CN_{2d}=2$ SCFT at $c=9$ can be split into two decoupled sectors, each of which enjoys a $U(1)$ symmetry. The first one carries central charge $c=1$ and can be completely bosonized; let us denote the $U(1)$ occurring in this sector by $U(1)_1$ and its $h=1$ current by ${\cal J}_1$. In addition, there exists a second decoupled sector with $c=8$ which involves conformal primiaries $g^{\pm}$ of weight $\frac{4}{3}$, see e.g. \cite{Feng:2012bb}. It enjoys an independent $U(1)_2$ under which the $g^\pm$ have opposite charges. The $c=9$ supercurrent can be split into two components $G_{\te{int}}^{\pm}$ that carry opposite charges under both $U(1)_1$ and $U(1)_2$ and factorize into conformal primaries of both sectors. The following figure \ref{N=1SCFT} summarizes the decoupling SCFT ingredients.

\begin{figure}[h]
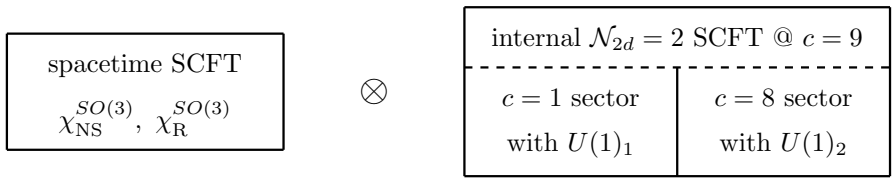

\centerline{
\tikzpicture [scale=1.4,line width=0.30mm]
\draw (0.7,-0.8) -- (0.7,0.3);
\draw (3.3,-0.8) -- (3.3,0.3);
\draw (0.7,-0.8) -- (3.3,-0.8);
\draw (0.7,0.3) -- (3.3,0.3);
\draw (2,0) node{spacetime SCFT};
\draw (2,-0.5) node{$\chi_{\te{NS}}^{SO(3)}, \ \chi_{\te{R}}^{SO(3)}$};
\draw (4.15,-0.25) node{\Large $\otimes$};
\begin{scope}[xshift=5cm]
\draw (0,-1.05) -- (0,0.55);
\draw (4,-1.05) -- (4,0.55);
\draw (0,-1.05) -- (4,-1.05);
\draw (0,0.55) -- (4,0.55);
\draw (2,0.25) node{internal ${\cal N}_{2d}=2$ SCFT @ $c=9$};
\draw (1,-0.3) node{$c=1$ sector};
\draw (1,-0.75) node{with $U(1)_1$};
\draw (3,-0.3) node{$c=8$ sector};
\draw (3,-0.75) node{with $U(1)_2$};
\draw[dashed] (0 ,-0.02) -- (4,-0.02);
\draw (2,-0.02) -- (2,-1.05);
\end{scope}
\endtikzpicture
}
\caption{Universal SCFT ingredients of ${\cal N}_{4d}=1$ scenarios.}
\label{N=1SCFT}
\end{figure}

Spacetime symmetries are generated by BRST invariant $h=1$ SCFT operators, and it turns out that only the current ${\cal J}_1+{\cal J}_2$ associated with the diagonal subgroup $S(U(1)_1 \times U(1)_2)$ is BRST closed. Hence, only $S(U(1)_1 \times U(1)_2)$ can take the role of the $U(1)_R$ symmetry of the spectrum. Accordingly, we have to define the charged internal character with respect to the diagonal current ${\cal J}_1+{\cal J}_2$ to see the $U(1)_R$ at the level of the partition function\footnote{We cannot give a local representation of ${\cal J}_2$ in terms of the $g^{\pm}$ fields from the $c=8$ sector, but we can make its existence plausible through an analogy: The currents of the $SO(d)$ Lorentz symmetry schematically read $\psi^\mu \psi^\nu + X^{[\mu} \partial X^{\nu]}$. Even though $X^{\mu}$ itself is not a conformal field involved in the construction of the spectrum, the product $X^{[\mu} \partial X^{\nu]}$ is inevitable to form a BRST invariant completion of the $h=1$ primary $\psi^\mu \psi^\nu$. The addition of $X^{[\mu} \partial X^{\nu]}$ for the sake of BRST closure is the spacetime SCFT analogue of the ${\cal J}_2$ current.}.

We denote the fugacity for charge under the $S(U(1)_1 \times U(1)_2) \cong U(1)_R$ subgroup by $s$.  On the level of the charged characters, this leads to a different fugacity dependence compared to (3.15)\footnote{The R sector analogue of the NS character (3.15) is not explicitly displayed in \cite{Odake:1989dm} but must be inferred through spectral flow.} of \cite{Odake:1989dm} where the internal charge is defined through the ${\cal J}_1$ eigenvalue rather than the ${\cal J}_1+{\cal J}_2$ eigenvalue\footnote{The author of \cite{Odake:1989dm} denotes by $z$ the fugacity of charge under $U(1)_1$. For us, it makes sense to rescale the units of internal charge by $3/2$ which amounts to the correspondence $s \leftrightarrow z^{3/2}$ (in addition to the aforementioned inclusion of ${\cal J}_2$). Moreover, the character in (3.15) of \cite{Odake:1989dm} is defined as the trace over $q^{L_0 - c/24}$, with $c=9$, instead of $q^{L_0}$.  The reason we consider the latter is because we are dealing with critical string theories, and so the total central charge of all matter and (super) ghost sectors taken together vanishes; this explains the presence of $q^{-9/24}$ factor in (3.15) of \cite{Odake:1989dm} but not in  \eref{N1NS}.}. For instance, the supercurrent components are products of operators from both sectors, so $G_{\textrm{int}}^{\pm}$ are charged under both $U(1)_1$ and $U(1)_2$ but neutral under the diagonal subgroup $S(U(1)_1 \times U(1)_2)$. The $\chi^{SO(3)}_{\te{NS,R}}$ factors in the following character formulae are due to the oscillator modes of the stress energy tensor, the internal current and the supercurrents. $U(1)_R$ neutrality of the latter forbids an $s$ dependence at this point and sets the second argument of the $\chi^{SO(3)}_{\te{NS,R}}$ characters to unity.
\subsubsection*{The NS-sector}  
The internal character in this sector is given by
\bea
\chi^{{\cal N}_{2d}=2,c=9}_{\te{NS},h=0,\ell=0} (q; s) &= (1-q)  \chi^{SO(3)}_{\NS}(q,1) \sum_{p\in \ZZ}  \frac{ q^{p^2 + p - \frac{1}{2}} \, s^{2p}  }{(1 \, + \,  q^{p-\frac{1}{2}}) \, (1 \, + \, q^{p+\frac{1}{2}})} \nn\\
&= (q;q)^{-3}_\infty (1-q) \vartheta_3 (1, q) \sum_{p\in \ZZ}  \frac{ q^{p^2 + p - \frac{1}{2}} \, s^{2p}  }{(1 \, + \,  q^{p-\frac{1}{2}}) \, (1 \, + \, q^{p+\frac{1}{2}})}  \label{N1NS} \\
%&= (q;q)^{-3}_\infty (1-q) \vartheta_3 (0, \tau) \sum_{p=0}^{\infty} s_{2p} \; \frac{ q^{p^2+p-\frac{1}{2}} }{(1+q^{p+\frac{1}{2}}) (1+q^{p-\frac{1}{2}})} \nn \\
&=  1+q+(2+s_2)q^{3/2}+( 3+s_2)q^2+(4+s_2)q^{5/2} +(6+2s_2)q^3 \nn \\
&\quad +(10+4s_2)q^{7/2}+(15+6s_2)q^4 +(20+8s_{2})q^{9/2}+(28+12s_2)q^5  \nn \\
&\quad +(42+19s_2+s_4)q^{11/2} +(59+27s_2+2s_4)q^6 \nn \\
&\quad +(78+36s_2+2s_4)q^{13/2}   +(107+51s_2+3s_4)q^7+O(q^{15/2})~, \nn
\eea
where we have introduced the notation
\bea
s_n = \left\{ \begin{array}{cl} s^n + s^{-n} &: \ n > 0 \\ 1 &: \ n=0 \end{array} \right.
\label{comp}
\eea
to compactly represent the fugacity dependence. 

The {\it unrefined} internal character (\ie~ setting $s$ to unity) can be rewritten in terms of modular functions as follows:
\bea
\chi^{{\cal N}_{2d}=2,c=9}_{\te{NS},h=0,\ell=0} (q; s=1) = q^{1/8} \frac{\vartheta_3 (1, q)}{\eta(q)^3} \left[ \vartheta_3(1,q^2) - q^{1/4} \vartheta_2(1,q^2) \right]~. \label{NSunrc9}
\eea

\subsubsection*{The R-sector}  
The internal character in this sector is given by
\bea
\chi^{{\cal N}_{2d} =2,c=9}_{\te{R},h=3/8,\ell=3/2} (q;s) 
&=  (1-q)  \chi^{SO(3)}_{\Ra}(q,1)  \sum_{p\in \ZZ}   \frac{ q^{p^2-1} \, s^{2p-1} }{(1 \, +  \, q^{p}) \, (1 \, + \,q^{p-1})} \nn \\
&=   (q;q)^{-3}_\infty (1-q) \vartheta_2 (1, q) \sum_{p \in \ZZ}   \frac{ q^{p^2-\frac{9}{8}} \, s^{2p-1} }{(1 \, +  \, q^{p}) \, (1 \, + \,q^{p-1})} \label{N1R} \\
%&= (q;q)^{-3}_\infty (1-q) q^{-\frac{3}{8}} \vartheta_2 (0, \tau) \sum_{p=0}^{\infty} s_{2p+1} \; \frac{ q^{(p+\frac{1}{2})^2} }{(1+q^{p+1}) (1+q^{-p})}   \nn \\
&= s_1+2s_1q+6s_1q^2  +(2s_3+14s_1)q^3+(4s_3+30s_1)q^4  \nn \\
&\quad   +(10s_3+62s_1)q^5+(24s_3+122s_1)q^6+(50s_3+230s_1)q^7+O(q^8)~.   \nn
\eea
The unrefined internal character can be rewritten in terms of modular functions as
\bea
\chi^{{\cal N}_{2d} =2,c=9}_{\te{R},h=3/8,\ell=3/2} (q;s=1) = q^{-1/4} \frac{\vartheta_2(1, q)}{\eta(q)^3} \left[ \vartheta_2(1,q^2) - q^{1/4} \vartheta_3(1, q^2) \right]~. \label{Runrc9}
\eea

\subsubsection*{Some features} 
Let us discuss some properties of the above internal characters.
\begin{itemize}
\item The units of $U(1)_R$ charge are normalized such that all integer powers of $s$ occur. According to the infinite sums within \eref{N1NS} and \eref{N1R}, even powers $s_{2p}$ firstly occur along with $q^{p^2+p-1/2}$, i.e. in the NS sector at mass level $p^2+p-1$. Odd powers $s_{2p-1}$ of the $U(1)_R$ fugacity, on the other hand, firstly show up at power $q^{p^2  - 5/8}$, i.e. in the R sector at mass level $p^2-1$
%\item The smallest units $\pm \frac{3}{2}$ of $U(1)$ charge correspond to powers $s_1 = s^1 + s^{-1}$. According to the infinite sums within \eref{N1NS} and \eref{N1R}, even powers $s_{2m}$ representing $U(1)$ charges $\pm 3m$ firstly occur along with $q^{m^2+m-1/2}$, i.e. in the NS sector at mass level $m^2+m-1$. Odd powers $s_{2m-1}$ of the $U(1)$ fugacity, on the other hand, firstly show up at power $q^{m^2  - 5/8}$, i.e. in the R sector at mass level $m^2-1$
\footnote{The onset of the $s_{m}$ at $q$ power $q^{\frac{1}{4}m^2+\frac{1}{2}m+const}$ might seem counterintuitive in view of the bosonized operators $e^{\pm \frac{i}{2}\sqrt{3}m H}$ (with $H$ a free boson) which contribute $s_m q^{\frac{3}{8} m^2}$ to the character. The mismatch between the $q$ exponents $\frac{1}{4}m^2+\frac{1}{2}m$ and $\frac{3}{8} m^2$ is caused by the fact that generic contributions to $s_{m}$ at fixed $U(1)_R \cong S(U(1)_1 \times U(1)_2)$ charge stem from composite fields with both $U(1)_1$ and $U(1)_2$ charges. The operator of lowest conformal weight along with some $s_{m \geq 3}$ is charged under both $U(1)_1$ and $U(1)_2$. Since the internal fugacities in \cite{Odake:1989dm} only count $U(1)_1$ charges and are insensitive to $U(1)_2$, the leading $q$ power associated with some $s_{m}$ in the characters of the reference can be directly traced back to the aforementioned operators $e^{\pm \frac{i}{2}\sqrt{3}m H}$.}.
\item The unrefined internal R character \eref{Runrc9} can be derived from the NS counterpart \eref{NSunrc9} by exchanging $\vartheta_2$ and $\vartheta_3$ and multiplying by an overall factor $q^{-3/8}$.
\item In contrast to their cousins in \cite{Odake:1989dm}, the charged characters (\ref{N1NS}) and (\ref{N1R}) of the NS- and R sector are not related by spectral flow because the internal fugacity $s$ is defined through the $U(1)_R$ symmetry current ${\cal J}_1+ {\cal J}_2$ and not through the bosonizable $U(1)_1$ current ${\cal J}_1$.
\item Both of the unrefined internal characters \eref{NSunrc9} and \eref{Runrc9} are {\it not} modular invariant.  This can be seen from the modular transformation $q \mapsto \tilde q = e^{-2\pi i/\tau}$,
\bea
\vartheta_2 \left( 1, \tilde q \right) = \vartheta_4(1,q) \sqrt{-i \tau}~, \qquad
\eta\left( \tilde q \right) = \eta(q) \sqrt{-i \tau} ~, \qquad
\vartheta_3(1, \tilde q) =\vartheta_3(1, q) \sqrt{-i \tau }~.
\eea
\end{itemize}
%%%%%%%%%%%%%
\subsection[${\cal N}_{2d}=4$ worldsheet superconformal algebra at $c=6$]{$\bm{{\cal N}_{2d}=4}$ worldsheet superconformal algebra at $\bm{c=6}$}
\label{sec:42}

The existence of eight supercharges in four or six dimensional spacetime implies that the universal part of the internal SCFT contains a sector with central charge $c=6$, enhanced ${\cal N}_{2d}=4$ worldsheet SUSY and $SU(2)$ Kac Moody symmetry at level 1. The $c=6$ representations contributing to the NS sector and R sector of ${\cal N}_{4d}=2$ and ${\cal N}_{6d}=(1,0)$ spectra are characterized by values $(h,\ell) = \left( 0,0 \right)$ and $(h,\ell) = \left( \frac{1}{4},\frac{1}{2} \right)$, respectively, of the conformal weight $h$ and the spin $\ell$ with respect to the $SU(2)$ Kac Moody symmetry.

The $c=6$ SCFT is governed by ${\cal N}_{2d}=4$ worldsheet SUSY and $SU(2)$ Kac Moody symmetry at level $k=1$. In the notation of \cite{Feng:2012bb}, the supercurrent components are built from two spin fields $\lambda^{1,2}$ of conformal weight $\frac{1}{4}$ which form a doublet under the $SU(2)_1$ Kac Moody currents ${\cal J}^{A=1,2,3}$ and additional weight $\frac{5}{4}$ fields $g_{1,2}$ which decouple from the ${\cal J}^A$. The $g_{1,2}$ form a doublet under another $SU(2)_2$ which is embedded into the SCFT sector decoupling from ${\cal J}^A$. Figure \ref{N=2SCFT} summarizes the mutually decoupling SCFT sectors involved in ${\cal N}_{6d}=(1,0)$ compactifications:

\begin{figure}[h]
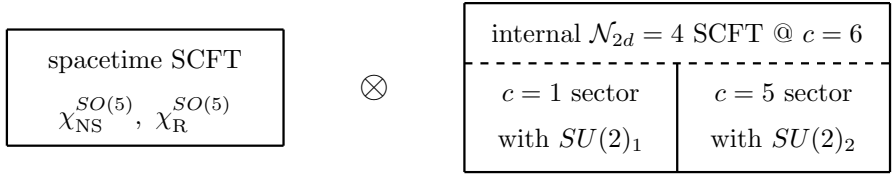

\centerline{
\tikzpicture [scale=1.4,line width=0.30mm]
\draw (0.7,-0.8) -- (0.7,0.3);
\draw (3.3,-0.8) -- (3.3,0.3);
\draw (0.7,-0.8) -- (3.3,-0.8);
\draw (0.7,0.3) -- (3.3,0.3);
\draw (2,0) node{spacetime SCFT};
\draw (2,-0.5) node{$\chi_{\te{NS}}^{SO(5)}, \ \chi_{\te{R}}^{SO(5)}$};
\draw (4.15,-0.25) node{\Large $\otimes$};
\begin{scope}[xshift=5cm]
\draw (0,-1.05) -- (0,0.55);
\draw (4,-1.05) -- (4,0.55);
\draw (0,-1.05) -- (4,-1.05);
\draw (0,0.55) -- (4,0.55);
\draw (2,0.25) node{internal ${\cal N}_{2d}=4$ SCFT @ $c=6$};
\draw (1,-0.3) node{$c=1$ sector};
\draw (1,-0.75) node{with $SU(2)_1$};
\draw (3,-0.3) node{$c=5$ sector};
\draw (3,-0.75) node{with $SU(2)_2$};
\draw[dashed] (0 ,-0.02) -- (4,-0.02);
\draw (2,-0.02) -- (2,-1.05);
\end{scope}
\endtikzpicture
}
\caption{Universal SCFT ingredients of ${\cal N}_{6d}=(1,0)$ scenarios.}
\label{N=2SCFT}
\end{figure}

We shall use charged characters in the following where the fugacity $r$ is defined with respect to the diagonal subgroup within the two decoupling $SU(2)$'s acting on the $\lambda^{1,2}$ and $g_{1,2}$ doublets. In other words, the insertion into the character trace is the BRST invariant sum of the two $SU(2)_{1,2}$ Cartan generators associated with the $SU(2)_R$ symmetry of the spectrum. This makes sure that the diagonal component $\lambda^{1}g_1 + \lambda^{2}g_2$ of the supercurrent is a singlet of the diagonal $SU(2)$, as required by the BRST invariance. The character formulae (21) and (22) in \cite{Eguchi:1988af}\footnote{Note that the sign in the second pair of brackets in the numerator of Eq. (24) of \cite{Eguchi:1988af} should be $+$.} are therefore slightly modified in their $r$ dependence.

\subsubsection*{The NS-sector}  
The internal character in this sector is given by
\bea
\chi^{{\cal N}_{2d}=4,c=6}_{\te{NS},h=0,\ell=0} (q; r) &=  \chi^{SO(3)}_{\NS}(q,1)   \sum_{m\in \ZZ} q^{\frac{1}{2}m^2+\frac{1}{4}} \, r^{2m} \; \frac{ q^{m-\frac{1}{2}} \, - \,r^{-2} }{1 \, +  \, q^{m-\frac{1}{2}}} \nn \\
%&=  \sum_{k=0}^\infty [2k]_x q^{k^2/2+k-1/2} \left( \frac{1-q-q^{k+1/2}+q^{k+3/2}}{(1+q^{k-1/2})(1+q^{k+3/2})} \right)  \nn \\
%&=  [0]_x+[2]_xq+([2]_x+[0]_x)q^{3/2}+([2]_x+2[0]_x)q^2+(2[2]_x+2[0]_x)q^{5/2} \nn \\
&=  (q;q)^{-3}_{\infty} \vartheta_3(1, q)  \sum_{k=0}^\infty [2k]_r   \frac{(1-q)(1-q^{k+\frac{1}{2}})}{(1+q^{k-\frac{1}{2}})(1+q^{k+\frac{3}{2}})} q^{\frac{1}{2}k^2+k-\frac{1}{2}}   \label{intNSN4} \\
&=  [0]_r+[2]_rq+([2]_r+[0]_r)q^{3/2}+([2]_r+2[0]_r)q^2+(2[2]_r+2[0]_r)q^{5/2} \nn \\
&\quad  +(4[2]_r+2[0]_r)q^3+([4]_r+5[2]_r+4[0]_r)q^{7/2}+(2[4]_r+6[2]_r+7[0]_r)q^4  \nn \\
&\quad +(2[4]_r+10[2]_r+8[0]_r)q^{9/2}+(3[4]_r+16[2]_r+9[0]_r)q^5 \nn \\
&\quad +(6[4]_r+21[2]_r+15[0]_r)q^{11/2}+(9[4]_r+27[2]_r+23[0]_r)q^6 \nn \\
&\quad +(12[4]_r+39[2]_r+27[0]_r)q^{13/2}+([6]_r+17[4]_r+56[2]_r+33[0]_r)q^7 \nn \\
&\quad +O(q^{15/2})~ . \nn
\eea
The unrefined internal character for the NS-sector can be written as
\bea
\chi^{{\cal N}_{2d}=4,c=6}_{\te{NS},h=0,\ell=0} (q; r=1) &= q^{1/8} \frac{\vartheta_3(1, q)^2}{\eta(q)^3}   \left[ 1- 2 i q^{1/8} \mu \left(\frac{1+\tau}{2} ,\tau \right) \right] ~,
\eea
where $\mu(u,\tau)$ is an Appell-Lerch sum defined in \eref{ALdef}; for our purpose, we have\footnote{This function is also closely related to the function $h_3(q)$ introduced in \cite{Eguchi:1988af, Eguchi:2008gc, Eguchi:2009cq}.}
\bea
\mu \left(\frac{1+\tau}{2},\tau \right) &=- \frac{i}{\vartheta_3(1,q)} \sum_{m \in \BZ} \frac{q^{\frac{1}{2}m^2-\frac{1}{8}}}{1+q^{m - \frac{1}{2}}}~, \label{ALhalf1}
\eea
where we have used the fact that $\vartheta_1\left(e^{2\pi i (1+\tau)/2}, q \right) =  q^{-1/8} \vartheta_3(1, q)$.

\subsubsection*{The R-sector}  
The internal character in this sector is given by
\bea
\chi^{{\cal N}_{2d}=4,c=6}_{\te{R},h=\frac{1}{4},\ell=\frac{1}{2}} (q; r) &= \chi^{SO(3)}_{\Ra}(q,1)  \sum_{m\in \ZZ}  \, r^{2m+1} \; \frac{  q^{m} \, - \, r^{-2} }{1 \, +  \, q^{m}} q^{\frac{1}{2}m^2 + \frac{1}{2}m } \nn \\
%&=  \sum_{k=0}^\infty [2k+1]_x q^{k^2/2+3k/2+1/4} \left( \frac{1-q-q^{k+1}+q^{k+2}}{(1+q^k)(1+q^{k+2})} \right) \nn \\
&= q^{-\frac{1}{8}} (q;q)^{-3}_\infty \vartheta_2 (1, q) \sum_{k=0}^\infty [2k+1]_r   \frac{(1-q)(1-q^{k+1})}{(1+q^k)(1+q^{k+2})}  q^{\frac{1}{2}k^2+\frac{3}{2}k} \label{intRN4} \\
&=  [1]_r+2[1]_rq+(2[3]_r+4[1]_r)q^2+(4[3]_r+10[1]_r)q^3\nn \\
&\quad +(10[3]_r+20[1]_r)q^4+(2[5]_r+22[3]_r+38[1]_r)q^5 \nn \\
&\quad +(6[5]_r+44[3]_r+72[1]_r)q^6+(14[5]_r+86[3]_r+130[1]_r)q^7+O(q^8)~ . \nn
\eea
The unrefined internal character for the R-sector can be written as
\bea
\chi^{{\cal N}_{2d}=4,c=6}_{\te{R},h=\frac{1}{4},\ell=\frac{1}{2}} (q; r=1)  
&= \frac{\vartheta_2(1, q)}{\eta(q)^3}  \sum_{m \in \BZ} \left( \frac{q^m -1}{1+q^m} \right) q^{\frac{1}{2}m(m+1)} \nn \\
&= \frac{\vartheta_2(1, q)}{\eta(q)^3}   \sum_{m \in \BZ} \left[ \left(1- \frac{2}{1+q^m} \right) q^{\frac{1}{2}m(m+1)} \right] \nn \\
&= q^{-1/8} \frac{\vartheta_2(1, q)^2}{\eta(q)^3}\left[  1- 2 i q^{1/8} \mu \left(1/2,\tau \right) \right]~,
\eea
where we have\footnote{This function is also closely related to the function $h_2(q)$ introduced in \cite{Eguchi:1988af, Eguchi:2008gc, Eguchi:2009cq}.}
\bea
\mu \left(1/2,\tau \right) &=- \frac{i}{\vartheta_2(1,q)} \sum_{m \in \BZ} \frac{q^{\frac{1}{2}m(m+1)}}{1+q^m}~, \label{ALhalf}
\eea
where we have used the fact that $\vartheta_1(-1,q) = \vartheta_2(1, q)$.

\paragraph{Some features} 
\begin{itemize}
\item According to appendix \ref{sec:not}, characters $[n]_r$ of $SU(2)_R$ follow the same highest weight notation as for $SO(3)$, i.e. we have $[1]_r = r+r^{-1}$ for the fundamental representation and $[n]_r = \sum_{k=-n/2}^{+n/2} r^{2k}$ in the general spin $n/2$ case. Again, the infinite sum representations allow to read off the lowest level where individual $SU(2)_R$ representations contribute: Integer spin representations $[2k]_r$ firstly occur at power $q^{k^2/2+k-1/2}$, i.e. at mass level\footnote{The floor function $\lfloor \cdot \rfloor$ picks out the nearest integer smaller than or equal to its argument.} $\lfloor k^2/2+k-1/2 \rfloor$. Spinorial representations $[2k+1]_r$, on the other hand, firstly show up at $q^{k^2/2+3k/2-1/4}$, i.e. at mass level $k(k+3)/2$.\footnote{The lowest $q$ power along with some $SU(2)_R$ representation $[n]_r$ is generically caused by an operator charged under both $SU(2)_1$ and $SU(2)_2$. That is why one cannot identify these leading $q$ exponents with the conformal dimension of a simple CFT operator such as an exponential $e^{\pm i q H}$, see the footnote at the end of subsection \ref{sec:41}.
}

\item Observe that the unrefined internal characters in both NS and R sectors involve Appell-Lerch sums, which are mock modular forms.  Since the characters and are holomorphic in $q$, it is immediate that they are {\it not} modular invariant. Also, as before in the ${\cal N}_{2d}=2$ SCFT, the relation between NS and R characters through spectral flow is absent due to the adaption of the internal fugacity to the $SU(2)_R$ symmetry.
\end{itemize}

\section[Spectrum in ${\cal N}_{4d}=1$ supersymmetric compactifications]{Spectrum in $\bm{{\cal N}_{4d}=1}$ supersymmetric compactifications}
\label{sec:4d}

This section opens up the main body of this work where the SCFT ingredients introduced so far are applied to counting universal super Poincar\'e multiplets in the perturbative string spectrum\footnote{The methods within this work are adapted to the representatives of physical states in the canonical superghost pictures: After stripping off the superghost contributions $e^{q\phi}$ from the $h=1$ vertex operators (with $q=-1$ and $h[e^{-\phi}] =\frac{1}{2}$ in the NS sector as well as $q=-\frac{1}{2}$ and $h[e^{-\phi/2}] =\frac{3}{8}$ in the R sector), this amounts to counting operators in the matter part of the SCFTs with weight $h=\frac{1}{2}$ in the NS sector and $h=\frac{3}{8}$ in the R sector.}. We start with the phenomenologocially relevant and mathematically most tractable ${\cal N}_{4d}=1$ supersymmetric scenario. Its SCFT description requires the internal sector with enhanced ${\cal N}_{2d} =2$ worldsheet SUSY introduced in subsection \ref{sec:41}, independently on the compactification details. The BRST invariant completion of the internal current takes the role of the $U(1)_R$ symmetry generator. Lorentz quantum numbers enter through the partition functions (\ref{exactchNSSO3}) and (\ref{exactchRSO3}) of the spacetime SCFT for the $\partial X^\mu$ and $\psi^\mu$ oscillators, expressed in terms of characters of the massive little group $SO(3)$ in four dimension.

The universal part of the ${\cal N}_{4d}=1$ spectrum is built from both spacetime oscillators and internal operators.  On the level of its partition function $\chi^{{\cal N}_{4d}=1} (q; y, s)$, this amounts to forming a GSO projected product of NS- and R characters from the spacetime- and internal SCFT, see (\ref{N1NS}) and (\ref{N1R}) for the latter. In a power series expansion in $q$, the coefficient of the $n$'th power $q^n$ comprises characters for the ${\cal N}_{4d}=1$ super Poincar\'e multiplets occurring at the $n$'th mass level with $m^2 = n/\ap$. The aforementioned massive supercharacters are functions of $SO(3)$ fugacity $y$ and $U(1)_R$ fugacity $s$.

The fundamental ${\cal N}_{4d}=1$ multiplet\footnote{As we shall see below, the fundamental multiplet does not appear on its own in both massless and massive spectra.  Representations appearing in the massive spectrum arise from certain non-trivial products with the fundamental multiplet.} consists of 2 real bosonic degrees of freedom and a Majorana fermion with 2 real fermonic on-shell degrees of freedom after taking the Dirac equation into account, see e.g. \cite{Boels:2011zz}.  The two real bosonic degrees of freedom can be complexified to yield a complex scalar and its complex conjugate; they transform as a singlet under the little group $SO(3)$ and each of them carries opposite $R$-charges $+1$ and $-1$.  On the other hand, the two real fermonic degrees of freedom transform as a doublet under the little group $SO(3)$ and each of them carries zero $R$-charge.  Thus, these 2+2 states yield the character
\beq
Z({\cal N}_{4d}=1) \eq [1]_y+ (s+s^{-1}) ~.
\label{fundn4d1}
\eeq
Any other massive representation of ${\cal N}_{4d}=1$ super Poincar\'e is specified by the little group $SO(3)$ quantum number $n$ and the $U(1)_R$ charge $Q$ of its highest weight state or Clifford vacuum. Its $SO(3) \times U(1)_R$ constituents follow from a tensor product:
\bea \label{N1rep}
\llbracket n,Q\rrbracket   \ \ := \ \ &Z({\cal N}_{4d}=1) \cdot s^Q [n]_y \eq s^Q [n]_y \left( [1]_y+ (s+s^{-1}) \right)  \\
= \ \ & \begin{cases} \ s^Q \, \big( \, [n+1] \ + \ (s + s^{-1}) \,[n] \ + \ [n-1] \, \big) & \qquad \text{for $n \geq 1$} \\
\ s^Q \, \big( \, [1] \ + \ (s + s^{-1})\, [0] \, \big) & \qquad \text{for $n=0$}  \end{cases} 
    \nn
\eea
The super-Poincar\'e character $\llbracket n,Q\rrbracket$ corresponds to $4(n+1)$ states of spin $\frac{n+1}{2}$, $\frac{n}{2}$ and (if $n\neq 0$) $\frac{n-1}{2}$ that can be generated from a Clifford vacuum with spin $n/2$ and $U(1)_R$ charge $Q+1$\footnote{In this terminology, the first label of $\llbracket n,Q\rrbracket$ refers to the average spin of the $SO(3)$ irreducibles. We deviate from the common practice that supermultiplets are referred to through the highest spin therein. The supercharacter $\llbracket 3,0\rrbracket = [4]+[2]+(s+s^{-1})[3]$, for instance, describes $U(1)_R$ neutral bosons of spin two and one, and two massive gravitinos of opposite $U(1)_R$ charges.}. Note that $Q$ is even whenever the maximum spin quantum number $n+1$ is.

In this setting, we find the (GSO projected) ${\cal N}_{4d}=1$ partition function
\bea
\chi^{{\cal N}_{4d}=1} (q; y, s)
 \ \ := \ \  \chi^{{\cal N}_{4d}=1}_{\NS} \mid_{\GSO} (q; y, s) + \chi^{{\cal N}_{4d}=1}_{\Ra} \mid_{\GSO} (q;y,s)~,
 \label{def:GSOedd}
\eea
where GSO projection removes half odd integer mass level $\ap m^2 \in \ZZ - \frac{1}{2}$ from the NS sector and interlocks spacetime chirality with $U(1)_R$ charges in the R sector. We can capture this projection through\footnote{The formula for the GSO projected R sector is reliable for positive powers $q^{\geq 1}$ only and inaccurate at the massless level: The coefficient of $q^0$ in $\chi^{{\cal N}_{4d}=1}_{\Ra} \mid_{\GSO}$ is $\frac{1}{2}(y+y^{-1}) (s+s^{-1})$ instead of the desired value $ys + (ys)^{-1}$. One can just add to the former $\frac{1}{2}(y-y^{-1})(s-s^{-1})$ to compensate this mismatch.  This artifact of the mismatch between massive and massless little groups does not affect the main focus our analysis -- the massive particle content.  Indeed, the character $ys$ corresponds to the left-handed gaugino and the character $(ys)^{-1}$ corresponds to the right-handed gaugino; they carry opposite $R$-charge $+1$ and $-1$ and opposite helicities $+1/2$ and $-1/2$.}:
\bea
 \chi^{{\cal N}_{4d}=1}_{\NS} \mid_{\GSO} (q) \eq & \frac{1}{2} \, q^{-\frac{1}{2}} \, \Bigg[ \chi_{\te{NS}}^{SO(3)} (q; y) \, \chi^{{\cal N}_{2d}=2,c=9}_{\te{NS},h=0,\ell=0} (q; s) \, - \, \chi_{\te{NS}}^{SO(3)} (e^{2\pi i}q; y) \, \chi^{{\cal N}_{2d}=2,c=9}_{\te{NS},h=0,\ell=0} (e^{2\pi i}q; s) \Bigg]~, \nn \\
\chi^{{\cal N}_{4d}=1}_{\Ra} \mid_{\GSO} (q) \eq &  \frac{1}{2} \ \chi_{\te{R}}^{SO(3)} (q; y)  \, \chi^{{\cal N}_{2d}=2,c=9}_{\te{R},h=3/8,\ell=3/2} (q; s)  ~. \label{def:GSOed}
\eea
In order to compactly represent the leading terms in a power series expansion of the partition function $\chi^{{\cal N}_{4d}=1}$, let us introduce the shorthand
\bea
\llbracket n,\pm Q \rrbracket \ \ := \ \ \left\{ \begin{array}{cl} \ \llbracket n,+ Q \rrbracket \ + \  \llbracket n, - Q \rrbracket \ &: \  Q \neq 0  \\
 \ \llbracket n, 0 \rrbracket  &: \ Q = 0  \end{array} \right.
\eea
which exploits that $U(1)_R$ charges always appear on symmetric footing $Q \leftrightarrow -Q$. The pairing of supermultiplets with opposite (nonzero) $U(1)_R$ charges combines Majorana fermions as they appear in the fundamental multiplet (\ref{fundn4d1}) to Dirac fermions.
\bea
\chi^{{\cal N}_{4d}=1} (q; y, s)
&\eq \underbrace{\left(y^2 + y^{-2} \, + \, \frac{1}{2} \, (y+y^{-1}) \, (s+s^{-1}) \right) \, q^0}_{4 \ \te{massless states}} \ +  \ \underbrace{ \big( \, \llbracket 3,0 \rrbracket \ + \ \llbracket 0,\pm 1 \rrbracket \, \big) \, q}_{24 \ \te{states at level} \ 1}  \notag \\
& \quad + \ \underbrace{\big( \,\llbracket 5,0 \rrbracket \ + \ \llbracket 3,0 \rrbracket \ + \ 2 \, \llbracket 2,\pm 1 \rrbracket \ + \ 2 \, \llbracket 1,0 \rrbracket \, \big) \, q^{2}}_{104 \ \te{states at level} \ 2} \notag \\
& \quad+ \ \big( \, \llbracket 7,0 \rrbracket \ + \ \llbracket 5,0 \rrbracket \ + \ 3 \, \llbracket 4,\pm 1 \rrbracket \ + \ 5 \, \llbracket 3,0 \rrbracket \ + \ 2 \, \llbracket 2,\pm 1 \rrbracket \notag \\
& \quad \ \ \ \ \ + \ \llbracket 1,\pm 2 \rrbracket \ + \ 5 \, \llbracket 1,0 \rrbracket \ + \ 3 \, \llbracket 0,\pm 1 \rrbracket\, \big) \, q^3 \ + \ {\cal O}(q^4)
\eea
The content of the first eight ${\cal N}_{4d}=1$ levels is summarized in \tref{tab:repsN1}.  The explicit form of the vertex operators at mass level one\footnote{Let us discuss about the states at the first mass level.   The 24 total states consist of the following multiplets:

 (1) {\bf the massive spin 3/2 multiplet $\llbracket 3,0 \rrbracket$}: it contains a massive spin 2 field with 5 on-shell degrees of freedoms (OSDOFs), a massive spin 1 field with 3 OSDOFs, a massive spin 3/2 field with 4 OSDOFs, and a Dirac fermion with 4 OSDOFs; so we have 8+8 real OSDOFs in total
 
 (2) {\bf the massive spin 0 multiplet $\llbracket 0,\pm1 \rrbracket$}:  the two constituents $\llbracket 0,1 \rrbracket$ and $\llbracket 0,-1 \rrbracket$ of the massive scalar multiplet correspond to two massless chiral fields, $\Phi$ and $\widetilde{\Phi}$ (not complex conjugate to each other) at $Q = \pm 1$. The opposite $Q$-charges are necessary to form an invariant mass term $ \Phi \widetilde{\Phi}$ in the superpotential. This multiplet contains 4 + 4 real OSDOFs coming from two complex scalars plus two Majorana fermions; the latter are equivalent to one massive Dirac fermion.  Note that the spin 0 multiplet is also referred to as two spin 1/2 multiplets in \cite{Feng:2012bb}.} can be found in section 5 of \cite{Feng:2012bb} (equations (5.3) to (5.6) for bosons and equations (5.14) to (5.18) for fermions) in the RNS framework, and references \cite{Berkovits:1997zd,Berkovits:1998ua} provide their superspace description.

\begin{table}[htdp]
\begin{center}
\begin{tabular}{|l|l|}
\hline  $\ap m^2$ &Representations of ${\cal N}_{4d}=1$ super Poincar\'e \\ \hline \hline   
  1 & $\llbracket 3,0 \rrbracket \, + \, \llbracket 0,\pm 1 \rrbracket$  \\\hline
   2 & $\llbracket 5,0 \rrbracket \, + \, \llbracket 3,0 \rrbracket \, + \, 2 \, \llbracket 2,\pm 1 \rrbracket \, + \, 2 \, \llbracket 1,0 \rrbracket$  \\\hline
   3 & $\llbracket 7,0 \rrbracket \, + \, \llbracket 5,0 \rrbracket \, + \,3 \, \llbracket 4,\pm 1 \rrbracket \, + \, 5 \, \llbracket 3,0 \rrbracket \, + \, 2 \, \llbracket 2,\pm 1 \rrbracket \, + \, \llbracket 1,\pm 2 \rrbracket \, + \, 5 \, \llbracket 1,0 \rrbracket \, + \, 3 \, \llbracket 0,\pm 1 \rrbracket$  \\\hline
   4 & $\llbracket 9,0 \rrbracket \, + \, \llbracket 7,0 \rrbracket \, + \, 3 \, \llbracket 6,\pm 1 \rrbracket \, + \, 7 \, \llbracket 5,0 \rrbracket \, + \, 4 \, \llbracket 4,\pm 1 \rrbracket \, + \, 2 \, \llbracket 3,\pm 2 \rrbracket \, + \, 12 \, \llbracket 3,0 \rrbracket$
   \\
   &$ \, + \, 11 \, \llbracket 2,\pm 1 \rrbracket \, + \, 2 \, \llbracket 1,\pm 2 \rrbracket \, + \, 12 \, \llbracket 1,0 \rrbracket \, + \, 3 \, \llbracket 0,\pm 1 \rrbracket$ \\\hline
   5 &$ \llbracket 11,0 \rrbracket \, + \,  \llbracket 9,0 \rrbracket \, + \, 3 \, \llbracket 8,\pm 1 \rrbracket \, + \, 7 \, \llbracket 7,0 \rrbracket \, + \, 5 \, \llbracket 6,\pm 1 \rrbracket \, + \, 2 \, \llbracket 5,\pm 2 \rrbracket \, + \, 17 \, \llbracket 5,0 \rrbracket\, + \, 18 \, \llbracket 4,\pm 1 \rrbracket$
     \\
     &$ \, + \, 6 \, \llbracket 3,\pm 2 \rrbracket \, + \, 31 \, \llbracket 3,0 \rrbracket \, + \, 20 \, \llbracket 2,\pm 1 \rrbracket \, + \, 6 \, \llbracket 1,\pm 2 \rrbracket \, + \, 28 \, \llbracket 1,0 \rrbracket \, + \,  \llbracket 0,\pm 3 \rrbracket \, + \, 15 \, \llbracket 0,\pm 1 \rrbracket$ \\\hline
   6 &$ \llbracket 13,0 \rrbracket \, + \, \llbracket 11,0 \rrbracket \, + \, 3 \, \llbracket 10,\pm 1 \rrbracket \, + \, 7 \, \llbracket 9,0 \rrbracket \, + \, 5 \, \llbracket 8,\pm 1 \rrbracket \, + \, 2 \, \llbracket 7,\pm 2 \rrbracket \, + \, 19 \, \llbracket 7,0 \rrbracket $ \\
   &$\, + \, 21 \, \llbracket 6,\pm 1 \rrbracket \, + \, 8 \, \llbracket 5,\pm 2 \rrbracket \, + \, 45 \, \llbracket 5,0 \rrbracket \, + \, 39 \, \llbracket 4,\pm 1 \rrbracket \, + \, 15 \, \llbracket 3,\pm 2 \rrbracket \, + \, 72 \, \llbracket 3,0 \rrbracket $ \\
   &$\, + \, 3 \, \llbracket 2,\pm 3 \rrbracket \, + \, 58 \, \llbracket 2,\pm 1 \rrbracket \, + \, 17 \, \llbracket 1,\pm 2 \rrbracket \, + \, 64 \, \llbracket 1,0 \rrbracket \, + \, 21 \, \llbracket 0,\pm 1 \rrbracket$
   \\\hline
   7& $\llbracket15,0\rrbracket \, + \,
\llbracket13,0\rrbracket \, + \,
3 \, \llbracket12,1\rrbracket \, + \,
7\, \llbracket11,0\rrbracket \, + \,
5\, \llbracket10,1\rrbracket \, + \,
2\, \llbracket9,2\rrbracket \, + \,
19\, \llbracket9,0\rrbracket \, + \,
22\, \llbracket8,1\rrbracket $ \\
   &$ \, + \,
8 \, \llbracket7,2\rrbracket \, + \,
51\, \llbracket7,0\rrbracket \, + \,
49\, \llbracket6,1\rrbracket \, + \,
22\, \llbracket5,2\rrbracket \, + \,
108 \, \llbracket5,0\rrbracket \, + \,
4\, \llbracket4,3\rrbracket \, + \,
105\, \llbracket4,1\rrbracket $ \\
   &$ \, + \,
43\, \llbracket3,2\rrbracket \, + \,
166 \,\llbracket3,0\rrbracket \, + \,
5\, \llbracket2,3\rrbracket \, + \,
115 \,\llbracket2,1\rrbracket \, + \,
38 \,\llbracket1,2\rrbracket \, + \,

136 \, \llbracket1,0\rrbracket \, + \,
6\, \llbracket0,3\rrbracket$ \\
   &$ \, + \,
66 \, \llbracket0,1\rrbracket $
   \\\hline
8&$
\, \llbracket17,0\rrbracket \, + \,
\, \llbracket15,0\rrbracket \, + \,
3\, \llbracket14,1\rrbracket \, + \,
7\, \llbracket13,0\rrbracket \, + \,
5\,\llbracket12,1\rrbracket \, + \,
2\,\llbracket11,2\rrbracket \, + \,
19\, \llbracket11,0\rrbracket $ \\
   &$ \, + \,
22\,\llbracket10,1\rrbracket \, + \,
8\,\llbracket9,2\rrbracket \, + \,
53\,\llbracket9,0\rrbracket \, + \,
52\,\llbracket8,1\rrbracket \, + \,
24\,\llbracket7,2\rrbracket \, + \,
125\,\llbracket7,0\rrbracket \, + \,
4\,\llbracket6,3\rrbracket $ \\
   &$ \, + \,
135\,\llbracket6,1\rrbracket \, + \,
62\,\llbracket5,2\rrbracket \, + \,
254\,\llbracket5,0\rrbracket \, + \,
10\,\llbracket4,3\rrbracket \, + \,
223\, \llbracket4,1\rrbracket \, + \,
101\,\llbracket3,2\rrbracket \, + \,
357 \, \llbracket3,0\rrbracket $ \\
   &$ \, + \,
21\, \llbracket2,3\rrbracket \, + \,
274 \, \llbracket2,1\rrbracket \, + \,
\llbracket1,4\rrbracket \, + \,
89 \, \llbracket1,2\rrbracket \, + \,
289 \, \llbracket1,0\rrbracket \, + \,
7\, \llbracket0,3\rrbracket \, + \,
112\, \llbracket0,1\rrbracket 
   $   \\\hline
\end{tabular}
\caption{The content of the first eight $\CN_{4d}=1$ levels.}
\label{tab:repsN1}
\end{center}
\end{table}
Character multiplicities up to mass level $\ap m^2 = 25$ are gathered in table \ref{tab4d0} and in the tables of appendix \ref{app4}.

\subsection{The total number of states at a given mass level} \label{sec:numberstates4SUSYs}

In this subsection, we focus on the total number of states present at a given mass level.  These numbers can indeed be obtained by adding up the dimensions of representations presented in table \ref{tab:repsN1}.  Our aim here is to compute such numbers analytically and asymptotically for large mass levels.

The starting point is the unrefined partition function obtained by setting the fugacities $y$ and $s$ in \eref{def:GSOedd} to unity.  The total number of states $N_m$ at the mass level $m$ can be read off from the coefficient of $q^m$ in the power series of $\chi^{{\cal N}_{4d}=1} (q; y=1, s=1)$.

Supersymmetry implies that 
\bea
 \chi^{{\cal N}_{4d}=1}_{\NS} \mid_{\GSO} (q; y=1, s=1)  = \chi^{{\cal N}_{4d}=1}_{\Ra} \mid_{\GSO} (q; y=1, s=1)~.
\eea
which can, of course, be checked directly using \eref{def:GSOed}, \eref{unrefchNSSO3}, \eref{unrefchRSO3}, \eref{NSunrc9} and \eref{Runrc9}. Since the formula for the R sector is simpler, we proceed from there.
\bea
\chi^{{\cal N}_{4d}=1} (q; y=1, s=1) &= 2  \chi^{{\cal N}_{4d}=1}_{\Ra} \mid_{\GSO} (q; y=1, s=1) \nn \\
&= \chi_{\te{R}}^{SO(3)} (q,y=1) \chi^{{\cal N}_{2d} =2,c=9}_{\te{R},h=3/8,\ell=3/2} (q;s=1) \nn \\
&= q^{-1/4} \frac{\vartheta_2(1, q)^2}{\eta(q)^6} \left[ \vartheta_2(1,q^2) - q^{1/4} \vartheta_3(1,q^2) \right]~. \label{exactunref4d}
\eea
Indeed, the power series of $\chi^{{\cal N}_{4d}=1} (q; y=1, s=1)$ in $q$ reproduces the numbers presented in the first column of \tref{numberofstates}.  We mention in passing that $\chi^{{\cal N}_{4d}=1} (q; y=1, s=1)$ is {\it not} a modular form.

\subsubsection*{The number of states at each mass level and its asymptotics} 
The number of states at the mass level $m$ can be computed from
\bea
N_m = \frac{1}{2 \pi i} \oint_{\CC} \frac{\ud q}{q^{m+1}}~ \chi^{{\cal N}_{4d}=1}(q; y=1, s=1 )~, \label{numstates4d}
\eea
where $\CC$ is a contour around the origin. 

%%%%%%%%%%%%%%%

Let us compute the number of states $N_m$ in the limit $m \rightarrow \infty$.  
Since the integrand of \eref{numstates4d} is sharply peaked near $q=1$, we need to examine the behaviour of $\chi^{{\cal N}_{4d}=1}(q;y=1,s=1)$ as $q \rightarrow 1^-$. The $q \rightarrow 1^-$ regime in question is related to the easily accessible $q \rightarrow 0$ limit
\bea
\eta(q) \sim q^{1/24} \co \vartheta_3(1,q) \sim 1 \co \vartheta_4(1,q) \sim 1~, \qquad \quad q \rightarrow 0
\label{obvious}
\eea
through modular transformation $q = e^{2\pi i \tau} \mapsto \widetilde{q} = e^{-2\pi i/\tau} $:
\bea
& \text{{\bf $\vartheta_2$-function}}: \qquad   \vartheta_4 \left( 1, \tilde q \right) = \vartheta_2(1, q ) \sqrt{-i \tau} \sim \frac{1}{\sqrt{ 2\pi}} (1-q)^{1/2} \vartheta_2(1,q) \nn \\
&\Rightarrow  \hspace{2.2cm}  \vartheta_2(1, q) \sim \sqrt{2 \pi} (1-q)^{-1/2},\qquad q \rightarrow 1^-  ~,\label{theta2approx} \\
& \text{{\bf $\vartheta_3$-function}}: \qquad   \vartheta_3 \left( 1, \tilde q \right) = \vartheta_3(1, q ) \sqrt{-i \tau} \sim \frac{1}{\sqrt{ 2\pi}} (1-q)^{1/2} \vartheta_3(1,q) \nn \\
&\Rightarrow  \hspace{2.2cm}  \vartheta_3(1, q) \sim \sqrt{2 \pi} (1-q)^{-1/2},\qquad q \rightarrow 1^-  ~,\label{theta3approx} \\
&\text{{\bf $\eta$-function}}: \qquad  \eta\left(\tilde q \right) = \eta(q) \sqrt{-i \tau } \sim \frac{1}{\sqrt{2 \pi}} (1-q)^{1/2} \eta(q)  \nn \\
&\Rightarrow \hspace{2.2cm} \eta(q) \sim \sqrt{2 \pi} (1-q)^{-1/2} \exp \left( {\frac{\pi^2}{6 \log q}} \right), \qquad q \rightarrow 1^-~,  \label{etaapprox}
\eea

%%%%%%%%%%%%%%%

Hence, we have
\bea
 \vartheta_2(1, q^{2}) \sim  \vartheta_3(1, q^{2}) \sim \sqrt{2 \pi} (1-q^2)^{-1/2}  ~, \quad q \rightarrow 1^-~,
\eea
and so as $q \rightarrow 1^{-}$,
\bea
\chi^{{\cal N}_{4d}=1} (q; y=1, s=1) \sim (2 \pi)^{-3/2} (1-q)^2 (1-q^{1/4}) (1-q^2)^{-1/2} \exp \left( -\frac{\pi^2}{\log q} \right)~.
\eea
Hence, as $m \rightarrow \infty$,
\bea
N_m \sim (2 \pi)^{-3/2} \frac{1}{2 \pi i} \oint_{\CC} \frac{\ud q}{q} ~(1-q)^2 (1-q^{1/4}) (1-q^2)^{-1/2}  \exp \left( -\frac{\pi^2}{\log q}  -m \log q\right)~.
\eea

%%%%%

Observe that the argument of the exponential function has a critical value at $q_0= \exp( -\pi/ \sqrt{m})$; this is the saddle point. The direction of steepest descent at this point is the imaginary direction in $q$.  We deform the contour $\CC$ such that it passes through $q=q_0$ and tangent to this direction. The leading contribution comes from expansions around $q=q_0$ in the steepest descent direction.  Writing $q = q_0 e^{i \theta}$, we have 
\bea
N_m 
&\sim (2 \pi)^{-3/2}  (1-q_0)^2 (1-q_0^{1/4}) (1-q_0^2)^{-1/2}  \times \nn \\
& \hspace{3cm} \frac{1}{2 \pi} \int_{-\epsilon}^{\epsilon} \ud \theta \exp \left(-\frac{ \pi^2}{i \theta+ \log q_0 } - m (i \theta + \log q_0)  \right), \quad \epsilon >0 \nn \\
&\sim (2 \pi)^{-3/2}  (1-q_0)^2 (1-q_0^{1/4}) (1-q_0^2)^{-1/2}  \times \nn \\
& \hspace{3cm} e^{2 \pi \sqrt{m}} \frac{1}{2 \pi} \int_{-\epsilon}^{\epsilon} \ud \theta \exp \left( -\frac{m^{3/2} }{\pi } \theta^2 + O(\theta^3) \right), \qquad \epsilon >0 \nn \\
&\sim (2 \pi)^{-3/2} (1-q_0)^2 (1-q_0^{1/4}) (1-q_0^2)^{-1/2} e^{2 \pi \sqrt{m}} \frac{1}{2 \pi} \int_{-\infty}^\infty \ud \theta \exp \left( -\frac{m^{3/2} }{\pi } \theta^2\right) \nn \\
&\sim \frac{\pi}{32} m^{-2} \exp\left( 2 \pi \sqrt{m} \right)~, \qquad \quad m\rightarrow \infty~. \label{asympNm4d}
\eea

\begin{figure}[htbp]
\begin{center}
\includegraphics[scale=1]{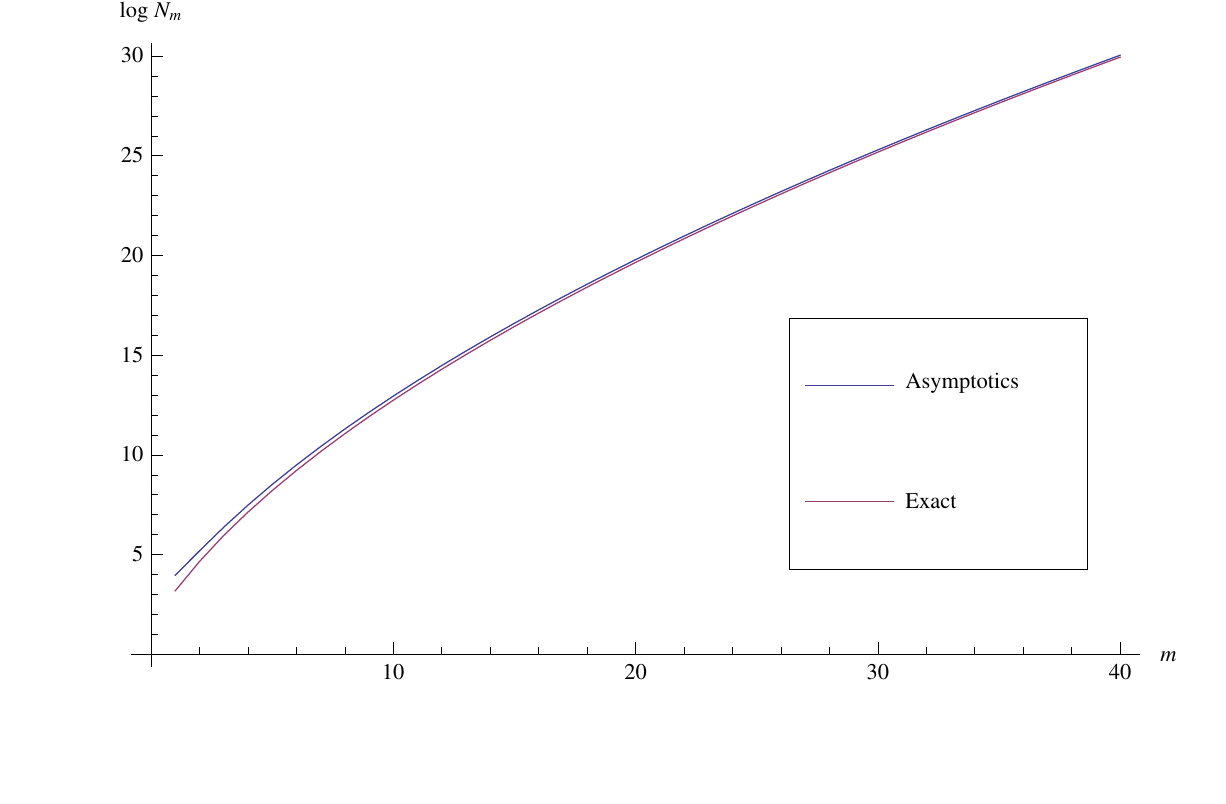}
\vspace{-1.2cm}
\caption{The plot of the exact and asymptotic values of  $\log N_m$ against the mass level $m$ for the case of 4 supercharges.}
\label{fig:compareexactasymp4d}
\end{center}
\end{figure}

The plot of the exact and asymptotic values for $N_m$ against $m$ is depicted in \fref{fig:compareexactasymp4d}.

\subsection{The GSO projected NS- and R sectors} 
In what follows, we compute analytic expressions of the refined partition function $\chi^{{\cal N}_{4d}=1}  (q; y, s)$ and discuss its asymptotic behaviour.

\subsubsection*{The NS sector} Let us write the partition function $\chi^{{\cal N}_{4d}=1}_{\NS} \mid_{\GSO} (q; y, s)$, defined in \eref{def:GSOed}, as 
\bea \label{GSONS4dN1}
\chi^{{\cal N}_{4d}=1}_{\NS} \mid_{\GSO} (q; y, s) = \sum_{k=0}^\infty \sum_{p=- \infty}^\infty [2k]_y  s^{2p}~F^{\NS}_{k,p} (q)~,
\eea
where the function $F^{\NS}_{k,p} (q)$ follows from \eref{exactchNSSO3}, \eref{N1NS} and \eref{def:GSOed}:
\bea
F^{\NS}_{k,p} (q) &=  (q;q)^{-6}_\infty (1-q) q^{p^2+p-1} \sum_{n=1}^\infty (-1)^{n+1} (1-q^n) q^{n \choose 2} \sum_{m=0}^\infty (q^{n|k-m|} -q^{n(k+m+1)})  \nn \\
& \qquad \times \frac{1}{2} q^{\frac{1}{2}m^2} \Bigg[ \frac{(1-q^{m+\frac{1}{2}}) \vartheta_3(1, q) }{ (1+q^{p-\frac{1}{2}})(1+q^{p+\frac{1}{2}}) }  + (-1)^{m^2} \frac{(1+q^{m+\frac{1}{2}}) \vartheta_4(1, q) }{ (1-q^{p-\frac{1}{2}})(1-q^{p+\frac{1}{2}}) }\Bigg]~.
\eea

This expression can be simplified further in the asymptotic limit $k \rightarrow \infty$.  In this limit, $q^{n|k-m|} \sim q^{n(k-m)}$ and the dominant contribution in the summation over $n$ comes from $n=1$.  The summation over $n$ can be asymptotically evaluated as follows (assume that $m$ is finite):
\bea
\sum_{n=1}^\infty (-1)^{n+1} (1-q^n) q^{n \choose 2} (q^{n|k-m|} -q^{n(k+m+1)}) &\sim \sum_{n=1}^\infty (-1)^{n+1} (1-q^n) q^{n(k-m)} (1-q^{n(2m+1)})\nn\\
&\sim \frac{ q^{k} (1-q) \left(1-q^{2 k}\right)}{\left(1+q^k\right)^4} \left \{  q^{-m} \left(1-q^{2 m+1}\right) \right \}~. \label{sumn4dN1}
\eea
The summation over $m$ can be evaluated by considering
\bea
 \sum_{m=0}^\infty q^{\frac{1}{2} m^2- m} \left(1-q^{m+\frac{1}{2}}\right) \left(1-q^{2 m+1}\right)
&= q^{-\frac{1}{2}} (1-q) \vartheta_3 (1, q)~,  \\
 \sum_{m=0}^\infty (-1)^{m^2} q^{\frac{1}{2} m^2- m} \left(1+q^{m+\frac{1}{2}}\right) \left(1-q^{2 m+1}\right)
&= -q^{-\frac{1}{2}} (1-q) \vartheta_4 (1, q)~. \label{summ4dN1}
\eea
In such a limit, the function $F^{\NS}_{k,p} (q)$ becomes
\bea
F^{\NS}_{k,p} (q) &\sim \frac{1}{2} (q;q)^{-6}_\infty (1-q)^3 q^{p^2+p+k-\frac{3}{2}} \frac{1-q^{2k}}{(1+q^k)^4} \left[ \frac{\vartheta_3(1, q)^2}{(1+q^{p-\frac{1}{2}})(1+q^{p+\frac{1}{2}}) } - \frac{\vartheta_4(1,q)^2}{(1-q^{p-\frac{1}{2}})(1-q^{p+\frac{1}{2}}) } \right] \nn \\
&\sim \frac{1}{2} (q;q)^{-6}_\infty (1-q)^3 q^{p^2+p+k-\frac{3}{2}} \left[ \frac{\vartheta_3(1, q)^2}{(1+q^{p-\frac{1}{2}})(1+q^{p+\frac{1}{2}}) } - \frac{\vartheta_4(1,q)^2}{(1-q^{p-\frac{1}{2}})(1-q^{p+\frac{1}{2}}) } \right]~, \qquad k \rightarrow \infty~. \label{asympFNSkp}
\eea

\subsubsection*{The R sector}   Similarly the partition function $\chi^{{\cal N}_{4d}=1}_{\Ra} \mid_{\GSO} (q; y, s)$, defined in \eref{def:GSOed}, can be written as
\bea  \label{GSOR4dN1}
\chi^{{\cal N}_{4d}=1}_{\Ra} \mid_{\GSO} (q; y, s) = \sum_{k=0}^\infty \sum_{p=- \infty}^\infty [2k+1]_y  s^{2p-1}~F^{\Ra}_{k,p} (q)~,
\eea
where the function $F^{\Ra}_{k,p} (q)$ follows from \eref{exactchRSO3}, \eref{N1R} and \eref{def:GSOed}:
\bea
F^{\Ra}_{k,p} (q) &= \frac{1}{2} (q;q)^{-6}_\infty (1-q) \frac{q^{p^2-\frac{5}{4}}}{(1+q^p)(1+q^{p-1})} \vartheta_2(1, q) \times \nn \\
& \quad \sum_{n=1}^\infty (-1)^{n+1} (1-q^n) q^{n \choose 2} \sum_{m=0}^\infty  q^{\frac{1}{2}(m+\frac{1}{2})^2}(1-q^{m+1}) (q^{n|k-m|} -q^{n(k+m+2)})~.
\eea
In the limit $k \rightarrow \infty$,  this function can be simplified further.
The summation over $n$ can be asymptotically evaluated as follows (assume that $m$ is finite):
\bea
\sum_{n=1}^\infty (-1)^{n+1} (1-q^n) q^{n \choose 2} (q^{n|k-m|} -q^{n(k+m+2)}) &\sim \sum_{n=1}^\infty (-1)^{n+1} (1-q^n) q^{n(k-m)} (1-q^{n(2m+2)})\nn\\
&\sim \frac{ q^{k} (1-q) \left(1-q^{2 k}\right)}{\left(1+q^k\right)^4} \left \{  q^{-m} \left(1-q^{2 m+2}\right) \right \}~, \label{sumnRa4dN1}
\eea
and the summation over $m$ can be computed as follows:
\bea
 \sum_{m=0}^\infty q^{\frac{1}{2} (m+\frac{1}{2})^2- m} \left(1-q^{m+1}\right) \left(1-q^{2 m+2}\right)
&= (1-q) \vartheta_2(1, q)~.
\eea
Therefore, we have the following asymptotic formula:
\bea \label{asympFRkp}
F^{\Ra}_{k,p} (q) &\sim \frac{1}{2} (q;q)^{-6}_\infty \frac{q^{p^2+k-\frac{5}{4}}(1-q)^3\left(1-q^{2 k}\right)}{\left(1+q^p\right)\left(1+q^{p-1}\right)\left(1+q^k\right)^4} \vartheta_2(1, q)^2 \nn \\ 
&\sim \frac{1}{2} (q;q)^{-6}_\infty \frac{q^{p^2+k-\frac{5}{4}}(1-q)^3}{\left(1+q^p\right)\left(1+q^{p-1}\right)} \vartheta_2(1, q)^2~, \qquad \qquad k \rightarrow \infty~.
\eea

%%%%%%%%

\subsubsection*{Combining both sectors} 

Combining the NS- and R contributions from the previous subsections gives rise to the following $SO(3) \times U(1)_R$ covariant partition function
\bea \label{N1partition1}
\chi^{{\cal N}_{4d}=1} (q; y, s) 
%&= \chi^{{\cal N}_{4d}=1}_{\NS} \mid_{\GSO} (q; y, s) + \chi^{{\cal N}_{4d}=1}_{\Ra} \mid_{\GSO} (q; y, s) \nn \\
&=  \sum_{k=0}^\infty \sum_{p=- \infty}^\infty \left( [2k]_y  s^{2p}~F^{\NS}_{k,p} (q) + [2k+1]_y  s^{2p-1}~F^{\Ra}_{k,p} (q) \right) \nn \\
&= \sum_{k=0}^\infty \left \{ [2k] \left( F^{\NS}_{k,0} (q) + \sum_{p=1}^\infty s_{2p} F^{\NS}_{k,p} (q) \right)  + [2k+1]  \sum_{p=1}^\infty s_{2p-1} F^{\Ra}_{k,p} (q) \right \} ~,
\eea
where $s_m$ is defined by \eref{comp}. Even though the $F^{\NS}_{k,p}$ and $F^{\Ra}_{k,p}$ functions are known, the representation (\ref{N1partition1}) of the overall partition function does not make ${\cal N}_{4d}=1$ SUSY manifest to all mass levels. In order to do so, we have to combine $SO(3) \times U(1)_R$ representations to supermultiplets (\ref{N1rep}) and rewrite (\ref{N1partition1}) as\footnote{The symmetry of \eref{N1partition1} under $s \rightarrow s^{-1}$ guarantees that $
M(\chi^{{\cal N}_{4d}=1}, \lb n,Q \rb, q)=M(\chi^{{\cal N}_{4d}=1}, \lb n,-Q \rb, q)$, so we shall henceforth assume that $Q\geq0$.}
\bea
\chi^{{\cal N}_{4d}=1} (q; y, s) = \sum_{n=0}^\infty  \sum_{Q=0}^\infty \lb n, \pm Q \rb \, M(\chi^{{\cal N}_{4d}=1}, \lb n ,Q \rb, q)  ~ .\label{N1partition2}
\eea
This introduces a {\it multiplicity generating function} $M(\chi^{{\cal N}_{4d}=1}, \lb n ,Q \rb, q)$ for the supermultiplet $\lb n ,Q \rb$ appearing in the partition function $\chi^{{\cal N}_{4d}=1}$. To lighten our notation in the subsequent steps, we shall use the shorthand 
\bea
G_{n,Q} (q) := M(\chi^{{\cal N}_{4d}=1}, \lb n ,Q \rb, q)~.
\eea
By comparing \eref{N1partition1} with \eref{N1partition2}, it is immediate that
\bea
G_{2n, 2Q}(q) = G_{2n+1, 2Q+1} (q)= 0~, \qquad \text{for all $n \geq0$ and $Q \geq 0$} ~.
\eea

%%%%%%%%

\subsubsection*{Recurrence relations} 

In order to relate the supersymmetric multiplicity generating functions $G_{n,Q}$ to their $SO(3) \times U(1)_R$ relatives $F^{\NS}_{k,p}$ and $F^{\Ra}_{k,p}$, we use (\ref{N1rep}) to rewrite \eref{N1partition2} in terms of characters of irreducible $SO(3)$ characters and the fugacity $s$ as 
\bea \label{N1partition3}
& \chi^{{\cal N}_{4d}=1} (q; y, s) =  [0] \Big[ (G_{1,0}+2G_{0,1}) + \sum_{Q=1}^\infty s_{2Q} \left( G_{0,2Q-1}+ G_{1,2Q} + G_{0,2Q+1} \right) \Big]  \nn \\
&\quad + \sum_{k=1}^\infty [2k] \Big[ \left( G_{2k-1,0} + 2G_{2k,1}+ G_{2k+1,0}\right)+ \sum_{Q=1}^\infty s_{2Q} \left( G_{2k-1,2Q} + G_{2k,2Q-1} + G_{2k,2Q+1} + G_{2k+1,2Q} \right) \Big]  \nn \\
&\quad + \sum_{k=0}^\infty [2k+1]   \sum_{Q=1}^\infty s_{2Q-1} \left( G_{2k,2Q-1} + G_{2k+1,2Q-2} + G_{2k+1,2Q} + G_{2k+2,2Q-1} \right)  ~,
\eea
where $G_{n,Q}$ is a shorthand notation for $G_{n,Q}(q)$.

Comparing \eref{N1partition1} with \eref{N1partition3}, we have the following relations:
\bea
2G_{0,1}(q)+G_{1,0}(q)  &= F^{\NS}_{0,0} (q)~,  \label{recrel1}\\ 
G_{2k-1,0} (q) + 2G_{2k,1} (q) + G_{2k+1,0} (q) &= F^{\NS}_{k,0} (q)~, \quad k \geq 1  \label{recrel2}\\
G_{0,2Q-1}(q)+G_{0,2Q+1}(q)+G_{1,2Q}(q) &= F^{\NS}_{0,Q} (q)~, \quad Q \geq 1 \label{recrel3} \\
G_{2k-1,2Q} (q) + G_{2k,2Q-1} (q) + G_{2k,2Q+1} (q) + G_{2k+1,2Q} (q) &= F^{\NS}_{k,Q} (q)~, \quad k,Q \geq 1  \label{recrel4} \\
G_{2k,2Q-1}(q) + G_{2k+1,2Q-2} (q)+ G_{2k+1,2Q}(q) + G_{2k+2,2Q-1} (q) &= F^{\Ra}_{k,Q} (q)~, \quad k \geq 0, Q \geq 1~. \label{recrel5}
\eea
These relations are useful for computing a multiplicity generating function for a representation $\lb \text{odd}, \text{even} \rb$ (or $\lb \text{even}, \text{odd} \rb$) when the one for opposite parity is known. However, the recursion is not powerful enough to directly determine all the $G_{n,Q}$ in terms of $F^{\NS}_{k,p}$ and $F^{\Ra}_{k,p}$. The following subsection follows an alternative approach to determine the $G_{n,Q}$.

%%%%%%%%

\subsection[Multiplicities of representations in the $\CN_{4d} =1$ partition function]{Multiplicities of representations in the $\bm{\CN_{4d} =1}$ partition function}  

Our aim in this subsection is to factor out the fundamental $\CN_{4d} =1$ super Poincar\'e character $Z({\cal N}_{4d}=1)=[1]_y+s+s^{-1}$ and to compute explicitly the multiplicity generating functions $G_{n,Q}(q)$ for $\lb n, Q \rb$ in
\bea
\chi^{{\cal N}_{4d}=1} (q; y, s) 
%= \sum_{n=0}^\infty \sum_{Q=-\infty}^\infty {\lb n, Q \rb} M(\chi^{{\cal N}_{4d}=1}, \lb n,Q \rb, q)
= \sum_{n=0}^\infty \sum_{Q=-\infty}^\infty {\lb n, Q \rb} G_{n,Q}(q) \ . \label{defMgenN1}
\eea
Using the second equality of \eref{N1rep} and orthogonality of $SO(3) \times U(1)_R$ representations, we have
\bea
G_{n,Q}(q) = M(\chi^{{\cal N}_{4d}=1}, \lb n ,  Q \rb, q) 
&= \frac{1}{2 \pi i} \oint_{\CC} \frac{\ud s}{s}  \int \ud \mu_{SO(3)} (y)~ [n]_y s^{-Q} \frac{\chi^{{\cal N}_{4d}=1}(q; y, s)}{[1]_y + (s+ s^{-1})} \ ,
\eea
where $\CC$ is a contour in the complex $s$-plane enclosing the origin. In order to proceed, we use the geometric series expansion of the inverse $Z({\cal N}_{4d}=1)$,\footnote{Note that $\frac{1}{[1]_y + (s+ s^{-1})} $ can also be written in another way as follows:
\bea
\frac{1}{[1]_y + (s+ s^{-1})} = \sum_{m=0}^\infty (-1)^m s^{m+1} [m]_y~.
\eea
However, we shall not take this approach, since otherwise this would lead to tensor products in \eref{orthogcase1} and \eref{orthogcase2} which are harder to evaluate in comparison with our current approach.}
\bea
\frac{1}{[1]_y + (s+ s^{-1})} = \frac{1}{s+s^{-1}} \frac{1}{ 1+ \frac{[1]_y}{s+s^{-1}}} = \sum_{m=0}^\infty (-1)^m \frac{[1]_y^m}{(s+s^{-1})^{m+1}}~.
\label{geoser}
\eea
In what follows, we consider the contributions from $\chi^{{\cal N}_{4d}=1}_{\NS} \mid_{\GSO} (q; y, s)$ and $\chi^{{\cal N}_{4d}=1}_{\Ra} \mid_{\GSO} (q; y, s)$ separately and then add up these results to yield the overall multiplicity generating function defined by \eref{defMgenN1},
\bea
M(\chi^{{\cal N}_{4d}=1}, \lb n ,  Q \rb, q) =M(\chi^{{\cal N}_{4d}=1}_{\NS} \mid_{\GSO}, \lb n , Q \rb, q)  + M(\chi^{{\cal N}_{4d}=1}_{\Ra} \mid_{\GSO}, \lb n , Q \rb, q)~,
\eea
where $\chi^{{\cal N}_{4d}=1}_{\NS, \Ra} \mid_{\GSO}$ are given by \eref{GSONS4dN1} and \eref{GSOR4dN1}.
%and the first and second equalities of \eref{N1partition1a},
%\bea
%\chi^{{\cal N}_{4d}=1} (q; y, s) &= \chi^{{\cal N}_{4d}=1}_{\NS} \mid_{\GSO} (q; y, s) + \chi^{{\cal N}_{4d}=1}_{\Ra} \mid_{\GSO} (q; y, s) \nn \\
%&=\sum_{k=0}^\infty \sum_{p=- \infty}^\infty \left( [2k]_y  s^{2p}~F^{\NS}_{k,p} (q) + [2k+1]_y  s^{2p-1}~F^{\Ra}_{k,p} (q) \right)~.
%\eea

\subsubsection*{Multiplicities in the NS-sector}

The series expansion of $( Z({\cal N}_{4d}=1) )^{-1}$ leads to the following NS sector contribution to the multiplicity generating function of the supermultiplet $ \lb n , Q \rb$ 
% \bea
%M(\chi^{{\cal N}_{4d}=1}_{\NS} \mid_{\GSO}, \lb n , Q \rb, q) 
%&:= \frac{1}{2 \pi i} \oint_{\CC} \frac{\ud s}{s} \int \ud \mu_{SO(3)} (y)~ \frac{[n]_y}{s^Q}  \times \frac{\chi^{{\cal N}_{4d}=1}_{\NS} \mid_{\GSO} (q; y, s) }{[1]_y + (s+ s^{-1})} \nn \\
%  &=  \sum_{m=0}^\infty \sum_{k=0}^\infty \sum_{p=- \infty}^\infty (-1)^m F^{\NS}_{k,p} (q) \left \{ \frac{1}{2 \pi i} \oint_{|s|=1-\epsilon} \frac{\ud s}{s} \frac{s^{2p}}{s^{Q} (s+s^{-1})^{m+1}}  \right \} \times \nn \\
%& \hspace{3cm} \left \{ \int \ud \mu_{SO(3)} (y)~ [n]_y  [1]_y^m   [2k]_y \right\} ~,
%\eea
\bea
M&(\chi^{{\cal N}_{4d}=1}_{\NS} \mid_{\GSO}, \lb n , Q \rb, q) 
:= \frac{1}{2 \pi i} \oint_{\CC} \frac{\ud s}{s} \int \ud \mu_{SO(3)} (y)~ \frac{[n]_y}{s^Q}  \times \frac{\chi^{{\cal N}_{4d}=1}_{\NS} \mid_{\GSO} (q; y, s) }{[1]_y + (s+ s^{-1})} \nn \\
  &=  \sum_{m=0}^\infty \sum_{k=0}^\infty \sum_{p=- \infty}^\infty (-1)^m F^{\NS}_{k,p} (q) \ \frac{1}{2 \pi i} \oint_{|s|=1-\epsilon} \frac{\ud s}{s} \frac{s^{2p}}{s^{Q} (s+s^{-1})^{m+1}}  \  \int \ud \mu_{SO(3)} (y)~ [n]_y  [1]_y^m   [2k]_y  ~,
\eea
We shall henceforth take $\CC$ to be a circle centred at the origin with the radius $1-\epsilon$, with $0<\epsilon <1$. The quantities in the curly brackets can be computed as follows:
\bea \label{residueS}
& \frac{1}{2 \pi i}  \oint_{|s|=1-\epsilon} \frac{\ud s}{s} \frac{s^{2p}}{s^{Q} (s+s^{-1})^{m+1}} \nn \\
 & = \begin{cases} (-1)^{\frac{1}{2}(Q-m-2p-1)} {\frac{1}{2}(Q+m-2 p-1) \choose m} &\quad \text{for $Q-m$ odd and $Q+m \geq 2p+1$}\\ 0 & \quad \text{otherwise~,}  \end{cases}
\eea 
and 
\bea
\int \ud \mu_{SO(3)}  (y)~ [2n]_y  [1]_y^m   [2k]_y  &= \begin{cases} T_{2n+1} \left( m, \frac{1}{2}m + n -k \right) \quad & \text{if $m$ is even} \\
0 & \text{if $m$ is odd}~, \end{cases} \label{orthogcase1} \\
\int \ud \mu_{SO(3)}  (y)~ [2n+1]_y  [1]_y^m   [2k]_y  &= \begin{cases} T_{2n+2} \left( m, \frac{1}{2}m + n +\frac{1}{2} -k \right) \quad & \text{if $m$ is odd} \label{orthogcase2}\\
0 & \text{if $m$ is even}~, \end{cases}
\eea
where
\bea
T_p (m,k) = {m \choose k} - {m \choose k-p}~.
\eea
Note that \eref{residueS}, \eref{orthogcase1} and \eref{orthogcase2} are in perfect agreement with the selection rule
\bea
M(\chi^{{\cal N}_{4d}=1} , \lb 2n , 2Q \rb, q) = M(\chi^{{\cal N}_{4d}=1} , \lb 2n+1 , 2Q+1 \rb, q) =0~.
\label{select}
\eea
The nonzero multiplicities of $\lb 2n , 2Q+1 \rb$ and $\lb 2n+1 , 2Q\rb$ receive the following NS sector contributions:
\begin{align}
 M(&\chi^{{\cal N}_{4d}=1}_{\NS} \mid_{\GSO}, \lb 2n , 2Q+1 \rb, q) \notag \\
&= \sum_{k= 0}^\infty \sum_{m= 0}^\infty \sum_{p=-\infty}^{Q+m}  (-1)^{Q-m-p} F^{\NS}_{k,p} (q) {Q+m-p \choose 2m} T_{2n+1} (2m, m+n-k)~, \\
 M(&\chi^{{\cal N}_{4d}=1}_{\NS} \mid_{\GSO}, \lb 2n+1 , 2Q \rb, q) \notag \\
&= \sum_{k= 0}^\infty \sum_{m= 0}^\infty \sum_{p=-\infty}^{Q+m}  (-1)^{Q-m-p} F^{\NS}_{k,p} (q) {Q+m-p \choose 2m+1} T_{2n+2} (2m+1, m+n+1-k)~.
\end{align}

\subsubsection*{Multiplicities in the R-sector}
Similarly to the NS-sector, the generating function for the multiplicity of the representation $ \lb n , Q \rb$ in the function $\chi^{{\cal N}_{4d}=1}_{\Ra} \mid_{\GSO} (q; y, s)$ is given by
\bea
& M(\chi^{{\cal N}_{4d}=1}_{\Ra} \mid_{\GSO}, \lb n , Q \rb, q) := \frac{1}{2 \pi i} \oint_{|s|=1-\epsilon} \frac{\ud s}{s} \int \ud \mu_{SO(3)} (y)~ \frac{[n]_y}{s^Q}  \times \frac{\chi^{{\cal N}_{4d}=1}_{\Ra} \mid_{\GSO} (q; y, s) }{[1]_y + (s+ s^{-1})} \nn \\
 &=  \sum_{m=0}^\infty \sum_{k=0}^\infty \sum_{p=- \infty}^\infty (-1)^m F^{\Ra}_{k,p} (q) \ \frac{1}{2 \pi i} \oint_{|s|=1-\epsilon} \frac{\ud s}{s} \frac{s^{2p-1}}{s^{Q} (s+s^{-1})^{m+1}}  \  \int \ud \mu_{SO(3)} (y)~ [n]_y  [1]_y^m   [2k+1]_y  ~,
 %&= \sum_{k= 0}^\infty \sum_{m= 0}^\infty \sum_{p=-\infty}^{Q+m}  (-1)^{Q-m-p} F^{\NS}_{k,p} (q) {Q+m-p \choose 2m} T_{2n+1} (2m, m+n-k)~.
\eea
with $0 < \epsilon <1$,
\bea
& \frac{1}{2 \pi i}  \oint_{|s|=1-\epsilon} \frac{\ud s}{s} \frac{s^{2p-1}}{s^{Q} (s+s^{-1})^{m+1}} \nn \\
 & = \begin{cases} (-1)^{\frac{1}{2}(Q-m-2p)} {\frac{1}{2}(Q+m-2 p) \choose m} &\quad \text{for $Q-m$ even and $Q+m \geq 2p$}\\ 0 & \quad \text{otherwise~,}  \end{cases}
\eea 
and 
\bea
\int \ud \mu_{SO(3)}  (y)~ [2n]_y  [1]_y^m   [2k+1]_y  &= \begin{cases} T_{2n+1} \left( m, \frac{1}{2}m + n -k -\frac{1}{2} \right) \quad & \text{if $m$ is odd} \\
0 & \text{if $m$ is even}~, \end{cases} \\
\int \ud \mu_{SO(3)}  (y)~ [2n+1]_y  [1]_y^m   [2k+1]_y  &= \begin{cases} T_{2n+2} \left( m, \frac{1}{2}m + n -k \right) \quad & \text{if $m$ is even}  \\
0 & \text{if $m$ is odd}~, \end{cases}
\eea
where $T_p (m,k)$ is defined as above and the zeros once again confirm the selection rule (\ref{select}).

%It follows immediately that 
%\bea
%M(\chi^{{\cal N}_{4d}=1}_{\Ra} \mid_{\GSO}, \lb 2n , 2Q \rb, q) = M(\chi^{{\cal N}_{4d}=1}_{\Ra} \mid_{\GSO}, \lb 2n+1 , 2Q+1 \rb, q) =0~.
%\eea
The multiplicities of $\lb 2n , 2Q+1 \rb$ are given by
\bea
& M(\chi^{{\cal N}_{4d}=1}_{\Ra} \mid_{\GSO}, \lb 2n , 2Q+1 \rb, q) \nn \\
&= \sum_{k= 0}^\infty \sum_{m= 0}^\infty \sum_{p=-\infty}^{Q+m}  (-1)^{Q-m-p+1} F^{\Ra}_{k,p} (q) {Q+m-p+1 \choose 2m+1} T_{2n+1} (2m+1, m+n-k)~.
\eea
The multiplicities of $\lb 2n+1 , 2Q \rb$ are given by
\bea
& M(\chi^{{\cal N}_{4d}=1}_{\Ra} \mid_{\GSO}, \lb 2n+1 , 2Q \rb, q) \nn \\
&= \sum_{k= 0}^\infty \sum_{m= 0}^\infty \sum_{p=-\infty}^{Q+m}  (-1)^{Q-m-p} F^{\Ra}_{k,p} (q) {Q+m-p\choose 2m} T_{2n+2} (2m, m+n-k)~.
\eea

\subsubsection*{Combining the NS and R sectors}
Now we can assemble the NS- and R sector results to obtain the full multiplicities of the representation $\lb n, Q \rb$ in $\chi^{{\cal N}_{4d}=1} (q; y, s)$. First, it is clear that
\bea
G_{2n,2Q} (q)= G_{2n+1,2Q+1} (q) =0~. \label{sameparityzero}
\eea
The nonzero multiplicities of $\lb 2n , 2Q+1 \rb$ and $\lb 2n+1 , 2Q \rb$ are most conveniently presented in terms of the shorthands
\bea
&\mathfrak{M}_{\lb 2n,2Q+1 \rb} (m,p,k; q) \ \ := \ \ (-1)^{Q-m-p} \Bigg[ F^{\NS}_{k,p} (q) {Q+m-p \choose 2m}  T_{2n+1} (2m, m+n-k) \nn \\
& \hspace{5cm} - F^{\Ra}_{k,p} (q) {Q+m-p+1 \choose 2m+1} T_{2n+1} (2m+1, m+n-k) \Bigg] \label{shorthd1} \\ 
&\mathfrak{M}_{\lb 2n+1,2Q \rb} (m,p,k; q) \ \ := \ \ (-1)^{Q-m-p} \Bigg[ F^{\NS}_{k,p} (q) {Q+m-p \choose 2m+1} T_{2n+2} (2m+1, m+n+1-k)  \nn \\
& \hspace{5cm} + F^{\Ra}_{k,p} (q) {Q+m-p\choose 2m} T_{2n+2} (2m, m+n-k) \Bigg]  \label{shorthd2} 
\eea
for the contributions $\mathfrak{M}_{\lb \cdot,\cdot \rb} (m,p,k; q)$ of individual terms in the $m,p,k$ triple sum to the multiplicity generating function. The result for $\lb 2n , 2Q+1 \rb$ supermultiplets is
\bea
 G_{2n,2Q+1} (q) &= \sum_{k= 0}^\infty \sum_{m= 0}^\infty \sum_{p=-\infty}^{Q+m} \mathfrak{M}_{\lb 2n,2Q+1 \rb} (m,p,k; q) \nn\\
&= \ \  \sum_{k= 0}^\infty \sum_{m= 0}^\infty \Bigg[ \sum_{p=0}^{\infty}  \Big \{ \frM_{\lb 2n,2Q+1 \rb} (m, -p-1, k;q) + \frM_{\lb 2n,2Q+1 \rb} (m+p, p, k;q) \Big \} \nn \\
& \hspace{2.5cm} + \sum_{p=0}^{Q-1}  \frM_{\lb 2n,2Q+1 \rb} (m, m+p+1, k;q) \Bigg]~.
\label{mulsumoddeven0}
\eea
whereas the multiplicities of $\lb 2n+1 , 2Q \rb$ are given by
\bea
 G_{2n+1,2Q} (q) &=\sum_{k= 0}^\infty \sum_{m= 0}^\infty \sum_{p=-\infty}^{Q+m} \mathfrak{M}_{\lb 2n+1,2Q \rb} (m,p,k; q) \nn\\
&= \ \ \sum_{k= 0}^\infty \sum_{m= 0}^\infty  \Bigg[ \sum_{p=0}^{\infty} \Big\{ \frM_{\lb 2n+1,2Q \rb} (m, -p-1, k;q) +\frM_{\lb 2n+1,2Q \rb} (m+p, p, k;q) \Big \} \nn \\
& \hspace{2.5cm} + \sum_{p=0}^{Q-1}  \frM_{\lb 2n+1,2Q \rb} (m, m+p+1, k;q) \Bigg]~.
\label{mulsumoddeven}
\eea

%\bea
%& M(\chi^{{\cal N}_{4d}=1}, \lb 2n , 2Q+1 \rb, q)  \nn \\
%&=  \sum_{k= 0}^\infty \sum_{m= 0}^\infty \sum_{p=-\infty}^{Q+m}  (-1)^{Q-m-p} \Bigg[ F^{\NS}_{k,p} (q) {Q+m-p \choose 2m}  T_{2n+1} (2m, m+n-k) \nn \\
%& \hspace{4cm} - F^{\Ra}_{k,p} (q) {Q+m-p+1 \choose 2m+1} T_{2n+1} (2m+1, m+n-k) \Bigg] \nn \\
%&=:  \sum_{k= 0}^\infty \sum_{m= 0}^\infty \sum_{p=-\infty}^{Q+m} \mathfrak{M}_{\lb 2n,2Q+1 \rb} (m,p,k; q) \nn\\
%&= \sum_{k= 0}^\infty \sum_{m= 0}^\infty \Bigg[ \sum_{p=0}^{\infty}  \Big \{ \frM_{\lb 2n,2Q+1 \rb} (m, -p-1, k;q) + \frM_{\lb 2n,2Q+1 \rb} (m+p, p, k;q) \Big \} \nn \\
%& \hspace{1.5cm} + \sum_{p=0}^{Q-1}  \frM_{\lb 2n,2Q+1 \rb} (m, m+p+1, k;q) \Bigg]~.
%\eea
%and the multiplicities of $\lb 2n+1 , 2Q \rb$ are given by
%\bea
%& M(\chi^{{\cal N}_{4d}=1}, \lb 2n+1 , 2Q \rb, q)  \nn \\
%&=  \sum_{k= 0}^\infty \sum_{m= 0}^\infty \sum_{p=-\infty}^{Q+m}  (-1)^{Q-m-p} \Bigg[ F^{\NS}_{k,p} (q) {Q+m-p \choose 2m+1} T_{2n+2} (2m+1, m+n+1-k)  \nn \\
%& \hspace{4cm} + F^{\Ra}_{k,p} (q) {Q+m-p\choose 2m} T_{2n+2} (2m, m+n-k) \Bigg] \nn \\
%&=:  \sum_{k= 0}^\infty \sum_{m= 0}^\infty \sum_{p=-\infty}^{Q+m} \mathfrak{M}_{\lb 2n+1,2Q \rb} (m,p,k; q) \nn\\
%&= \sum_{k= 0}^\infty \sum_{m= 0}^\infty  \Bigg[ \sum_{p=0}^{\infty} \Big\{ \frM_{\lb 2n+1,2Q \rb} (m, -p-1, k;q) +\frM_{\lb 2n+1,2Q \rb} (m+p, p, k;q) \Big \} \nn \\
%& \hspace{1.5cm} + \sum_{p=0}^{Q-1}  \frM_{\lb 2n+1,2Q \rb} (m, m+p+1, k;q) \Bigg]~.
%\eea

\subsection{Asymptotic analysis for the multiplicities}
\label{asympt4dim}

This subsection is devoted to the multiplicity generating function $G_{n,Q}(q)$ in the limit $n \rightarrow \infty$. We shall present analytic expressions for their $n \rightarrow \infty$ asymptotics whose derivation is defered to appendix \ref{deriv}. The method essentially relies on identifying the dominant contribution to the triple sums in (\ref{mulsumoddeven0}) and (\ref{mulsumoddeven}). The end result for multiplicity generating functions $G_{n,Q}(q)$ reads
\bea
G_{2n+1,2Q}(q) &\sim \frac{(1-q)^2 q^{n-\frac{3}{2}} }{2 (q;q)_\infty^6} \CF(q,Q)~,\quad n \rightarrow \infty ~, \label{asympmuloddeven} \\
G_{2n,2Q+1}(q) &\sim  \frac{(1-q)^2 q^{n-\frac{3}{2}}}{2(q;q)^{6}_\infty (1+q)} \times  \left[ \frac{q^{(Q+1)^2+\frac{1}{4}}(1-q)}{\left(1+q^{Q}\right) \left(1+q^{Q+1} \right)} \vartheta_2(1, q)^2  - \CF(q,Q)- \CF(q,Q+1) \right]
\label{mainferm}
\eea
with the function ${\cal F}(q,Q)$ given by
\bea
{\cal F}(q,Q) &= \vartheta_2(1,q)^2 \Big[ q^{1-Q} u_1(\sqrt{q},Q)+(-1)^Q (1-q)  (v_1(\sqrt{q},Q)+q^{-1/4} w_1(\sqrt{q},Q)) \Big] \nn \\
& \quad +\vartheta_3(1,q)^2 \Big[ -q^{1-Q} u_2(\sqrt{q},Q)+(-1)^Q (1-q)  (v_2(\sqrt{q},Q)+ q^2 w_2(\sqrt{q},Q)) \Big] \nn \\
& \quad +\vartheta_4(1,q)^2 \Big [ q^{1-Q} u_2(-\sqrt{q},Q)-(-1)^Q (1-q)  (v_2(-\sqrt{q},Q)+q^2 w_2(-\sqrt{q},Q)) \Big] \ .
\label{defF}
\eea
The three pairs of functions $u_i, v_i$ and $w_i$ correspond to the three summations in (\ref{mulsumoddeven0}) and (\ref{mulsumoddeven}):
\bea
u_1(q,Q) &= \sum_{p=0}^\infty q^{2\left(p+\frac{3}{2} \right)^2} \frac{1-q^{4p+4Q+6}}{(1+q^{2p+2})(1+q^{2p+4})}~, \nn \\
u_2(q,Q) &=  \sum_{p=0}^\infty q^{2\left(p+1 \right)^2} \frac{1-q^{4p+4Q+4}}{(1+q^{2p+1})(1+q^{2p+3})}~. \\
v_1(q,Q) &= \sum_{p=0}^{\lfloor Q/2 \rfloor} \frac{q^{2(p-\frac{1}{2})^2}(1+q^2)^{2p}}{(1+q^{2p-2})(1+q^{2p})} {Q \choose 2 p} {}_3F_2\left[{1,\ Q+1,\ 2p-Q \atop p+1/2,\ p+1} ; \ \frac{(1+q)^2}{4q}\right] ~,  \nn \\
v_2(q,Q) &= \sum_{p=0}^{\lfloor Q/2 \rfloor} \frac{ (1+q) q^{2 p^2}(1+q^2)^{2p}}{(1+q^{2p-1})(1+q^{2p+1})}  {Q \choose 2 p+1} {}_3F_2\left[{1,\ Q+1,\ 2p+1-Q \atop p+1,\ p+3/2};\ \frac{(1+q)^2}{4q} \right] ~,  \\
w_1 (q,Q) &= \sum_{m= 0}^\infty   \sum_{p=0}^{Q-1} \frac{ (-1)^{p+1}q^{1+2(1+m+p)^2-2m}\left(1+q^2\right)^{2m} {Q-1-p \choose 2 m}}{ \left(1+q^{2(m+p)}\right) \left(1+q^{2(1+m+p)}\right) }~, \nn \\
w_2 (q,Q) &= q^{-\frac{9}{2}} \sum_{m= 0}^\infty   \sum_{p=0}^{Q-1} \frac{(-1)^{p+1}q^{2\left(m+p+\frac{3}{2}\right)^2-2m}\left(1+q^2\right)^{2m+1} {Q-1-p \choose 1+2 m}}{\left(1+q^{1+2m+2p}\right) \left(1+q^{3+2m+2p}\right)}~.
\eea
Note that the leading orders in the power series are
\bea
G_{2n+1,2Q}(q) \ \sim \ q^{n+Q(Q+2)} \co G_{2n,2Q+1}(q) \ \sim \ q^{n+Q^2+3Q+1} \co q \rightarrow 0  \ ,
\eea
i.e. the supermultiplet $\lb 2n+1,2Q \rb$ firstly occurs at mass level $n+Q(Q+2)$ whereas the $\lb 2n,2Q+1 \rb$ multiplet firstly occurs at mass level $n+Q^2+3Q+1$.

For reference, we list the leading $q$ powers for the $G_{n\rightarrow \infty,Q}$ regime for some small values of the $U(1)_R$ charge, obtained by expansion of (\ref{asympmuloddeven}) and (\ref{mainferm}): firstly for even values $Q \in 2\NN_0$
%\bea
%G_{2n+1,0}(q) &\sim q^n( 1+q+7 q^2+19 q^3+53 q^4+133 q^5+328 q^6+752 q^7+1689 q^8+3635 q^9 \nn \\
%& \qquad \quad +7642 q^{10}+ O(q^{11}) )~, \nn \\
%G_{2n+1,2}(q) &\sim q^{n+3} (  2+8 q+24 q^2+73 q^3+187 q^4+467 q^5+1090 q^6+2457 q^7+5314 q^8+11197 q^9 \nn \\
%& \qquad \quad +22888 q^{10}+ O(q^{11}))~, \nn \\
%G_{2n+1,4}(q) &\sim q^{n+8} (2+10 q+36 q^2+110 q^3+306 q^4+773 q^5+1861 q^6+4245 q^7+9327 q^8+19774 q^9 \nn \\
%& \qquad \quad  +40775 q^{10}+ O(q^{11}))~, \nn \\
%G_{2n+1,6}(q) &\sim  q^{n+15} ( 2+10 q+38 q^2+124 q^3+352 q^4+928 q^5+2282 q^6+5335 q^7+11929 q^8 \nn \\
%& \qquad \quad +25747 q^9+53791 q^{10} + O(q^{11}))~,
%\eea
\bea
G_{2n+1,0}(q) &\sim q^n( 1+q+7 q^2+19 q^3+53 q^4+133 q^5+328 q^6+752 q^7+1689 q^8+3635q^9+O(q^{10}) )~, \nn \\
G_{2n+1,2}(q) &\sim q^{n+3} (  2+8 q+24 q^2+73 q^3+187 q^4+467 q^5+1090 q^6+2457 q^7+ 5314q^8+O(q^{9}))~, \nn \\
G_{2n+1,4}(q) &\sim q^{n+8} (2+10 q+36 q^2+110 q^3+306 q^4+773 q^5+1861 q^6+4245 q^7+9327 q^8+O(q^{9}))~, \nn \\
G_{2n+1,6}(q) &\sim  q^{n+15} ( 2+10 q+38 q^2+124 q^3+352 q^4+928 q^5+2282 q^6+5335 q^7+ O(q^{8}))~,
\eea
and secondly for odd values $Q \in 2\NN-1$
\bea
G_{2n,1}(q) &\sim q^{n+1} (3+5 q+22 q^2+53 q^3+150 q^4+345 q^5+836 q^6+1824 q^7+4011 q^8+O(q{9}))~, \nn \\
G_{2n,3}(q) &\sim q^{n+5} (4+11 q+46 q^2+117 q^3+331 q^4+784 q^5+1876 q^6+4133 q^7+O(q^{8}))~, \nn \\
G_{2n,5}(q) &\sim q^{n+11} (4+12 q+55 q^2+150 q^3+437 q^4+1078 q^5+2640 q^6+5951 q^7+O(q^{8}))~, \nn \\
G_{2n,7}(q) &\sim  q^{n+19} (4+12 q+56 q^2+159 q^3+474 q^4+1197 q^5+2994 q^6+6882 q^7+O(q^{8}))~.
\eea
Note that the general formula greatly simplifies at $U(1)_R$ charges $Q=0$ and $Q=1$,
\bea
G_{2n+1,0}&(q) \sim \frac{q^n}{(q;q)_\infty^6} \Bigg \{ \frac{1}{2} (1-q)^2 q^{-\frac{1}{2}} \Big ( u_1(\sqrt{q}) \vartheta_2 (1, q)^2 - \left[ u_2(\sqrt{q}) \vartheta_3(1, q)^2-u_2(-\sqrt{q}) \vartheta_4(1,q)^2 \right] \Big ) \nn \\
& \qquad + \frac{1}{4} \frac{(1-q)^3}{1+q} q^{-\frac{1}{4}} \vartheta_2 (1,q)^2 \Bigg \}~, \qquad n \rightarrow \infty \label{simpleeven} \\
G_{2n,1}&(q) \sim  \frac{(1-q)^3 q^{n+1} }{4 (q;q)^{6}_\infty}\Bigg[ q^{-\frac{5}{2}}   \left \{ \frac{\vartheta_3(1,q)^2}{(1+q^{-\frac{1}{2}})(1+q^{\frac{1}{2}}) } - \frac{\vartheta_4(1,q)^2}{(1-q^{-\frac{1}{2}})(1-q^{\frac{1}{2}}) } \right \} - \frac{1}{2} q^{-\frac{9}{4}}  \vartheta_2 (1,q)^2 \nn \\
& \qquad -   q^{-\frac{5}{2}} \frac{1+q}{1-q} \Big ( u_1(\sqrt{q}) \vartheta_2 (1, q)^2 - \left[ u_2(\sqrt{q}) \vartheta_3(1, q)^2-u_2(-\sqrt{q}) \vartheta_4(1,q)^2 \right] \Big )  \Bigg] \label{simpleodd} 
\eea
where $u_i(q) \equiv u_i(q;0)$, see the first subsection of appendix \ref{deriv}.

%%%

\subsection[Empirical approach to ${\cal N}_{4d}=1$ asymptotic patterns]{Empirical approach to $\bm{{\cal N}_{4d}=1}$ asymptotic patterns}
\label{emp4dim}

In the previous subsection, we have derived the large spin asymptotics for multiplicity generating functions $G_{k,Q}(q)$ of individual ${\cal N}_{4d}=1$ multiplets (at finite $Q$ while $k\rightarrow \infty$), the main results being (\ref{asympmuloddeven}) and (\ref{mainferm}). The asymptotic formulae can be viewed as the supersymmetric generalization of truncating the infinite sum expression (\ref{genfnmul}) for the $SO(3)$ multiplicity generating function in the $d=4$ bosonic partition function to its $n=1$ term. In \cite{Curtright:1986di}, this $n=1$ term is interpreted as the leading (additive) Regge trajectory of unit slope, followed by an infinite tower of sister trajectories of fractional slope and alternating sign. Let us borrow the $\tau$ notation from equation (6.2) of \cite{Curtright:1986di} and expand the $G_{k,Q}(q)$ in an infinite series of trajectories:
%% RHS without differentiating even/odd:
%%  q^{n} \, \tau_1^Q(q) \ - \ q^{2n} \, \tau_2^Q(q) \ + \ q^{3n} \, \tau_3^Q(q) \ - \ \ldots \eq \sum_{\ell=1}^{\infty} (-1)^{\ell-1} \, q^{  \ell n} \, \tau_\ell^Q(q)
\begin{align}
G_{2n+1,2Q}(q) \ = \ &q^{n} \, \tau_1^{2Q}(q) \ - \ q^{2n} \, \tau_2^{2Q}(q) \ + \ q^{3n} \, \tau_3^{2Q}(q) \ - \ \ldots \ = \ \sum_{\ell=1}^{\infty} (-1)^{\ell-1} \, q^{  \ell n} \, \tau_\ell^{2Q}(q)
\label{tau} \\ 
G_{2n,2Q+1}(q) \ = \ &q^{n} \,  \tau_1^{2Q+1}(q) \ - \ q^{2n} \,  \tau_2^{2Q+1}(q) \ + \ q^{3n} \,  \tau_3^{2Q+1}(q) \ - \ \ldots \ = \ \sum_{\ell=1}^{\infty} (-1)^{\ell-1} \, q^{  \ell n} \, \tau_\ell^{2Q+1}(q)
\notag
\end{align}
All our ${\cal N}_{4d}=1$ data suggests that both of $\tau_\ell^{2Q}(q)$ and $\tau_\ell^{2Q+1}(q)$ are power series in $q$ with non-negative coefficients. Our analytic results (\ref{asympmuloddeven}) and (\ref{mainferm}) identify the first coefficient functions $\tau_1(q)$ in (\ref{tau}):
\begin{align}
\tau_1^{2Q}(q) \eq& \frac{(1-q)^2 q^{-\frac{3}{2}} }{2 (q;q)_\infty^6} \CF(q,Q)
\label{tau3} \\
 \tau_1^{2Q+1}(q) \eq& \frac{(1-q)^2 q^{-\frac{3}{2}}}{2(q;q)^{6}_\infty (1+q)} \times \, \left[ \frac{q^{(Q+1)^2+\frac{1}{4}}(1-q)}{\left(1+q^{Q}\right) \left(1+q^{Q+1} \right)} \vartheta_2(1, q)^2  - \CF(q,Q)- \CF(q,Q+1) \right]
\label{tau4}
\end{align}
The methods presented in appendix \ref{deriv} and applied in the previous subsection are not suitable to extract subleading Regge trajectories $\tau_{\ell \geq 2}(q)$, i.e. ${\cal N}_{4d}=1$ analogues of $n \geq 2$ terms in the sum (\ref{genfnmul}). Instead, we shall rely on an empirical approach, more specifically on explicit results obtained from a supercharacter expansion of the partition function (\ref{def:GSOed}) up to the 25th mass level. 

As an illustrative example, let us first of all investigate the family of $Q=0$ supermultiplets: The following table \ref{tab4d0} gathers $\llbracket2n+1,0\rrbracket$ multiplicities in the first 25 levels. Numbers marked in red directly correspond to the leading trajectory $\tau_1^0(q)$ whereas those in blue are additionally affected by the subleading trajectory $\tau_2^0(q)$. Given the leading trajectories (\ref{tau3}), our data from table \ref{tab4d0} can be used to determine the following subleading behaviour for $Q=0$ multiplets:

\begin{table}
\begin{scriptsize}
\begin{center}
%\begin{sidewaysfigure}
\begin{tabular}{|c|| c|c|c|c|c| c|c|c|c|c| c|}
%\begin{longtable}{|c|| c|c|c|c|c| c|c|c|c|c| c|c|c|c|c| c|c|c|c|c| c|c|c|c|c| c|} % \small
 \hline
\begin{turn}{-90}$ \ \ \ \ \ \ap m^2$  \end{turn}
&\begin{turn}{-90}\# $\llbracket1,0\rrbracket$ \end{turn}
&\begin{turn}{-90}\# $\llbracket3,0\rrbracket$ \end{turn}
&\begin{turn}{-90}\# $\llbracket5,0\rrbracket$ \end{turn}
&\begin{turn}{-90}\# $\llbracket7,0\rrbracket$ \end{turn}
&\begin{turn}{-90}\# $\llbracket9,0\rrbracket$ \end{turn}
&\begin{turn}{-90}\# $\llbracket11,0\rrbracket \ $ \end{turn}
&\begin{turn}{-90}\# $\llbracket13,0\rrbracket$ \end{turn}
&\begin{turn}{-90}\# $\llbracket15,0\rrbracket$ \end{turn}
&\begin{turn}{-90}\# $\llbracket17,0\rrbracket$ \end{turn}
&\begin{turn}{-90}\# $\llbracket19,0\rrbracket$ \end{turn}
&\begin{turn}{-90}\# $\llbracket21,0\rrbracket$ \end{turn}
\\\hline \hline
1 &0 &\textcolor{red}{1} &\textcolor{red}{0} & &
& & & & &
&
  \\\hline
2 &2 &\textcolor{red}{1} &\textcolor{red}{1} &\textcolor{red}{0} &
& & & & &
&
  \\\hline
3 &5 &\textcolor{blue}{5} &\textcolor{red}{1} &\textcolor{red}{1} &\textcolor{red}{0}
& & &
& & &    \\\hline
4 &12 &12 &\textcolor{red}{7} &\textcolor{red}{1} &\textcolor{red}{1}
&\textcolor{red}{0} & & & &
& 
  \\\hline
5 &28 &31 &\textcolor{blue}{17} &\textcolor{red}{7} &\textcolor{red}{1}
&\textcolor{red}{1} &\textcolor{red}{0} & & &
& 
  \\\hline
6 &64 &72 &\textcolor{blue}{45} &\textcolor{red}{19} &\textcolor{red}{7}
&\textcolor{red}{1} &\textcolor{red}{1} &\textcolor{red}{0} & &
&
  \\\hline
7 &136 &166 &108 &\textcolor{blue}{51} &\textcolor{red}{19}
&\textcolor{red}{7} &\textcolor{red}{1} &\textcolor{red}{1} &\textcolor{red}{0} &
&
  \\\hline
8 &289 &357 &254 &\textcolor{blue}{125} &\textcolor{red}{53}
&\textcolor{red}{19} &\textcolor{red}{7} &\textcolor{red}{1} &\textcolor{red}{1} &\textcolor{red}{0}
&
  \\\hline
9 &588 &757 &557 &\textcolor{blue}{302} &\textcolor{blue}{131}
&\textcolor{red}{53} &\textcolor{red}{19} &\textcolor{red}{7} &\textcolor{red}{1} &\textcolor{red}{1}
&\textcolor{red}{0}
 \\\hline
10 &1175 &1548 &1200 &675 &\textcolor{blue}{320}
&\textcolor{red}{133} &\textcolor{red}{53} &\textcolor{red}{19} &\textcolor{red}{7} &\textcolor{red}{1}
&\textcolor{red}{1} 
 \\\hline
11 &2293 &3100 &2482 &1479 &\textcolor{blue}{726}
&\textcolor{blue}{326} &\textcolor{red}{133} &\textcolor{red}{53} &\textcolor{red}{19} &\textcolor{red}{7}
&\textcolor{red}{1} 
  \\\hline
12 &4399 &6053 &5028 &3106 &\textcolor{blue}{1611}
&\textcolor{blue}{744} &\textcolor{red}{328} &\textcolor{red}{133} &\textcolor{red}{53} &\textcolor{red}{19}
&\textcolor{red}{7}
  \\\hline
13 &8267 &11620 &9910 &6373 &3422
&\textcolor{blue}{1663} &\textcolor{blue}{750} &\textcolor{red}{328} &\textcolor{red}{133} &\textcolor{red}{53}
&\textcolor{red}{19}
  \\\hline
14 &15325 &21855 &19173 &12713 &7098
&\textcolor{blue}{3557} &\textcolor{blue}{1681} &\textcolor{red}{752} &\textcolor{red}{328} &\textcolor{red}{133}
&\textcolor{red}{53} \\\hline
15 &27949 &40496 &36322 &24856 &14297
&\textcolor{blue}{7428} &\textcolor{blue}{3609} &\textcolor{blue}{1687} &\textcolor{red}{752} &\textcolor{red}{328}
&\textcolor{red}{133}
  \\\hline
16 &50306 &73846 &67720 &47539 &28216
&15061 &\textcolor{blue}{7564} &\textcolor{blue}{3627} &\textcolor{red}{1689} &\textcolor{red}{752}
&\textcolor{red}{328} 
  \\\hline
17 &89367 &132860 &124161 &89401 &54430
&29909 &\textcolor{blue}{15394} &\textcolor{blue}{7616} &\textcolor{blue}{3633} &\textcolor{red}{1689}
&\textcolor{red}{752}
  \\\hline
18 &156930 &235871 &224479 &165210 &103182
&58054 &\textcolor{blue}{30687} &\textcolor{blue}{15530} &\textcolor{blue}{7634} &\textcolor{red}{3635}
&\textcolor{red}{1689}
  \\\hline
19 &272424 &413879 &400257 &300837 &192109
&110702 &59786 &\textcolor{blue}{31021} &\textcolor{blue}{15582} &\textcolor{blue}{7640} 
&\textcolor{red}{3635}
  \\\hline
20 &468130 &717909 &705032 &539962 &352279
&207282 &114437 &\textcolor{blue}{60567} &\textcolor{blue}{31157} &\textcolor{blue}{15600} 
&\textcolor{red}{7642} 
  \\\hline
21 &796410 &$\!$1232463$\!$ &1227214 &956883 &636445
&382179 &215074 &\textcolor{blue}{116183} &\textcolor{blue}{60901} &\textcolor{blue}{31209} 
&\textcolor{blue}{15606}  
  \\\hline
22 &$\!$1342531$\!$&$\!$2094716$\!$&2113394&$\!$1674933$\!$&$\!$1134836$\!$
&694090 &398007 &218848 &$\!$\textcolor{blue}{116965}$\!$&\textcolor{blue}{61037} 
&\textcolor{blue}{31227} 
  \\\hline
23 &$\!$2243232$\!$&$\!$3527456$\!$ &3602086 &$\!$2899342$\!$&$\!$1997955$\!$
&$\!$1243836$\!$ &725457 &405910 &$\!$\textcolor{blue}{220597}$\!$ &$\!$\textcolor{blue}{117299}$\!$
&\textcolor{blue}{61089} 
  \\\hline
24 &$\!$3717405$\!$ &$\!$5887668$\!$ &6081317 &$\!$4965411$\!$&$\!$3477396$\!$
&$\!$2200438$\!$ &$\!$1304682$\!$ &741559 &$\!$\textcolor{blue}{409698}$\!$&$\!$\textcolor{blue}{221379}$\!$
&$\!$\textcolor{blue}{117435}$\!$ 
  \\\hline
25 &$\!$6111615$\!$ &$\!$9745995$\!$ &$\!$10173766$\!$ &$\!$8420331$\!$ &$\!$5986079$\!$
&$\!$3847540$\!$ &$\!$2316123$\!$ &$\!$1336712$\!$ &$\!$749501$\!$ &$\!$\textcolor{blue}{411448}$\!$
&$\!$\textcolor{blue}{221713}$\!$  
  \\\hline
\end{tabular}
%\end{longtable}
%\end{sidewaysfigure}
\end{center}
\end{scriptsize}
\caption{${\cal N}_{4d}=1$ multiplets at $U(1)_R$ charge $Q=0$}
\label{tab4d0}
\end{table}

\begin{align}
  &G_{2n+1,0}(q) \ \ \sim \ \  q^n\, ( 1 + q + 7 q^2 + 19 q^3 + 53 q^4 + 133 q^5 + 328 q^6 + 
   752 q^7 + 1689 q^8 + 3635 q^9 \notag \\
   & \ \ \ \ \ \ \ \ + 7642 q^{10} + 15608 q^{11} + 
   31235 q^{12} + 61115 q^{13} + 117513 q^{14} + 221927 q^{15} + 
   412778 q^{16} \notag \\
   & \ \ \ \ \ \ \ \ + 756372 q^{17} + 1367753 q^{18}  + 2441849 q^{19} + 
   4309132 q^{20} + 7520092 q^{21} \notag \\
   & \ \ \ \ \ \ \ \ + 12989357 q^{22} + 22216885 q^{23} + 
   37651970 q^{24} + 63252874 q^{25} + \ldots) \notag \\
   & - \  q^{2 n + 1}\, ( 
  2 + 8 q + 26 q^2 + 78 q^3 + 214 q^4 + 548 q^5 + 1330 q^6 + 
   3080 q^7 + 6872 q^8 + 14832 q^9 \notag \\
   & \ \ \ \ \ \ \ \  + 31102 q^{10} + 63574 q^{11} + 
   127020 q^{12} + 248590 q^{13} +  477504 q^{14} +\ldots) \notag \\
   &+q^{3 n + 1} \, ( 
  1 + 4 q + 19 q^2 + 61 q^3 + 187 q^4 + 503 q^5 + 1294 q^6 + 
   3113 q^7 \notag \\
   & \ \ \ \ \ \ \ \ + 7217 q^8  +16036 q^9 + 34584 q^{10} + \ldots) \notag \\ 
   &- \ q^{4 n + 2} \, ( 
  2 + 10 q + 38 q^2 + 124 q^3  + 364 q^4 + 978 q^5 + 2476 q^6 +\ldots) \notag \\
  & +\ q^{5 n + 2}  \, (1 + 4 q + 21 q^2 + 72 q^3+\ldots) \ + \ \ldots \co n \rightarrow \infty
  \label{tau5}
\end{align}
The first term linear in $q^n$ simply reproduces (\ref{simpleeven}) for $\tau_1^{Q=0}(q)$ whereas higher powers of $q^n$ allow to read off subleading $\tau^{Q=0}_{\ell \geq 2}(q)$ to certain order in $q$:
\begin{align}
\tau^{Q=0}_2(q) \eq &q \, ( 
  2 + 8 q + 26 q^2 + 78 q^3 + 214 q^4 + 548 q^5 + 1330 q^6 + 
   3080 q^7 + 6872 q^8+ 14832 q^9 \notag \\
   & \ \ \ \ \ \ \ \  + 31102 q^{10} + 63574 q^{11} + 
   127020 q^{12} + 248590 q^{13} +  477504 q^{14} +\ldots) \label{tau6} \\
\tau^{Q=0}_3(q) \eq &q \, ( 
  1 + 4 q + 19 q^2 + 61 q^3 + 187 q^4 + 503 q^5 + 1294 q^6 + 
   3113 q^7 \notag \\
   & \ \ \ \ \ \ \ \ + 7217 q^8  +16036 q^9 + 34584 q^{10} + \ldots) \label{tau7} \\
\tau^{Q=0}_4(q) \eq &q^2 \,  ( 
  2 + 10 q + 38 q^2 + 124 q^3  + 364 q^4 + 978 q^5 + 2476 q^6 +\ldots) \label{tau8} \\
\tau^{Q=0}_5(q) \eq &q^2  \, (1 + 4 q + 21 q^2 + 72 q^3+\ldots)  \label{tau9}
\end{align}
Determining higher order terms in the $\tau^{Q=0}_{\ell \geq 2}(q)$ would require ${\cal O}(q^{26})$ parts of (\ref{def:GSOed}), this is where we stopped the explicit evaluation.
% b14 c10 d6 e3
%The $q^{14}$ coefficient of the $q^{2n+1}$ trajectory, the $q^{10}$ coefficient of the $q^{3n+1}$ trajectory, the $q^{6}$ coefficient of the $q^{4n+2}$ trajectory and the $q^{3}$ coefficient of the $q^{5n+2}$ trajectory additionally require Omega's analytic all-order expression for the leading asymptotics.

Similarly, the $\llbracket 2n+1,2Q\rrbracket$ multiplicities up to level $q^{25}$ as tabulated in appendix \ref{app4} determine the associated $\tau_\ell^{2Q}(q)$ coefficients for low charges $Q$ to the following orders:
\begin{itemize}
\item $U(1)_R$ charge $Q=2$:
\begin{align}
\tau^{Q=2}_{2}(q)  \eq & q^{3} \, ( 
  2 + 11 q + 37 q^2 + 114 q^3 + 319 q^4 + 822 q^5 + 2000 q^6 + 
   4645 q^7 + 10354 q^8 \notag \\
   & \ \ \ \ \ \ \ \ + 22317 q^9 + 46702 q^{10} + 95210 q^{11} +  189656 q^{12} + \ldots) \notag \\
\tau^{Q=2}_{3}(q)  \eq & q^{3}  \, (2 + 8 q + 33 q^2 + 104 q^3 + 310 q^4 + 826 q^5 +   2093 q^6 + 4991 q^7 +  11454  q^8 + \ldots) \notag \\
\tau^{Q=2}_{4}(q)  \eq & q^{3} \, ( 1 + 5 q + 22 q^2 + 77 q^3 + 237 q^4 + 664 q^5 + \ldots) \notag \\
  \tau^{Q=2}_{5}(q)  \eq &  q^{4}\, (3 + 12 q + 49 q^2 + \ldots) \label{tau10} 
  \end{align}
%%%%%%%%%%%%
\item $U(1)_R$ charge $Q=4$:
\begin{align}
\tau^{Q=4}_{2}(q) \eq &q^{8} \, ( 
  2 + 14 q + 57 q^2 + 187 q^3 + 542 q^4 + 1438 q^5 + 3563 q^6 \notag \\
   & \ \ \ \ \ \ \ \ + 
   8376 q^7 +  18846 q^8 + 40866 q^9 + \ldots ) \notag \\
\tau^{Q=4}_{3}(q) \eq &q^{8} \, ( 2 + 14 q + 58 q^2 + 200 q^3 + 591 q^4 + 1612 q^5 + \ldots) \notag \\
\tau^{Q=4}_{4}(q) \eq & q^{8}\, ( 2 + 13 q + 53 q^2 + \ldots) 
\label{tau11} 
\end{align}
%%%%%%%%%%%%
\item $U(1)_R$ charge $Q=6$:
\begin{align}
\tau^{Q=6}_{2}(q) \eq & q^{15}\,  (2 + 14 q + 60 q^2 + 209 q^3 + 633 q^4 +\ldots) \notag \\
\tau^{Q=6}_{3}(q) \eq &q^{15}\, (2 + 14 q + 64 q^2 +\ldots) \label{tau12}
\end{align}
\end{itemize}
Note that the analytic result (\ref{asympmuloddeven}) for $\tau_1^{2Q}(q)$ was used as an extra input, in addition to the explicit results for the first 25 mass level, to make a few more orders of the subleading $\tau_{\ell \geq 2}^{2Q}(q)$ accessible.

Also in the $\llbracket 2n,2Q+1\rrbracket$ sector, we can use the data from appendix \ref{app4} to expand the subleading trajectories $\tau_{\geq 2}^{2Q+1}(q)$:
\begin{itemize}
\item $U(1)_R$ charge $Q=1$:
\begin{align}
\tau_2^{Q=1}(q) \eq & 1 + 4 q + 15 q^2 + 50 q^3 + 143 q^4 + 379 q^5 + 947 q^6 + 
   2244 q^7 + 5103 q^8 + 11196 q^9 \notag \\
   & \ \  + 23804 q^{10} + 49252 q^{11} + 
   99465 q^{12} + 196522 q^{13} + 380719 q^{14} + \ldots  \notag \\
\tau_3^{Q=1}(q) \eq &1 + 5 q + 22 q^2 + 70 q^3 + 212 q^4 + 568 q^5 + 1458 q^6 + 
   3496 q^7 \notag \\
   & \ \ + 8093 q^8 + 17936 q^9 + \ldots \notag \\
\tau_4^{Q=1}(q) \eq &1 + 6 q + 24 q^2 + 83 q^3 + 252 q^4 + 698 q^5 + \ldots \notag \\
\tau_5^{Q=1}(q) \eq &1 + 6 q + 25 q^2 + \ldots
\label{tau13}
\end{align}
%%%
\item $U(1)_R$ charge $Q=3$:
\begin{align}
\tau_2^{Q=3}(q) \eq & q^ 4 \, (1 + 9 q + 37 q^2 + 120 q^3 + 347 q^4 + 922 q^5 + 
   2287 q^6 +   5385 q^7 +   12142 q^8 \notag \\
   & \ \ + 26395 q ^9 + 55605 q^{10} + 
   113973 q^{11} + \ldots) \notag \\
\tau_3^{Q=3}(q) \eq &q^4 \, (4 + 17 q + 68 q^2 + 208 q^3 + 603 q^4 + 1573 q^5 + 
   3919 q^6 + 9195 q^7 + \ldots ) \notag \\
\tau_4^{Q=3}(q) \eq &q^3 \, (1 + 7 q + 28 q^2 + 99 q^3 + 304 q^4 + 851 q^5 + \ldots ) \notag \\
\tau_5^{Q=3}(q) \eq &q^3 \, (2 + 9 q  + 38 q^2+ \ldots) \label{tau14}
\end{align}
%%%
\item $U(1)_R$ charge $Q=5$:
\begin{align}
\tau_2^{Q=5}(q) \eq & q^{10}\, (1 + 9 q + 43 q^2 + 151 q^3 + 462 q^4 + 1277 q^5 + 
   3264 q^6 + 7865 q^7 + \ldots) \notag \\
\tau_3^{Q=5}(q) \eq &q^{10} \, (4 + 20 q + 89 q^2 + 292 q^3 + \ldots) \notag \\
\tau_4^{Q=5}(q) \eq &q^9 \, (1 + 9 q + \ldots ) \label{tau15}
\end{align}
%%%
\item $U(1)_R$ charge $Q=7$:
\begin{align}
\tau_2^{Q=7}(q) \eq & q^{18}\, (1 + 9 q  + \ldots )
\label{tau16}
\end{align}
\end{itemize}
Again, the set of accessible coefficients in $\tau^{2Q+1}_{\ell \geq 2}(q)$ could be slightly improved by making use of the $\tau_1^{2Q+1}(q)$ expression (\ref{mainferm}) to order $q^{25}$. 

Note that for all values of the $U(1)_R$ charge $Q$ considered here, the leading $q$ powers of the $\tau_\ell^Q(q)$ at fixed $Q$ hardly vary with $\ell$ (at $Q=2$, for instance, we can read off $\tau_1^{2},\tau_2^{2},\tau_3^{2}, \tau_4^{2} \sim {\cal O}(q^3)$ and $\tau_5^{2} \sim {\cal O}(q^4)$ from (\ref{tau10})). In particular, the approximate agreement of the leading $q$ powers of $\tau_1(q)$ and $\tau_2(q)$ supports our claim in the introduction that half of the nonzero multiplicities exactly match with the stable patterns.

\section{Spectra in compactifications with 8 supercharges}
\label{sec:6d}

In six dimensional Minkowski space, the minimal realization of SUSY involves eight supercharges. They form two left- handed Weyl spinors of $SO(6)$ which are related through an $SU(2)_R$ R symmetry. Our notation for such minimally supersymmetric theories in $d=6$ is ${\cal N}_{6d}=(1,0)$. Superstring compactification subject to ${\cal N}_{6d}=(1,0)$ SUSY are described by a universal SCFT sector with $c=6$ and ${\cal N}_{2d}=4$ SUSY on the worldsheet, see subsection \ref{sec:42} for details. In addition, the SCFT introduces $SO(5)$ quantum numbers for the massive string states through a six dimensional spacetime sector for which the methods of subsections \ref{sec:highdim} and \ref{sec:highdimm} are applicable.

The fundamental multiplet of ${\cal N}_{6d}=(1,0)$ theories consists of 8+8 states
\bea
Z(\CN_{6d}=(1,0)) \ \ := \ \  [1,0]  \ + \  [2]_R \ + \  [1]_R \,  [0,1] ~.  \label{fundmassive6d}
\eea
where $[p]_R$ is the character of the $p+1$ dimensional representation of $SU(2)_R$. Generic multiplets follow through the tensor product with some $SO(5) \times SU(2)_R$ representation with little group quantum numbers $[n_1,n_2]$ and $R$ symmetry content $[k]_R$. This leads to the general supercharacter
\beq
\llbracket n_1,n_2; p\rrbracket \ \ := \ \ Z(\CN_{6d}=(1,0)) \cdot [p]_R \, [n_1,n_2] \ . \label{ch6d}
\eeq
The partition function capturing the universal spectrum of six dimensional ${\cal N}_{6d}=(1,0)$ compactifications is obtained thorugh a GSO projected product of internal $\chi_{\ldots}^{{\cal N}_{2d}=4,c=6}(q;r)$ characters (with $SU(2)_R$ fugacity $r$) defined by (\ref{intNSN4}) as well as (\ref{intRN4}) and $SO(5)$ spacetime characters (\ref{BnNS}) and (\ref{BnR}). The GSO projection removes half odd integer mass leves from the NS sector and enforces the R spin field to be a left handed $SO(6)$ spinor, therefore:
\begin{align} \label{NSandR6d1}
\chi^{{\cal N}_{6d}=(1,0)}(q;\vec y,r) &\eq \chi^{{\cal N}_{6d}=(1,0)} _{\te{NS}} \mid_{\GSO} (q;\vec y,r) \ + \ \chi^{{\cal N}_{6d}=(1,0)}_{\te{R}} \mid_{\GSO} (q;\vec y,r) \notag \\
\chi^{{\cal N}_{6d}=(1,0)} _{\te{NS}} \mid_{\GSO} (q;\vec y,r) &\eq
\frac{1}{2} \, q^{-\frac{1}{2}} \, \big[ \, \chi_{\te{NS}}^{SO(5)}(q;\vec y)  \,  \chi^{{\cal N}_{2d}=4,c=6}_{\te{NS},h=0,\ell=0}(q;r)  \notag \\
& \ \ \ \ \ \ \ \ \ \ \ \ \ \ \ - \ \chi_{\te{NS}}^{SO(5)}(e^{2\pi i}q;\vec y)  \,  \chi^{{\cal N}_{2d}=4,c=6}_{\te{NS},h=0,\ell=0}(e^{2\pi i}q;r) \, \big] \notag \\
\chi^{{\cal N}_{6d}=(1,0)}_{\te{R}} \mid_{\GSO} (q;\vec y,r) &\eq \frac{1}{2} \,   \chi_{\te{R}}^{SO(5)}(q;\vec y)  \, \chi^{{\cal N}_{2d}=4,c=6}_{\te{R},h=1/4,\ell=1/2}(q;r)~.
\end{align}

The power series expansion of (\ref{NSandR6d1}) starts as\footnote{Again, there is a subtlety in applying (\ref{NSandR6d1}) to the massless R sector, see the footnote before \eref{def:GSOed}. However, this can be fixed easily: one can simply add to it $\frac{1}{2}(y_1-y_1^{-1})(y_2-y_2^{-1})(r-r^{-1})$ to get the correct massless character in R sector.}
\begin{align}
\chi^{{\cal N}_{6d}=(1,0)}(q;\vec y,r)&\eq \underbrace{\left( \, y^2_1 + y_1^{-2} + y^2_2 + y_2^{-2} \ + \ \frac{1}{2}[1]_{r} {\prod_{i=1}^2 (y_i +y_i^{-1}) } \, \right) \, q^0}_{8 \ \te{massless states}}  \ + \ \underbrace{ \llbracket 1,0;0\rrbracket \, q}_{80 \ \te{states at level} \ 1}  \notag \\
&  \hskip-2cm + \ \underbrace{\big( \, \llbracket 2,0;0\rrbracket \ + \  \llbracket 0,2;0\rrbracket \ + \ \llbracket 0,1;1\rrbracket\, \big) \, q^2}_{512 \ \te{states at level} \ 2} \ + \ \big( \, \llbracket 3,0;0\rrbracket \, + \, 2 \, \llbracket 1,0;0\rrbracket \, + \, \llbracket 0,0;0\rrbracket \notag \\
 &  \hskip-0.7cm + \,  \llbracket 1,2;0 \rrbracket \, + \,  \llbracket 0,2;0\rrbracket \, + \, \llbracket 0,0;2\rrbracket \, + \, 2\, \llbracket 1,1;1 \rrbracket \, + \,  \llbracket 0,1;1 \rrbracket \, \big) \, q^3 \ + \ {\cal O}(q^4)  \ .
 \label{part6dim}
\end{align}
The $q^{\leq 6}$ coefficients are listed in table \ref{table6,2}, further information on the particle content up to level 25 is tabulated in appendix \ref{app6}.

\bigskip
\noindent
\begin{table}[htdp]
\begin{tabular}{|l|l|}\hline  $\ap m^2$ &representations of ${\cal N}_{6d}=(1,0)$ super Poincar\'e \\ \hline \hline   
  1 & $\llbracket 1,0;0 \rrbracket $  \\\hline
   2 & $\llbracket 2,0;0\rrbracket \, + \,  \llbracket 0,2;0\rrbracket \, + \, \llbracket 0,1; 1\rrbracket $  \\\hline
   3 & $\llbracket 3,0;0\rrbracket \, + \, 2 \, \llbracket 1,0;0\rrbracket\, + \, \llbracket 0,0;0\rrbracket \, + \,  \llbracket 1,2;0 \rrbracket \, + \,  \llbracket 0,2;0\rrbracket \, + \, \llbracket 0,0;2\rrbracket \, + \, 2\, \llbracket 1,1;1 \rrbracket \, + \,  \llbracket 0,1;1 \rrbracket $ \\ \hline
   4 & $ \llbracket 4,0;0\rrbracket \, + \, 3\, \llbracket 2,0;0\rrbracket \, + \, 2\, \llbracket 1,0;0\rrbracket \, + \, 2 \, \llbracket 0,0;0\rrbracket \, + \,  \llbracket 2,2;0\rrbracket   \, + \, 2 \, \llbracket 1,2;0 \rrbracket  \, + \, 4 \, \llbracket 0,2;0\rrbracket $
   \\
   &$ \, + \, 2 \, \llbracket 1,0;2 \rrbracket  \, + \,  \llbracket 0,2;2 \rrbracket  \, + \, 3 \, \llbracket 1,1;1 \rrbracket   \, + \, 4 \, \llbracket 0,1;1 \rrbracket   \, + \, 2 \, \llbracket 2,1;1 \rrbracket      $ 
   \\\hline        
   5 &$\llbracket 5,0;0 \rrbracket \, + \, 3 \, \llbracket 3,0;0\rrbracket \, + \, 4 \, \llbracket 2,0;0 \rrbracket\, + \, 9 \, \llbracket 1,0;0 \rrbracket \, + \, 3 \, \llbracket 0,0;0 \rrbracket \, + \, \llbracket 3,2;0 \rrbracket \, + \, 2 \, \llbracket 2,2;0 \rrbracket $
     \\
     &$\, + \, 7 \, \llbracket 1,2;0 \rrbracket \, + \, 6 \, \llbracket 0,2;0 \rrbracket \, + \,  \llbracket 0,4;0 \rrbracket \, + \, 3\, \llbracket 2,0;2 \rrbracket \, + \, 3 \, \llbracket 1,0;2\rrbracket \, + \, 3 \, \llbracket 0,0;2 \rrbracket \, + \, \llbracket 1,2;2 \rrbracket $
     \\
     &$ \, + \, 3 \, \llbracket 0,2;2 \rrbracket \, + \, 2 \, \llbracket 3,1;1\rrbracket \, + \, 4 \, \llbracket 2,1;1 \rrbracket \, + \, 9 \, \llbracket 1,1;1 \rrbracket \, + \, 8 \, \llbracket 0,1;1 \rrbracket \, + \,  \llbracket 1,3;1 \rrbracket \, + \, 4 \, \llbracket 0,3;1 \rrbracket  $  \\
     &$\, + \,  \llbracket 0,1;3\rrbracket$
     \\\hline
6 &$\llbracket6,0;0\rrbracket \, + \,
\llbracket4,2;0\rrbracket \, + \, 
2\, \llbracket4,1;1\rrbracket \, + \, 
3\, \llbracket4,0;0\rrbracket \, + \, 
2\, \llbracket3,2;0\rrbracket \, + \, 
4\, \llbracket3,1;1\rrbracket \, + \, 
3\, \llbracket3,0;2\rrbracket $
     \\
     &$ \, + \, 
5\, \llbracket3,0;0\rrbracket \, + \, 
\llbracket2,3;1\rrbracket \, + \, 
\llbracket2,2;2\rrbracket \, + \, 
8\, \llbracket2,2;0\rrbracket \, + \, 
12\, \llbracket2,1;1\rrbracket \, + \, 
4\, \llbracket2,0;2\rrbracket \, + \, 
14\, \llbracket2,0;0\rrbracket $
     \\
     &$ \, + \, 
\llbracket1,4;0\rrbracket \, + \, 
5\, \llbracket1,3;1\rrbracket \, + \, 
6\, \llbracket1,2;2\rrbracket \, + \, 
13\, \llbracket1,2;0\rrbracket \, + \, 
2\, \llbracket1,1;3\rrbracket \, + \, 
23\,\llbracket1,1;1\rrbracket \, + \, 
9\, \llbracket1,0;2\rrbracket $
     \\
     &$ \, + \, 
12\, \llbracket1,0;0\rrbracket \, + \, 
4\, \llbracket0,4;0\rrbracket \, + \, 
9\, \llbracket0,3;1\rrbracket \, + \, 
9\, \llbracket0,2;2\rrbracket \, + \, 
19 \,\llbracket0,2;0\rrbracket \, + \, 
3\, \llbracket0,1;3\rrbracket $
     \\
     &$ \, + \, 
18 \, \llbracket0,1;1\rrbracket \, + \, 
4\, \llbracket0,0;2\rrbracket \, + \, 
8\, \llbracket0,0;0\rrbracket
$\\\hline     
\end{tabular}
\caption{${\cal N}_{6d}=(1,0)$ multiplets occurring up to mass level 6}
\label{table6,2}
\end{table}

%%%%%%%%%%%%%%%%%%%%%%%

\subsection{The total number of states at a given mass level}

In this subsection, we compute the total number of states present at a given mass level through the unrefined partition function, \ie~ by setting the fugacities $y_1, y_2$ and $r$ in \eref{part6dim} to unity.  The total number of states $N_m$ at the mass level $m$ can be read off from the coefficient of $q^m$ in the power series of $\chi^{{\cal N}_{6d}=(1,0)} (q; \{ y_i=1, r=1\})$.

We follow the analysis presented in subsection \ref{sec:numberstates4SUSYs}.  The unrefined partition function is given by
\bea
\chi^{{\cal N}_{6d}=(1,0)} (q; \{ y_i=1, r=1\})
&= 2  \chi^{{\cal N}_{6d}=(1,0)}_{\Ra} \mid_{\GSO} (q; y=1, s=1) \nn \\
&= \chi_{\te{R}}^{SO(5)}(q;\{y_i=1\})  \, \chi^{{\cal N}_{2d}=4,c=6}_{\te{R},h=1/4,\ell=1/2}(q;r=1) \nn \\
%&= \chi_{\te{R}}^{SO(3)}(q;\{y=1\})^2  \, \chi^{{\cal N}_{2d}=4,c=6}_{\te{R},h=1/4,\ell=1/2}(q;r=1) \nn \\
&= q^{-1/8} \frac{\vartheta_2(1, q)^4}{\eta(q)^9}\left[  1- 2 i q^{1/8} \mu \left(1/2,\tau \right) \right]~. \label{exactunref6d}
\eea
Indeed, the power series of $\chi^{{\cal N}_{6d}=(1,0)} (q; \{ y_i=1, r=1\})$ in $q$ reproduces the numbers presented in the second column of \tref{numberofstates}.  Note that $\chi^{{\cal N}_{6d}=(1,0)} (q; \{ y_i=1, r=1\})$ is {\it not} a modular form, since the Appell-Lerch sum is a mock modular form and it is not added by a suitable non-holomorphic component to be modular.

\subsubsection*{The number of states at each mass level and its asymptotics} 
The number of states at the mass level $m$ can also be computed from
\bea
N_m = \frac{1}{2 \pi i} \oint_{\CC} \frac{\ud q}{q^{m+1}}~\chi^{{\cal N}_{6d}=(1,0)} (q; \{ y_i=1, r=1\})~, \label{numstates6d}
\eea
where $\CC$ is a contour around the origin. 

Now let us compute an asymptotic formula for the number of states $N_m$ at a mass level $m$ when $m \rightarrow \infty$.   We focus on the limit $q \rightarrow 1^{-}$ and proceed in a similar way to subsection \ref{sec:numberstates4SUSYs}.   

Let us first examine the leading behaviour of $\mu \left(1/2,\tau \right)$ as $q \rightarrow 1^-$ or $\tau \rightarrow 0$.  Using the second point of Proposition 1.5 of \cite{Zwegers2002}, we find that
\bea \label{invtauAL}
\frac{1}{\sqrt{- i \tau}} \mu \left(\frac{1}{2 \tau}, -\frac{1}{\tau} \right) + \mu \left( \frac{1}{2} ,\tau \right) =\frac{1}{2 i}~.
\eea
Let us consider $ \mu \left(\frac{1}{2 \tau}, -\frac{1}{\tau} \right)$ as $q \rightarrow 1^-$ or equivalently $\tau =i \epsilon$ as $\epsilon \rightarrow 0^+$.   It follows from the definition of Appell-Lerch sum that
\bea
 \mu \left(\frac{1}{2 \tau}, -\frac{1}{\tau} \right)  &= - \frac{e^{i \pi/ (2 \tau)}}{\vartheta_1(e^{2 \pi i/(2 \tau)},e^{-2 \pi i/\tau})} \sum_{m \in \BZ} (-1)^m \frac{e^{-i \pi m^2/ \tau  }}{1-e^{-2 \pi i m/\tau+ \pi i/\tau }} \nn \\
 &\sim - \frac{e^{\pi/ (2\epsilon)} }{-i e^{\pi/(4 \epsilon)}} \times(- 2e^{-\pi/ \epsilon} )~, \qquad \tau = i \epsilon,~ \epsilon \rightarrow 0^+ \nn\\
&= 2i \exp \left( -\frac{3\pi}{4\epsilon} \right) ~,
\eea
where in the second `equality' only $m=0, 1$ in the infinite sum contribute to the leading behaviour and we have used the fact that $\vartheta_1(e^{2 \pi i/(2 \tau)},e^{-2 \pi i/\tau}) = -i e^{\pi/(4 \epsilon)}$, as $\tau = i \epsilon, ~ \epsilon \rightarrow 0^+$.  Hence, to the leading order, one can neglect the first term in \eref{invtauAL} in comparison with $1/(2i)$ on the right hand side and so
\bea
 \mu \left( \frac{1}{2} ,\tau \right) \sim \frac{1}{2 i}~, \qquad q \rightarrow 1^-~.
\eea

\begin{figure}[htbp]
\begin{center}
\includegraphics[scale=0.8]{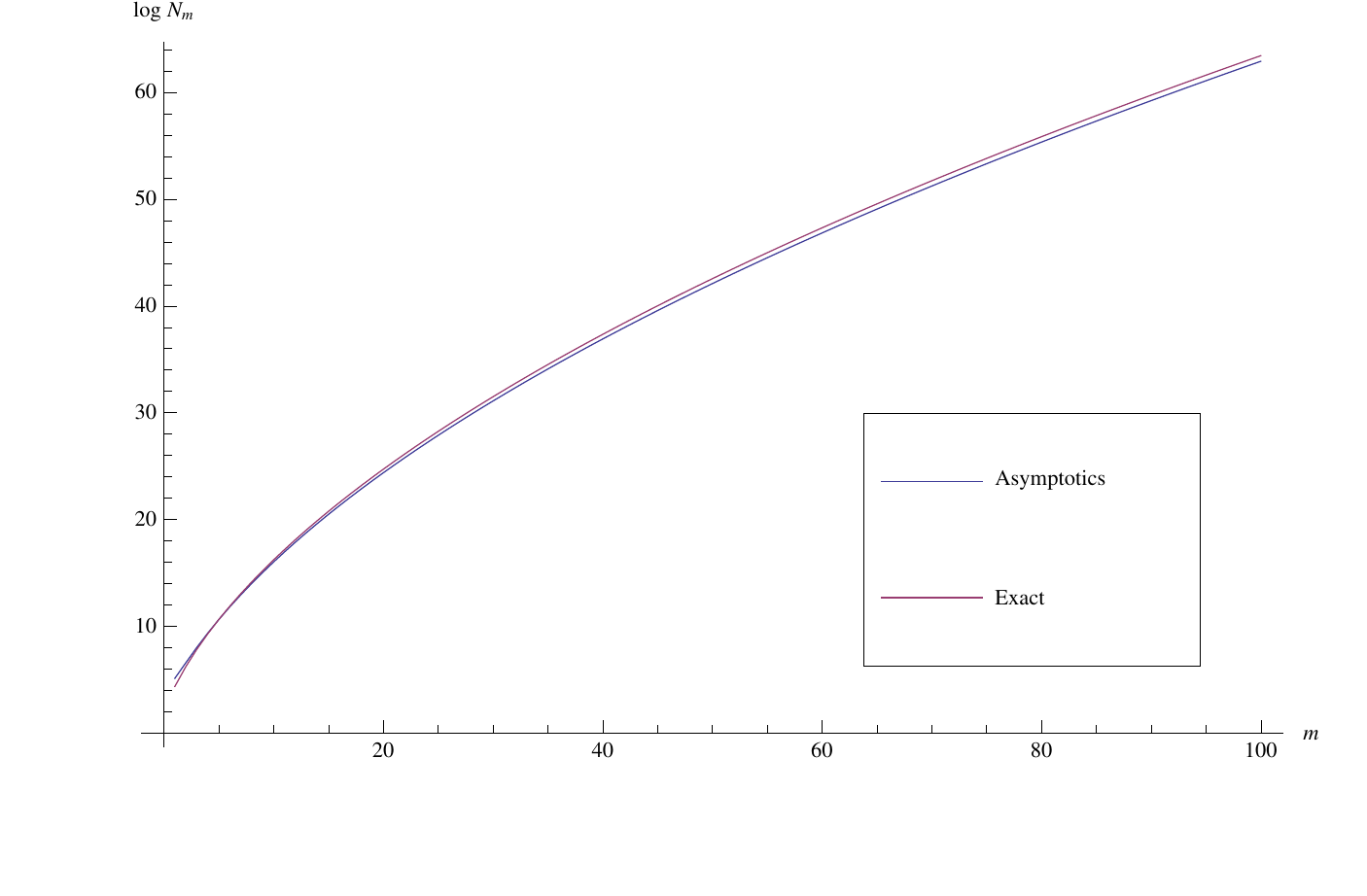}
\vspace{-1.2cm}
\caption{The plot of the exact and asymptotic values of  $\log N_m$ against the mass level $m$ for the case of 8 supercharges.}
\label{fig:compareexactasymp6d}
\end{center}
\end{figure}

Therefore it follows from \eref{exactunref6d} that, as $q \rightarrow 1^-$,
\bea
\chi^{{\cal N}_{6d}=(1,0)} (q; \{ y_i=1, r=1\}) &\sim q^{-1/8} \frac{\vartheta_2(1, q)^4}{\eta(q)^9}\left(  1- q^{1/8}  \right) \nn \\
&\sim  (2 \pi)^{-5/2} (1-q^{1/8}) (1-q)^{5/2} \exp \left( - \frac{3 \pi^2}{2 \log q} \right)~,
\eea
where we have used \eref{etaapprox} and \eref{theta2approx}.  Hence, as $m \rightarrow \infty$,
\bea
N_m \sim (2 \pi)^{-5/2} \frac{1}{2 \pi i} \oint_{\CC} \frac{\ud q}{q} ~(1-q^{1/8}) (1-q)^{5/2} \exp \left( - \frac{3 \pi^2}{2 \log q} - m \log q \right)~.
\eea
The saddle point is at $q_0 = \exp\left( - \pi \sqrt{3}/\sqrt{2 m} \right)$ and the steepest descent direction is the imaginary direction in $q$.  We proceed in a similar way to \eref{asympNm4d} by writing $q=q_0e^{i\theta}$ and using Laplace's method to obtain
\bea
N_m &\sim (2 \pi)^{-5/2} (1-q_0^{1/8}) (1-q_0)^{5/2}  e^{\pi \sqrt{6m}} \frac{1}{2 \pi} \int_{-\infty}^\infty \ud \theta \exp \left( -\frac{1}{\pi} \sqrt{\frac{2}{3}} m^{3/2} \theta^2\right) \nn \\
&\sim \frac{9\pi}{2^{17/2}} m^{-5/2} \exp \left( \pi \sqrt{6m} \right) ~, \qquad m \rightarrow \infty~. \label{asympNm6d}
\eea
The plot of the exact and asymptotic values for $N_m$ against $m$ is depicted in \fref{fig:compareexactasymp6d}.

\subsection{The GSO projected NS and R sectors}
\subsubsection*{The NS sector} 
From \eref{NSandR6d}, the partition function of the GSO projected NS sector is 
\bea
\chi^{\CN_{6d}=(1,0)}_{\NS} \mid_{\GSO} (q; y, s) = \sum_{k_1,k_2, p=0}^\infty [2k_1]_{y_1} [2k_2]_{y_2}  [2 p]_r   ~F^{\NS}_{k_1,k_2,p} (q)~,
\eea
where the function $F^{\NS}_{k,p} (q)$ is given by 
\bea
F^{\NS}_{k_1,k_2,p} (q) &=  (q;q)^{-9}_\infty (1-q)   q^{\frac{1}{2}p^2+p-1} \nn \\
& \quad \times \sum_{\vec n \in \BZ^2_+} \sum_{\vec m \in \BZ^2_{\geq 0}} 
\prod_{A=1}^2  (-1)^{n_A+1} (1-q^{n_A}) q^{\frac{1}{2}m_A^2+{n_A \choose 2}} (q^{n_A|k_A-m_A|} -q^{n_A(k_A+m_A+1)})  \nn \\
& \quad \times \frac{1}{2}  \Bigg[ \frac{(1-q^{p+\frac{1}{2}}) \vartheta_3(1, q) }{ (1+q^{p-\frac{1}{2}})(1+q^{p+\frac{3}{2}}) } \prod_{A=1}^2 (1-q^{m_A+\frac{1}{2}})  \nn \\
& \hspace{3cm} + (-1)^{m_1^2+m_2^2+p^2} \frac{(1+q^{p+\frac{1}{2}}) \vartheta_4(1, q) }{ (1-q^{p-\frac{1}{2}})(1-q^{p+\frac{3}{2}}) } \prod_{A=1}^2 (1+q^{m_A+\frac{1}{2}})\Bigg]~.
\eea

\paragraph{Asymptotics.} This expression can be simplified further in the asymptotic limit $k_1, k_2 \rightarrow \infty$.  Using \eref{sumn4dN1}, we have
\bea
& \sum_{\vec n \in \BZ^2_+} \prod_{A=1}^2  (-1)^{n_A+1} (1-q^{n_A}) q^{{n_A \choose 2}} (q^{n_A|k_A-m_A|} -q^{n_A(k_A+m_A+1)})  \nn \\
&\sim  (1-q)^2 \prod_{A=1}^2 \frac{ q^{k_A} \left(1-q^{2 k_A+2}\right)}{\left(1+q^{k_A}\right)^4} \left \{  q^{-m_A} \left(1-q^{2 m_A+1}\right) \right \}~,
\eea
and using \eref{summ4dN1} we have
\bea
  \sum_{\vec m \in \BZ^2_{\geq 0}}  \prod_{A=1}^2 q^{\frac{1}{2} m_A^2- m_A} \left(1-q^{m_A+\frac{1}{2}}\right) \left(1-q^{2 m_A+1}\right)
&= q^{-1} (1-q)^2 \vartheta_3 (1, q)^2~,  \\
  \sum_{\vec m \in \BZ^2_{\geq 0}} \prod_{A=1}^2  (-1)^{m_A^2} q^{\frac{1}{2} m_A^2- m_A} \left(1+q^{m_A+\frac{1}{2}}\right) \left(1-q^{2 m_A+1}\right)
&= q^{-1} (1-q)^2 \vartheta_4 (1, q)^2~, 
\eea
Therefore we arrive at an asymptotic formula for $F^{\NS}_{k_1,k_2,p} (q)$ when $k_1, k_2 \rightarrow \infty$:
\bea
F^{\NS}_{k_1,k_2,p} (q) &\sim \frac{1}{2} (q;q)^{-9}_\infty (1-q)^5 q^{\frac{1}{2} p^2+p+k_1+k_2-2} \Bigg[ \frac{(1-q^{p+\frac{1}{2}})  }{ (1+q^{p-\frac{1}{2}})(1+q^{p+\frac{3}{2}}) } \vartheta_3(1, q)^3 \nn \\
& \hspace{3cm} + (-1)^{p^2} \frac{(1+q^{p+\frac{1}{2}})}{ (1-q^{p-\frac{1}{2}})(1-q^{p+\frac{3}{2}}) }  \vartheta_4(1,q)^3  \Bigg]~, \qquad k_1, k_2 \rightarrow \infty~.
\eea

\subsubsection*{The R sector}  
The partition function of the GSO projected R sector is 
\bea
\chi^{\CN_{6d}=(1,0)}_{\Ra} \mid_{\GSO} (q; y, s) = \sum_{k_1,k_2,p=0}^\infty  [2k_1+1]_{y_1} [2k_2+1]_{y_2}  [2 p+1]_r   ~F^{\Ra}_{k_1,k_2,p} (q)~,
\eea
where $F^{\Ra}_{k_1,k_2,p} (q)$ is given by
%{\small
%\bea
%F^{\Ra}_{k_1,k_2,p} (q) &= (q;q)^{-9}_\infty (1-q) q^{\frac{1}{2} p^2 + \frac{3}{2} p - \frac{3}{8}}  \nn \\
%& \qquad \times \sum_{\vec n \in \BZ^2_+} \sum_{\vec m \in \BZ^2_{\geq 0}}
%\prod_{A=1}^2 (-1)^{n_A} (1-q^{n_A}) q^{ \frac{1}{2}(m_A+\frac{1}{2})^2+{n_A \choose 2}}  (q^{n_A |k_A-m_A|}- q^{n_A(k_A+m_A+2)} ) \nn \\
%& \qquad \times \frac{1}{2}  \frac{(1-q^{p+1}) \vartheta_2 (1, q)}{(1+q^p)(1+q^{p+2})} \prod_{A=1}^2 (1-q^{m_A+1})~.
%\eea}
{\small
\bea
&F^{\Ra}_{k_1,k_2,p} (q) = (q;q)^{-9}_\infty (1-q) q^{\frac{1}{2} p^2 + \frac{3}{2} p - \frac{3}{8}} \times \frac{1}{2}  \frac{(1-q^{p+1}) \vartheta_2 (1, q)}{(1+q^p)(1+q^{p+2})}  \nn \\
& \times \sum_{\vec n \in \BZ^2_+} \sum_{\vec m \in \BZ^2_{\geq 0}}
\prod_{A=1}^2 (-1)^{n_A} (1-q^{n_A}) q^{ \frac{1}{2}(m_A+\frac{1}{2})^2+{n_A \choose 2}}  (q^{n_A |k_A-m_A|}- q^{n_A(k_A+m_A+2)} ) (1-q^{m_A+1}) 
\eea}
Similarly to the NS sector, an asymptotic formula for $F^{\NS}_{k_1,k_2,p} (q)$ when $k_1, k_2 \rightarrow \infty$ is given by
\bea
F^{\Ra}_{k_1,k_2,p} (q) &\sim \frac{1}{2} (q;q)^{-9}_\infty (1-q)^5 q^{\frac{1}{2} p^2 + \frac{3}{2} p +k_1+k_2- \frac{3}{8}}  \frac{(1-q^{p+1})}{(1+q^p)(1+q^{p+2})}  \vartheta_2 (1, q)^3~.
\eea

\subsection[Multiplicities of representations in the $\CN_{6d} = (1,0)$ partition function]{Multiplicities of representations in the $\bm{\CN_{6d} = (1,0)}$ partition function}
Combining the contributions from the NS and R sectors, we have
\bea
\chi^{{\cal N}_{6d}=(1,0)}(q;\vec y,r) &= \chi^{{\cal N}_{6d}=(1,0)} _{\te{NS}} \mid_{\GSO} (q;\vec y,r) \ + \ \chi^{{\cal N}_{6d}=(1,0)}_{\te{R}} \mid_{\GSO} (q;\vec y,r) \nn \\
&= \sum_{k_1,k_2,p=0}^\infty \Big( [2k_1]_{y_1} [2k_2]_{y_2}  [2 p]_r   ~F^{\NS}_{k_1,k_2,p} (q)  \nn \\
& \qquad \qquad +  [2k_1+1]_{y_1} [2k_2+1]_{y_2}  [2 p+1]_r   ~F^{\Ra}_{k_1,k_2,p} (q) \Big)~. \label{chi6d10sumboth}
\eea
Making SUSY manifest amounts to rewriting the partition function as
\bea
\chi^{{\cal N}_{6d}=(1,0)}(q;\vec y,r) &= \sum_{n_1, n_2 \geq 0} \sum_{p=0}^\infty \lb n_1, n_2; p \rb ~ G_{n_1,n_2,p} (q) ~,
\label{GGG6}
\eea
and the aim is to compute explicitly a {\it multiplicity generating function} $G_{n_1,n_2,p} (q)$.

Before proceeding further, we observe the selection rule
\bea
G_{n_1, 2n_2, 2p+1}(q) =0~, \qquad G_{n_1, 2n_2+1, 2p} (q)= 0~. \label{6diden1}
\eea
%It follows from \eref{chi6d10sumboth} that odd (respectively even) values of $p$ only come with a product of two representations with both odd (resp. even) $k_1$ and $k_2$.  According to \eref{decompeven} and \eref{decompodd}, a decomposition of product two representations with both odd (resp. even) $k_1$ and $k_2$ contains only spin (resp. non-spin) representations of $SO(5)$.  In other words, a spin (resp. non-spin) representation only comes with an even (resp. odd) value of $p$, and hence \eref{6diden1} follows.
It follows from \eref{chi6d10sumboth} that $[k_1]_{y_1} [k_2]_{y_2}  [p]_r$ with odd (respectively even) values of $p$ only enter with a product of two representations with both odd (resp. even) $k_1$ and $k_2$.  According to \eref{decompeven} and \eref{decompodd}, the product $[k_1]_{y_1} [k_2]_{y_2}$ with both odd (resp. even) $k_1$ and $k_2$ decomposes into only spin (resp. non-spin) representations of $SO(5)$.  In other words, a spin (resp. non-spin) representation only comes with an odd (resp. even) value of $p$, and hence \eref{6diden1} follows.

The multiplicity of $\lb n_1,n_2;p \rb$ appearing in $\chi^{{\cal N}_{6d}=(1,0)}(q; \vec y,r)$ can be determined as follows:
\bea
G_{n_1,n_2,p} (q) &= \int \ud \mu_{SU(2)} (r) [p]_r \int \ud \mu_{SO(5)}(\vec y) [n_1, n_2]_{\vec y} ~\frac{\chi^{{\cal N}_{6d}=(1,0)}(q;\vec y,r) }{Z(\CN_{6d}=(1,0)) (\vec y,r)}~, \nn \\
&= G^{\NS}_{n_1,n_2,p} (q) + G^{\Ra}_{n_1,n_2,p} (q)~, \label{sumboth6d}
%&= \int \ud \mu_{SU(2)} (r) [p]_r  \Delta_{n_1+\frac{1}{2}n_2, \frac{1}{2}n_2; 2k_1, 2k_2}  \frac{\chi^{{\cal N}_{6d}=(1,0)}(q;\vec y,r) }{Z(\CN_{6d}=(1,0)) (\vec y,r)} \nn \\
\eea
where
\bea
& G^{\NS}_{n_1,n_2,p} (q) =  \int \ud \mu_{SU(2)} (r) [p]_r  ~ \int \ud \mu_{SO(5)}(\vec y) [n_1, n_2]_{\vec y} \times \nn \\
& \hspace{3cm} \sum_{k_1,k_2,p' \geq 0}  \frac{[2k_1]_{y_1} [2k_2]_{y_2}  [2 p']_r}{{Z(\CN_{6d}=(1,0)) (\vec y,r)}}   ~F^{\NS}_{k_1,k_2,p'} (q)~, \\
& G^{\Ra}_{n_1,n_2,p} (q) = \int \ud \mu_{SU(2)} (r) [p]_r  ~  \int \ud \mu_{SO(5)}(\vec y) [n_1, n_2]_{\vec y} \times \nn \\
& \hspace{3cm} \sum_{k_1,k_2,p' \geq 0}  \frac{[2k_1+1]_{y_1} [2k_2+1]_{y_2}  [2 p'+1]_r}{{Z(\CN_{6d}=(1,0)) (\vec y,r)}}   ~F^{\Ra}_{k_1,k_2,p'} (q)~,
\eea
and the inverse of the character of the fundamental multiplet in \eref{fundmassive6d} can be written as a geometric series\footnote{Note that this can also be rewritten as 
\bea
\left[ Z(\CN_{6d}=(1,0)) (\vec y,r) \right]^{-1} &= r^2 \PE \left[ s [0,1]_{\vec y} \right] \quad \text{with~$s= -r$} \nn \\
&= \sum_{m=0}^\infty (-1)^m r^{m+2} [0,m]_{\vec y}~.
\eea} similar to (\ref{geoser})
\bea
\left[ Z(\CN_{6d}=(1,0)) (\vec y,r) \right]^{-1} 
&= \frac{r^2}{\left(1+\frac{r}{y_1 y_2}\right) \left(1+\frac{r y_1}{y_2}\right) \left(1+\frac{r y_2}{y_1}\right) \left(1+r y_1 y_2\right)} \nn \\
&= \sum_{m_1, \ldots, m_4 \geq 0} (-1)^{m_1+m_2+m_3+m_4} r^{2+m_1+m_2+m_3+m_4}  \times \nn \\
& \hspace{2cm} y_1^{-m_1+m_2-m_3+m_4} y_2^{-m_1-m_2+m_3+m_4}~.
\label{geoser6d}
\eea
%where in the last equality we have used the fact that $\Sym^m [0,1] = [0,m]$.

\subsubsection{Some useful identities}
Before we proceed further, let us derive some useful identities for the elementary building blocks of $G_{n_1,n_2,p}$.  The first one follows from \eref{idenortho}:
\bea \label{iden6d1}
& \CI_0(w; p_1, p_2) := \int \ud \mu_{SO(3)} (r)~  r^w [p_1]_r  [p_2]_r  \nn \\
&= \begin{cases} 
\delta_{p_1,p_2} &\quad \text{for $w =  0$}  \\
\frac{1}{2} \sum_{p= 0}^{\frac{1}{2}(p_1+p_2- |p_1-p_2|)}  \left( \delta_{|w|,2p+|p_1-p_2|} - \delta_{|w|,2p+2+|p_1-p_2|} \right) &\quad \text{for $w \neq 0$}  
\end{cases}
\eea
Next, we are interested in the following integral:
\bea
%\CI(m_1, m_2, \vec k; \vec n) 
%:= \int \ud \mu_{SO(5)} (\vec y)~  [0,m]_{\vec y} [k_1]_{y_1} [k_2]_{y_2} [n_1,n_2]_{\vec y}
\CI(\vec w; \vec k; \vec n) := \int \ud \mu_{SO(5)} (\vec y)~  y_1^{w_1} y_2^{w_2} [k_1]_{y_1} [k_2]_{y_2} [n_1,n_2]_{\vec y}~.
\eea
There are four cases to be considered.  Each of them can be computed using the decomposition formula \eref{decompnonspin} or \eref{decompspin}, together with \eref{iden6d1}.  In what follows, we assume that $\vec k, \vec n \in \BZ_{\geq 0}^2$ and $\vec w \in \BZ^2$.
\bea
&\CI(\vec w; 2k_1,2k_2; n_1,2n_2)   \nn \\
&\hspace{0.5cm}  =  \sum_{\vec k' \in \BZ_{\geq 0}^2} \Delta( n_1 +n_2, n_2; 2k'_1, 2k'_2) \prod_{A =1}^2 \CI_0 (w_A; 2k_A, 2k'_A)~,   \\
&\CI(\vec w; 2k_1+1,2k_2+1; n_1,2n_2)  \nn \\
&\hspace{0.5cm} = \sum_{\vec k' \in \BZ_{\geq 0}^2} \Delta( n_1 +n_2, n_2; 2k'_1, 2k'_2) \prod_{A =1}^2 \CI_0 (w_A; 2k_A+1, 2k'_A)~, \\
&\CI(\vec w; 2k_1,2k_2; n_1,2n_2+1) \nn \\
& \hspace{0.5cm}  = \sum_{\vec k' \in \BZ_{\geq 0}^2} \Delta( n_1 +n_2+ \frac{1}{2}, n_2+ \frac{1}{2}; 2k'_1+1, 2k'_2+1) \prod_{A =1}^2 \CI_0 (w_A; 2k_A, 2k'_A+1)~,  \\
&\CI(\vec w; 2k_1+1,2k_2+1; n_1,2n_2+1) \nn \\
& \hspace{0.5cm} = \sum_{\vec k' \in \BZ_{\geq 0}^2} \Delta( n_1 +n_2+ \frac{1}{2}, n_2+ \frac{1}{2}; 2k'_1+1, 2k'_2+1) \prod_{A =1}^2 \CI_0 (w_A; 2k_A+1, 2k'_A+1)~,
\eea
where from \eref{CTorthoeven} and \eref{CTorthoodd} 
\bea
\Delta(\lambda_1, \lambda_2; 2k_1, 2k_2) &= \frac{1}{2} \sum_{\sigma \in S_2} \det \left(\theta_{|\lambda_A-A+B|}^{4+\lambda_A-A-B} \left(k_{\sigma (A)} \right)  \right)_{A,B=1}^2~, \\
\Delta(\lambda_1, \lambda_2; 2k_1+1, 2k_2+1) &= \frac{1}{2} \sum_{\sigma \in S_2} \det \left(\theta_{|\lambda_A-A+B|}^{4+\lambda_A-A-B} \left(k_{\sigma (A)}+\frac{1}{2} \right)  \right)_{A,B=1}^2~.
\eea

\subsubsection{Multiplicity generating function}
The NS- and R sector contributions to the multiplicity generating function for the representation $\lb n_1,n_2; p \rb$ can be rewritten as
\begin{align}
 G^{\NS}_{n_1,n_2,p} (q) &= \sum_{m_1, \dots, m_4 \geq 0} (-1)^{\sum_{j=1}^4 m_j} \sum_{p' \geq 0}  \CI_0(W_1(\vec m), p, 2p') \times  \\
&\hspace{3cm}   \sum_{k_1,k_2 \geq 0}  \CI(\vec W_2( \vec m); 2k_1,2k_2; n_1,n_2)~F^{\NS}_{k_1,k_2,p'} (q) ~\notag \\
G^{\Ra}_{n_1,n_2,p} (q) &= \sum_{m_1, \dots, m_4 \geq 0}  (-1)^{\sum_{j=1}^4 m_j}\sum_{p' \geq 0}   \CI_0(W_1(\vec m), p, 2p'+1) \times \nn \\
&  \hspace{3cm} \sum_{k_1,k_2 \geq 0} \CI(\vec W_2( \vec m); 2k_1+1,2k_2+1; n_1,n_2)~F^{\Ra}_{k_1,k_2,p'} (q)~,
\end{align}
where we define
\bea
W_1(\vec m) &= 2+m_1+m_2+m_3+m_4~, \nn \\
\vec W_2( \vec m) &= (-m_1+m_2-m_3+m_4, -m_1-m_2+m_3+m_4)~.
\eea

As stated in \eref{sumboth6d}, the multiplicity of the representation $\lb n_1,n_2; p \rb$ in the $\CN_{6d} = (1,0)$ partition function is given by
{\small
\bea
G_{n_1,n_2,p}(q) &= G^{\NS}_{n_1,n_2,p} (q) + G^{\Ra}_{n_1,n_2,p} (q) \nn \\
&= \sum_{m_1, \dots, m_4 \geq 0} \! \! (-1)^{\sum_{j=1}^4 m_j} \sum_{p' \geq 0}  \Big [  \CI_0(W_1(\vec m); p, 2p') \sum_{k_1,k_2 \geq 0}  \CI(\vec W_2( \vec m); 2k_1,2k_2; n_1,n_2)~F^{\NS}_{k_1,k_2,p'} (q)  \nn \\
& \quad +  \CI_0(W_1(\vec m), p, 2p'+1) \sum_{k_1,k_2 \geq 0}  \CI(\vec W_2( \vec m); 2k_1+1,2k_2+1; n_1,n_2)~F^{\Ra}_{k_1,k_2,p'} (q) \Big ]~.
\label{end6dim}
\eea}

%\subsubsection{Recurrence relations}
%Let us write the the partition function $\chi^{\CN_{6d}=(1,0)} (q; \vec y,r)$ as
%\bea
%\chi^{\CN_{6d}=(1,0)} (q; \vec y,r) = \sum_{k_1=0}^\infty \sum_{k_2=0}^\infty \sum_{p=0}^\infty \lb n_1, n_2; r \rb G_{n_1,n_2,p}~.
%\eea 
%Now we rewrite \eref{fundmassive6d} and \eref{ch6d} in terms of characters of representation of $SU(2)^3$:
%\bea
%Z(\CN_{6d} =(1,0)) (\vec y, r) &= ([2]_{y_1}+ [2]_{y_2} -1) + [2]_r + [1]_{y_1} [1]_{y_2} [1]_r~,  \\
%[n_1,2n_2]_{\vec y} &= \frac{1}{\rho(\vec y)} \sum_{k_1,k_2 = 0}^\infty [2k_1]_{y_1} [2k_2]_{y_2} ~\Delta(n_1+n_2,n_2; 2k_1, 2k_2)~, \nn \\
%&=  \frac{1}{\rho(\vec y)}  \Big[ \delta_{n_1,0} \delta_{n_2,0} + \sum_{k=1}^\infty ( [2k]_{y_1} + [2k]_{y_2} ) \Delta(n_1+n_2,n_2; 2k, 0) \nn \\
%& \hspace{2cm}+  \sum_{k_1,k_2 = 1}^\infty [2k_1]_{y_1} [2k_2]_{y_2} ~\Delta(n_1+n_2,n_2; 2k_1, 2k_2) \Big]~, \\
%[n_1,2n_2+1]_{\vec y} &= \frac{1}{\rho(\vec y)} \sum_{k_1,k_2 = 0}^\infty [2k_1]_{y_1} [2k_2]_{y_2} ~\Delta(n_1+n_2+ \frac{1}{2},n_2+ \frac{1}{2}; 2k_1+1, 2k_2+1) \nn \\
%&=  \frac{1}{\rho(\vec y)}  \Big[ \delta_{n_1,0} \delta_{n_2,0} + \sum_{k=1}^\infty ( [2k+1]_{y_1} + [2k+1]_{y_2} ) \Delta(n_1+n_2+ \frac{1}{2},n_2+ \frac{1}{2}; 2k+1, 0) \nn \\
%& \hspace{0.5cm}+  \sum_{k_1,k_2 = 1}^\infty [2k_1+1]_{y_1} [2k_2+1]_{y_2} ~\Delta(n_1+n_2+ \frac{1}{2},n_2+ \frac{1}{2}; 2k_1+1, 2k_2+1) \Big]~.
%\eea
%Hence,
%\bea
%\lb n_1,2n_2; p \rb = 
%\eea
%\subsubsection{Asymptotic analysis for the multiplicities}
%In this subsection, we focus on the limit $n_1,n_2 \rightarrow \infty$ whereas $p$ kept fixed and finite.  

\subsection[Empirical approach to ${\cal N}_{6d}=(1,0)$ asymptotic patterns]{Empirical approach to $\bm{{\cal N}_{6d}=(1,0)}$ asymptotic patterns}
\label{emp6dim}

In this subsection, we follow the lines of subsection \ref{emp4dim} and investigate the large spin asymptotics of multiplicity generating functions $G_{n,k,p}(q)$ for universal ${\cal N}_{6d}=(1,0)$ supermultiplets $\llbracket n,k;p\rrbracket$. Similar to the ${\cal N}_{4d}=1$ strategy, the $G_{n,k,p}(q)$ are expanded in powers of $q^n$ where $n$ denotes the first Dynkin label that we loosely identify with the spin. The coefficients $\tau^{k,p}_{\ell}(q)$ of $(q^n)^{\ell}$ turn out to be power series with non-negative coefficients which enter with alternating sign $(-1)^{\ell-1}$:
\begin{align}
G_{n,k,p}(q) \eq &q^n \, \tau_1^{k,p}(q) \ - \ q^{2n} \, \tau_2^{k,p}(q) \ + \ q^{3n} \, \tau_3^{k,p}(q) \ - \ \ldots \notag \\
\eq &\sum_{\ell=1}^{\infty} (-1)^{\ell-1} \, q^{\ell n} \, \tau_\ell^{k,p}(q)
\label{6dimtau}
\end{align}
In spacetime dimensions higher than four, the analytic methods of subsection \ref{asympt4dim} are no longer efficiently applicable. We could not find an asymptotic formula for (\ref{end6dim}) resembling (\ref{asympmuloddeven}) and (\ref{mainferm}) for the large spin regime of the ${\cal N}_{4d}=1$ multiplicity generating functions. Hence, we determine the $\tau^{k,p}_{\ell}(q)$ including the leading trajectory $\tau^{k,p}_1(q)$ from our data found by expanding the partition function (\ref{NSandR6d1}) up to mass level 25. The multiplicities of $\llbracket n,0;0\rrbracket$ multiplets are shown in the following table \ref{tab6d00}, data for nonzero values of $k$ or $p$ can be found in appendix \ref{app6}. Table entries marked in red are only affected by the stable pattern $\tau_{\ell=1}^{k,p}(q)$ whereas the blue numbers arise from $q^n \tau_{1}^{k,p}(q)-q^{2n}\tau_{2}^{k,p}(q)$, i.e. by including the (subtractive) subleading trajectory.

\begin{table}
\begin{scriptsize}
\begin{center}
%\begin{sidewaysfigure}
\begin{tabular}{|c|| c|c|c|c|c| c|c|c|c|c| c|c|}
%\begin{longtable}{|c|| c|c|c|c|c| c|c|c|c|c| c|c|c|c|c| c|c|c|c|c| c|c|c|c|c| c|} % \small
 \hline
\begin{turn}{-90}$ \ \ \ \ \ \ap m^2$  \end{turn}
&\begin{turn}{-90}\# $\llbracket0,0;0\rrbracket$ \end{turn}
&\begin{turn}{-90}\# $\llbracket1,0;0\rrbracket$ \end{turn}
&\begin{turn}{-90}\# $\llbracket2,0;0\rrbracket$ \end{turn}
&\begin{turn}{-90}\# $\llbracket3,0;0\rrbracket$ \end{turn}
&\begin{turn}{-90}\# $\llbracket4,0;0\rrbracket$ \end{turn}
&\begin{turn}{-90}\# $\llbracket5,0;0\rrbracket$ \end{turn}
&\begin{turn}{-90}\# $\llbracket6,0;0\rrbracket$ \end{turn}
&\begin{turn}{-90}\# $\llbracket7,0;0\rrbracket$ \end{turn}
&\begin{turn}{-90}\# $\llbracket8,0;0\rrbracket$ \end{turn}
&\begin{turn}{-90}\# $\llbracket9,0;0\rrbracket$ \end{turn}
&\begin{turn}{-90}\# $\llbracket10,0;0\rrbracket\ $ \end{turn}
&\begin{turn}{-90}\# $\llbracket11,0;0\rrbracket$ \end{turn}
\\\hline \hline
1 &0 &\textcolor{red}{1} &\textcolor{red}{0} & &
& & & & &
& &  \\\hline
2 &0 &\textcolor{red}{0} &\textcolor{red}{1} &\textcolor{red}{0} &
& & & & &
& &  \\\hline
3 &1 &\textcolor{blue}{2} &\textcolor{red}{0} &\textcolor{red}{1} &\textcolor{red}{0}
& & & & &
& & \\\hline
4 &2 &2 &\textcolor{red}{3} &\textcolor{red}{0} &\textcolor{red}{1}
&\textcolor{red}{0} & & & &
& & \\\hline
5 &3 &9 &\textcolor{blue}{4} &\textcolor{red}{3} &\textcolor{red}{0}
&\textcolor{red}{1} &\textcolor{red}{0} & & &
& & \\\hline
6 &8 &12 &\textcolor{blue}{14} &\textcolor{red}{5} &\textcolor{red}{3}
&\textcolor{red}{0} &\textcolor{red}{1} &\textcolor{red}{0} & &
& & \\\hline
7 &13 &35 &24 &\textcolor{blue}{17} &\textcolor{red}{5}
&\textcolor{red}{3} &\textcolor{red}{0} &\textcolor{red}{1} &\textcolor{red}{0} &
& & \\\hline
8 &30 &58 &63 &\textcolor{blue}{29} &\textcolor{red}{18}
&\textcolor{red}{5} &\textcolor{red}{3} &\textcolor{red}{0} &\textcolor{red}{1} &\textcolor{red}{0}
& & \\\hline
9 &53 &135 &116 &\textcolor{blue}{82} &\textcolor{blue}{32}
&\textcolor{red}{18} &\textcolor{red}{5} &\textcolor{red}{3} &\textcolor{red}{0} &\textcolor{red}{1}
&\textcolor{red}{0} &   \\\hline
10 &107 &243 &265 &153 &\textcolor{blue}{88}
&\textcolor{red}{33} &\textcolor{red}{18} &\textcolor{red}{5} &\textcolor{red}{3} &\textcolor{red}{0}
&\textcolor{red}{1} &\textcolor{red}{0}   \\\hline
11 &193 &505 &503 &358 &\textcolor{blue}{172}
&\textcolor{blue}{91} &\textcolor{red}{33} &\textcolor{red}{18} &\textcolor{red}{5} &\textcolor{red}{3}
&\textcolor{red}{0} &\textcolor{red}{1}  \\\hline
12 &376 &918 &1044 &696 &\textcolor{blue}{403}
&\textcolor{blue}{178} &\textcolor{red}{92} &\textcolor{red}{33} &\textcolor{red}{18} &\textcolor{red}{5}
&\textcolor{red}{3} &\textcolor{red}{0}   \\\hline
13 &670 &1803 &1975 &1474 &801
&\textcolor{blue}{423} &\textcolor{blue}{181} &\textcolor{red}{92} &\textcolor{red}{33} &\textcolor{red}{18}
&\textcolor{red}{5} &\textcolor{red}{3} \\\hline
14 &1246 &3269 &3887 &2839 &1711
&\textcolor{blue}{846} &\textcolor{blue}{429} &\textcolor{red}{182} &\textcolor{red}{92} &\textcolor{red}{33}
&\textcolor{red}{18} &\textcolor{red}{5}   \\\hline
15 &2220 &6136 &7235 &5687 &3355
&\textcolor{blue}{1824} &\textcolor{blue}{866} &\textcolor{blue}{432} &\textcolor{red}{182} &\textcolor{red}{92}
&\textcolor{red}{33} &\textcolor{red}{18} \\\hline
16 &4005 &11015 &13691 &10754 &6784
&3605 &\textcolor{blue}{1870} &\textcolor{blue}{872} &\textcolor{red}{433} &\textcolor{red}{182}
&\textcolor{red}{92} &\textcolor{red}{33}  \\\hline
17 &7025 &20052 &25041 &20649 &13021
&7348 &\textcolor{blue}{3718} &\textcolor{blue}{1890} &\textcolor{blue}{875} &\textcolor{red}{433}
&\textcolor{red}{182} &\textcolor{red}{92}   \\\hline
18 &12407 &35469 &45971 &38304 &25243
&14213 &\textcolor{blue}{7606} &\textcolor{blue}{3764} &\textcolor{blue}{1896} &\textcolor{red}{876}
&\textcolor{red}{433} &\textcolor{red}{182}  \\\hline
19 &21469 &63030 &82532 &71226 &47411
&27774 &14790 &\textcolor{blue}{7720} &\textcolor{blue}{3784} &\textcolor{blue}{1899}
&\textcolor{red}{876} &\textcolor{red}{433}   \\\hline
20 &37182 &109838 &147906 &129443 &89013
&52547 &29015 &\textcolor{blue}{15048} &\textcolor{blue}{7766} &\textcolor{blue}{3790}
&\textcolor{red}{1900} &\textcolor{red}{876}   \\\hline
21 &63492 &191293 &260818 &234646 &163536
&99387 &55177 &\textcolor{blue}{29600} &\textcolor{blue}{15162} &\textcolor{blue}{7786}
&\textcolor{blue}{3793} &\textcolor{red}{1900}  \\\hline
22 &$\!$108142$\!$&328527 &457957 &418298 &299140
&183903 &$\!$104797$\!$&56431 &\textcolor{blue}{29859} &\textcolor{blue}{15208}
&\textcolor{blue}{7792} &\textcolor{red}{3794}   \\\hline
23 &$\!$182254$\!$&562391 &794256 &741961 &538495
&338749 &$\!$194850$\!$&$\!$107476$\!$&\textcolor{blue}{57016} &\textcolor{blue}{29973}
&$\!$\textcolor{blue}{15228}$\!$&\textcolor{blue}{7795}   \\\hline
24 &$\!$306007$\!$&952431 &$\!$1369976$\!$&$\!$1299438$\!$&963344
&613928 &$\!$360467$\!$&$\!$200360$\!$&$\!$\textcolor{blue}{108738}$\!$&\textcolor{blue}{57275}
&$\!$\textcolor{blue}{30019}$\!$&$\!$\textcolor{blue}{15234}$\!$  \\\hline
25 &$\!$509309$\!$&$\!$1605996$\!$&$\!$2339762$\!$&$\!$2261945$\!$&$\!$1702039$\!$
&$\!$1105604$\!$&$\!$656324$\!$&$\!$371692$\!$&$\!$203052$\!$ &$\!$\textcolor{blue}{109324}$\!$
&$\!$\textcolor{blue}{57389}$\!$&$\!$\textcolor{blue}{30039}$\!$  \\\hline
\end{tabular}
%\end{longtable}
%\end{sidewaysfigure}
\end{center}
\end{scriptsize}
\caption{${\cal N}_{6d}=(1,0)$ multiplets with $SO(5)$ quantum numbers $[n,0]$ and $SU(2)_R$ spin 0}
\label{tab6d00}
\end{table}

\subsubsection{Levels of first appearance}

Let us firstly determine the level of first appearance for various families $\{ \llbracket n,k;p\rrbracket, \ n=0,1,\ldots\}$ of ${\cal N}_{6d}=(1,0)$ supermultiplets with second $SO(5)$ Dynkin label $k$ and R symmetry quantum number $p$ fixed. It is identical to the leading $q$ power of the multiplicity generaing function $G_{0,k,p}(q)$ or its expansion coefficients $\tau_\ell^{k,p}(q)$ defined by (\ref{6dimtau}). The following table \ref{tab:appearance} gathers the mass levels $\ap m^2 \leq 25$ where the first instance of a $\{ \llbracket n,k;p\rrbracket, \ n=0,1,\ldots\}$ member can be found:

\begin{table}[htdp]
\begin{center}
\begin{tabular}{|c||c|  c|c|c|c|c|  c|c|c|c|c|  c|c|c|c|c|  c|c|}
\hline  $\! \! \downarrow p \Big. ,\ \overrightarrow{ k} \! \!$
&0 &1 &2 &3 &4 &5 &6 &7 &8 &9 &10 &11 &12 &13 &14 &15 &16 &17 \\ \hline \hline   
   %%%%
0 &1 
& &2 & &5 &
&8 & &11 & &14
& &17 & &20 &
&23 &     \\\hline
1 & 
&2 & &4 & &7
& &10 & &13 &
&16 & &19 & &22
& &25     \\\hline
2 &3 
& &4 & &7 &
&10 & &13 & &16
& &19 & &22 &
&25 &     \\\hline
3 & 
&5 & &7 & &10
& &13 & &16 &
&19 & &22 & &25
& &     \\\hline
4 &7 
& &8 & &10 &
&13 & &16 & &19
& &22 & &25 &
& &     \\\hline
5 & 
&9 & &11 & &14
& &17 & &20 &
&23 & & & &
& &     \\\hline
6 & 11
& &12 & &15 &
&18 & &21 & &24
& & & & &
& &     \\\hline
7 & 
&14 & &16 & &19
& &22 & &25 &
& & & & &
& &     \\\hline
8 & 17
& &18 & &20 &
&23 & & & &
& & & & &
& &     \\\hline
9 & 
&20 & &22 & &25
& & & & &
& & & & &
& &     \\\hline
10 & 23
& &24 & & &
& & & & &
& & & & &
& &     \\\hline
\end{tabular}
\caption{Mass level where the $\llbracket0,k;p\rrbracket$ multiplet of ${\cal N}_{6d}=(1,0)$ firstly occurs. Empty spaces indicate that the representations in question do not occur at levels $\leq 25$.}
\label{tab:appearance}
\end{center}
\end{table}

We observe that, roughly speaking, the level of first appearance for supermultiplets $\llbracket n,k;p\rrbracket$ depends linearly\footnote{The linear $k$ dependence can be partially understood from the $\la_{1,2}$ dependence in (\ref{GBSO(2n+1)}). However, the bosonic string suggests that an $SO(5)$ representation $[n,k]$ is delayed by {\it two} levels under $k \mapsto k+1$ whereas the observations from table \ref{tab:appearance} clearly show a delay of {\it three} levels per $k \mapsto k+1$. Even though we cannot give a detailed explanation on analytical grounds, it is clear that this extra delay in mass level must be due to the worldsheet fermions, see e.g. (\ref{BnNS}) and (\ref{BnR}).} on the $SO(5)$ Dynkin label $k$ (with slope $\frac{3}{2}$) but quadratically on the R symmetry spin $p/2$, in agreement with the final remark in subsection \ref{sec:42}. 

\subsubsection{Explicit formulae for the $\tau_{\ell}^{k,p}(q)$}

Let us now list the leading terms in various $\tau_{\ell}^{k,p}(q)$, obtained through the entries of table \ref{tab6d00} and its $(k,p) \neq (0,0)$ relatives displayed in appendix \ref{app6}. This allows to reconstruct the large spin asymptotics of the multiplicity generating functions $G_{n,k,p}(q)$ via (\ref{6dimtau}). 

At zero $SU(2)_R$ charge, we have
\begin{itemize}
\item $SO(5)$ Dynkin labels $[n \rightarrow \infty, 0]$ and $SU(2)_R$ representation $[0]$
\begin{align}
\tau_1^{0,0}(q) \ \ = \ \ &  1 + 0 q + 3 q^2 + 5 q^3 + 18 q^4 + 33 q^5 + 92 q^6 + 182 q^7 + 
 433 q^8 + 876 q^9 + 1900 q^{10} \notag \\
 & \ \ + 3794 q^{11} + 7796 q^{12} + 15238 q^{13} + 
 30049 q^{14} + 57465 q^{15} + 109773 q^{16} \notag \\
 & \ \ + 205349 q^{17} + 382249 q^{18} + 
 700520 q^{19} + \ldots \notag \\
\tau_2^{0,0}(q) \ \ = \ \ &q\, ( 1 + 4 q^1 + 10 q^2 + 30 q^3 + 76 q^4 + 190 q^5 + 449 q^6 + 
  1035 q^7 + 2298 q^8 + 4999 q^9 \notag \\
  & \ \ + 10580 q^{10} + 21976 q^{11} + 
  44727 q^{12} + 89543 q^{13}+\ldots) \notag\\
\tau_3^{0,0}(q) \ \ = \ \ & q \,(1 + q + 10 q^2 + 23 q^3 + 81 q^4 + 194 q^5  + 531 q^6 + 1232 q^7 + 
  2967 q^8 +  6586 q^9+\ldots ) \notag\\
\tau_4^{0,0}(q) \ \ = \ \ & q^2 \,(1 + 5 q + 16 q^2 + 53 q^3 + 153 q^4 + 417 q^5 + \ldots) \notag \\
 \tau_5^{0,0}(q) \eq &q^2 \, (1 + q + 11 q^2 + \ldots)
\end{align}
\item $SO(5)$ Dynkin labels $[n \rightarrow \infty, 2]$ and $SU(2)_R$ representation $[0]$
\begin{align}
\tau_1^{2,0}(q) \eq& q^{2} \, (1 + 2 q + 8 q^2 + 17 q^3 + 52 q^4 + 117 q^5 + 293 q^6 + 645 q^7 + 
 1468 q^8 \notag \\
 & \ \ + 3119 q^9 + 6667 q^{10} + 13674 q^{11} + 27913 q^{12} + 
 55446 q^{13} + 109165 q^{14} \notag \\
 & \ \  + 210717 q^{15} + 402714 q^{16} + 757889 q^{17} + 
 1412208 q^{18}+\ldots) \notag \\
 \tau_2^{2,0}(q) \eq& q^{3} \, (1 + 4 q + 14 q^2 + 41 q^3 + 118 q^4 + 306 q^5 + 764 q^6 + 1818 q^7 + 
 4191 q^8 \notag \\
 & \ \ + 9344 q^9 + 20318 q^{10} + 43083 q^{11} + 89493 q^{12} + 
 182239 q^{13}+\ldots) \notag \\
 \tau_3^{2,0}(q) \eq& q^{5} \, ( 3 + 9 q + 40 q^2 + 114 q^3 + 345 q^4 + 890 q^5 + 2297 q^6 + 
 5481 q^7 + 12871 q^8  +\ldots) \notag \\
 \tau_4^{2,0}(q) \eq& q^{6} \, (1 + 5 q + 23 q^2 + 79 q^3 + 251 q^4 + 717 q^5+\ldots) \notag \\
 \tau_5^{2,0}(q) \eq& q^{8} \, (3 + 10 q + 48 q^2+\ldots) 
 \end{align}
\item $SO(5)$ Dynkin labels $[n \rightarrow \infty, 4]$ and $SU(2)_R$ representation $[0]$
\begin{align}
\tau_1^{4,0}(q) \eq& q^{5} \,(1 + 5 q + 14 q^2 + 43 q^3 + 113 q^4 + 294 q^5 + 698 q^6 + 1648 q^7 + 
 3677 q^8 \notag \\
 & \ \ + 8090 q^9 + 17182 q^{10} + 35919 q^{11} + 73211 q^{12} + 
 147036 q^{13} +  289598 q^{14} \notag \\
 & \ \ + 562694 q^{15} + 1076373 q^{16}+\ldots) \notag \\
\tau_2^{4,0}(q) \eq& q^{6} \,(1 + 5 q + 18 q^2 + 56 q^3 + 166 q^4 + 446 q^5 + 1143 q^6 + 2787 q^7 + 
 6549 q^8 \notag \\
 & \ \ + 14864 q^9 + 32811 q^{10} + 70532 q^{11} + 148268 q^{12}+\ldots) \notag \\
 \tau_3^{4,0}(q) \eq& q^{9} \,(4 + 14 q + 61 q^2 + 184 q^3 + 561 q^4 + 1495 q^5 + 3896 q^6 + 9478 q^7+\ldots) \notag \\
 \tau_4^{4,0}(q) \eq& q^{11} \,(1 + 8 q + 36 q^2 + 131 q^3+\ldots) 
\end{align}
\item $SO(5)$ Dynkin labels $[n \rightarrow \infty, 6]$ and $SU(2)_R$ representation $[0]$
\begin{align}
\tau_1^{6,0}(q) \eq& q^{8} \, (1 + 5 q + 18 q^2 + 53 q^3 + 158 q^4 + 407 q^5 + 1033 q^6 + 2452 q^7 + 
 5686 q^8 \notag \\
 & \ \ + 12640 q^9 + 27521 q^{10} + 58151 q^{11} + 120616 q^{12} + 
 244647 q^{13}+\ldots)\notag \\
 \tau_2^{6,0}(q) \eq& q^{9} \, (1 + 5 q + 18 q^2 + 57 q^3 + 173 q^4 + 473 q^5 + 1234 q^6 + 3060 q^7 \notag \\
 & \ \ + 
 7308 q^8 + 16835 q^9+\ldots)\notag \\
 \tau_3^{6,0}(q) \eq& q^{13} \, (4  + 15 q + 67 q^2 + 209 q^3+\ldots)
  \end{align}
  \end{itemize}
Observe that the leading $q$ powers of $\tau_{1}^{k,p},\tau_{2}^{k,p},\tau_{3}^{k,p},\tau_{4}^{k,p},\ldots$ increase more rapidly at higher values of $k$. In other words, the series (\ref{6dimtau}) converges more quickly as the second Dynkin label $k$ increases.

The simplest examples with vanishing second $SO(5)$ Dynkin label and nonzero R symmetry charge are the following:
\begin{itemize}
\item $SO(5)$ Dynkin labels $[n \rightarrow \infty, 0]$ and $SU(2)_R$ representation $[2]$
\begin{align}
\tau_1^{0,2}(q) \eq &q^3\, ( 
  3 + 5 q + 20 q^2 + 46 q^3 + 128 q^4 + 288 q^5 + 696 q^6 + 
   1513 q^7 + 3354 q^8  \notag \\
   & \ \ + 7025 q^9 + 14707 q^{10} + 29736 q^{11} + 
   59679 q^{12} + 116933 q^{13} \notag \\
   & \ \ +  226900 q^{14} + 432515 q^{15} + 816089 q^{16} + \ldots) \notag\\
   \tau_2^{0,2}(q) \eq &q^2\, (1 + 3 q^1 + 13 q^2 + 37 q^3 + 109 q^4 + 285 q^5 + 727 q^6 + 
  1737 q^7 + 4050 q^8 + 9075 q^9 \notag \\
  & \ \ + 19868 q^{10} + 42302 q^{11} + 
  88278 q^{12}+\ldots)\notag \\
      \tau_3^{0,2}(q) \eq &q^2\, ( 
  1 + 2 q + 13 q^2 + 37 q^3 + 124 q^4 + 331 q^5 + 906 q^6 + 2233 q^7 + 
 5456 q^8+ \ldots)\notag \\
   \tau_4^{0,2}(q) \eq &q^3\, ( 2 + 7 q + 29 q^2 + 92 q^3 + 282 q^4+ \ldots) \notag \\
   \tau_5^{0,2}(q) \eq &q^3\, ( 1 + 3 q + 18 q^2 + \ldots) 
   \end{align}
%%%%%%%%%%%%%%%%%%%%%%%%%
\item $SO(5)$ Dynkin labels $[n \rightarrow \infty, 0]$ and $SU(2)_R$ representation $[4]$
\begin{align}
\tau_1^{0,4}(q) \eq& q^{6} \,(1 + 4 q + 18 q^2 + 47 q^3 + 142 q^4 + 353 q^5 + 887 q^6 + 2049 q^7 + 
 4692 q^8 \notag \\
 & \ \ + 10215 q^9 + 21942 q^{10} + 45608 q^{11} + 93377 q^{12} + 
 186790 q^{13} + 368341 q^{14}+\ldots) \notag \\
 \tau_2^{0,4}(q) \eq& q^{6} \,(3 + 10 q + 41 q^2 + 124 q^3 + 362 q^4 + 952 q^5 + 2424 q^6 + 
 5811 q^7 \notag \\
 & \ \ + 13526 q^8 + 30317 q^9+\ldots) \notag \\
 \tau_3^{0,4}(q) \eq& q^{5} \,(1 + 3 q + 17 q^2 + 53 q^3 + 179 q^4 + 501 q^5 + 1392 q^6+\ldots) \notag \\
 \tau_4^{0,4}(q) \eq& q^{5} \,(1 + 3 q + 16 q^2 + 53 q^3+\ldots) 
  \end{align}
%%%%%%%%%%%%%
\item $SO(5)$ Dynkin labels $[n \rightarrow \infty, 0]$ and $SU(2)_R$ representation $[6]$
\begin{align}
\tau_1^{0,6}(q) \eq & q^{11} \,(3+ 8 q+ 35 q^2+ 98 q^3+ 291 q^4 + 733 q^5 +1856 q^6  + 4339 q^7 \notag \\
& \ \ + 
 9987 q^8 + 21954 q^9+\ldots) \notag \\
 \tau_2^{0,6}(q) \eq & q^{10} \,(1 + 5 q^1 + 27 q^2 + 88 q^3 + 286 q^4 + 804 q^5 + 2171 q^6+\ldots) \notag \\
 \tau_3^{0,6}(q) \eq & q^{10} \,(3 + 10 q + 46 q^2 + 148 q^3+\ldots) \notag \\
 \tau_4^{0,6}(q) \eq & q^{9} \,(1 + 3 q+\ldots) 
\end{align}
\end{itemize}
Observe that the leading $q$ powers of $\tau_{1}^{k,p},\tau_{2}^{k,p},\tau_{3}^{k,p},\tau_{4}^{k,p},\ldots$ tend to decrease at higher values of $p$. In other words, the series (\ref{6dimtau}) converges more slowly as the $SU(2)_R$ spin $p$ increases.

Finally, we shall display asymptotic data for the simplest family of fermionic multiplets with $SO(5)$ Dynkin labels $[n \rightarrow \infty, 1]$ and $SU(2)_R$ representation $[1]$
\begin{align}
\tau_1^{1,1}(q) \eq& q^{2} \,( 2 + 4  q +13q^2  +35q^3 +89 q^4 + 216 q^5 + 508 q^6 + 1145 q^7 + 
 2521 q^8  \notag \\
 & \ \ + 5402 q^9+ 11320 q^{10} + 23238 q^{11} + 46856 q^{12} + 
 92850 q^{13}  \notag \\
 & \ \ + 181217 q^{14} +  348612 q^{15} + 661792 q^{16} + 1240786 q^{17} + \ldots ) \notag \\
\tau_2^{1,1}(q) \eq& q^{2} \,( 1 + 4 q + 13 q^2 + 43 q^3 + 122 q^4 + 323 q^5 + 814 q^6 + 
 1962 q^7 + 4550 q^8 \notag \\
 &\ \ + 10233 q^9 + 22370 q^{10} + 47718 q^{11} + 
 99574 q^{12} + \ldots ) \notag \\
 \tau_3^{1,1}(q) \eq& q^{3} \,(1 + 5 q + 21 q^2 + 70 q^3 + 211 q^4 + 584 q^5 + 1529 q^6 + 3798 q^7+ \ldots ) \notag \\
 \tau_4^{1,1}(q) \eq& q^{4} \,( 1 + 6 q + 24 q^2 + 85 q^3+ \ldots ) \notag \\
 \tau_5^{1,1}(q) \eq& q^{5} \,( 1+ \ldots ) 
\end{align}
Further $\tau_{\ell}^{k,p}(q)$ for $\llbracket n, k;p \rrbracket$ multiplets at $(k,p)=(2,2), (4,2), (2,4), (3,1), (1,3), (5,1), (3,3), (1,5)$ are listed in appendix \ref{app:data6}. They confirm our observations that the $\tau_{\ell}^{k,p}(q)$ expansion (\ref{6dimtau}) coverges more quickly with larger values of $k$ and smaller values of $p$.

%%%%%%%%%

\subsection[Four dimensional ${\cal N}_{4d}=2$ spectra]{Four dimensional $\bm{{\cal N}_{4d}=2}$ spectra}
\label{sec:4d,N2}

In order to determine universal string spectra with ${\cal N}_{4d}=2$ SUSY, we shall now compactify two dimensions of minimally supersymmetric ${\cal N}_{6d}=(1,0)$ theories on a $T^2$. This preserves all the eight supercharges and the internal rotation symmetry becomes an R symmetry factor of $SO(2)_R \cong U(1)_R$. Hence, the dimensionally reduced theory in $d=4$ spacetime dimensions enjoys ${\cal N}_{4d}=2$ SUSY and R symmetry $SU(2)_R \times U(1)_R$. The fundamental ${\cal N}_{4d}=2$ super Poincar\'e multiplet encompasses 8+8 states,
\bea
Z({\cal N}_{4d}=2) \eq [2]_y \ + \ [2]_r \, [0]_y \ +\ (z^2 + z^{-2}) \, [0]_y \ + \ (z+z^{-1}) \, [1]_r \, [1]_y
\label{4dim2susy}
\eea
where $z$ denotes the $U(1)_R$ fugacity. The tensor product of (\ref{4dim2susy}) with a Clifford vacuum in some $SO(3) \times SU(2)_R \times U(1)_{R}$ representation yields a family of supermultiplets characterized by three quantum numbers -- $n$ for $SO(3)$ spin, $m$ for $SU(2)_R$ spin and $p$ for $U(1)_R$ charge. The resulting $16(n+1)(m+1)$ states are described by the supercharacter
\footnote{
The simplicity of the $SO(3)$ tensor product $[2m] \cdot [2k] = \sum_{l=|k-m|}^{k+m} [2l]$ allows for compact closed formulae for the $SO(3) \times SU(2)_R \times U(1)_{R}$ decomposition of a general ${\cal N}_{4d}=2$ supercharacter:
\begin{align}
\llbracket n;m&,p\rrbracket \ \ = \ \ z^p \, \big\{ \, [m]_r \, [n+2] \ + \ [m]_r \, [n-2] \ + \ [m+2]_r \, [n] \ + \ [m-2]_r \, [n] \ + \ 2 \, [m]_r \, [n]  \notag \\
& + \ (z^2+z^{-2}) \, [m]_r \, [n] \ + \ (z+z^{-1}) \, \big( \, [m+1]_r \, + \, [m-1]_r \, \big) \, \big( \, [n+1] \, + \, [n-1] \, \big) \, \big\}
\label{genchar}
\end{align}
This generic character formula \eref{genchar} holds for values $n,m \geq 2$ of the Clifford vacuum's $SO(3) \times SU(2)_R$ spin quantum numbers and specializes otherwise:
\begin{align}
\llbracket n;0,p\rrbracket \eq &z^p \, \big\{ \, [n+2] \ + \ [n-2] \ +\ [2]_r \, [n] \ + \ (1+z^2+z^{-2}) \, [n] \notag \\
& \ \ \ \ \ +\ (z+z^{-1}) \, [1]_r \, \big( \, [n+1] \, + \, [n-1] \, \big) \, \big\} \co  \ \ \ \ \ \ \ \ \ \ n \geq 2
\\
\llbracket 0;m,p\rrbracket \eq &z^p \, \big\{ \,[m]_r \, [2] \ + \ [m]_r\, [0] \ + \ [m+2]_r \, [0] \ + \ [m-2]_r \, [0] \ + \ (z^2 + z^{-2}) \, [m]_r \, [0] \notag \\
& \ \ \ \ \ +\ (z+z^{-1}) \, \big( \, [m+1]_r \, + \, [m-1]_r\, \big) \, [1]\, \big\} \co  \ \ \ \ \ \ m \geq 2
\\
\llbracket 0;0,p\rrbracket \eq &z^p \, \big\{ \, [2] \ + \ [2]_r \, [0] \ +\ (z^2 + z^{-2}) \, [0] \ + \ (z+z^{-1}) \, [1]_r \, [1] \, \big\}
\\
\llbracket 1;1,p\rrbracket \eq &z^p \, \big\{ \, [1]_r \, [3] \ + \ [3]_r \, [1] \ + \ (2+z^2 + z^{-2}) \,[1]_r \, [1] \notag \\
& \ \ \ \ \ (z+z^{-1}) \, \big( \, [2]_r \, [2] \ + \ [2]_r \, [0] \ + \ [2] \ + \ [0] \, \big) \, \big\}
\end{align}
We observe the general selection rule that either none or all of $n,m,p$ are odd, hence, there is no need to consider $\llbracket 1;0,p\rrbracket $ or $\llbracket 0;1,p\rrbracket $.
}
\bea
\llbracket n;m,p\rrbracket \ \ := \ \ Z({\cal N}_{4d}=2)  \cdot z^p \, [m]_r \, [n]_y \ .
\eea
The position of the semicolon in the arguments of the supercharacter allows to distinguish ${\cal N}_{4d}=2$ multiplets $\llbracket \cdot ;\cdot ,\cdot \rrbracket$ from ${\cal N}_{6d}=(1,0)$ multiplets $\llbracket \cdot ,\cdot ;\cdot \rrbracket$.

The universal partition function of ${\cal N}_{4d}=2$ scenarios is obtained through GSO projection of the following character products%\footnote{Again, our GSO projection in the R sector is appropriate at orders $q^{\geq 1}$ only and leads to the massless character $\frac{1}{2} [1]_y [1]_z [1]_R$  instead of the desired coefficient $(yz + (yz)^{-1}) (1)_r$ along with $q^0$ in $\chi^{{\cal N}_{4d}=2}_{\Ra} \mid_{\GSO}$. We will neglect this subtlety because we are mainly interested in the massive particle content.}
:
\begin{align}
\chi^{{\cal N}_{4d}=2}(q;y,r,z) &\eq \chi^{{\cal N}_{4d}=2}_{\te{NS}} \mid_{\GSO} (q;y,r,z) \ + \ \chi^{{\cal N}_{4d}=2}_{\te{R}} \mid_{\GSO} (q;y,r,z) \notag \\
%
%\chi^{{\cal N}_{4d}=2}_{\te{NS}} \mid_{\GSO}(q;y,r,z) &\eq
% \frac{1}{2 } \, q^{-\frac{1}{2}} \, \big[ \, \chi_{\te{NS}}^{SO(3)}(q;y)  \,  \chi^{{\cal N}_{2d}=4,c=6}_{\te{NS},h=0,q=0}(q;r) \, \chi^{{\cal N}_{2d}=2,c=3}_{\te{NS},h=0,q=0}(q;z)  \notag \\
%& \ \ \ \ \ \ \ \ \ \ \ \ \ \ \ - \ \chi_{\te{NS}}^{SO(3)}(e^{2\pi i}q;y)  \,  \chi^{{\cal N}_{2d}=4,c=6}_{\te{NS},h=0,q=0}(e^{2\pi i}q;r) \, \chi^{{\cal N}_{2d}=2,c=3}_{\te{NS},h=0,q=0}(e^{2\pi i}q;z) \, \big] \notag \\
%%
%\chi^{{\cal N}_{4d}=2}_{\te{R}} \mid_{\GSO}(q;y,r,z) &\eq  \frac{1}{2 } \, q^{-\frac{3}{8}} \, \chi_{\te{R}}^{SO(3)}(q;y)  \, \chi^{{\cal N}_{2d}=4,c=6}_{\te{R},h=1/4,q=1/2}(q;r) \, \chi^{{\cal N}_{2d}=2,c=3}_{\te{R},h=1/8,q=1/2}(q;z) 
\chi^{{\cal N}_{4d}=2}_{\te{NS}} \mid_{\GSO}(q;y,r,z) &\eq
 \frac{1}{2 } \, q^{-\frac{1}{2}} \, \big[ \, \chi_{\te{NS}}^{SO(3)}(q;y)  \,  \chi^{{\cal N}_{2d}=4,c=6}_{\te{NS},h=0,\ell=0}(q;r) \, \chi^{SO(3)}_{\te{NS} }(q;z)  \notag \\
& \ \ \ \ \ \ \ \ \ \ \ \ \ \ \ - \ \chi_{\te{NS}}^{SO(3)}(e^{2\pi i}q;y)  \,  \chi^{{\cal N}_{2d}=4,c=6}_{\te{NS},h=0,\ell=0}(e^{2\pi i}q;r) \, \chi^{SO(3)}_{\te{NS} }(e^{2\pi i}q;z) \, \big] \notag \\
\chi^{{\cal N}_{4d}=2}_{\te{R}} \mid_{\GSO}(q;y,r,z) &\eq  \frac{1}{2 } \, \chi_{\te{R}}^{SO(3)}(q;y)  \, \chi^{{\cal N}_{2d}=4,c=6}_{\te{R},h=1/4,\ell=1/2}(q;r) \, \chi^{SO(3)}_{\te{R}}(q;z) 
\label{N2part0}
\end{align}
Its symmetry under reversal $p \mapsto -p$ of $U(1)_R$ charges motivates the definition
\beq
\llbracket n;m,\pm p\rrbracket  \ \ := \ \ \left\{ \begin{array}{cl} \ \llbracket n;m,p\rrbracket \ + \ \llbracket n;m,-p\rrbracket \ &: \ p \neq 0 \\ \llbracket n;m,0\rrbracket &: \ p=0 \end{array} \right. \ ,
\eeq
then the power series expansion of (\ref{N2part0}) starts like\footnote{Again, there is a subtlety in applying the above formula to the massless R sector; see the footnote before \eref{def:GSOed}. However, this can be fixed easily: one can simply add to it $\frac{1}{2}(y-y^{-1})(z-z^{-1})(r-r^{-1})$ to get the correct massless character in R sector.  }
\begin{align}
&\chi^{{\cal N}_{4d}=2}(q;y,r,z) \eq \underbrace{\left( \, y^2 + y^{-2} + z^2 + z^{-2} \ + \ \frac{1}{2} (y+y^{-1})  [1]_z [1]_r  \, \right) \, q^0}_{8 \ \te{massless states}}  \ + \ \underbrace{ \big( \, \llbracket 2;0,0\rrbracket \ + \ \llbracket 0;0,\pm 2\rrbracket \, \big) \, q}_{80 \ \te{states at level} \ 1}  \notag \\
& \ \ + \ \underbrace{ \big(  \llbracket 4;0,0\rrbracket \, + \, 2 \, \llbracket 2;0,\pm 2\rrbracket \, + \, \llbracket 2;0,0\rrbracket \, + \, \llbracket 1;1,\pm 1\rrbracket \, + \, \llbracket 0;0,\pm 4\rrbracket \, + \, 2 \, \llbracket 0;0,0\rrbracket \, \big) \, q^2}_{512 \ \te{states at level} \ 2} \notag \\
& \ \ + \ \big( \llbracket 6;0,0\rrbracket \, + \, 2 \, \llbracket 4;0,\pm 2\rrbracket\, + \, \llbracket 4;0,0\rrbracket \, + \, 2 \, \llbracket 3;1,\pm 1\rrbracket \, + \, 2 \, \llbracket 2;0,\pm 4\rrbracket  \, + \, 2 \, \llbracket 2;0,\pm 2\rrbracket \, + \, 6 \, \llbracket 2;0,0\rrbracket \notag \\
& \hskip0.5cm  \, + \, 2 \, \llbracket 1;1,\pm 3\rrbracket \, + \,3 \, \llbracket 1;1,\pm 1\rrbracket \, + \, \llbracket 0;2,0\rrbracket \, + \, \llbracket 0;0,\pm 6\rrbracket \, + \, 4 \, \llbracket 0;0,\pm 2\rrbracket \, + \, 2 \, \llbracket 0;0,0\rrbracket \,\big) \, q^3
 \ + \  {\cal O}(q^4) 
\label{N2part}
\end{align}
The vertex operators occurring in the three multiplets of the first mass level have been constructed in \cite{Feng:2012bb}, see equations (6.3) to (6.11) of that reference for bosons and equations (6.22) to (6.30) for fermions. The content of the first five levels is summarized in table \ref{table4,2}:

\bigskip
\noindent
\begin{table}[htdp]
\begin{tabular}{|l|l|}\hline  $\ap m^2$ &representations of ${\cal N}_{4d}=2$ super Poincar\'e \\ \hline \hline   
  1 & $\llbracket 2;0,0 \rrbracket \, + \, \llbracket 0;0,\pm 2 \rrbracket$  \\\hline
   2 & $\llbracket 4;0,0\rrbracket \, + \, 2 \, \llbracket 2;0,\pm 2\rrbracket \, + \, \llbracket 2;0,0\rrbracket \, + \, \llbracket 1;1,\pm 1\rrbracket \, + \, \llbracket 0;0,\pm 4\rrbracket \, + \, 2 \, \llbracket 0;0,0\rrbracket$  \\\hline
   3 & $\llbracket 6;0,0\rrbracket \, + \, 2 \, \llbracket 4;0,\pm 2\rrbracket\, + \, \llbracket 4;0,0\rrbracket \, + \, 2 \, \llbracket 3;1,\pm 1\rrbracket \, + \, 2 \, \llbracket 2;0,\pm 4\rrbracket \, + \, 2 \, \llbracket 2;0,\pm 2\rrbracket \, + \, 6 \, \llbracket 2;0,0\rrbracket $ \\
   &$\, + \,2 \, \llbracket 1;1,\pm 3\rrbracket \, + \,3 \, \llbracket 1;1,\pm 1\rrbracket \, + \, \llbracket 0;2,0\rrbracket \, + \, \llbracket 0;0,\pm 6\rrbracket \, + \, 4 \, \llbracket 0;0,\pm 2\rrbracket \, + \, 2 \, \llbracket 0;0,0\rrbracket$  \\\hline
   4 & $ \llbracket 8;0,0\rrbracket \, + \, 2\, \llbracket 6;0,\pm 2\rrbracket \, + \, \llbracket 6;0,0\rrbracket \, + \, 2 \, \llbracket 5;1,\pm 1\rrbracket \, + \, 2 \, \llbracket 4;0,\pm 4\rrbracket   \, + \, 3 \, \llbracket 4;0,\pm 2 \rrbracket  \, + \, 8 \, \llbracket 4;0,0\rrbracket $
   \\
   &$ \, + \, 3 \, \llbracket 3;1,\pm 3 \rrbracket  \, + \, 6 \, \llbracket 3;1,\pm 1 \rrbracket  \, + \, \llbracket 2;2,\pm 2 \rrbracket   \, + \, 3 \, \llbracket 2;2,0 \rrbracket   \, + \, 2 \, \llbracket 2;0,\pm 6 \rrbracket   \, + \, 3 \, \llbracket 2;0,\pm 4 \rrbracket      $ 
   \\
&$  \, + \, 12 \, \llbracket 2;0,\pm 2 \rrbracket  \, + \, 11 \, \llbracket 2;0,0 \rrbracket   \, + \, 2 \, \llbracket 1;1,\pm5 \rrbracket  \, + \, 5 \, \llbracket 1;1,\pm 3 \rrbracket  \, + \, 10 \, \llbracket 1;1,\pm 1 \rrbracket   \, + \, 2 \, \llbracket 0;2, \pm 2 \rrbracket   $ \\
&$\, + \,\llbracket 0;2,0 \rrbracket  \, + \,  \llbracket 0;0,\pm 8 \rrbracket  \, + \, 5 \, \llbracket 0;0,\pm 4 \rrbracket   \, + \, 4 \, \llbracket 0;0,2 \rrbracket  \, + \, 11 \, \llbracket 0;0,0 \rrbracket $
   \\\hline        
   5 &$\llbracket 10;0,0 \rrbracket \, + \, 2 \, \llbracket 8;0,\pm 2 \rrbracket \, + \, \llbracket 8;0,0 \rrbracket\, + \, 2 \, \llbracket 7;1,\pm 1 \rrbracket \, + \, 2 \, \llbracket 6;0,\pm 4 \rrbracket \, + \, 3 \, \llbracket 6;0,\pm 2 \rrbracket \, + \, 8 \, \llbracket 6;0,0 \rrbracket $
     \\
     &$\, + \, 3 \, \llbracket 5;1,\pm 3 \rrbracket \, + \, 7 \, \llbracket 5;1,\pm 1 \rrbracket \, + \,  \llbracket 4;2,\pm 2 \rrbracket \, + \, 4 \, \llbracket 4;2,0\rrbracket \, + \, 2 \, \llbracket 4;0,\pm 6 \rrbracket \, + \, 4 \, \llbracket 4;0,\pm 4 \rrbracket$
     \\
     &$\, + \, 16 \, \llbracket 4;0,\pm 2 \rrbracket \, + \, 17 \, \llbracket 4;0,0 \rrbracket \, + \, 3 \, \llbracket 3;1,\pm 5 \rrbracket \, + \, 11 \, \llbracket 3;1,\pm 3 \rrbracket \, + \, 21 \, \llbracket 3;1,\pm 1 \rrbracket \, + \,  \llbracket 2;2,\pm 4 \rrbracket $ \\
     &$\, + \, 7 \, \llbracket 2;2,\pm 2 \rrbracket \, + \, 8 \, \llbracket 2;2,0 \rrbracket \, + \, 2 \, \llbracket 2;0,\pm 8 \rrbracket \, + \, 3 \, \llbracket 2;0,\pm 6 \rrbracket \, + \, 15 \, \llbracket 2;0,\pm 4 \rrbracket \, + \, 23 \, \llbracket 2;0,\pm 2 \rrbracket$ \\
     &$\, + \, 38 \, \llbracket 2;0,0 \rrbracket \, + \,  \llbracket 1;3,\pm 1 \rrbracket \, + \, 2 \, \llbracket 1;1,\pm 7 \rrbracket \, + \, 6 \, \llbracket 1;1,\pm 5 \rrbracket \, + \, 16 \, \llbracket 1;1,\pm 3 \rrbracket \, + \, 28 \, \llbracket 1;1,\pm 1 \rrbracket$ \\
     &$\, + \, 3 \, \llbracket 0;2,\pm 4 \rrbracket \, + \, 4 \, \llbracket 0;2,\pm 2 \rrbracket \, + \, 9 \, \llbracket 0;2,0 \rrbracket \, + \, \llbracket 0;0,\pm 10 \rrbracket \, + \, 5 \, \llbracket 0;0,\pm 6 \rrbracket \, + \, 6 \, \llbracket 0;0,\pm 4 \rrbracket $ \\
     &$ \, + \, 21 \, \llbracket 0;0,\pm 2 \rrbracket \, + \, 16 \, \llbracket 0;0,0 \rrbracket $
     \\\hline
\end{tabular}
\caption{${\cal N}_{4d}=2$ multiplets occurring up to mass level 5}
\label{table4,2}
\end{table}

Comparison with the partition function \eref{part6dim} of the ${\cal N}_{6d}=(1,0)$ ancestor theory (and table \ref{table6,2}) clearly demonstrates that the six dimensional viewpoint gives a more streamlined handle on the spectrum in terms of fewer supermultiplets. This is why we do not provide an asymptotic analysis and data tables for the universal ${\cal N}_{4d}=2$ spectrum like we did for the $d=6$ ancestor in subsection \ref{emp6dim} and appendix \ref{app6}.

%%%%%%%%%%
%%%%%%%%%%
%%%%%%%%%%

\section{Spectra in compactifications with 16 supercharges}
\label{sec:10d}

This section is devoted to maximally supersymmetric type I superstring compactifications on even dimensional tori where all the sixteen supercharges are preserved \cite{Narain:1986am}. The methods introduced in subsections \ref{sec:highdim} and \ref{sec:highdimm} are applied to decompose the partition function of the $(\partial X^i,\psi^i)$ CFT describing $d=10,8,6,4$ spacetime dimensions into characters of the little group $SO(d-1)$. According to figure \ref{roadmap}, the $d=10$ case takes the role of the ancestor theory for 16 supercharges, so its spectrum will be analyzed in particular detail. In the remaining cases $d=8,6,4$, dimensional reduction converts part of the higher dimensional Lorentz symmetry into an internal R symmetry, i.e. we branch the ten dimensional little group into $SO(9) \rightarrow SO(d-1) \times SO(10-d)_R$. In this process, individual Lorentz fugacities $y_k$ with $k > \frac{1}{2}(d-2)$ are reinterpreted as R symmetry fugacities $r_k$.

Before looking at individual dimensionalities in detail, let us fix the notation for describing supersymmetric spectra with R symmetries: Characters of the spacetime little group $SO(d-1)$ are denoted by $[a_1,\ldots,a_n]$ with fugacities $y_1,\ldots,y_n$ and $n = \frac{1}{2}(d-2)$ whereas those of the R symmetry $SO(10-d)_R$ receive an extra subscript $[b_1,\ldots,b_\ell]_R$ with fugacities $r_1,\ldots,r_\ell$ and $\ell = 5-\frac{d}{2}$. Our notation for supercharacters makes use of double brackets $\llbracket a_1, \ldots, a_n;b_1,\ldots,b_\ell\rrbracket$ enclosing the $SO(d-1) \times SO(10-d)_R$ quantum numbers of the highest weight state. The semicolon between $a_n$ and $b_1$ separates spacetime from R symmetry Dynkin labels and eliminates any ambiguity about the spacetime dimension under consideration.

\subsection[Ten dimensional ${\cal N}_{10d}=1$ spectra]{Ten dimensional $\bm{{\cal N}_{10d}=1}$ spectra}
In this subsection, we want to revisit the results of \cite{Hanany:2010da} on $SO(9)$ covariant partition functions for ten dimensional open string excitations and examine further symmetry patterns. The minimal massive ${\cal N}_{10d}=1$ SUSY multiplet encompasses $SO(9)$ representations of a spin two tensor, a three-form and a massive gravitino\footnote{Note that $Z({\cal N}_{10d}=1)$ is denoted by $Z_Q$ in \cite{Hanany:2010da}.}
\beq
Z({\cal N}_{10d}=1) \ \ := \ \ [2,0,0,0] \ + \ [0,0,1,0] \ + \ [1,0,0,1] ~ .
\label{Z10}
\eeq
This is precisely the particle content of the first mass level, its vertex operators can for instance be found in equations (2.8), (2.9) and (2.22) of \cite{Feng:2012bb}.

The generic multiplet is obtained as a tensor product of $Z({\cal N}_{10d}=1)$ with some $SO(9)$ representation and therefore described by the following ${\cal N}_{10d}=1$ supercharacter:
\beq
\llbracket a_1, a_2, a_3, a_4\rrbracket \ \ := \ \ Z({\cal N}_{10d}=1) \cdot [a_1, a_2, a_3, a_4] 
\label{ZZ10}
\eeq
This is the basic building blocks of the refined ten dimensional partition function. The latter can be obtained through standard GSO projection of the spacetime CFT
\bea \label{NSandR10d}
\chi^{{\cal N}_{10d}=1}(q;\vec y) &\eq \chi^{{\cal N}_{10d}=1}_{\te{NS}} \mid_{\GSO} (q;\vec y) \ + \ \chi^{{\cal N}_{10d}=1}_{\te{R}} \mid_{\GSO} (q;\vec y) \notag \\
\chi^{{\cal N}_{10d}=1}_{\te{NS}} \mid_{\GSO} (q;\vec y,r) &\eq
\frac{1}{2} \, q^{-\frac{1}{2}} \, \big[ \, \chi_{\te{NS}}^{SO(9)}(q;\vec y) \ - \ \chi_{\te{NS}}^{SO(9)}(e^{2\pi i}q;\vec y) \, \big] \notag \\
\chi^{{\cal N}_{10d}=1}_{\te{R}} \mid_{\GSO} (q;\vec y,r) &\eq \frac{1}{2} \,  \chi_{\te{R}}^{SO(9)}(q;\vec y)  \ ,
\eea
where $\chi_{\te{NS}}^{SO(9)}(q;\vec y)$ and $\chi_{\te{R}}^{SO(9)}(q;\vec y)$ are given by \eref{BnNS} and \eref{BnR}.

In a power series expansion in $q$, the coefficient of the $n$'th power $q^n$ comprises the super Poincar\'e characters of the $n$'th mass level $m^2 = n /\ap$:\footnote{Note the usual subtlety about the massless R sector which was explained in the footnote before (\ref{def:GSOed}). One can simply fix this by adding $\frac{1}{2}([0,0,0,1]_{SO(8)}-[0,0,1,0]_{SO(8)}) = \frac{1}{2} \prod_{i=1}^4 (y_i - y_i^{-1})$ to the present result and obtain the correct answer; see also (3.16) of \cite{Hanany:2010da}. The $\frac{1}{2}[1,0,0,0]_9$ factor in the massive sector of the aforementioned (3.16) exactly matches our formula at any positive $q$ power.}
\begin{align} \label{chi10dref}
\chi^{{\cal N}_{10d}=1}(q;\vec y)&\eq \underbrace{\left( \, \sum_{j=1}^4 (y_j^2 + y_j^{-2}) \ + \ \frac{1}{2} \, \prod_{j=1}^4 (y_j + y_j^{-1}) \, \right) \, q^0}_{16 \ \te{massless states}}  \ + \ \underbrace{ \llbracket 0,0,0,0\rrbracket  \, q}_{256 \ \te{states at level} \ 1}   \notag \\
&   + \ \underbrace{ \llbracket 1,0,0,0\rrbracket  \, q^2}_{2304 \ \te{states at level} \ 2} \ + \ \underbrace{\big( \, \llbracket 2,0,0,0 \rrbracket  \, + \, \llbracket 0,0,0,1 \rrbracket  \, \big) \, q^3}_{15360 \ \te{states at level} \ 3} \notag \\
&   + \ \big( \, \llbracket 3,0,0,0 \rrbracket  \, + \, \llbracket 1,0,0,1 \rrbracket  \, + \, \llbracket 1,0,0,0 \rrbracket  \, + \, \llbracket 0,1,0,0 \rrbracket  \, \big) \, q^4 \ + \ {\cal O}(q^5)  \ .
\end{align}
The supermultiplets up to level eight are listed in table \ref{table10,1} and the complete first 25 mass levels can be found in table \ref{tab10d000} and appendix \ref{app10}.

\bigskip
\noindent
\begin{table}[htdp]
\begin{tabular}{|l|l|}\hline  $\ap m^2$ &representations of ${\cal N}_{10d}=1$ super Poincar\'e \\ \hline \hline   
  1 & $\llbracket 0,0,0,0\rrbracket  $  \\\hline
   2 & $\llbracket 1,0,0,0\rrbracket  $  \\\hline
   3 & $\llbracket 2,0,0,0 \rrbracket  \, + \, \llbracket 0,0,0,1 \rrbracket  $ \\ \hline
   4 & $\llbracket 3,0,0,0 \rrbracket  \, + \, \llbracket 1,0,0,1 \rrbracket  \, + \, \llbracket 1,0,0,0 \rrbracket  \, + \, \llbracket 0,1,0,0 \rrbracket $
   \\\hline        
   5 &$\llbracket 4,0,0,0 \rrbracket  \, + \,\llbracket 2,0,0,1 \rrbracket  \, + \, \llbracket 2,0,0,0 \rrbracket  \, + \, \llbracket 1,1,0,0 \rrbracket  \, + \,\llbracket 1,0,0,1 \rrbracket  \, + \, \llbracket 0,1,0,0 \rrbracket  $
     \\
     &$\, + \, \llbracket 0,0,1,0 \rrbracket  \, + \,  \llbracket 0,0,0,1 \rrbracket  \, + \, \llbracket 0,0,0,0 \rrbracket  $
     \\\hline
   6 &$\llbracket 5,0,0,0 \rrbracket  \, + \,\llbracket 3,0,0,1 \rrbracket  \, + \, \llbracket 3,0,0,0 \rrbracket  \, + \, \llbracket 2,1,0,0 \rrbracket  \, + \,\llbracket 2,0,0,1 \rrbracket \, + \, \llbracket 2,0,0,0 \rrbracket \, + \,2\, \llbracket 1,1,0,0 \rrbracket    $
     \\
     &$  \, + \, \llbracket 1,0,1,0 \rrbracket \, + \, 2\, \llbracket 1,0,0,1 \rrbracket    \, + \, 2 \, \llbracket 1,0,0,0 \rrbracket  \, + \, \llbracket 0,1,0,1 \rrbracket  \, + \, \llbracket 0,1,0,0 \rrbracket  \, + \, \llbracket 0,0,0,2 \rrbracket   \, + \, 2 \, \llbracket 0,0,0,1 \rrbracket  $
     \\\hline
     7 &$\llbracket 6,0,0,0 \rrbracket \, + \,  
\llbracket4,0,0,1 \rrbracket \, + \, 
\llbracket4,0,0,0 \rrbracket \, + \,  
\llbracket3,1,0,0 \rrbracket \, + \,  
\llbracket3,0,0,1 \rrbracket \, + \,  
\llbracket3,0,0,0 \rrbracket \, + \,  
2 \, \llbracket2,1,0,0 \rrbracket $ \\
&$ \, + \,  
\llbracket2,0,1,0 \rrbracket \, + \,  
3\,\llbracket2,0,0,1 \rrbracket \, + \,  
3\,\llbracket2,0,0,0 \rrbracket \, + \,  
\llbracket1,1,0,1 \rrbracket \, + \,  
2\,\llbracket1,1,0,0 \rrbracket \, + \,  
\llbracket1,0,1,0 \rrbracket \, + \,  
\llbracket1,0,0,2 \rrbracket $ \\
&$ \, + \,  
4\,\llbracket1,0,0,1 \rrbracket \, + \,  
2\,\llbracket1,0,0,0 \rrbracket \, + \,  
\llbracket0,2,0,0 \rrbracket \, + \,  
2\,\llbracket0,1,0,1 \rrbracket \, + \,  
2\,\llbracket0,1,0,0 \rrbracket \, + \, 
3\,\llbracket0,0,1,0 \rrbracket $ \\
&$ \, + \,  
\llbracket0,0,0,2 \rrbracket \, + \,  
2\,\llbracket0,0,0,1 \rrbracket \, + \,  
2\,\llbracket 0,0,0,0 \rrbracket   $
     \\\hline
8 &$\llbracket 7,0,0,0\rrbracket \, + \,  
 \llbracket5,0,0,1\rrbracket \, + \,  
 \llbracket5,0,0,0\rrbracket \, + \,  
 \llbracket4,1,0,0\rrbracket \, + \, 
 \llbracket4,0,0,1\rrbracket \, + \, 
 \llbracket4,0,0,0\rrbracket \, + \, 
2\, \llbracket3,1,0,0\rrbracket $ \\
&$ \, + \,  
 \llbracket3,0,1,0\rrbracket \, + \, 
3\, \llbracket3,0,0,1\rrbracket \, + \,  
4\, \llbracket3,0,0,0\rrbracket \, + \,  
 \llbracket2,1,0,1\rrbracket \, + \,  
3\, \llbracket2,1,0,0\rrbracket \, + \,  
 \llbracket2,0,1,0\rrbracket \, + \, 
 \llbracket2,0,0,2\rrbracket $ \\
&$\, + \, 
5\, \llbracket2,0,0,1\rrbracket \, + \, 
3\, \llbracket2,0,0,0\rrbracket \, + \, 
 \llbracket1,2,0,0\rrbracket \, + \, 
3\, \llbracket1,1,0,1\rrbracket \, + \, 
5\, \llbracket1,1,0,0\rrbracket \, + \, 
4\, \llbracket1,0,1,0\rrbracket $ \\
&$ \, + \, 
2\, \llbracket1,0,0,2\rrbracket \, + \, 
7\, \llbracket1,0,0,1\rrbracket \, + \, 
5\, \llbracket1,0,0,0\rrbracket \, + \, 
 \llbracket0,2,0,0\rrbracket \, + \, 
 \llbracket0,1,1,0\rrbracket \, + \,  
4\, \llbracket0,1,0,1\rrbracket $ \\
&$\, + \, 
5\, \llbracket0,1,0,0\rrbracket \, + \, 
 \llbracket0,0,1,1\rrbracket \, + \, 
2\, \llbracket0,0,1,0\rrbracket \, + \,
3\, \llbracket0,0,0,2\rrbracket \, + \,  
4\, \llbracket0,0,0,1\rrbracket \, + \,  
 \llbracket0,0,0,0\rrbracket   $
     \\\hline
\end{tabular}
\caption{${\cal N}_{10d}=1$ multiplets occurring up to mass level eight}
\label{table10,1}
\end{table}

\subsubsection{The total number of states at a given mass level} \label{sec:totalnumstates16susys}

The total number of states at a given mass level $m$ can be read off from the coefficient of $q^m$ in the partition function $\chi^{{\cal N}_{10d}=1}(q;\vec y)$ when the $SO(9)$ fugacities $y_1, \ldots, y_4$ are set to unity.  The function $\chi^{{\cal N}_{10d}=1}(q;\{ y_i =1 \} )$ is referred to as the {\it unrefined partition function}.  From \eref{unrefBnNS},  \eref{NSandR10d} and SUSY\footnote{The agreement of GSO projected partition functions for NS and R sectors follows from Jacobi's abstruse identity:
\bea
\vartheta_3(1, q)^4 - \vartheta_4(1, q)^4 - \vartheta_2(1, q)^4  = 0~.
\eea
}, we have
%\bea
%\chi^{{\cal N}_{10d}=1} _{\te{NS}} \mid_{\GSO} (q; \{y_i =1\}) &= \frac{1}{2} q^{-\frac{1}{2}} \left[ q^{\frac{1}{2}} \frac{\vartheta_3(1, q)^4}{\eta(q)^{12}} - q^{\frac{1}{2}} \frac{\vartheta_4(1, q)^4}{\eta(q)^{12}} \right]  =\frac{\vartheta_2(1, q)^4}{2\eta(q)^{12}}~, \nn \\
%\chi^{{\cal N}_{10d}=1}_{\te{R}} \mid_{\GSO} (q;\{y_i =1\})  &= \frac{\vartheta_2(1, q)^4}{2\eta(q)^{12}}~,
%\eea
%where in the first line we have use Jacobi's abstruse identity:
%\bea
%\vartheta_3(1, q)^4 - \vartheta_4(1, q)^4 - \vartheta_2(1, q)^4  = 0~.
%\eea
%Observe that GSO projected partition functions for NS and R sectors are identical.  Hence, the numbers of states at a given mass level in the NS and R sectors are equal; this is indeed an expected result of supersymmetry.
%
%Thus, the unrefined partition function is given by
\bea
\chi^{{\cal N}_{10d}=1}(q;\{ y_i =1 \} ) = 2\chi^{{\cal N}_{10d}=1}_{\te{R}} \mid_{\GSO} (q;\{y_i =1\}) = \frac{\vartheta_2(1, q)^4}{\eta(q)^{12}} = 16 \prod_{n=1}^\infty  \left( \frac{1+q^n}{1-q^n} \right)^8~. \label{exactunref10d}
\eea
The coefficients in the power series of this formula reproduces the third column of \tref{numberofstates}.  It also agrees with (5.3.37) of \cite{Green:1987sp}. Note that $\chi^{{\cal N}_{10d}=1}(q;\{ y_i =1 \} )$ is {\it not} a modular form.

%... is not for the following reason.  Taking $q = \exp(2 \pi i \tau) \mapsto \exp (-2\pi i/\tau) =: \tilde q$, we have
%\bea
% \vartheta_2 \left( 1, \tilde q \right) = \vartheta_4(1, q) \sqrt{-i \tau}~, \qquad
%\eta\left(  \tilde q \right) = \eta(q) \sqrt{-i \tau} ~, \label{modulartheta2andeta}
%\eea
%and so
%\bea
%\chi^{{\cal N}_{10d}=1}(\tilde q ;\{ y_i =1 \} ) =  \frac{\vartheta_2 \left( 1, \tilde q \right)^4}{\eta\left( \tilde q \right)^{12}}  = \tau^4~\frac{\vartheta_4(1, q)^4}{\eta(q)^{12}}~;
%\eea
%since the right hand side cannot be arranged into the form $\tau^d ~ \frac{\vartheta_2(1, q)^4}{\eta(q)^{12}}  $,  it is then clear that the unrefined partition function is not modular.

\subsubsection*{The number of states at each mass level and its asymptotics} 
The number of states at the mass level $m$ can be determined by
\bea
N_m = \frac{1}{2 \pi i} \oint_{\CC} \frac{\ud q}{q^{m+1}}~ \chi^{{\cal N}_{10d}=1}(q;\{ y_i =1 \} )~, \label{numstates10d}
\eea
where $\CC$ is a contour around the origin. 

Now let us compute an asymptotic formula for the number of states $N_m$ at mass level $m$ when $m \rightarrow \infty$. Note that a similar discussion can be found in subsections 4.3.3 and 5.3.1 of \cite{Green:1987sp}. For completeness, let us go over some details here.
We focus on the limit $q \rightarrow 1^-$ and proceed in a similar way to subsection \ref{sec:numberstates4SUSYs}. The asymptotic behaviour (\ref{theta2approx}) and (\ref{etaapprox}) of $\vartheta_2(1,q)$ and $\eta(q)$, respectively, leads to

%Let us compute the number of states $N_m$ in the limit $m \rightarrow \infty$.  
%Since the integrand of \eref{numstates10d} is sharply peaked near $q=1$, we need to examine the behaviour of $\chi^{{\cal N}_{10d}=1}(q;\{ y_i =1 \} )$ as $q \rightarrow 1^-$.  This can be done using the modular property of \eref{exactunref10d} as follows (using $\tilde q = e^{-2\pi i/\tau}$).
%\bea
%& \text{{\bf $\vartheta$-function}}: \qquad   \vartheta_4 \left( 1, \tilde q \right) = \vartheta_2(1, q ) \sqrt{-i \tau} \sim \frac{1}{\sqrt{ 2\pi}} (1-q)^{1/2} \vartheta_2(1,q) \nn \\
%&\Rightarrow  \hspace{2.2cm}  \vartheta_2(1, q) \sim \sqrt{2 \pi} (1-q)^{-1/2},\qquad q \rightarrow 1^-  ~,\label{theta2approx}
%\eea
%where we have used the fact that $\vartheta_4(1, q) \rightarrow 1$ as $\tau \rightarrow i \infty$.
%\bea
%&\text{{\bf $\eta$-function}}: \qquad  \eta\left(\tilde q \right) = \eta(q) \sqrt{-i \tau } \sim \frac{1}{\sqrt{2 \pi}} (1-q)^{1/2} \eta(q)  \nn \\
%&\Rightarrow \hspace{2.2cm} \eta(q) \sim \sqrt{2 \pi} (1-q)^{-1/2} \exp \left( {\frac{\pi^2}{6 \log q}} \right), \qquad q \rightarrow 1^-~,  \label{etaapprox}
%\eea
%where we have used the fact that $\eta(q) \sim q^{1/24}$ as $q \rightarrow 0$ or $\tau \rightarrow i \infty$.
%
%Therefore,
\bea
\chi^{{\cal N}_{10d}=1}(q;\{ y_i =1 \} ) \sim \frac{1}{(2 \pi)^4} (1-q)^{4} \exp \left( -\frac{2 \pi^2}{\log q} \right) ~, \quad q \rightarrow 1^- ~. \label{asymp10dunrefq1} 
\eea
Let us now combine \eref{numstates10d} with \eref{asymp10dunrefq1}.  As $m \rightarrow \infty$,
\bea
N_m \sim \frac{1}{(2 \pi)^4} \frac{1}{2 \pi i} \oint_{\CC} \frac{\ud q}{q} ~(1-q)^4\exp \left( -\frac{2 \pi^2}{\log q} - m \log q \right)~.
\eea
%Observe that the argument of the exponential function has a critical value at $q_0 = \exp \left( - \pi \sqrt{\frac{2}{m}}  \right)$; this is the saddle point. The direction of steepest descent at this point is the imaginary direction in $q$.  We deform the contour $\CC$ such that it passes through $q=q_0$ and tangent to this direction. The leading contribution comes from expansions around $q=q_0$ in the steepest descent direction.  Writing $q = q_0 e^{i \theta}$, we have

\begin{figure}[htbp]
\begin{center}
\includegraphics[scale=1]{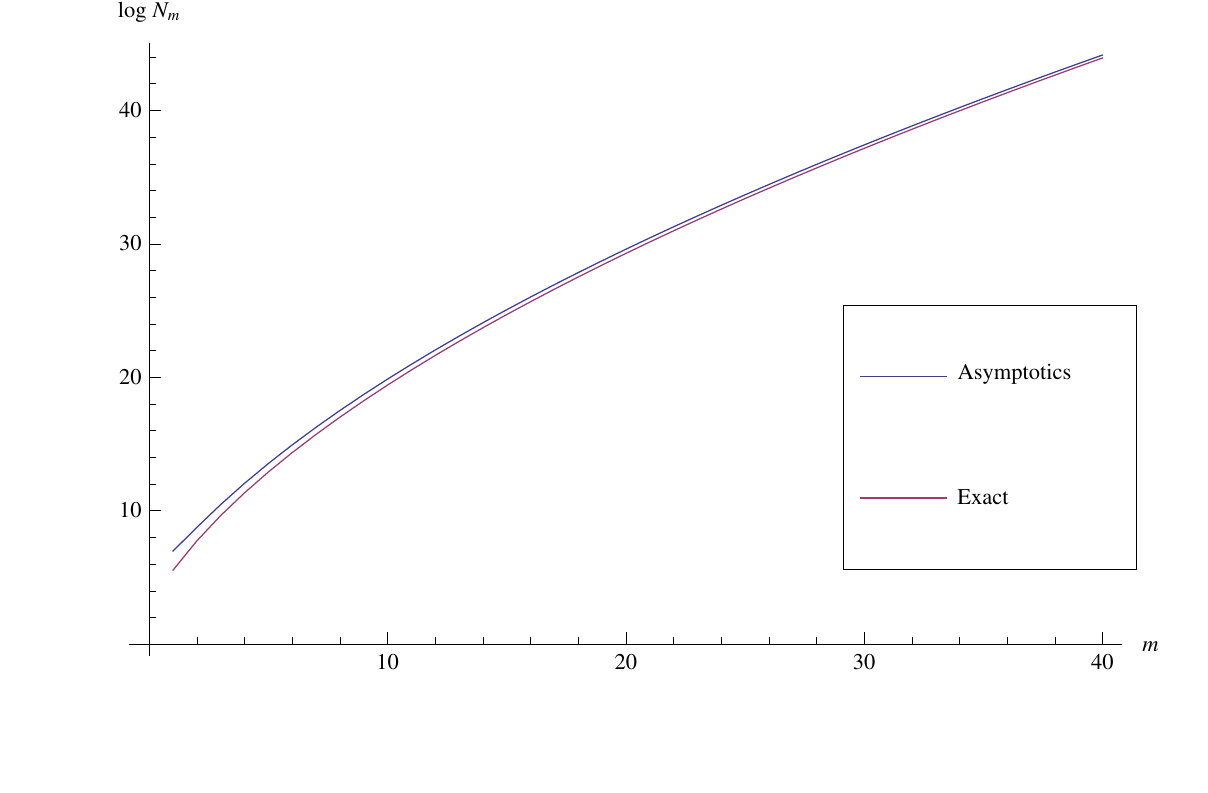}
\vspace{-1.2cm}
\caption{The plot of the exact and asymptotic values of  $\log N_m$ against the mass level $m$ for the case of 16 supercharges.}
\label{fig:compareexactasymp10d}
\end{center}
\end{figure}

The saddle point is at $q_0 = \exp \left( - \pi \sqrt{\frac{2}{m}}  \right)$ and the steepest descent direction is the imaginary direction in $q$.  We proceed in a similar way to \eref{asympNm4d} by writing $q=q_0e^{i\theta}$ and using Laplace's method to obtain
\bea
N_m 
%&\sim \frac{1}{(2 \pi)^4} (1-q_0)^4  \frac{1}{2 \pi} \int_{-\epsilon}^{\epsilon} \ud \theta \exp \left(-\frac{2 \pi^2}{i \theta+ \log q_0 } - m (i \theta + \log q_0)  \right), \quad \epsilon >0 \nn \\
& \sim \frac{1}{4} m^{-2} \exp \left( 2 \pi \sqrt{2m } \right) \frac{1}{2 \pi} \int_{-\infty}^{\infty} \ud \theta \exp \left( -\frac{m^{3/2}}{\pi \sqrt{2}  } \theta^2 \right) \nn \\
%& =  \frac{1}{4} m^{-2} \exp \left( 2 \pi \sqrt{2m } \right) \frac{1}{2 \pi} \frac{2^{1/4} \pi }{m^{3/4}} \nn \\
&\sim \frac{1}{2^{11/4}} m^{-11/4} e^{2 \pi \sqrt{2m}}~, \qquad m \rightarrow \infty~. \label{asympNm10d}
\eea
%For example, for $m=100$, the exact value for $N_{100}$ is $1.59 \times 10^{32}$ and the value from \eref{asympNm10d} is  $1.83 \times 10^{32}$ ; the error is approximately 15 \%.
The plot of the exact and asymptotic values for $N_m$ against $m$ is depicted in \fref{fig:compareexactasymp10d}.

\subsubsection{The GSO projected NS and R sectors}
In this section we compute the contributions from the NS and R sectors to the partition function given in \eref{NSandR10d}.  Here we consider the refined partition function, \ie~ the fugacities $y$'s are kept explicit. 
\paragraph{The NS sector.}
From \eref{NSandR10d} and \eref{BnNS}, the partition function of the GSO projected NS sector has the structure 
\bea
\chi^{\CN_{10d}=1}_{\NS} \mid_{\GSO} (q; y) 
&= \sum_{\vec k \in \BZ^4_{\geq 0}} F^{\NS}_{k_1, \ldots, k_4} (q) \prod_{A=1}^4 [2k_A]_{y_A}~,
\eea
where the functions $F^{\NS}_{k_1, \ldots, k_4} (q)$ are given by 
\bea
F^{\NS}_{k_1, \ldots, k_4} (q) &=  (q;q)^{-12}_\infty \sum_{\vec n \in \BZ^4_+} \sum_{\vec m \in \BZ^4_{\geq 0}} \prod_{A=1}^4  (-1)^{n_A+1} (1-q^{n_A}) q^{\frac{1}{2}m_A^2+{n_A \choose 2}} (q^{n_A|k_A-m_A|} -q^{n_A(k_A+m_A+1)}) \nn \\
& \quad \times \frac{1}{2} \Bigg[ \prod_{A=1}^4 (1-q^{m_A+\frac{1}{2}}) + (-1)^{m_1^2+m_2^2+m_3^2+m_4^2} \prod_{A=1}^4 (1+q^{m_A+\frac{1}{2}})\Bigg]~.
\eea

\paragraph{The R sector.}
From \eref{NSandR10d} and \eref{BnR}, the partition function of the GSO projected R sector is
\bea
\chi^{\CN_{10d}=1}_{\Ra} \mid_{\GSO} (q; y, s) 
&= \sum_{\vec k \in \BZ^4_{\geq 0}} F^{\Ra}_{k_1, \ldots, k_4} (q) \prod_{A=1}^4 [2k_A+1]_{y_A}~,
\eea
where the function $F^{\Ra}_{k_1, \ldots, k_4} (q)$ is given by
\bea
F^{\Ra}_{k_1, \ldots, k_4} (q) &= \frac{1}{2} q^{-\frac{1}{2}}  (q;q)_\infty^{-12}  \sum_{\vec m  \in\BZ_{\geq 0}^4} \sum_{\vec n \in \BZ_+^4}   \prod_{A=1}^4 (-1)^{n_A+1} (1-q^{m_A+1}) (1-q^{n_A}) q^{\frac{1}{2}\left(m_A+\frac{1}{2} \right)^2+ {n_A \choose 2}}   \nn \\
& \hspace{3cm}\times   \prod_{A=1}^4  ( q^{n_A|k_A-m_A|} - q^{n_A(k_A+m_A+2)} )~.
\eea

\subsubsection{Multiplicities of representations in the $\CN_{10d} = 1$ partition function}
Combining the contributions from the NS and R sectors, we have
\bea
\chi^{{\cal N}_{10d}=1}(q;\vec y) &= \chi^{{\cal N}_{10d}=1} _{\te{NS}} \mid_{\GSO} (q;\vec y) \ + \ \chi^{{\cal N}_{10d}=1}_{\te{R}} \mid_{\GSO} (q;\vec y) \nn \\
&= \sum_{\vec k \in \BZ^4_{\geq 0}} \Big(F^{\NS}_{\vec k} (q)  \prod_{A=1}^4 [2k_A]_{y_A}+ F^{\Ra}_{\vec k} \prod_{A=1}^4 [2k_A+1]_{y_A}  \Big)~.
\eea
Supersymmetry implies that this partition function can be rewritten as
\bea
\chi^{{\cal N}_{10d}=1}(q;\vec y) &= \sum_{\vec n \in \BZ^4_{\geq 0}}  \lb n_1, n_2, n_3, n_4 \rb ~ G_{n_1,n_2,n_3,n_4} (q) ~,
\label{GGG10}
\eea
and the aim is to compute explicitly a {\it multiplicity generating function} $G_{n_1,n_2,n_3,n_4} (q) $.

The multiplicity of $\lb n_1,n_2,n_3,n_4\rb$ appearing in $\chi^{{\cal N}_{10d}=1}(q; \vec y)$ can be determined as follows:
\bea
G_{n_1,n_2,n_3,n_4} (q) &= \int \ud \mu_{SO(9)}(\vec y) [n_1, n_2, n_3, n_4]_{\vec y} ~\frac{\chi^{{\cal N}_{10d}=1}(q;\vec y) }{Z(\CN_{10d}=1) (\vec y)}~, \nn \\
&= G^{\NS}_{n_1,n_2,n_3, n_4} (q) + G^{\Ra}_{n_1,n_2,n_3, n_4} (q)~, \label{sumboth10d}
\eea
where 
\bea
G^{\NS}_{n_1,n_2,n_3, n_4} (q) &=  \int \ud \mu_{SO(9)}(\vec y) [n_1, n_2, n_3, n_4]_{\vec y} ~\sum_{\vec k \in \BZ^4_{\geq 0}} \frac{\prod_{A=1}^4 [2k_A]_{y_A}}{Z(\CN_{10d}=1) (\vec y)} F^{\NS}_{k_1, \ldots, k_4} (q) ~, \\
G^{\Ra}_{n_1,n_2,n_3, n_4} (q) &=  \int \ud \mu_{SO(9)}(\vec y) [n_1, n_2, n_3, n_4]_{\vec y} ~\sum_{\vec k \in \BZ^4_{\geq 0}} \frac{\prod_{A=1}^4 [2k_A+1]_{y_A}}{Z(\CN_{10d}=1) (\vec y)} F^{\Ra}_{k_1, \ldots, k_4} (q) ~.
\eea
The inverse of the character of the fundamental multiplet in \eref{Z10} can be written as a geometric series\footnote{Note that this can also be rewritten as 
\bea
\left[ Z(\CN_{10d}=1) (\vec y) \right]^{-1} &=  \lim_{s \rightarrow -1} (\PE \left[ s [0,0,0,1]_{\vec y} \right] )^{1/2} = \left[ \sum_{m=0}^\infty (-1)^m  \Sym^m [0,0,0,1]_{\vec y} \right]^{1/2}~.
\eea} similar to (\ref{geoser}) and (\ref{geoser6d})
\bea
&\left[ Z(\CN_{10d}=1) (\vec y,r) \right]^{-1}  \nn \\
&= \frac{y_4^4}{\left(1+\frac{y_4}{y_1 y_2 y_3}\right) \left(1+\frac{y_1 y_4}{y_2 y_3}\right) \left(1+\frac{y_2 y_4}{y_1 y_3}\right) \left(1+\frac{y_1 y_2 y_4}{y_3}\right) \left(1+\frac{y_3 y_4}{y_1 y_2}\right) \left(1+\frac{y_1 y_3 y_4}{y_2}\right) \left(1+\frac{y_2 y_3 y_4}{y_1}\right) \left(1+y_1 y_2 y_3 y_4\right)} \nn \\
&= \sum_{\vec m \in \BZ^8_{\geq 0}} (-1)^{\sum_{j=1}^8 m_j} y_1^{\sum_{j=1}^8 (-1)^j m_j} y_2^{\sum_{j=1}^8 (-1)^{\lfloor (j+1)/2 \rfloor} m_j} y_3^{\sum_{j=1}^8 (-1)^{\lfloor (j+3)/4 \rfloor}  m_j}   y_4^{4+\sum_{j=1}^8 m_j}~.
\label{geoserr}
\eea

\subsubsection{Some useful identities}
In this section, we derive some useful identities that will be put into use later. Once we plug the series expansion (\ref{geoserr}) of the inverse $Z(\CN_{10d}=1)$ into the integrand of (\ref{sumboth10d}), the elementary contributions to multiplicity generating functions $G_{n_1,n_2,n_3,n_4}$ are integrals of type
%The first one follows from \eref{idenortho}: 
\bea \label{iden10d1}
& {\cal J}_0(w; \vec p) := \int \ud \mu_{SO(3)} (r)~  r^w \prod_{A=1}^4 [p_A]_{r}~.
%&= \begin{cases} 
%\delta_{p_1,p_2} &\quad \text{for $w =  0$}  \\
%\frac{1}{2} \sum_{p= 0}^{\frac{1}{2}(p_1+p_2- |p_1-p_2|)}  \left( \delta_{|w|,2p+|p_1-p_2|} - \delta_{|w|,2p+2+|p_1-p_2|} \right) &\quad \text{for $w \neq 0$}  
%\end{cases}
\eea
as well as
\bea
{\cal J} (\vec w; \vec k; \vec n) :=\int \ud \mu_{SO(9)}(\vec y) [n_1,n_2,n_3,n_4]_{\vec y} \prod_{A=1}^4 y_A^{w_A}  [k_A]_{y_A} ~.
\eea
There are four cases to be considered, namely spin/non-spin representations of $SO(9)$ and for each of these cases $k_1, \ldots, k_4$ can be all even or all odd.  In what follows, we assume that $\vec k, \vec n \in \BZ_{\geq 0}^4$ and $\vec w \in \BZ^4$. For non-spin representations,
{\small
\bea
{\cal J}(\vec w; ~2 k_1, \ldots, 2k_4; ~n_1,\ldots,2n_4)  &= \sum_{\vec k' \in \BZ_{\geq 0}^4} \Delta(\vec \lambda_{ns}; 2k'_1, \ldots, 2k'_4) \prod_{A =1}^4 {\cal J}_0 (w_A; 2k_A, 2k'_A)~,   \\
{\cal J}(\vec w; ~2 k_1+1, \ldots, 2k_4+1; ~n_1,\ldots,2n_4) 
& = \sum_{\vec k' \in \BZ_{\geq 0}^4} \Delta( \vec \lambda_{ns} ; 2k'_1, \ldots, 2k'_4) \prod_{A =1}^4 {\cal J}_0 (w_A; 2k_A+1, 2k'_A)~,
\eea}
where $ \vec \lambda_{ns} = (n_1 + n_2 +n_3+ n_4, n_2 +n_3 + n_4, n_3+n_4, n_4)$. 
For spin representations,
{\small
\bea
{\cal J}(\vec w; ~2 k_1, \ldots, 2k_4;~ n_1,\ldots,2n_4+1) 
&  = \sum_{\vec k' \in \BZ_{\geq 0}^4} \Delta(  \vec \lambda_{s}; 2k'_1+1, \ldots,2k'_4+1) \prod_{A =1}^4 {\cal J}_0 (w_A; 2k_A, 2k'_A+1)~,  \\
{\cal J}(\vec w; ~2 k_1+1, \ldots, 2k_4+1; n_1,\ldots, 2n_4+1)
&= \sum_{\vec k' \in \BZ_{\geq 0}^4} \Delta( \vec \lambda_{s}; 2k'_1+1, \ldots, 2k'_4+1) \prod_{A =1}^4 {\cal J}_0 (w_A; 2k_A+1, 2k'_A+1)~,
\eea}
where $ \vec \lambda_{s} = \left(n_1 + n_2 +n_3+ n_4+\frac{1}{2}, n_2 +n_3 + n_4+\frac{1}{2}, n_3+n_4+\frac{1}{2}, n_4+\frac{1}{2} \right)$. 
Recall from \eref{CTorthoeven} and \eref{CTorthoodd} that
\bea
\Delta(\vec \lambda; 2k_1, \ldots, 2k_4) &= \frac{1}{4!} \sum_{\sigma \in S_4} \det \left(\theta_{|\lambda_A-A+B|}^{8+\lambda_A-A-B} \left(k_{\sigma (A)} \right)  \right)_{A,B=1}^4~, \\
\Delta(\vec \lambda; 2k_1+1, \ldots, 2k_4+1) &= \frac{1}{4!} \sum_{\sigma \in S_4} \det \left(\theta_{|\lambda_A-A+B|}^{8+\lambda_A-A-B} \left(k_{\sigma (A)}+\frac{1}{2} \right)  \right)_{A,B=1}^4~.
\eea

\subsubsection{Multiplicity generating function}
The NS- and R sector contributions to the multiplicity generating function for the representation $\lb n_1,n_2, n_3, n_4 \rb$ can be rewritten as
\bea
& G^{\NS}_{n_1,\ldots,n_4} (q) = \sum_{\vec m \in \BZ_{\geq 0}^8} (-1)^{\sum_{j=1}^8 m_j}   \sum_{\vec k \in \BZ_{\geq 0}^4}  {\cal J}(\vec W( \vec m); 2k_1, \ldots, 2k_4; \vec n)~F^{\NS}_{k_1,\ldots,k_4} (q)~, ~\\
& G^{\Ra}_{n_1,\ldots,n_4} (q) = \sum_{\vec m \in \BZ_{\geq 0}^8} (-1)^{\sum_{j=1}^8 m_j}  \sum_{\vec k \in \BZ_{\geq 0}^4}  {\cal J}(\vec W(\vec m); 2k_1+1,\ldots, 2k_4+1; \vec n)~F^{\Ra}_{k_1,\ldots,k_4} (q)~, 
\eea
where 
\bea
\vec W(\vec m) = \left( \sum_{j=1}^8 (-1)^j m_j, ~ \sum_{j=1}^8 (-1)^{\lfloor (j+1)/2 \rfloor} m_j, ~\sum_{j=1}^8 (-1)^{\lfloor (j+3)/4 \rfloor}  m_j, ~4+\sum_{j=1}^8 m_j \right)~.
\eea

As stated in \eref{NSandR10d}, the multiplicity of the representation $\lb n_1,n_2, n_3, n_4 \rb$ in the $\CN_{10d} = 1$ partition function is given by
\bea
G_{n_1,n_2,n_3,n_4} (q) &= \sum_{\vec m \in \BZ_{\geq 0}^8} (-1)^{\sum_{j=1}^8 m_j}   \sum_{\vec k \in \BZ_{\geq 0}^4}  \Big[ {\cal J}(\vec W( \vec m); 2k_1, \ldots, 2k_4; \vec n)~F^{\NS}_{k_1,\ldots,k_4} (q) \nn \\
& \hspace{1cm} + {\cal J}(\vec W( \vec m); 2k_1+1, \ldots, 2k_4+1; \vec n)~F^{\Ra}_{k_1,\ldots,k_4} (q)  \Big]~.
\label{end10dim}
\eea

\subsection[Empirical approach to ${\cal N}_{10d}=1$ asymptotic patterns]{Empirical approach to $\bm{{\cal N}_{10d}=1}$ asymptotic patterns}
\label{emp10dim}

In this subsection, we proceed like in subsections \ref{emp4dim} and \ref{emp6dim} to obtain large spin asymptotics of multiplicity generating functions $G_{n,x,y,z}(q)$ for ${\cal N}_{10d}=1$ supermultiplet $\llbracket n,x,y,z\rrbracket$. The supermultiplet content of the first 25 mass levels is used to determine the $q$ expansion of the leading coefficients $\tau^{x,y,z}_{\ell}(q)$ defined by:
\begin{align}
G_{n,x,y,z}(q) \eq &q^n \, \tau_1^{x,y,z}(q) \ - \ q^{2n} \, \tau_2^{x,y,z}(q) \ + \ q^{3n} \, \tau_3^{x,y,z}(q) \ - \ \ldots \notag \\
\eq &\sum_{\ell=1}^{\infty} (-1)^{\ell-1} \, q^{\ell n} \, \tau_\ell^{x,y,z}(q)
\label{10dimtau}
\end{align}
Again, the $\tau^{x,y,z}_{\ell}(q)$ are found to be power series in $q$ with non-negative coefficients.

Having $d > 4$ spacetime dimensions makes the analytic methods of subsection \ref{asympt4dim} inefficient, i.e. we did not find a manageable asymptotic formula for (\ref{end10dim}). Hence, we compute the $\tau_\ell^{x,y,z}(q)$ at $\ell \leq 5$ on the basis of an ${\cal O}(q^{25})$ expansion of the partition function (\ref{NSandR10d}). The multiplicities of $\llbracket n,0,0,0\rrbracket$ multiplets are shown in the following table \ref{tab10d000}, and analogous data tables for $\llbracket n,x,y,z\rrbracket$ at nonzero values of $x,y,z$ can be found in appendix \ref{app10}. The numbers marked in red match with the leading trajectory contribution $q^n\tau_{1}^{x,y,z}(q)$ whereas blue numbers correspond to $q^n \tau_{1}^{x,y,z}(q)-q^{2n}\tau_{2}^{x,y,z}(q)$ including one subleading trajectory.

\begin{table}
\begin{scriptsize}
\begin{center}
%\begin{sidewaysfigure}
\begin{tabular}{|c|| c|c|c|c|c| c|c|c|c|c| c|c|c|c|c|}
%\begin{longtable}{|c|| c|c|c|c|c| c|c|c|c|c| c|c|c|c|c| c|c|c|c|c| c|c|c|c|c| c|} % \small
 \hline
\begin{turn}{-90}$ \ \ \ \ \ \ap m^2$  \end{turn}
&\begin{turn}{-90}\# $\llbracket0,0,0,0\rrbracket$ \end{turn}
&\begin{turn}{-90}\# $\llbracket1,0,0,0\rrbracket$ \end{turn}
&\begin{turn}{-90}\# $\llbracket2,0,0,0\rrbracket$ \end{turn}
&\begin{turn}{-90}\# $\llbracket3,0,0,0\rrbracket$ \end{turn}
&\begin{turn}{-90}\# $\llbracket4,0,0,0\rrbracket$ \end{turn}
&\begin{turn}{-90}\# $\llbracket5,0,0,0\rrbracket$ \end{turn}
&\begin{turn}{-90}\# $\llbracket6,0,0,0\rrbracket$ \end{turn}
&\begin{turn}{-90}\# $\llbracket7,0,0,0\rrbracket$ \end{turn}
&\begin{turn}{-90}\# $\llbracket8,0,0,0\rrbracket$ \end{turn}
&\begin{turn}{-90}\# $\llbracket9,0,0,0\rrbracket$ \end{turn}
&\begin{turn}{-90}\# $\llbracket10,0,0,0\rrbracket\ $ \end{turn}
&\begin{turn}{-90}\# $\llbracket11,0,0,0\rrbracket$ \end{turn}
&\begin{turn}{-90}\# $\llbracket12,0,0,0\rrbracket$ \end{turn}
&\begin{turn}{-90}\# $\llbracket13,0,0,0\rrbracket$ \end{turn}
&\begin{turn}{-90}\# $\llbracket14,0,0,0\rrbracket$ \end{turn}
%&\begin{turn}{-90}\# $\llbracket15,0,0,0\rrbracket$ \end{turn}
%
\\\hline \hline
1 &\textcolor{red}{1} &\textcolor{red}{0} & & &
& & & & &
& & & & & \\\hline
2 &\textcolor{red}{0} &\textcolor{red}{1} &\textcolor{red}{0} & &
& & & & &
& & & & & \\\hline
3 &\textcolor{blue}{0} &\textcolor{red}{0} &\textcolor{red}{1} &\textcolor{red}{0} &
& & & & &
& & & & & \\\hline
4 &0 &\textcolor{red}{1} &\textcolor{red}{0} &\textcolor{red}{1} &\textcolor{red}{0}
& & & & &
& & & & & \\\hline
5 &1 &\textcolor{blue}{0} &\textcolor{red}{1} &\textcolor{red}{0} &\textcolor{red}{1}
&\textcolor{red}{0} & & & &
& & & & & \\\hline
6 &0 &\textcolor{blue}{2} &\textcolor{red}{1} &\textcolor{red}{1} &\textcolor{red}{0}
&\textcolor{red}{1} &\textcolor{red}{0} & & &
& & & & & \\\hline
7 &2 &2 &\textcolor{blue}{3} &\textcolor{red}{1} &\textcolor{red}{1}
&\textcolor{red}{0} &\textcolor{red}{1} &\textcolor{red}{0} & &
& & & & & \\\hline
8 &1 &5 &\textcolor{blue}{3} &\textcolor{red}{4} &\textcolor{red}{1}
&\textcolor{red}{1} &\textcolor{red}{0} &\textcolor{red}{1} &\textcolor{red}{0} &
& & & & & \\\hline
9 &3 &5 &\textcolor{blue}{9} &\textcolor{blue}{4} &\textcolor{red}{4}
&\textcolor{red}{1} &\textcolor{red}{1} &\textcolor{red}{0} &\textcolor{red}{1} &\textcolor{red}{0}
& & & & & \\\hline
10 &3 &12 &10 &\textcolor{blue}{11} &\textcolor{red}{5}
&\textcolor{red}{4} &\textcolor{red}{1} &\textcolor{red}{1} &\textcolor{red}{0} &\textcolor{red}{1}
&\textcolor{red}{0} & & & & \\\hline
11 &8 &15 &23 &\textcolor{blue}{14} &\textcolor{blue}{12}
&\textcolor{red}{5} &\textcolor{red}{4} &\textcolor{red}{1} &\textcolor{red}{1} &\textcolor{red}{0}
&\textcolor{red}{1} &\textcolor{red}{0} & & & \\\hline
12 &8 &30 &31 &\textcolor{blue}{31} &\textcolor{blue}{16}
&\textcolor{red}{13} &\textcolor{red}{5} &\textcolor{red}{4} &\textcolor{red}{1} &\textcolor{red}{1}
&\textcolor{red}{0} &\textcolor{red}{1} &\textcolor{red}{0} & & \\\hline
13 &19 &41 &61 &45 &\textcolor{blue}{36}
&\textcolor{blue}{17} &\textcolor{red}{13} &\textcolor{red}{5} &\textcolor{red}{4} &\textcolor{red}{1}
&\textcolor{red}{1} &\textcolor{red}{0} &\textcolor{red}{1} &\textcolor{red}{0} & \\\hline
14 &22 &77 &89 &87 &\textcolor{blue}{53}
&\textcolor{blue}{38} &\textcolor{red}{18} &\textcolor{red}{13} &\textcolor{red}{5} &\textcolor{red}{4}
&\textcolor{red}{1} &\textcolor{red}{1} &\textcolor{red}{0} &\textcolor{red}{1} &\textcolor{red}{0} \\\hline
15 &41 &109 &164 &132 &\textcolor{blue}{104}
&\textcolor{blue}{58} &\textcolor{blue}{39} &\textcolor{red}{18} &\textcolor{red}{13} &\textcolor{red}{5}
&\textcolor{red}{4} &\textcolor{red}{1} &\textcolor{red}{1} &\textcolor{red}{0} &\textcolor{red}{1} \\\hline
16 &57 &190 &245 &244 &162
&\textcolor{blue}{113} &\textcolor{blue}{60} &\textcolor{red}{40} &\textcolor{red}{18} &\textcolor{red}{13}
&\textcolor{red}{5} &\textcolor{red}{4} &\textcolor{red}{1} &\textcolor{red}{1} &\textcolor{red}{0} \\\hline
17 &100 &282 &426 &378 &299
&\textcolor{blue}{179} &\textcolor{blue}{118} &\textcolor{blue}{61} &\textcolor{red}{40} &\textcolor{red}{18}
&\textcolor{red}{13} &\textcolor{red}{5} &\textcolor{red}{4} &\textcolor{red}{1} &\textcolor{red}{1} \\\hline
18 &138 &471 &656 &657 &473
&\textcolor{blue}{332} &\textcolor{blue}{188} &\textcolor{blue}{120} &\textcolor{red}{62} &\textcolor{red}{40}
&\textcolor{red}{18} &\textcolor{red}{13} &\textcolor{red}{5} &\textcolor{red}{4} &\textcolor{red}{1} \\\hline
19 &235 &710 &1097 &1040 &830
&532 &\textcolor{blue}{350} &\textcolor{blue}{193} &\textcolor{blue}{121} &\textcolor{red}{62}
&\textcolor{red}{40} &\textcolor{red}{18} &\textcolor{red}{13} &\textcolor{red}{5} &\textcolor{red}{4} \\\hline
20 &336 &1153 &1699 &1751 &1333
&938 &\textcolor{blue}{565} &\textcolor{blue}{359} &\textcolor{blue}{195} &\textcolor{red}{122}
&\textcolor{red}{62} &\textcolor{red}{40} &\textcolor{red}{18} &\textcolor{red}{13} &\textcolor{red}{5} \\\hline
21 &544 &1750 &2778 &2769 &2263
&1523 &\textcolor{blue}{1000} &\textcolor{blue}{583} &\textcolor{blue}{364} &\textcolor{blue}{196}
&\textcolor{red}{122} &\textcolor{red}{62} &\textcolor{red}{40} &\textcolor{red}{18} &\textcolor{red}{13} \\\hline
22 &799 &2785 &4309 &4561 &3630
&2600 &1635 &\textcolor{blue}{1034} &\textcolor{blue}{592} &\textcolor{blue}{366}
&\textcolor{red}{197} &\textcolor{red}{122} &\textcolor{red}{62} &\textcolor{red}{40} &\textcolor{red}{18} \\\hline
23 &$\!$1261$\!$ &4237 &6907 &7201 &6025
&4212 &2803 &\textcolor{blue}{1697} &\textcolor{blue}{1052} &\textcolor{blue}{597}
&\textcolor{blue}{367} &\textcolor{red}{197} &\textcolor{red}{122} &\textcolor{red}{62} &\textcolor{red}{40}    \\\hline
24 &$\!$1860$\!$ &6634 &$\!$10700$\!$ &$\!$11637$\!$ &9629
&7034 &4567 &\textcolor{blue}{2918} &\textcolor{blue}{1731} &\textcolor{blue}{1061}
&\textcolor{blue}{599} &\textcolor{red}{368} &\textcolor{red}{197} &\textcolor{red}{122} &\textcolor{red}{62} \\\hline
25 &$\!$2895$\!$ &$\!$10082$\!$ &$\!$16893$\!$ &$\!$18301$\!$ &$\!$15694$\!$
&$\!$11337$\!$ &7662 &4774 &\textcolor{blue}{2981} &\textcolor{blue}{1749}
&\textcolor{blue}{1066} &\textcolor{blue}{600} &\textcolor{red}{368} &\textcolor{red}{197} &\textcolor{red}{122} \\\hline
\end{tabular}
%\end{longtable}
%\end{sidewaysfigure}
\end{center}
\end{scriptsize}
\caption{${\cal N}_{10d}=1$ multiplets with $SO(9)$ quantum numbers $[n,0,0,0]$}
\label{tab10d000}
\end{table}

%For a given family $\{ \llbracket n,k;p\rrbracket, \ n=0,1,\ldots\}$ of ${\cal N}_{6d}=(1,0)$ supermultiplets with second $SO(5)$ Dynkin label $k$ and R symmetry quantum number, the first quantity of interest is the level of first appearance. This determines the leading $q$ power of the $\tau_\ell^{k,p}$ coefficients defined by (\ref{6dimtau}). The following table \ref{tab:appearance} gathers the mass levels where the first instance of a $\{ \llbracket n,k;p\rrbracket, \ n=0,1,\ldots\}$ member can be found up to level 25:

\subsubsection{Levels of first appearance}

The mass level where some $\llbracket0,x,y,z\rrbracket$ multiplet firstly occurs can be studied by inspecting the leading power of the multiplicity generating function $G_{0,x,y,z} (q)$ and therefore $\tau_\ell^{x,y,z}(q)$. The following tables \ref{tab:appearance10} and \ref{tab:appearance10f} give an overview of this mass level threshold for various values of $x,y,z$.

\begin{table}[htdp]
\begin{center}
\[
\begin{array}{|c|c|c||c|} \hline
\ x \ & \ y \ & \ z \ & \, \te{level} \, \\\hline
0&0&0&1 \\
1&0&0&4\\
2&0&0&7\\
3&0&0&10\\
4&0&0&13\\
5&0&0&16\\
6&0&0&19\\ 
7&0&0&22 \\
8&0&0&25 \\
9&0&0& \\\hline
0&1&0 &5\\
1&1&0 &8\\
2&1&0 &11\\
3&1&0 &14\\
4&1&0 &17\\
5&1&0 &20\\
6&1&0 &23\\\hline
\end{array}  \ \ \ \ \ \ \ \ \ \
\begin{array}{|c|c|c||c|} \hline
\ x \ & \ y \ & \ z \ & \, \te{level} \, \\\hline
7&1&0 &\\
\hline
0&2&0 &11\\
1&2&0 &14\\
2&2&0 &17\\
3&2&0 &20\\
4&2&0 &23\\
5&2&0 &\\\hline
0&3&0 &17\\
1&3&0 &20\\
2&3&0 &23\\
3&3&0 &\\\hline
0&4&0 &23\\
1&4&0 &\\\hline 
0&5&0 &\\\hline 
0&0&2&6 \\
1&0&2&9\\\hline
\end{array}  \ \ \ \ \ \ \ \ \ \
\begin{array}{|c|c|c||c|} \hline
\ x \ & \ y \ & \ z \ & \, \te{level} \, \\\hline
%%%%%%%
2&0&2&12\\
3&0&2&15\\
4&0&2&18\\
5&0&2&21\\
6&0&2&24\\
7&0&2& \\\hline
0&1&2&12 \\
1&1&2&15\\
2&1&2&18\\
3&1&2&21\\
4&1&2&24\\
5&1&2& \\\hline
0&2&2&18 \\
1&2&2&21\\
2&2&2&24\\
3&2&2& \\\hline
\end{array}  \ \ \ \ \ \ \ \ \ \
\begin{array}{|c|c|c||c|} \hline
\ x \ & \ y \ & \ z \ & \, \te{level} \, \\\hline
0&3&2&24 \\
1&3&2& \\\hline
0&4&2& \\\hline
0&0&4 &14 \\
1&0&4 &17 \\
2&0&4 &20 \\
3&0&4 &23 \\
4&0&4 & \\ \hline
0&1&4 &20 \\
1&1&4 &23 \\
2&1&4 & \\ \hline
0&2&4 & \\ \hline
0&0&6 &24 \\ 
1&0&6 & \\ \hline
0&1&6 & \\ \hline
0&0&8 & \\ \hline
%%%%% now the fermions
\end{array}  \]
\caption{First mass level where bosonic supermultiplets $\llbracket0,x,y,z\rrbracket$ of ${\cal N}_{10d}=1$ firstly occur. Empty spaces indicate that the representations in question do not occur at levels $\leq 25$.}
\label{tab:appearance10}
\end{center}
\end{table}

\begin{table}[htdp]
\begin{center}
\[
\begin{array}{|c|c|c||c|} \hline
\ x \ & \ y \ & \ z \ &\, \te{level} \, \\\hline
0&0&1 &3 \\
1&0&1 &6 \\
2&0&1 &9 \\
3&0&1 &12 \\
4&0&1 &15 \\
5&0&1 &18 \\
6&0&1 &21 \\
7&0&1 &24 \\
8&0&1 & \\ \hline
0&1&1 &8 \\
1&1&1 &11 \\
2&1&1 &14 \\
3&1&1 &17 \\
 \hline
\end{array} 
\ \ \ \ \ \ \ \ \ \
\begin{array}{|c|c|c||c|} \hline
\ x \ & \ y \ & \ z \ &\, \te{level} \, \\\hline
4&1&1 &20 \\
5&1&1 &23 \\
6&1&1 &  \\\hline
0&2&1 &14 \\
1&2&1 &17 \\
2&2&1 &20 \\
3&2&1 &23 \\
4&2&1 &  \\ \hline
0&3&1 &20 \\
1&3&1 &23 \\
2&3&1 & \\ \hline
0&4&1 & \\ \hline
\end{array} 
\ \ \ \ \ \ \ \ \ \
\begin{array}{|c|c|c||c|} \hline
\ x \ & \ y \ & \ z \ &\, \te{level} \, \\\hline
0&0&3 &10 \\
1&0&3 &13 \\
2&0&3 &16 \\
3&0&3 &19 \\
4&0&3 &22 \\
5&0&3 &25 \\
6&0&3 & \\ \hline
0&1&3 &16 \\
1&1&3 &19 \\
2&1&3 &22 \\
3&1&3 &25 \\
4&1&3 & \\ \hline
\end{array} 
\ \ \ \ \ \ \ \ \ \
\begin{array}{|c|c|c||c|} \hline
\ x \ & \ y \ & \ z \ &\, \te{level} \, \\\hline
0&2&3 &22 \\
1&2&3 &25 \\
2&2&3 & \\ \hline
0&3&3 & \\ \hline
0&0&5 &19 \\ 
1&0&5 &22 \\ 
2&0&5 &25 \\ 
3&0&5 & \\ \hline
0&1&5 &25 \\ 
1&1&5 & \\ \hline
0&2&5 & \\ \hline
0&0&7 & \\ \hline
\end{array} 
\]
\caption{First mass level where fermionic supermultiplets $\llbracket0,x,y,z\rrbracket$ of ${\cal N}_{10d}=1$ firstly occur. Empty spaces indicate that the representations in question do not occur at levels $\leq 25$.}
\label{tab:appearance10f}
\end{center}
\end{table}

For all supermultiplets $\llbracket 0,x,y,z\rrbracket$ considered in tables \ref{tab:appearance10} and \ref{tab:appearance10f}, the level of first appearance is delayed by three whenever the second Dynkin label is incremented as $x \mapsto x+1$. This suggests to look for a similar linear effect of $y \mapsto y+1$ and $z \mapsto z+1$. Up to the two exceptions $\llbracket 0,0,0,0\rrbracket$ and $\llbracket 0,0,0,1\rrbracket$, the data in the tables shows that the value $y$ of the third Dynkin label increases the level of first appearance by $6y$. 

The influence of the last Dynkin label $z$ is much more difficult to probe without any explicit multiplicities beyond level 25 at hand. If an asymptotically linear relation between $z$ and the level of first appearance of $\llbracket 0,x,y,z\rrbracket$ exists, then it certainly admits even more exceptions than in the $y \mapsto y+1$ case. The onset of $\llbracket n,0,0,4\rrbracket$, $\llbracket n,0,0,5\rrbracket$ and $\llbracket n,0,0,6\rrbracket$ multiplets at levels 14, 19 and 24, respectively, suggests that an increment $z \mapsto z+1$ delays the $\llbracket 0,x,y,z\rrbracket$ multiplet by five levels -- at least in the regime of sufficiently large values of $x,y,z$.

On the basis of this reasonning, we conjecture that sufficiently high mass levels of first occurrence for general supermultiplets $\llbracket n,x,y,z\rrbracket$ are determined by the following overall prefactor in their multiplicity generating function:
\beq
G_{n,x,y,z}(q) \ \ \sim \ \ q^{n+3x+6y+5z-6} \, \times \, {\cal O}(1) \co x,y,z \ \te{large}
\eeq
Note that also the six dimensional ${\cal N}_{6d}=(1,0)$ spectrum exhibis an asymptotic linear relation between the second $SO(5)$ label $k$ and the level of first appearance: Table \ref{tab:appearance} shows that sufficiently high levels of first appearance for $\llbracket n,k;p\rrbracket$ are delayed by three under $k \mapsto k+2$.

\subsubsection{Explicit formulae for the $\tau_{\ell}^{x,y,z}(q)$}

We shall now give the explicit results for a large class of $\tau_{\ell}^{x,y,z}(q)$, obtained through the entries of table \ref{tab10d000} and its generalizations to $(x,y,z) \neq (0,0,0)$ gathered in appendix \ref{app10}. This reflects large spin information on the multiplicity generating functions $G_{n,x,y,z}(q)$ via (\ref{10dimtau}).

We firstly consider supermultiplet families of vanishing third and fourth Dynkin labels $y=z=0$ and vary the second Dynkin label through values $x=0,1,2,3$:
\begin{itemize}
\item $SO(9)$ Dynkin labels $[n \rightarrow \infty,0,0,0]$
\begin{align}
\tau_1^{0,0,0}(q) \eq& q^{1} \, (1 + 0 q + 1 q^2 + 1 q^3 + 4 q^4 + 5 q^5 + 13 q^6 + 18 q^7 + 40 q^8 + 
 62 q^9 + 122 q^{10}  \notag \\
 & \ \ + 197 q^{11} + 368 q^{12} + 601 q^{13} + 1070 q^{14} + 
 1767 q^{15} + 3051 q^{16} + 5022 q^{17} \notag \\
 & \ \ + 8489 q^{18} + 13897 q^{19}+\ldots) \notag \\
 \tau_2^{0,0,0}(q) \eq& q^{1} \, (1 + 2 q + 4 q^2 + 9 q^3 + 18 q^4 + 36 q^5 + 70 q^6 + 133 q^7 + 
 249 q^8 + 460 q^9   \notag \\
 & \ \ + 836 q^{10} + 1503 q^{11} + 2672 q^{12} + 4699 q^{13}+\ldots) \notag \\
 \tau_3^{0,0,0}(q) \eq& q^{1} \, (1 + 1 q + 5 q^2 + 9 q^3 + 26 q^4 + 48 q^5 + 112 q^6 + 211 q^7 + 
 439 q^8 + 818 q^9+\ldots) \notag \\
 \tau_4^{0,0,0}(q) \eq& q^{1} \, (1   +3 q + 8 q^2 + 20 q^3 + 48 q^4 + 106 q^5+\ldots) \notag \\
 \tau_5^{0,0,0}(q) \eq& q^{1} \, (1 + 1 q + 6 q^2+\ldots) 
\end{align}
\item $SO(9)$ Dynkin labels $[n \rightarrow \infty,1,0,0]$
\begin{align}
\tau_1^{1,0,0}(q) \eq& q^{4} \, (1 + 2 q + 3 q^2 + 7 q^3 + 14 q^4 + 28 q^5 + 53 q^6 + 103 q^7 + 
 189 q^8 + 352 q^9 + 634 q^{10} \notag \\
 &+ 1146 q^{11} + 2026 q^{12} + 3578 q^{13} + 
 6209 q^{14} + 10752 q^{15} + 18378 q^{16} + 31279 q^{17}+ \ldots) \notag \\
\tau_2^{1,0,0}(q) \eq& q^{5} \, (1 + 2 q + 5 q^2 + 11 q^3 + 26 q^4 + 54 q^5 + 114 q^6 + 227 q^7 + 
 449 q^8 + 863 q^9 \notag \\
 & + 1639 q^{10} + 3050 q^{11} + 5618 q^{12} + 10187 q^{13}+ \ldots) \notag \\
 \tau_3^{1,0,0}(q) \eq& q^{8} \, (2 + 5 q + 15 q^2 + 35 q^3 + 86 q^4 + 185 q^5 + 403 q^6 + 825 q^7+ \ldots) \notag \\
 \tau_4^{1,0,0}(q) \eq& q^{10} \, (1  + 3 q + 11 q^2 + 30 q^3+ \ldots)
\end{align}
\item $SO(9)$ Dynkin labels $[n \rightarrow \infty,2,0,0]$
\begin{align}
\tau_1^{2,0,0}(q) \eq& q^{7} \,(1 + 2 q + 6 q^2 + 11 q^3 + 27 q^4 + 52 q^5 + 112 q^6 + 212 q^7 + 
 423 q^8 \notag \\
 & \ \  + 787 q^9 + 1496 q^{10} + 2724 q^{11} + 5001 q^{12} + 8927 q^{13} + 
 15950 q^{14}+\ldots) \notag \\
 \tau_2^{2,0,0}(q) \eq& q^{8} \,(1 + 2 q^1 + 6 q^2 + 14 q^3 + 34 q^4 + 74 q^5 + 161 q^6 + 333 q^7 + 
 680 q^8 \notag \\
 & \ \ + 1346 q^9 + 2627 q^{10}+\ldots) \notag \\
 \tau_3^{2,0,0}(q) \eq& q^{12} \,(3 + 7 q + 23 q^2 + 54 q^3 + 138 q^4+\ldots) \notag \\
 \tau_4^{2,0,0}(q) \eq& q^{15} \,(1 +\ldots)  
 \end{align}
\item $SO(9)$ Dynkin labels $[n \rightarrow \infty,3,0,0]$
\begin{align}
\tau_1^{3,0,0}(q) \eq& q^{10} \, ( 1 + 2 q + 6 q^2 + 14 q^3 + 32 q^4 + 69 q^5 + 147 q^6 + 299 q^7 + 
 597 q^8 + 1168 q^9 \notag\\
 & \ \ + 2239 q^{10} + 4226 q^{11} + 7854 q^{12}+\ldots) \notag \\
 \tau_2^{3,0,0}(q) \eq& q^{11} \, ( 1 + 2 q + 6 q^2 + 14 q^3 + 35 q^4 + 77 q^5 + 172 q^6 + 361 q^7 + 
 752 q^8 + 1513 q^9+\ldots) \notag \\
 \tau_3^{3,0,0}(q) \eq& q^{16} \, ( 3 + 8 q + 25 q^2 + 63 q^3+\ldots) 
\end{align}
\end{itemize}
Let us next freeze the second and fourth Dynkin label to $x=z=0$ and vary the third Dynkin label to $y=1$ and $y=2$:
\begin{itemize}
\item $SO(9)$ Dynkin labels $[n \rightarrow \infty,0,1,0]$
\begin{align}
\tau_1^{0,1,0}(q) \eq& q^{5} \, (1 + 1 q + 5 q^2 + 8 q^3 + 22 q^4 + 40 q^5 + 90 q^6 + 165 q^7 + 
 338 q^8 + 619 q^9 + 1190 q^{10} \notag \\
 & \ \ + 2149 q^{11} + 3969 q^{12} + 7048 q^{13} + 
 12630 q^{14} + 22060 q^{15} + 38603 q^{16}+\ldots) \notag \\
 \tau_2^{0,1,0}(q) \eq& q^{6} \, (1 + 2 q + 7 q^2 + 17 q^3 + 41 q^4 + 91 q^5 + 199 q^6 + 412 q^7 + 
 841 q^8 + 1665 q^9 \notag \\
 & \ \  + 3241 q^{10} + 6178 q^{11} + 11611 q^{12}+\ldots) \notag \\
 \tau_3^{0,1,0}(q) \eq& q^{8} \, (1 + 2 q + 11 q^2 + 25 q^3 + 71 q^4 + 160 q^5 + 381 q^6 + 809 q^7+\ldots) \notag \\
 \tau_4^{0,1,0}(q) \eq& q^{11} \, (2  + 7 q + 23 q^2+\ldots) 
\end{align}
\item $SO(9)$ Dynkin labels $[n \rightarrow \infty,0,2,0]$
\begin{align}
\tau_1^{0,2,0}(q) \eq& q^{11} \, (2 + 4 q + 17 q^2 + 36 q^3 + 97 q^4 + 207 q^5 + 473 q^6 + 963 q^7 + 
 2016 q^8 \notag \\
 & \ \ + 3957 q^9 + 7809 q^{10} + 14838 q^{11}+\ldots) \notag \\
 \tau_2^{0,2,0}(q) \eq& q^{12} \, (2 + 6 q + 22 q^2 + 59 q^3 + 153 q^4 + 365 q^5 + 842 q^6 + 1842 q^7+\ldots) \notag \\
 \tau_3^{0,2,0}(q) \eq& q^{14} \, (2 + 5 q + 24 q^2 + 62 q^3+\ldots) 
\end{align}
\end{itemize}
Finally, at $x=y=0$ and nonzero fourth Dynkin label, we have the bosonic supermultiplets
\begin{align}
\tau_1^{0,0,2}(q) \eq& q^{6} \, (1 + 2 q + 7 q^2 + 13 q^3 + 33 q^4 + 66 q^5 + 143 q^6 + 277 q^7 + 
 559 q^8 + 1053 q^9 \notag \\
 & \ \ + 2019 q^{10} + 3715 q^{11} + 6859 q^{12} + 
 12338 q^{13} + 22156 q^{14} + 39043 q^{15}+\ldots) \notag \\
 \tau_2^{0,0,2}(q) \eq& q^{7} \, (1 + 4 q^1 + 11 q^2 + 28 q^3 + 68 q^4 + 155 q^5 + 339 q^6 + 716 q^7 + 
 1469 q^8 \notag \\
 & \ \ + 2938 q^9 + 5755 q^{10} + 11054 q^{11}+\ldots) \notag \\
 \tau_3^{0,0,2}(q) \eq& q^{9} \, (2 + 5 q + 19 q^2 + 48 q^3 + 130 q^4 + 301 q^5 + 703 q^6 + 1518 q^7+\ldots) \notag \\
 \tau_4^{0,0,2}(q) \eq& q^{11} \, (1  + 4 q + 16 q^2 + 49 q^3+\ldots)
\end{align}
and the simplest fermionic supermultiplets $\llbracket n, 0,0,1 \rrbracket$:
\begin{align}
\tau_1^{0,0,1}(q) \eq& q^{3} \, (1 + 1 q + 3 q^2 + 6 q^3 + 12 q^4 + 24 q^5 + 48 q^6 + 90 q^7 + 
 171 q^8 + 317 q^9 + 579 q^{10} \notag \\
 &  + 1045 q^{11} + 1870 q^{12} + 3299 q^{13} + 
 5777 q^{14} + 10017 q^{15} + 17222 q^{16} + 29370 q^{17}+\ldots) \notag \\
 \tau_2^{0,0,1}(q) \eq& q^{4} \, (1 + 2 q^1 + 5 q^2 + 13 q^3 + 29 q^4 + 62 q^5 + 130 q^6 + 263 q^7 + 
 520 q^8 + 1008 q^9 \notag \\
 & + 1916 q^{10} + 3583 q^{11} + 6609 q^{12}+\ldots) \notag \\
 \tau_3^{0,0,1}(q) \eq& q^{6} \, (1 + 3 q^1 + 10 q^2 + 26 q^3 + 63 q^4 + 143 q^5 + 315 q^6 + 664 q^7+\ldots) \notag \\
 \tau_4^{0,0,1}(q) \eq& q^{8} \, (1  + 4 q + 12 q^2 + 35 q^3+\ldots) \notag \\
 \tau_5^{0,0,1}(q) \eq& q^{10} \, (1 +\ldots) 
\end{align}
The leading $q$ powers of various $\tau_{\ell}^{x,y,z}(q)$ suggest that the $\tau_{\ell}^{x,y,z}(q)$ expansion (\ref{10dimtau}) converges more quickly at higher value of $x,y,z$. The rapid increase of leading $q$ powers of $\tau_{1}^{x,y,z},\tau_{2}^{x,y,z},\tau_{3}^{x,y,z},\ldots$ with growing $x$ becomes particular obvious from the $\tau_{\ell}^{x,y,z}(q)$ shown. These trends are supported by further $\tau_{\ell}^{x,y,z}(q)$ at $(x,y,z)=(1,1,0), (1,0,2), (2,1,0), (1,0,1), (0,1,1), (2,0,1), (0,0,3), (1,1,1)$, listed in appendix \ref{app:data10}.

\subsection[Eight dimensional ${\cal N}_{8d}=1$ spectra]{Eight dimensional $\bm{{\cal N}_{8d}=1}$ spectra}
\label{sec:8dim}

Starting from this subsection, we consider even dimensional type I superstring compactifications on $T^2$ tori preserving all the sixteen supercharges. The highest dimensional example is ${\cal N}_{8d}=1$ SUSY in eight spacetime dimensions. As explained in \cite{Boels:2012ie,Boels:2012if}, dimensional reduction of the open superstring from $d=10$ to $d=8$ paves the way towards powerful on-shell SUSY techniques to manifest hidden simplicity of scattering amplitudes among massive string modes (further examples following in \cite{Boels:progress}): One technical advantage of the eight dimensional setting is the possibility to covariantly single out a Clifford vacuum which is annihilated by half of the supercharges, say the right handed $SO(8)$ spinor of SUSY generators \cite{Boels:2012ie,Boels:2012if}. This is a particular motivation to focus on the covariant particle content of the maximally supersymmetric open superstring in $d=8$.

Let $r$ denote the fugacity with respect to the R symmetry $SO(2)_R \cong U(1)_R$ and $y_{i}$ the fugacities of the massive little group $SO(7)$, then the fundamental ${\cal N}_{8d}=1$ super Poincar\'e multiplet is described by the supercharacter
\begin{align}
Z&({\cal N}_{8d}=1) \ \ := \ \ (r^4+r^{-4}) \, [0,0,0] \ + \ (r^3 + r^{-3}) \, [0,0,1] \ + \ (r^2+r^{-2}) \, \big( \, [0,1,0] \, + \, [1,0,0] \, \big) \notag \\
& \ + \ (r+r^{-1}) \, \big( \,[1,0,1] \, + \, [0,0,1] \, \big) \ + \ [2,0,0] \, + \, [0,0,2] \, + \, [1,0,0] \, + \, [0,0,0]
\label{Z8}
\end{align}
which is obtained by branching the $SO(9)$ representations contributing to the ${\cal N}_{10d}=1$ analogue (\ref{Z10}) to $SO(7) \times U(1)_R$. The minimal multiplet (\ref{Z8}) can be generated from a scalar Clifford vacuum of $U(1)_R$ charge $+4$, and the generic ${\cal N}_{8d}=1$ multiplet follows from a Clifford vacuum with nontrivial $SO(7) \times U(1)_R$ quantum numbers\footnote{Recall that the semicolon in $\llbracket a_1,a_2,a_3;b \rrbracket$ separating the $U(1)_R$ quantum number $b$ from the $SO(7)$ Dynkin labels $a_1,a_2,a_3$ eliminates potential confusion with ${\cal N}_{10d}=1$ supercharacters (\ref{ZZ10}).}. This gives rise to the supercharacter
\beq
\llbracket a_1,a_2,a_3;Q\rrbracket \ \ := \ \ Z({\cal N}_{8d}=1) \cdot r^Q \, [a_1,a_2,a_3] \ .
\label{ZZ8}
\eeq
The eight dimensional partition function is obtained from its ten dimensional ancestor (\ref{NSandR10d}) by singling out an internal factor $\chi^{SO(3)}_{\te{NS,R}}$ within $\chi^{SO(9)}_{\te{NS,R}}(\vec y) = \prod_{k=1}^4 \chi^{SO(3)}_{\te{NS,R}}(y_k)$ and reinterpreting its argument as an R-symmetry fugacity:
\bea \label{NSandR8d}
\chi^{{\cal N}_{8d}=1}(q;\vec y,r) &\eq \chi^{{\cal N}_{8d}=1}_{\te{NS}} \mid_{\GSO} (q;\vec y,r) \ + \ \chi^{{\cal N}_{8d}=1}_{\te{R}} \mid_{\GSO} (q;\vec y,r) \notag \\
\chi^{{\cal N}_{8d}=1}_{\te{NS}} \mid_{\GSO} (q;\vec y,r) &\eq
\frac{1}{2} \, q^{-\frac{1}{2}} \, \big[ \, \chi_{\te{NS}}^{SO(7)}(q;\vec y) \, \chi_{\te{NS}}^{SO(3)}(q;r) \ - \ \chi_{\te{NS}}^{SO(7)}(e^{2\pi i}q;\vec y) \,  \chi_{\te{NS}}^{SO(3)}(e^{2\pi i}q;r) \, \big] \notag \\
\chi^{{\cal N}_{8d}=1}_{\te{R}} \mid_{\GSO} (q;\vec y,r) &\eq \frac{1}{2} \,  \chi_{\te{R}}^{SO(7)}(q;\vec y) \, \chi_{\te{R}}^{SO(3)}(q; r) 
\eea
Let us display the first four coefficients of the power series expansion in $q$:\footnote{Again, there is a subtlety in applying the above formula to the massless R sector; see the footnote before \eref{def:GSOed}. However, this can be fixed easily: one can simply add to it $\frac{1}{2} \, \prod_{j=1}^3 (y_j - y_j^{-1}) \, (r-r^{-1})$ to get the correct massless character in R sector.}
\begin{align}
\chi^{{\cal N}_{8d}=1}(q;\vec y,r)&\eq \underbrace{\left( \, \sum_{j=1}^3 (y_j^2 + y_j^{-2}) \ + \ r^2 \ + \ r^{-2} \ + \ \frac{1}{2} \, \prod_{j=1}^3 (y_j + y_j^{-1}) \, (r+r^{-1}) \, \right) \, q^0}_{16 \ \te{massless states}}  \notag  \\
&   + \ \underbrace{ \llbracket 0,0,0;0\rrbracket  \, q}_{256 \ \te{states at level} \ 1}   \  + \ \underbrace{\big( \,\llbracket 0,0,0;\pm 2\rrbracket \, + \, \llbracket 1,0,0;0\rrbracket \,\big) \, q^2}_{2304 \ \te{states at level} \ 2}  \notag \\
&+ \  \big( \, \llbracket 0,0,0;\pm 4 \rrbracket \, + \, \llbracket 1,0,0;\pm 2 \rrbracket \, + \, \llbracket 0,0,1;\pm 1\rrbracket \, \notag \\
& \ \ \ \ \ \ \ \ \ \ \ \ \ \ \ \ \ \ \ \ \ \ \ \ \ \ \ \ \ \ + \, \llbracket 2,0,0;0\rrbracket \, + \, \llbracket 0,0,0;0\rrbracket \, \big) \, q^3  \ + \ {\cal O}(q^4)  \ .
\end{align}
The pairing of opposite $U(1)_R$ charges $\pm Q$ motivates the following shorthand:
\beq
\llbracket a_1,a_2,a_3;\pm Q\rrbracket   \ \ := \ \ \begin{cases} \ \llbracket a_1,a_2,a_3; Q\rrbracket + \llbracket a_1,a_2,a_3;- Q\rrbracket  & \text{for $Q \neq 0$~, } \\
 \ \llbracket a_1,a_2,a_3;0\rrbracket  & \text{for $Q = 0$~. } \end{cases}
\eeq
The supermultiplets up to level six are listed in table \ref{table8,1}, some of their scattering amplitudes are discussed in \cite{Boels:2012if, Boels:progress}. The branching process obviously increases the number and diversity of multiplets compared to the ten dimensional analogue, cf. table \ref{table10,1}. This is why we do not repeat the higher level analysis carried out for the $d=10$ ancestor in dimensionally reduced settings.

Note that this partition function can also be obtained by branching the $SO(9)$ representations appearing in the the $\CN_{10d}=1$ partition function \eref{chi10dref} into $SO(7)  \times U(1)_R$ representations.  In terms of characters, one simply maps $SO(9)$ fugacities into  $SO(7)  \times U(1)_R$ fugacities; a possible fugacity map is as follows:
\bea
z_1 = y_1, \qquad z_2 = y_2, \qquad z_3 = y_3, \qquad z_4= s~, \label{B4toA1xD3}
\eea
where $z_1, \ldots, z_4$ are fugacities of $SO(9)$, $y_1, y_2, y_3$ are fugacities of $SO(7)$ and $s$ is a fugacity of $U(1)_R$.
For example,
\bea
[1,0,0,0]_{\vec z} &= 1+\frac{1}{z_1^2}+z_1^2+\frac{1}{z_2^2}+z_2^2+\frac{1}{z_3^2}+z_3^2+\frac{1}{z_4^2}+z_4^2\nn\\
&=1+\frac{1}{y_1^2}+y_1^2+\frac{1}{y_2^2}+y_2^2+\frac{1}{y_3^2}+y_3^2+\frac{1}{s^2}+s^2 \nn \\
&= [1,0,0;0]_{\vec y; s} + [0,0,0; +2] _{\vec y; s}+ [0,0,0; -2]_{\vec y; s} ~,
\eea
where the notation $[b_1, b_2, b_3; Q]$ denotes the $SO(7) \times U(1)_R$ representation.

\bigskip
\noindent
\begin{table}[htdp]
\begin{tabular}{|l|l|}\hline  $\ap m^2$ &representations of ${\cal N}_{8d}=1$ super Poincar\'e \\ \hline \hline   
  1 & $\llbracket 0,0,0;0\rrbracket  $  \\\hline
   2 & $\llbracket 0,0,0;\pm 2\rrbracket \, + \, \llbracket 1,0,0;0\rrbracket  $  \\\hline
   3 & $  \llbracket 0,0,0;\pm 4 \rrbracket \, + \, \llbracket 1,0,0;\pm 2 \rrbracket \, + \, \llbracket 0,0,1;\pm 1\rrbracket \, + \, \llbracket 2,0,0;0\rrbracket \, + \, \llbracket 0,0,0;0\rrbracket  $ \\ \hline
   4 & $\llbracket0,0,0;\pm6 \rrbracket \, + \,
\llbracket1,0,0;\pm 4 \rrbracket \, + \,
\llbracket0,0,1;\pm 3 \rrbracket \, + \,
\llbracket2,0,0;\pm 2 \rrbracket \, + \,
\llbracket1,0,0;\pm 2 \rrbracket \, + \,
2 \, \llbracket0,0,0;\pm 2 \rrbracket  $ \\ &$\, + \,
\llbracket1,0,1;\pm 1 \rrbracket \, + \,
\llbracket0,0,1;\pm 1 \rrbracket \, + \,
\llbracket3,0,0;0 \rrbracket \, + \,
2 \, \llbracket1,0,0;0 \rrbracket \, + \,
\llbracket0,1,0;0 \rrbracket \, + \,
\llbracket0,0,0;0 \rrbracket  $
   \\\hline        
   5 &$\llbracket0,0,0;\pm 8 \rrbracket \, + \,
\llbracket1,0,0;\pm 6 \rrbracket \, + \,
\llbracket0,0,1;\pm 5 \rrbracket \, + \,
\llbracket2,0,0;\pm 4 \rrbracket \, + \,
\llbracket1,0,0 ;\pm 4\rrbracket \, + \,
2\, \llbracket0,0,0;\pm 4 \rrbracket $ \\ &$ \, + \,
\llbracket1,0,1;\pm 3 \rrbracket \, + \,
2\, \llbracket0,0,1;\pm 3 \rrbracket \, + \,
\llbracket3,0,0;\pm 2 \rrbracket \, + \,
\llbracket2,0,0;\pm 2 \rrbracket \, + \,
3\, \llbracket1,0,0;\pm 2 \rrbracket \, + \,
2\, \llbracket0,1,0;\pm 2 \rrbracket $ \\ &$\, + \,
\llbracket0,0,0;\pm 2 \rrbracket \, + \,
\llbracket2,0,1;\pm 1 \rrbracket \, + \,
2\, \llbracket1,0,1;\pm 1 \rrbracket \, + \,
3\, \llbracket0,0,1;\pm 1 \rrbracket \, + \,
\llbracket4,0,0;0 \rrbracket \, + \,
2\, \llbracket2,0,0;0 \rrbracket $ \\ &$ \, + \,
\llbracket1,1,0;0 \rrbracket \, + \,
3\, \llbracket1,0,0;0 \rrbracket \, + \,
\llbracket0,1,0;0 \rrbracket \, + \,
\llbracket0,0,2;0 \rrbracket \, + \,
4\, \llbracket0,0,0;0 \rrbracket  $
     \\\hline
   6 &$\llbracket0,0,0;\pm 10\rrbracket \, + \,
\llbracket1,0,0;\pm 8\rrbracket \, + \,
\llbracket0,0,1;\pm 7\rrbracket \, + \,
\llbracket2,0,0;\pm 6\rrbracket \, + \,
\llbracket1,0,0;\pm 6\rrbracket \, + \,
2 \, \llbracket0,0,0;\pm 6\rrbracket $ \\ &$ \, + \,
\llbracket1,0,1;\pm 5\rrbracket \, + \,
2 \, \llbracket0,0,1;\pm 5\rrbracket \, + \,
\llbracket3,0,0;\pm 4\rrbracket \, + \,
\llbracket2,0,0;\pm 4\rrbracket \, + \,
3 \, \llbracket1,0,0;\pm 4\rrbracket \, + \,
2 \, \llbracket0,1,0;\pm 4\rrbracket $ \\ &$ \, + \,
2 \, \llbracket0,0,0;\pm 4\rrbracket \, + \,
\llbracket2,0,1;\pm 3\rrbracket \, + \,
3 \, \llbracket1,0,1;\pm 3\rrbracket \, + \,
3 \, \llbracket0,0,1;\pm 3\rrbracket \, + \,
\llbracket4,0,0;\pm 2\rrbracket \, + \,
\llbracket3,0,0;\pm 2\rrbracket $ \\ &$ \, + \,
3 \, \llbracket2,0,0;\pm 2\rrbracket \, + \,
2 \, \llbracket1,1,0;\pm 2\rrbracket \, + \,
5 \, \llbracket1,0,0;\pm 2\rrbracket \, + \,
\llbracket0,1,0;\pm 2\rrbracket \, + \,
2 \, \llbracket0,0,2;\pm 2\rrbracket $ \\ &$ \, + \,
4 \, \llbracket0,0,0;\pm 2\rrbracket \, + \,
\llbracket3,0,1;\pm 1\rrbracket \, + \,
2 \, \llbracket2,0,1;\pm 1\rrbracket \, + \,
4 \, \llbracket1,0,1;\pm 1\rrbracket \, + \,
\llbracket0,1,1;\pm 1\rrbracket $ \\ &$ \, + \,
5 \, \llbracket0,0,1;\pm 1\rrbracket \, + \,
\llbracket5,0,0;0\rrbracket \, + \,
2 \, \llbracket3,0,0;0\rrbracket \, + \,
\llbracket2,1,0;0\rrbracket \, + \,
4 \, \llbracket2,0,0;0\rrbracket \, + \,
\llbracket1,1,0;0\rrbracket  $ \\ &$ \, + \,
\llbracket1,0,2;0\rrbracket \, + \,
5 \, \llbracket1,0,0;0\rrbracket \, + \,
5 \, \llbracket0,1,0;0\rrbracket \, + \,
\llbracket0,0,2;0\rrbracket \, + \,
3 \, \llbracket0,0,0;0\rrbracket$ 
     \\\hline
\end{tabular}
\caption{${\cal N}_{8d}=1$ multiplets occurring up to mass level six}
\label{table8,1}
\end{table}

\subsection[Six dimensional ${\cal N}_{6d}=(1,1)$ spectra]{Six dimensional $\bm{{\cal N}_{6d}=(1,1)}$ spectra}
\label{sec:6dims}

Six dimensional type I compactifications with sixteen supercharges are said to possess ${\cal N}_{6d}=(1,1)$ SUSY. The spacetime symmetry branches to $SO(9) \rightarrow SO(5) \times SO(4)_R$, i.e. two Cartan generators of ten dimensional Lorentz group take the role of R symmetry generators probing fugacities $r_1,r_2$ of $SO(4)_R \cong SU(2)_R \times SU(2)_R$. The fundamental supermultiplet of the ${\cal N}_{6d}=(1,1)$ super Poincar\'e group has the following $SO(5) \times SU(2)_R \times SU(2)_R$ particle content:
\begin{align}
Z&({\cal N}_{6d}=(1,1)) \ \ := \ \ [2,0] \cdot [0,0]_R \ + \ [0,2] \cdot [0,0]_R \ + \ [0,2] \cdot [1,1]_R \ + \ [1,0] \cdot [1,1]_R \notag \\
& \ \ + \ [1,0] \cdot \big( \, [2,0]_R \ + \ [0,2]_R \, \big) \ + \ [0,0] \cdot [2,2]_R \ + \ [0,0] \cdot [1,1]_R \ + \ [0,0] \cdot [0,0]_R \notag \\
& \ \ + \ [1,1] \cdot \big( \, [1,0]_R \ + \ [0,1]_R \, \big) \ + \ [0,1] \cdot\big( \, [2,1]_R \ + \ [1,2]_R \ + \ [1,0]_R \ + \ [0,1]_R \, \big)
\label{Z6}
\end{align}
Note that the R-symmetry characters $[\ldots ]_R$ carry a subscript to avoid confusion with the Lorentz symmetry of identical rank.

The most general multiplet follows from (\ref{Z6}) by taking tensor products with $SO(5) \times SU(2)_R \times SU(2)_R$ representations, this leads to the supercharacter
\beq
\llbracket a_1,a_2;b_1,b_2\rrbracket \ \ := \ \ Z({\cal N}_{6d}=(1,1)) \cdot  [a_1,a_2] \cdot [b_1,b_2]_R
\label{ZZ6}
\eeq
The six dimensional partition function is obtained from its ten dimensional ancestor (\ref{NSandR10d}) by singling out two internal factor $\chi^{SO(3)}_{\te{NS,R}}$ within $\chi^{SO(9)}_{\te{NS,R}}(\vec y) = \prod_{k=1}^4 \chi^{SO(3)}_{\te{NS,R}}(y_k)$ and reinterpreting their second argument as an R-symmetry fugacity:
\bea \label{NSandR6d}
\chi^{{\cal N}_{6d}=(1,1)}(q;\vec y,\vec r) &\eq \chi^{{\cal N}_{6d}=(1,1)}_{\te{NS}} \mid_{\GSO} (q;\vec y,\vec r) \ + \ \chi^{{\cal N}_{6d}=(1,1)}_{\te{R}} \mid_{\GSO} (q;\vec y,\vec r) \notag \\
\chi^{{\cal N}_{6d}=(1,1)}_{\te{NS}} \mid_{\GSO} (q;\vec y, \vec r) &\eq
\frac{1}{2} \, q^{-\frac{1}{2}} \, \big[ \, \chi_{\te{NS}}^{SO(5)}(q;\vec y) \, \chi_{\te{NS}}^{SO(5)}(q;\vec r) \ - \ \chi_{\te{NS}}^{SO(5)}(e^{2\pi i}q;\vec y) \,  \chi_{\te{NS}}^{SO(5)}(e^{2\pi i}q;\vec r) \, \big] \notag \\
\chi^{{\cal N}_{6d}=(1,1)}_{\te{R}} \mid_{\GSO} (q;\vec y, \vec r) &\eq \frac{1}{2} \,  \chi_{\te{R}}^{SO(5)}(q;\vec y) \, \chi_{\te{R}}^{SO(5)}(q; \vec r) 
\eea
Its $q$ expansion starts like\footnote{Again, there is a subtlety in applying the above formula to the massless R sector; see the footnote before \eref{def:GSOed}. However, this can be fixed easily: one can simply add to it $\frac{1}{2} \, \prod_{j=1}^2 (y_j - y_j^{-1}) \, \prod_{j=1}^2 (r_j - r_j^{-1})$ to get the correct massless character in R sector.}
\begin{align}
\chi^{{\cal N}_{6d}=(1,1)}(q;\vec y,\vec r)&\eq \underbrace{\left( \, \sum_{j=1}^2 (y_j^2 + y_j^{-2}) \ + \ \sum_{j=1}^2 (r_j^2 + r_j^{-2}) \ + \ \frac{1}{2} \, \prod_{j=1}^2 (y_j + y_j^{-1}) \, \prod_{j=1}^2 (r_j + r_j^{-1}) \, \right) \, q^0}_{16 \ \te{massless states}}  \notag  \\
&   + \ \underbrace{ \llbracket 0,0;0,0\rrbracket  \, q}_{256 \ \te{states at level} \ 1}   \  + \ \underbrace{\big( \,\llbracket 0,0;1,1\rrbracket \, + \, \llbracket 1,0;0,0\rrbracket \,\big) \, q^2}_{2304 \ \te{states at level} \ 2}  \notag \\
&+ \  \big( \, \llbracket 0,0;2,2 \rrbracket \, + \, \llbracket 1,0;1,1 \rrbracket \, + \, \llbracket 0,1;1,0\rrbracket \notag \\
& \ \ \ \ \ \ \ \ \ \ \ \ +   \, \llbracket 0,1;0,1\rrbracket \, + \, \llbracket 2,0;0,0\rrbracket\, + \, \llbracket 0,0;0,0\rrbracket \, \big) \, q^3  \ + \ {\cal O}(q^4)  \ ,
\end{align}
and supermultiplets at higher levels $\leq 5$ are listed in table \ref{table6,1}.

Note that this partition function can also be obtained by branching the $SO(9)$ representations appearing in the the $\CN_{10d}=1$ partition function \eref{chi10dref} into $SO(5)  \times SU(2)_R \times SU(2)_R$ representations.  In terms of characters, one simply maps $SO(9)$ fugacities into $SO(5)  \times SU(2)_R \times SU(2)_R$ fugacities; a possible fugacity map is as follows:
\bea
z_1 = y_1, \qquad z_2 = y_2, \qquad z_3 = r_1 r_2, \qquad z_4= r_1 r_2^{-1}~, \label{B4toB2A1A1}
\eea
where $z_1, \ldots, z_4$ are fugacities of $SO(9)$, $y_1, y_2$ are fugacities of $SO(5)$, and $r_1, r_2$ are fugacities for the two $SU(2)_R$ factors.
For example,
\bea
[1,0,0,0]_{\vec z} &= 1+\frac{1}{z_1^2}+z_1^2+\frac{1}{z_2^2}+z_2^2+\frac{1}{z_3^2}+z_3^2+\frac{1}{z_4^2}+z_4^2\nn\\
&=1+\frac{1}{y_1^2}+y_1^2+\frac{1}{y_2^2}+y_2^2+ (r_1 +r_1^{-1}) (r_2 +r_2^{-1})  \nn \\
&= [1,0; 0 , 0]_{\vec y; \vec r} + [0,0; 1, 1] _{\vec y; \vec r}~,
\eea
where the notation $[a_1, a_2; b_1, b_2]$ denotes the $SO(5) \times SU(2)_R \times SU(2)_R$ representation.

\bigskip
\noindent
\begin{table}[htdp]
\begin{tabular}{|l|l|}\hline  $\ap m^2$ &representations of ${\cal N}_{6d}=(1,1)$ super Poincar\'e \\ \hline \hline   
  1 & $\llbracket 0,0;0,0\rrbracket  $  \\\hline
   2 & $\llbracket 0,0;1,1\rrbracket \, + \, \llbracket 1,0;0,0\rrbracket  $  \\\hline
   3 & $ \llbracket0,0;2,2\rrbracket \, + \,
\llbracket1,0;1,1\rrbracket \, + \,
\llbracket0,1;1,0\rrbracket \, + \,
\llbracket0,1;0,1\rrbracket \, + \,
\llbracket2,0;0,0\rrbracket \, + \,
\llbracket0,0;0,0\rrbracket  $ \\ \hline
   4 & $\llbracket0,0;3,3\rrbracket \, + \,
\llbracket1,0;2,2\rrbracket \, + \,
\llbracket0,1;2,1\rrbracket \, + \,
\llbracket0,0;2,0\rrbracket \, + \,
\llbracket0,1;1,2\rrbracket \, + \,
\llbracket2,0;1,1\rrbracket  \, + \,
\llbracket1,0;1,1\rrbracket $ \\ &$ \, + \,
2 \, \llbracket0,0;1,1\rrbracket \, + \,
\llbracket1,1;1,0\rrbracket \, + \,
\llbracket0,1;1,0\rrbracket \, + \,
\llbracket0,0;0,2\rrbracket \, + \,
\llbracket1,1;0,1\rrbracket \, + \,
\llbracket0,1;0,1\rrbracket $ \\ &$ \, + \,
\llbracket3,0;0,0\rrbracket \, + \,
2 \, \llbracket1,0;0,0\rrbracket \, + \,
\llbracket0,2;0,0\rrbracket $
   \\\hline        
   5 &$\llbracket0,0;4,4\rrbracket \, + \,
\llbracket1,0;3,3\rrbracket \, + \,
\llbracket0,1;3,2\rrbracket \, + \,
\llbracket0,0;3,1\rrbracket \, + \,
\llbracket0,1;2,3\rrbracket \, + \,
\llbracket2,0;2,2\rrbracket \, + \,
\llbracket1,0;2,2\rrbracket $ \\ &$ \, + \,
2 \, \llbracket0,0;2,2\rrbracket  \, + \,
\llbracket1,1;2,1\rrbracket \, + \,
2\, \llbracket0,1;2,1\rrbracket \, + \,
2\, \llbracket1,0;2,0\rrbracket \, + \,
\llbracket0,0;2,0\rrbracket \, + \,
\llbracket0,0;1,3\rrbracket  $ \\ &$ \, + \,
\llbracket1,1;1,2\rrbracket \, + \,
2\, \llbracket0,1;1,2\rrbracket \, + \,
\llbracket3,0;1,1\rrbracket \, + \,
\llbracket2,0;1,1\rrbracket \, + \,
3\, \llbracket1,0;1,1\rrbracket \, + \,
2\, \llbracket0,2;1,1\rrbracket $ \\ &$ \, + \,
2\, \llbracket0,0;1,1\rrbracket \, + \,
\llbracket2,1;1,0\rrbracket \, + \,
2\, \llbracket1,1;1,0\rrbracket \, + \,
3\, \llbracket0,1;1,0\rrbracket \, + \,
2\, \llbracket1,0;0,2 \rrbracket \, + \,
\llbracket0,0;0,2\rrbracket $ \\ &$ \, + \,
\llbracket2,1;0,1\rrbracket \, + \,
2\, \llbracket1,1;0,1\rrbracket \, + \,
3\, \llbracket0,1;0,1\rrbracket \, + \,
\llbracket4,0;0,0\rrbracket \, + \,
2\, \llbracket2,0;0,0\rrbracket \, + \,
\llbracket1,2;0,0\rrbracket $ \\ &$ \, + \,
\llbracket1,0;0,0\rrbracket \, + \,
2\, \llbracket0,2;0,0\rrbracket \, + \,
3\, \llbracket0,0;0,0\rrbracket $
     \\\hline
\end{tabular}
\caption{${\cal N}_{6d}=(1,1)$ multiplets occurring up to mass level five}
\label{table6,1}
\end{table}

\subsection[Four dimensional ${\cal N}_{4d}=4$ spectra]{Four dimensional $\bm{{\cal N}_{4d}=4}$ spectra}
\label{sec:4dims}

Finally, four dimensional theories with maximal ${\cal N}_{4d}=4$ SUSY follow from the ten dimensional ancestor through compactification on $T^6$. The internal rotation group is identified with the $R$ symmetry $SO(6)_R$, its characters are denoted by $[b_1,b_2,b_3]_R$. The universal partition function decomposes into characters of the ${\cal N}_{4d}=4$ super Poincar\'e algebra, the fundamental one being
\begin{align}
&Z({\cal N}_{4d}=4) \ = \  [0] \,\big( \, [0, 0, 2]_R  \, + \,   [0, 2, 0]_R \, + \,[2, 0, 0]_R  \, + \, 2 \, \big)  \ + \ [2] \, [0, 1, 1]_R \  + \ 2 \,[2] \,[1, 0, 0]_R \ + \ [4] 
 \notag \\
&+ \ [1] \, \big( \, [0, 0, 1]_R  \ + \ [0, 1, 0]_R \  + \ [1, 0, 1]_R \ + \ [1, 1,0]_R \, \big)  \ + \ [3]  \,\big( \, [0,0,1]_R\ + \ [0, 1, 0]_R \, \big)  \, .
\label{Z4}
\end{align}
Any other supermultiplet follows by taking a tensor product of (\ref{Z4}) with the $SO(3) \times SO(6)_R$ representation $[n] \, [b_1,b_2,b_3]_R$ of the the Clifford vacuum,
\bea
\llbracket n;b_1,b_2,b_3 \rrbracket & \ \ := \ \ Z({\cal N}_{4d}=4) \cdot  [n] \, [b_1,b_2,b_3]_R ~.
\label{chN=4}
\eea
% According to the following dimension formula for the representations of $SO(6)_R$
%\bea
%\dim~(k,p,q) = \frac{1}{12}(k+p+q+3)(k+p+2)(k+q+2)(k+1)(p+1)(q+1)~,
%\eea
%we have the dimension formula of representations of the $\CN_{4d}=2$ super-Poincar\'e algebra:
%\bea
%\te{dim}\llbracket n;k,p,q \rrbracket_4 &= 256 \, (n+1) \, \frac{1}{12} \, (k+p+q+3) \, (k+p+2) \, (k+q+2) \times \nn \\
%&\qquad  (k+1)\,(p+1) \,(q+1) \ .
%\eea
%%%%
The four dimensional partition function is obtained through the usual procedure from the ten dimensional ancestor (\ref{NSandR10d}), this time we have to interpret three factors of $\chi^{SO(3)}_{\te{NS,R}}$ as carrying R-symmetry fugacities $r_j$:
\bea \label{NSandR4d}
\chi^{{\cal N}_{4d}=4}(q;y,\vec r) &\eq \chi^{{\cal N}_{4d}=4}_{\te{NS}} \mid_{\GSO} (q; y,\vec r) \ + \ \chi^{{\cal N}_{4d}=4}_{\te{R}} \mid_{\GSO} (q; y,\vec r) \notag \\
\chi^{{\cal N}_{4d}=4}_{\te{NS}} \mid_{\GSO} (q;  y, \vec r) &\eq
\frac{1}{2} \, q^{-\frac{1}{2}} \, \big[ \, \chi_{\te{NS}}^{SO(3)}(q; y) \, \chi_{\te{NS}}^{SO(7)}(q;\vec r) \ - \ \chi_{\te{NS}}^{SO(3)}(e^{2\pi i}q; y) \,  \chi_{\te{NS}}^{SO(7)}(e^{2\pi i}q;\vec r) \, \big] \notag \\
\chi^{{\cal N}_{4d}=4}_{\te{R}} \mid_{\GSO} (q; y, \vec r) &\eq \frac{1}{2} \,  \chi_{\te{R}}^{SO(3)}(q; y) \, \chi_{\te{R}}^{SO(7)}(q; \vec r) 
\eea
The power series in $q$ starts with\footnote{Again, there is a subtlety in applying the above formula to the massless R sector; see the footnote before \eref{def:GSOed}. However, this can be fixed easily: one can simply add to it $\frac{1}{2}(y -y^{-1})\prod_{j=1}^3 (r_j - r_j^{-1})$ to get the correct massless character in R sector.}
\begin{align}
\chi^{{\cal N}_{4d}=4}(q;y,r_j)  \ \
&= \ \ \underbrace{\left(y^2 + y^{-2} \, + \, \sum_{j=1}^3 (r_j^2 + r_j^{-2}) \ + \ \frac{1}{2}  \, [1]_y \, \prod_{j=1}^3 (r_j + r_j^{-1}) \right) \, q^0}_{16 \ \te{massless states}} \ + \ \underbrace{ \llbracket 0;0,0,0 \rrbracket \, q}_{256 \ \te{states at level} \ 1}  \notag \\
& \ \ \ \ \ \ \ \ + \ \underbrace{ \big( \, \llbracket0; 1, 0, 0\rrbracket + \llbracket2; 0, 0, 0\rrbracket \, \big) \, q^2}_{2304 \ \te{states at level} \ 2} \ + \ \big( \, \llbracket0; 0, 0, 0\rrbracket + \llbracket0; 2, 0, 0\rrbracket + \llbracket1; 0, 0, 1\rrbracket \notag \\
& \ \ \ \ \ \ \ \ \ \ \ \ \ \ \ \ \ \ + \llbracket1; 0, 1, 0\rrbracket + \llbracket2; 1, 0, 0\rrbracket + \llbracket4; 0, 0, 0\rrbracket \, \big) \, q^3\
+ \ {\cal O}(q^4) \ ,
\end{align}
the coefficients of $q^4$ and $q^5$ can be found in table \ref{table4,4}. The explicit vertex operators from the first level are listed in section 4 of \cite{Feng:2012bb}.

Note that this partition function can also be obtained by branching the $SO(9)$ representations appearing in the the $\CN_{10d}=1$ partition function \eref{chi10dref} into $SO(3)  \times SO(6)_R$ representations.  In terms of characters, one simply maps $SO(9)$ fugacities into  $SO(3)  \times SO(6)_R$ fugacities; a possible fugacity map is as follows:
\bea
z_1 = r_1, \qquad z_2 = r_2, \qquad z_3 = r_3, \qquad z_4= y~, \label{B4toD1D3}
\eea
where $z_1, \ldots, z_4$ are fugacities of $SO(9)$, $r_1, r_2, r_3$ are fugacities of $SO(6)_R$ and $y$ is a fugacity of $SO(3)$.
For example,
\bea
[1,0,0,0]_{\vec z} &= 1+\frac{1}{z_1^2}+z_1^2+\frac{1}{z_2^2}+r_2^2+\frac{1}{z_3^2}+z_3^2+\frac{1}{z_4^2}+z_4^2 \nn\\
&=\frac{1}{r_1^2}+r_1^2+\frac{1}{r_2^2}+r_2^2+\frac{1}{r_3^2}+r_3^2 + \left( 1+\frac{1}{y^2}+y^2 \right) \nn \\
&= [0; 1,0,0]_{\vec r; y} + [2; 0,0,0]_{\vec r; y}  ~,
\eea
where the notation $[a; b_1, b_2, b_3]$ denotes the $SO(3) \times SO(6)_R$ representation for which the $SO(3)$ representation is $[a]$ and $SO(6)_R$ representation is $[b_1, b_2, b_3]_R$. 

\bigskip
\noindent
\begin{table}[htdp]
\begin{tabular}{|l|l|}\hline  $\ap m^2$ &representations of ${\cal N}_{4d}=4$ super Poincar\'e \\ \hline \hline   
  1 & $ \llbracket 0;0,0,0 \rrbracket $  \\\hline
   2 & $\llbracket0; 1, 0, 0\rrbracket   \, + \, \llbracket2; 0, 0, 0\rrbracket $  \\\hline
   3 & $\llbracket0; 0, 0, 0\rrbracket   \, + \, \llbracket0; 2, 0, 0\rrbracket   \, + \, \llbracket1; 0, 0, 1\rrbracket   \, + \, \llbracket1; 0, 1, 0\rrbracket   \, + \, \llbracket2; 1, 0, 0\rrbracket   \, + \, \llbracket4; 0, 0, 0\rrbracket $  \\\hline
   4 & $ \llbracket0; 0, 1, 1\rrbracket   \, + \, 2 \llbracket0; 1, 0, 0\rrbracket   \, + \, \llbracket0; 3, 0, 0\rrbracket   \, + \, \llbracket1; 0, 0, 1\rrbracket   \, + \, \llbracket1; 0, 1, 0\rrbracket   \, + \, \llbracket1; 1, 0, 1\rrbracket  $ \\
   &  $ \, + \, \llbracket1; 1, 1, 0\rrbracket   \, + \, 3 \llbracket2; 0, 0, 0\rrbracket   \, + \, \llbracket2; 1, 0, 0\rrbracket   \, + \, \llbracket2; 2, 0, 0\rrbracket   \, + \, \llbracket3; 0, 0, 1\rrbracket   \, + \, \llbracket3; 0, 1, 0\rrbracket $ \\
   & $ \, + \, \llbracket4; 1, 0, 0\rrbracket   \, + \, \llbracket6; 0, 0, 0\rrbracket $
   \\\hline        
   5 &$4 \llbracket0; 0, 0, 0\rrbracket   \, + \, \llbracket0; 0, 0, 2\rrbracket   \, + \, \llbracket0; 0, 1, 1\rrbracket   \, + \, \llbracket0; 0, 2, 0\rrbracket   \, + \, \llbracket0; 1, 0, 0\rrbracket   \, + \, \llbracket0; 1, 1, 1\rrbracket  $ \\
   & $ \, + \, 2 \llbracket0; 2, 0, 0\rrbracket   \, + \, \llbracket0; 4, 0, 0\rrbracket   \, + \, 3 \llbracket1; 0, 0, 1\rrbracket   \, + \, 3 \llbracket1; 0, 1, 0\rrbracket   \, + \, 2 \llbracket1; 1, 0, 1\rrbracket $ \\
   & $ \, + \, 2 \llbracket1; 1, 1, 0\rrbracket   \, + \, \llbracket1; 2, 0, 1\rrbracket   \, + \, \llbracket1; 2, 1, 0\rrbracket   \, + \, 2 \llbracket2; 0, 0, 0\rrbracket   \, + \, 2 \llbracket2; 0, 1, 1\rrbracket   \, + \, 5 \llbracket2; 1, 0, 0\rrbracket $ \\
   & $ \, + \, \llbracket2; 2, 0, 0\rrbracket   \, + \, \llbracket2; 3, 0, 0\rrbracket   \, + \, 2 \llbracket3; 0, 0, 1\rrbracket   \, + \, 2 \llbracket3; 0, 1, 0\rrbracket   \, + \, \llbracket3; 1, 0, 1\rrbracket   \, + \, \llbracket3; 1, 1, 0\rrbracket  $ \\
   & $ \, + \, 3 \llbracket4; 0, 0, 0\rrbracket    \, + \, \llbracket4; 1, 0, 0\rrbracket   \, + \, \llbracket4; 2, 0, 0\rrbracket   \, + \, \llbracket5; 0, 0, 1\rrbracket   \, + \, \llbracket5; 0, 1, 0\rrbracket   \, + \, \llbracket6; 1, 0, 0\rrbracket $ \\
   & $ \, + \, \llbracket8; 0, 0, 0\rrbracket  $
     \\\hline
\end{tabular}
\caption{${\cal N}_{4d}=4$ multiplets occurring up to mass level 5}
\label{table4,4}
\end{table}

\section{Conclusion}

We have investigated model independent superstring states common to all type I compactifications that preserve ${\cal N}_{4d}=1$ and ${\cal N}_{6d}=(1,0)$ SUSY, respectively, and identified the underlying super Poincar\'e multiplets at individual mass levels. Part of our results are the associated unrefined partition functions together with their asymptotics for large mass levels, see \eref{exactunref4d}--\eref{asympNm4d} and \eref{exactunref6d}--\eref{asympNm6d}. The refined versions of the universal partition functions are given by (\ref{def:GSOed}) and (\ref{NSandR6d1}) and rewritten in terms of super Poincar\'e characters in (\ref{defMgenN1}), (\ref{mulsumoddeven0}), (\ref{mulsumoddeven}), (\ref{GGG6}) and (\ref{end6dim}). Moreover, we have presented dimensional reductions of the universal ${\cal N}_{6d}=(1,0)$ and ${\cal N}_{10d}=1$ spectra to even dimensions $d\geq 4$ in subsections \ref{sec:4d,N2}, \ref{sec:8dim}, \ref{sec:6dims} and \ref{sec:4dims}.

Multiplicity generating functions for individual supermultiplets tend to stabilize in the regime where the spin $j$ (or more generally the first $SO(d-1)$ Dynkin label) is comparable to the mass level $M=\ap m^2$. More specifically, the validity for the stable pattern roughly ranges between $\frac{1}{2}(M-M_0) \lessapprox j \lessapprox M-M_0$ where the offset $M_0$ depends on the remaining super Poincar\'e quantum numbers of the multiplets beyond the spin. In the mathematically most tractable ${\cal N}_{4d}=1$ case, we have derived closed formulae (\ref{asympmuloddeven}) and (\ref{mainferm}) for the leading Regge trajectory. In the highest dimensional scenarios with given number of supercharges -- ${\cal N}_{4d}=1$, ${\cal N}_{6d}=(1,0)$ and ${\cal N}_{10d}=1$ -- we extracted both leading and subleading Regge trajectories from explicitly computed multiplicities up to level $\ap m^2 =25$, see subsections \ref{emp4dim}, \ref{emp6dim} and \ref{emp10dim}.
 
Identifying the super Poincar\'e covariant spectrum in scenarios with different numbers of supercharges provides a significant step towards a better understanding of the string S-matrix. As pointed out in \cite{Boels:2012if}, cubic tree level vertices among all the massive states are the seeds for superstring amplitudes of higher multiplicity and genus. The results of this work appear inspiring to push this programme further, using on-shell superspace techniques in various dimenions \cite{Boels:2011zz,Boels:2012ie}. Refined partition functions as computed here serve as generating functions for helicitiy supertraces \cite{deWit:1999ir} which allow to disentangle contribution of individual supermultiplets to loop amplitudes.

Extending flat space results as presented in this work to curved spacetime provides an exciting direction of further research. Anti-de-Sitter backgrounds are of particular interest in view of their conjectured correspondence to conformal field theories \cite{Maldacena:1997re,Aharony:1999ti}. For instance, the model independent higher spin string spectrum at the first massive level in $AdS_3 \times S^3$ compactifications with pure NSNS background has been pioneered in \cite{Gerigk:2012cq}. This is a motivating starting point towards generalizations to nonzero RR flux and superstrings in $AdS_5 \times S^5$, see \cite{Mazzucato:2011jt} for a review. Also, we would like to mention reference \cite{Bianchi:2003wx} which extracts information on the $AdS_5 \times S^5$ Kaluza Klein excitations from the ${\cal N}_{10d}=1$ flat space spectrum, in particular from its large spin regime investigated in detail here. Finally it would be also very interesting to explore the extended symmetry structure
of the universal higher spin states in supersymmetric string compactifications, in analogy to the ${\cal W}_N$-symmetries in three-dimensional higher spin theories on  $AdS_3$
\cite{Gaberdiel:2010ar,Gaberdiel:2010pz,Gaberdiel:2011zw}.

\acknowledgments
We are very grateful to Stefan Hohenegger and Roberto Volpato for a number of useful discussions.  N.~M.~ would like to express his gratitude towards the following institutes and collaborators for the hospitality during the completion of this project: Max-Planck-Instituts f\"ur Gravitationsphysik (Albert-Einstein-Institut), Humboldt-Universit\"at zu Berlin, Imperial College London, 2012 Simons Summer Workshop at Simons Center for Geometry and Physics; Matthias Staudacher, Amihay Hanany, Theerasak Mingarcha and Aroonroj Mekareeya. O.~S.~ is grateful to the Max Planck Institute Munich for hospitality during various stages of the project and to Rutger Boels, Wan-Zhe Feng, Stephan Stieberger and Tomasz Taylor for collaboration on related projects. The work of N.~M.~ is supported by a research grant of the Max Planck Society.
\newpage
\section*{\huge Appendices}

\appendix

\section{Notation and conventions} \label{sec:not}
Unless stated otherwise, the following notation and conventions will be used throughout the paper.

%We apply the methods of \cite{Hanany:2010da, Curtright:1986di} and keep track of Lorentz- and R symmetry quantum numbers through $SO(n)$ characters. For this purpose, it is essential to have a clean and unambiguous notation:
\subsubsection*{Group and representation theoretic objects}
\begin{itemize}
\item The plethystic exponential of a multivariate function $f(t_1, \ldots, t_n)$ that vanishes at the origin, $f(0,\ldots, 0)=0$, is defined as
\bea
\PE \left[ f(t_1, \ldots, t_n) \right] = \exp \left( \sum_{k=1}^\infty \frac{1}{k} f(t_1^k, \ldots, t_n^k) \right)~.
\eea 
The ferminonic plethystic exponential is defined by
\bea
{\PE}_F \left[ f(t_1, \ldots, t_n) \right] = \exp \left( \sum_{k=1}^\infty \frac{(-1)^{k+1}}{k} f(t_1^k, \ldots, t_n^k) \right)~.
\eea
%%%%%
\item An irreducible representation of a simple group $G$ can be denoted by its highest weight vector.
\begin{itemize}
\item With respect to a basis consisting of the fundamental weights (the $\omega$-basis), we write the highest vector as $[a_1, \ldots, a_r]$ with $r = \mathrm{rank}~G$. This is also known as the \emph{Dynkin label}. 
\item With respect to a basis of the dual Cartan subalgebra (the $e$-basis), we write the the highest vector as $(\lambda_1, \ldots, \lambda_r)$.  
\item Note that we use the round brackets to distinguish the latter from the former for which the square brackets are used.
\end{itemize}
%%%%%
\item For $SO(2n+1)$, the label $[a_1, a_2, \ldots, a_n]$ is related to the label $(\lambda_1,  \lambda_2, \ldots, \lambda_n)$ by the formula 
\bea
\lambda_i &= a_i + a_{i+1} + \ldots+ a_{n-1} + \frac{1}{2}a_n~, \quad 1\leq i \leq n-1~, \nn \\
\lambda_n &= \frac{1}{2} a_n~, \label{lambdaanda}
\eea
or equivalently
\bea
a_i &= \lambda_i-\lambda_{i+1}~, \quad 1\leq i \leq n-1~, \nn \\
a_n &= 2 \lambda_n~, \label{aandlambda}
\eea
%%%%%%
\item Note that a representation is uniquely specified by its character.
We use the notation $[a_1, a_2, \ldots, a_r]_{\vec y}$ (resp. $(\lambda_1, \ldots, \lambda_r)_{\vec y}$) to denote the character of the representation $[a_1, a_2, \ldots, a_r]$ (resp. $(\lambda_1, \ldots, \lambda_r)$) written in terms of the variables $\vec y = (y_1, \ldots, y_r)$. Whenever there is no potential confusion, we drop the subscript $\vec y$ to avoid cluttered notation.
%%%%%%%
\item A representation of a product group $G_1 \times G_2$ is denoted by $[a_1, \ldots, a_{r_1}; b_1, b_2, \ldots, b_{r_2}]$ where $[a_1, \ldots, a_{r_1}]$ is an irreducible representation of $G_1$ and $[b_1, \ldots, b_{r_2}]$ is that of $G_2$. We use a semi-colon (;) to separate each representation.
%%%%%%%%%
\item We use the notation $[n]$ to denote the $(n+1)$-dimensional irreducible representation of $SU(2)$ and $SO(3)$.  Its character is given by
\begin{align}
[n]_y  = \sum_{k=-n/2}^{+n/2} \, y^{2k}~. \label{char1}
\end{align}
%%%%%%%
\item The character of the vector representation of $SO(2n+1)$, with $n >1$, is taken to be
\bea
(1,0,\ldots, 0)_{y} = [1,0,\ldots, 0]_y = 1+ \sum_{k=1}^n \left(y^2_k + y_k^{-2} \right)~. \label{chaBn}
\eea  
In general, the character of the irreducible representation $(\lambda_1,\ldots, \lambda_n)$ of $SO(2n+1)$ is given by the Weyl character formula:
\bea
(\lambda_1,\ldots, \lambda_n)_y = \frac{\det \left( y_j^{2(\lambda_{i}+n-i+\frac{1}{2})}- y_j^{-2(\lambda_{i}+n-i+\frac{1}{2})} \right)_{i,j=1}^n}{\det \left( y_j^{2(n-i+\frac{1}{2})}- y_j^{-2(n-i+\frac{1}{2})} \right)_{i,j=1}^n}~.
\eea
%%%%%%%%%

\item The choice of the character in \eref{chaBn} has a great advantage: One can relate the character of the vector representation of $SO(2n+1)$ to that of the vector representation of $SO(3)$ in a simple way:   
\bea
[1,0, \ldots, 0]^{SO(2n+1)}_y  = \sum_{k=1}^n [2]^{SO(3)}_{y_k} -(n-1) ~. \label{vecBn}
\eea
As we shall see in subsequent sections, this helps simplify a number of computations.

\item The Haar measures of $SO(3)$ and $SU(2)$ are taken to be
\bea
\int \ud \mu_{SO(3)} (y) &= \int \ud \mu_{SU(2)} (y) = \frac{1}{2} \frac{1}{2 \pi i} \oint_{|y|=1} \frac{\ud y}{y}  (1-y^2)(1-y^{-2})~. \label{HaarSO3}
%\int \ud \mu_{SO(5)} (y_1, y_2) &=  \frac{1}{8} \frac{1}{(2 \pi i)^2} \oint_{|y_1|=1} \frac{\ud y_1}{y_1} \oint_{|y_2|=1} \frac{\ud y_2}{y_2}~  (1-y_1^2)(1-y_2^2) (1-y_1^{-2})(1-y_2^{-2}) \times \nn \\
%& \qquad (1-y_1^2 y_2^2) (1-y_1^{-2} y_2^{-2})\left(1-y_1^2 y_2^{-2} \right)\left(1-y_1^{-2} y_2^{2} \right) \nn \\
%&= \int \ud \mu_{SO(3)} (y_1) \int \ud \mu_{SO(3)} (y_2) \times  \nn \\
%& \qquad  \frac{1}{2} (1-y_1^2 y_2^2) (1-y_1^{-2} y_2^{-2})\left(1-y_1^2 y_2^{-2} \right)\left(1-y_1^{-2} y_2^{2} \right)~. \label{HaarSO5}
\eea
In general, the Haar measure for $SO(2n+1)$ can be written as
\bea \label{HaarBn}
\int \ud \mu_{SO(2n+1)} (\vec y) = \int \ud \mu_{SO(3)} (y_1) \cdots \int \ud \mu_{SO(3)} (y_n)~\rho (\vec y)~,
\eea
where
\bea \label{defrho}
\rho(\vec y) = \frac{1}{n!} \prod_{1\leq i<j \leq n} (1-y_i^2 y_j^2) (1-y_i^{-2} y_j^{-2}) \left(1-y_i^2 y_j^{-2} \right)\left(1-y_i^{-2} y_j^{2} \right)~. 
\eea
\end{itemize}

\subsubsection*{Special functions}
\begin{itemize}
\item The $q$-Pochhammer symbols are defined as
\bea
(a;q)_n = \prod_{k=0}^{n-1} (1-a q^k)~, \qquad (a;q)_\infty = \prod_{k=0}^{\infty} (1-a q^k)~.
\eea

\item Our conventions for the Dedekind eta and the Jacobi theta functions are
\footnote{These conventions are related to, for example, those adopted in Appendix A of \cite{Kiritsis:1997hj} by $y = \exp( 2\pi i v),~q = \exp(2 \pi i \tau)$.   We refer the reader to this reference for further properties of such functions.}
%\bea
%\eta(\tau) & =  q^{\frac{1}{24}} \prod_{n=1}^{\infty} (1 - q^{n}) = q^{\frac{1}{24}} (q;q)_\infty~,  \\
%\vartheta_1(u,\tau) & = 2q^{\frac{1}{8}} \sin (\pi u) \, \prod_{n=1}^\infty(1-q^{n})(1-y q^{n})(1-y^{-1}q^{n})~, \\
%\vartheta_2(u,\tau) & =  2q^{\frac{1}{8}} \cos (\pi u) \, \prod_{n=1}^{\infty} (1-q^{n})\, (1+y q^{n})(1+y^{-1} q^{n})~,   \\
%\vartheta_3(u,\tau) & = \prod_{n=1}^{\infty} (1-q^{n}) \, (1+y q^{n-1/2})(1+y^{-1 }q^{n-1/2})~, \\
%\vartheta_4(u,\tau) & = \prod_{n=1}^{\infty} (1-q^{n}) \, (1-y q^{n-1/2}) (1-y^{-1} q^{n-1/2})~, \ \
%\eea
\bea
\eta(q) & =  q^{\frac{1}{24}} \prod_{n=1}^{\infty} (1 - q^{n}) = q^{\frac{1}{24}} (q;q)_\infty~,  \\
\vartheta_1(y,q) & = -i q^{\frac{1}{8}} (y^{\frac{1}{2}}-y^{-\frac{1}{2}}) \, \prod_{n=1}^\infty(1-q^{n})(1-y q^{n})(1-y^{-1}q^{n})~, \\
\vartheta_2(y,q) & =  q^{\frac{1}{8}} (y^{\frac{1}{2}}+y^{-\frac{1}{2}})\, \prod_{n=1}^{\infty} (1-q^{n})\, (1+y q^{n})(1+y^{-1} q^{n})~,   \\
\vartheta_3(y,q) & = \prod_{n=1}^{\infty} (1-q^{n}) \, (1+y q^{n-1/2})(1+y^{-1 }q^{n-1/2})~, \\
\vartheta_4(y,q) & = \prod_{n=1}^{\infty} (1-q^{n}) \, (1-y q^{n-1/2}) (1-y^{-1} q^{n-1/2})~, \ \
\eea
%where here and throughout the paper we define
%\bea
%y = \exp( 2\pi i u)~, \qquad q = \exp(2 \pi i \tau)~.
%\eea
In terms of an infinite sum, the Jacobi theta functions can be written as
\bea
\vartheta [^a_b] (y, q) = \sum_{m \in \BZ} q^{\frac{1}{2}(m-a/2)^2} (e^{-i\pi b} y)^{(m-a/2)}~,
\eea
where 
\bea
\th_1=\th[^1_1]~,\qquad  \th_2= \th[^1_0]~, \qquad  \th_3=\th[^0_0]~, \qquad  \th_4=\th[^0_1]~.
\eea

\item The Appell-Lerch sum is defined as follows \cite{Zwegers2002}:\footnote{The notation in this paper and that in Proposition 1.4 of \cite{Zwegers2002} can be related as follows. Our notation is on the left hand sides of the following equalities: $\mu(u,q) = \mu(u,u,q)$, and $\vartheta_1(u, \tau) = -\vartheta(u,\tau)$.}
\bea
\mu(u,\tau) = -\frac{e^{i \pi u} }{\vartheta_1(y,q)} \sum_{m \in \BZ}  (-1)^m \frac{e^{\pi i m(m+1)\tau + 2\pi i  m u}}{1- e^{2 \pi i m \tau + 2\pi  i u} }~, \label{ALdef}
\eea
where
\bea
y= \exp(2 \pi i u)~, \qquad q = \exp(2 \pi i \tau)~.
\eea
\end{itemize}

\cofeAm{0.4}{0.8}{90}{-3cm}{1cm}

\newpage
%%%%%%%%

\section{Data tables for super Poincar\'e multiplicities}

This appendix contains data tables for multiplicities of super Poincar\'e representations up to mass level $\ap m^2 = 25$. We only display tables for the ancestor theories with 4, 8 and 16 supercharges, respectively, since these highest dimensional theories organize the states in the most economic number of supermultiplets. Particular attention is paid to stable patterns, i.e. to the asymptotics of multiplicity generating functions for large spins and mass levels.

Each of the following tables is devoted to family of supermultiplets whose quantum numbers differ in the first $SO(d-1)$ Dynkin label and match in the remaining $SO(d-1)$ and $R$ symmetry quantum numbers. Rows are associated with mass levels, and columns are associated with the value of the first $SO(d-1)$ Dynkin label to which we loosely refer to as the spin. Independently of spacetime dimensions and supercharges, the multiplicity generating functions $G_{\ldots}(q)$ tend to stabilize for large values of the spin and the mass level in the limit where both of them are uniformly increased. This leading Regge trajectory (corresponding to the $\tau^{\ldots}_1(q)$ contribution in (\ref{tau}), (\ref{6dimtau}) and (\ref{10dimtau})) is exact when numbers occur repeatedly along diagonal lines in the tables, these entries are marked in \textcolor{red}{red}.

Moreover, once the asymptotic numbers in red are subtracted from the data outside the first stable region, further subleading trajectory emerge. The leftover after this subtraction tends to stabilize along lines where the mass level grows twice as fast as the spin. This can be understood as the second Regge trajectory (corresponding to the $\tau^{\ldots}_2(q)$ contribution in (\ref{tau}), (\ref{6dimtau}) and (\ref{10dimtau})) with slope $\frac{1}{2}$ and subtractive sign. Its region of exact validity is highlighted in \textcolor{blue}{blue}.

%%%%%%%%%%%%%%%%%%%%%%%%%%%%%%%%%%
%%%%%%%%%%%%%%%%%%%%%%%%%%%%%%%%%%

\subsection{4 supercharges in four dimensions}
\label{app4}

The tables in this subsection are based on the ${\cal N}_{4d}=1$ partition function (\ref{def:GSOed}), organized in terms of multiplicity generating functions $G_{n,Q}(q)$, see (\ref{defMgenN1}).

% Q=2 here

%\begin{table}
\begin{scriptsize}
\begin{center}
\begin{tabular}{|c|| c|c|c|c|c| c|c|c|c|c| c|c|c|c|c| c|c|c|c|c| c|c|c|}
%\begin{longtable}{|c|| c|c|c|c|c| c|c|c|c|c| c|c|c|c|c| c|c|c|c|c| c|c|c|c|c| c|} % \small
 \hline
\begin{turn}{-90}$ \  \ap m^2 \ $  \end{turn}
&\begin{turn}{-90} $\llbracket1;2\rrbracket$ \end{turn}
&\begin{turn}{-90} $\llbracket3;2\rrbracket$ \end{turn}
&\begin{turn}{-90} $\llbracket5;2\rrbracket$ \end{turn}
&\begin{turn}{-90} $\llbracket7;2\rrbracket$ \end{turn}
&\begin{turn}{-90} $\llbracket9;2\rrbracket$ \end{turn}
&\begin{turn}{-90} $\llbracket11;2\rrbracket$ \end{turn}
&\begin{turn}{-90} $\llbracket13;2\rrbracket$ \end{turn}
&\begin{turn}{-90} $\llbracket15;2\rrbracket$ \end{turn}
&\begin{turn}{-90} $\llbracket17;2\rrbracket$ \end{turn}
&\begin{turn}{-90} $\llbracket19;2\rrbracket$ \end{turn}
&\begin{turn}{-90} $\llbracket21;2\rrbracket$ \end{turn}
&\begin{turn}{-90} $\llbracket23;2\rrbracket$ \end{turn}
%&\begin{turn}{-90}\# $\llbracket25;2\rrbracket \ $ \end{turn}
%
\\\hline \hline
1 &\textcolor{red}{0} & & & &
& & & & &
& &  
  \\\hline
2 &\textcolor{red}{0} & & & &
& & & & &
& & 
  \\\hline
3 &1 &\textcolor{red}{0} & & &
& & & & &
& &  
  \\\hline
4 &2 &\textcolor{red}{2} &\textcolor{red}{0} & &
& & & & &
& &  
  \\\hline
5 &6 &\textcolor{blue}{6} &\textcolor{red}{2} &\textcolor{red}{0} &
& & & & &
& & 
  \\\hline
6 &17 &15 &\textcolor{red}{8} &\textcolor{red}{2} &\textcolor{red}{0}
& & & & &
& & 
  \\\hline
7 &38 &43 &\textcolor{blue}{22} &\textcolor{red}{8} &\textcolor{red}{2}
&\textcolor{red}{0} & & & &
& &   
  \\
\hline
\end{tabular}
%\end{longtable}
%\end{sidewaysfigure}
\end{center}
\end{scriptsize}
%\caption{${\cal N}_{4d}=1$ multiplets at $U(1)_R$ charge $Q=2$}
%\label{tab4d2}
%\end{table}

%%% Now the Q=4 stuff

%\begin{table}
\begin{scriptsize}
\begin{center}
%\begin{sidewaysfigure}
\begin{tabular}{|c|| c|c|c|c|c| c|c|c|c|c| c|c|c|c|c| c|c|c|c|c|}
%\begin{longtable}{|c|| c|c|c|c|c| c|c|c|c|c| c|c|c|c|c| c|c|c|c|c| c|c|c|c|c| c|} % \small
 \hline
\begin{turn}{-90}$ \  \ap m^2 \ $  \end{turn}
%&\begin{turn}{-90}\# $\llbracket1;4\rrbracket$ \end{turn}
%&\begin{turn}{-90}\# $\llbracket3;4\rrbracket$ \end{turn}
%&\begin{turn}{-90}\# $\llbracket5;4\rrbracket$ \end{turn}
%&\begin{turn}{-90}\# $\llbracket7;4\rrbracket$ \end{turn}
%&\begin{turn}{-90}\# $\llbracket9;4\rrbracket$ \end{turn}
%%
%&\begin{turn}{-90}\# $\llbracket11;4\rrbracket \ $ \end{turn}
%&\begin{turn}{-90}\# $\llbracket13;4\rrbracket$ \end{turn}
%&\begin{turn}{-90}\# $\llbracket15;4\rrbracket$ \end{turn}
%&\begin{turn}{-90}\# $\llbracket17;4\rrbracket$ \end{turn}
%&\begin{turn}{-90}\# $\llbracket19;4\rrbracket$ \end{turn}
%%
%&\begin{turn}{-90}\# $\llbracket21;4\rrbracket$ \end{turn}
%&\begin{turn}{-90}\# $\llbracket23;4\rrbracket$ \end{turn}
%&\begin{turn}{-90}\# $\llbracket25;4\rrbracket$ \end{turn}
%&\begin{turn}{-90}\# $\llbracket27;4\rrbracket$ \end{turn}
&\begin{turn}{-90} $\llbracket1;4\rrbracket$ \end{turn}
&\begin{turn}{-90} $\llbracket3;4\rrbracket$ \end{turn}
&\begin{turn}{-90} $\llbracket5;4\rrbracket$ \end{turn}
&\begin{turn}{-90} $\llbracket7;4\rrbracket$ \end{turn}
&\begin{turn}{-90} $\llbracket9;4\rrbracket$ \end{turn}
&\begin{turn}{-90} $\llbracket11;4\rrbracket \ $ \end{turn}
&\begin{turn}{-90} $\llbracket13;4\rrbracket$ \end{turn}
&\begin{turn}{-90} $\llbracket15;4\rrbracket$ \end{turn}
&\begin{turn}{-90} $\llbracket17;4\rrbracket$ \end{turn}
&\begin{turn}{-90} $\llbracket19;4\rrbracket$ \end{turn}
&\begin{turn}{-90} $\llbracket21;4\rrbracket$ \end{turn}
&\begin{turn}{-90} $\llbracket23;4\rrbracket$ \end{turn}
&\begin{turn}{-90} $\llbracket25;4\rrbracket$ \end{turn}
&\begin{turn}{-90} $\llbracket27;4\rrbracket$ \end{turn}
%&\begin{turn}{-90}\# $\llbracket29;4\rrbracket$ \end{turn}
%
%&\begin{turn}{-90}\# $\llbracket31;4\rrbracket$ \end{turn}
\\
\hline
\end{tabular}
%\end{longtable}
%\end{sidewaysfigure}

\vspace{0.4cm}

\begin{tabular}{|c|| c|c|c|c|c| c|c|c|c|c| c|c|c|c|c| c|c|c|c|c| c|c|c|c|c| c|}
%\begin{longtable}{|c|| c|c|c|c|c| c|c|c|c|c| c|c|c|c|c| c|c|c|c|c| c|c|c|c|c| c|} % \small
 \hline
\begin{turn}{-90}$ \ \ap m^2 \ $  \end{turn}
%&\begin{turn}{-90}\# $\llbracket1;6\rrbracket$ \end{turn}
%&\begin{turn}{-90}\# $\llbracket3;6\rrbracket$ \end{turn}
%&\begin{turn}{-90}\# $\llbracket5;6\rrbracket$ \end{turn}
%&\begin{turn}{-90}\# $\llbracket7;6\rrbracket$ \end{turn}
%&\begin{turn}{-90}\# $\llbracket9;6\rrbracket$ \end{turn}
%%
%&\begin{turn}{-90}\# $\llbracket11;6\rrbracket \ $ \end{turn}
%&\begin{turn}{-90}\# $\llbracket13;6\rrbracket$ \end{turn}
%&\begin{turn}{-90}\# $\llbracket15;6\rrbracket$ \end{turn}
%&\begin{turn}{-90}\# $\llbracket17;6\rrbracket$ \end{turn}
%&\begin{turn}{-90}\# $\llbracket19;6\rrbracket$ \end{turn}
%%
%&\begin{turn}{-90}\# $\llbracket21;6\rrbracket$ \end{turn}
%&\begin{turn}{-90}\# $\llbracket23;6\rrbracket$ \end{turn}
%&\begin{turn}{-90}\# $\llbracket25;6\rrbracket$ \end{turn}
&\begin{turn}{-90} $\llbracket1;6\rrbracket$ \end{turn}
&\begin{turn}{-90} $\llbracket3;6\rrbracket$ \end{turn}
&\begin{turn}{-90} $\llbracket5;6\rrbracket$ \end{turn}
&\begin{turn}{-90} $\llbracket7;6\rrbracket$ \end{turn}
&\begin{turn}{-90} $\llbracket9;6\rrbracket$ \end{turn}
&\begin{turn}{-90} $\llbracket11;6\rrbracket \ $ \end{turn}
&\begin{turn}{-90} $\llbracket13;6\rrbracket$ \end{turn}
&\begin{turn}{-90} $\llbracket15;6\rrbracket$ \end{turn}
&\begin{turn}{-90} $\llbracket17;6\rrbracket$ \end{turn}
&\begin{turn}{-90} $\llbracket19;6\rrbracket$ \end{turn}
&\begin{turn}{-90} $\llbracket21;6\rrbracket$ \end{turn}
&\begin{turn}{-90} $\llbracket23;6\rrbracket$ \end{turn}
&\begin{turn}{-90} $\llbracket25;6\rrbracket$ \end{turn}
\\\hline \hline
%13 & & & & &
%& & & & &
%& & &  \\\hline
14 &\textcolor{red}{0} & & & &
& & & & &
& & &   \\\hline
15 &1 &\textcolor{red}{0} & & &
& & & & &
& & &  \\\hline
16 &3 &\textcolor{red}{2} &\textcolor{red}{0} & &
& & & & &
& & &  \\
\hline
\end{tabular}
%\end{longtable}

\vspace{0.4cm}

%\begin{sidewaysfigure}
\begin{tabular}{|c|| c|c|c|c|c| c|c|c|c|c| c|}
%\begin{longtable}{|c|| c|c|c|c|c| c|c|c|c|c| c|c|c|c|c| c|c|c|c|c| c|c|c|c|c| c|} % \small
 \hline
\begin{turn}{-90}$ \ \ap m^2 \ $  \end{turn}
%&\begin{turn}{-90}\# $\llbracket0;1\rrbracket$ \end{turn}
%&\begin{turn}{-90}\# $\llbracket2;1\rrbracket$ \end{turn}
%&\begin{turn}{-90}\# $\llbracket4;1\rrbracket$ \end{turn}
%&\begin{turn}{-90}\# $\llbracket6;1\rrbracket$ \end{turn}
%&\begin{turn}{-90}\# $\llbracket8;1\rrbracket$ \end{turn}
%%
%&\begin{turn}{-90}\# $\llbracket10;1\rrbracket\ $ \end{turn}
%&\begin{turn}{-90}\# $\llbracket12;1\rrbracket$ \end{turn}
%&\begin{turn}{-90}\# $\llbracket14;1\rrbracket$ \end{turn}
%&\begin{turn}{-90}\# $\llbracket16;1\rrbracket$ \end{turn}
%&\begin{turn}{-90}\# $\llbracket18;1\rrbracket$ \end{turn}
%%
%&\begin{turn}{-90}\# $\llbracket20;1\rrbracket$ \end{turn}
&\begin{turn}{-90} $\llbracket0;1\rrbracket$ \end{turn}
&\begin{turn}{-90} $\llbracket2;1\rrbracket$ \end{turn}
&\begin{turn}{-90} $\llbracket4;1\rrbracket$ \end{turn}
&\begin{turn}{-90} $\llbracket6;1\rrbracket$ \end{turn}
&\begin{turn}{-90} $\llbracket8;1\rrbracket$ \end{turn}
&\begin{turn}{-90} $\llbracket10;1\rrbracket\ $ \end{turn}
&\begin{turn}{-90} $\llbracket12;1\rrbracket$ \end{turn}
&\begin{turn}{-90} $\llbracket14;1\rrbracket$ \end{turn}
&\begin{turn}{-90} $\llbracket16;1\rrbracket$ \end{turn}
&\begin{turn}{-90} $\llbracket18;1\rrbracket$ \end{turn}
&\begin{turn}{-90} $\llbracket20;1\rrbracket$ \end{turn}
%&\begin{turn}{-90}\# $\llbracket22;1\rrbracket$ \end{turn}
\\\hline \hline
1 &1 &\textcolor{red}{0} & & &
& & & & &
&   \\\hline
2 &0 &\textcolor{blue}{2} &\textcolor{red}{0} & &
& & & & &
&   \\\hline
3 &3 &2 &\textcolor{red}{3} &\textcolor{red}{0} &
& & & & &
&   \\\hline
4 &3 &11 &\textcolor{blue}{4} &\textcolor{red}{3} &\textcolor{red}{0}
& & & & &
&   \\\hline
5 &15 &20 &\textcolor{blue}{18} &\textcolor{red}{5} &\textcolor{red}{3}
&\textcolor{red}{0} & & & &
&  \\
\hline
\end{tabular}
%\end{longtable}
%\end{sidewaysfigure}
\end{center}
\end{scriptsize}
%\caption{${\cal N}_{4d}=1$ multiplets at $U(1)_R$ charge $Q=1$}
%\label{tab4d1}
%\end{table}

%%%

%\begin{table}
\begin{scriptsize}
\begin{center}
%\begin{sidewaysfigure}
\begin{tabular}{|c|| c|c|c|c|c| c|c|c|c|c| c|c|}
%\begin{longtable}{|c|| c|c|c|c|c| c|c|c|c|c| c|c|c|c|c| c|c|c|c|c| c|c|c|c|c| c|} % \small
 \hline
\begin{turn}{-90}$ \ \ap m^2 \ $  \end{turn}
&\begin{turn}{-90} $\llbracket0;3\rrbracket$ \end{turn}
&\begin{turn}{-90} $\llbracket2;3\rrbracket$ \end{turn}
&\begin{turn}{-90} $\llbracket4;3\rrbracket$ \end{turn}
&\begin{turn}{-90} $\llbracket6;3\rrbracket$ \end{turn}
&\begin{turn}{-90} $\llbracket8;3\rrbracket$ \end{turn}
&\begin{turn}{-90} $\llbracket10;3\rrbracket$ \end{turn}
&\begin{turn}{-90} $\llbracket12;3\rrbracket$ \end{turn}
&\begin{turn}{-90} $\llbracket14;3\rrbracket$ \end{turn}
&\begin{turn}{-90} $\llbracket16;3\rrbracket$ \end{turn}
&\begin{turn}{-90} $\llbracket18;3\rrbracket$ \end{turn}
&\begin{turn}{-90} $\llbracket20;3\rrbracket$ \end{turn}
&\begin{turn}{-90} $\llbracket22;3\rrbracket$ \end{turn}
%&\begin{turn}{-90}\# $\llbracket24;3\rrbracket \ $ \end{turn}
\\\hline \hline
4 &\textcolor{red}{0} & & & &
& & & & &
& &  \\\hline
5 &1 &\textcolor{red}{0} & & &
& & & & &
& &  \\\hline
6 &0 &\textcolor{blue}{3} &\textcolor{red}{0} & &
& & & & &
& &  \\\hline
7 &6 &5 &\textcolor{red}{4} &\textcolor{red}{0} &
& & & & &
& &  \\\hline
8 &7 &21 &\textcolor{blue}{10} &\textcolor{red}{4} &\textcolor{red}{0}
& & & & &
& &  \\\hline
9 &29 &44 &\textcolor{blue}{37} &\textcolor{red}{11} &\textcolor{red}{4}
&\textcolor{red}{0} & & & &
& &  \\\hline
10 &50 &122 &84 &\textcolor{blue}{45} &\textcolor{red}{11}
&\textcolor{red}{4} &\textcolor{red}{0} & & &
& & \\
\hline
\end{tabular}
%\end{longtable}
%\end{sidewaysfigure}

\vspace{0.7cm}

%\begin{sidewaysfigure}
\begin{tabular}{|c|| c|c|c|c|c| c|c|c|c|c| c|c|c|c|}
%\begin{longtable}{|c|| c|c|c|c|c| c|c|c|c|c| c|c|c|c|c| c|c|c|c|c| c|c|c|c|c| c|} % \small
 \hline
\begin{turn}{-90}$ \ \ap m^2 \ $  \end{turn}
&\begin{turn}{-90} $\llbracket0;5\rrbracket$ \end{turn}
&\begin{turn}{-90} $\llbracket2;5\rrbracket$ \end{turn}
&\begin{turn}{-90} $\llbracket4;5\rrbracket$ \end{turn}
&\begin{turn}{-90} $\llbracket6;5\rrbracket$ \end{turn}
&\begin{turn}{-90} $\llbracket8;5\rrbracket$ \end{turn}
&\begin{turn}{-90} $\llbracket10;5\rrbracket$ \end{turn}
&\begin{turn}{-90} $\llbracket12;5\rrbracket$ \end{turn}
&\begin{turn}{-90} $\llbracket14;5\rrbracket$ \end{turn}
&\begin{turn}{-90} $\llbracket16;5\rrbracket$ \end{turn}
&\begin{turn}{-90} $\llbracket18;5\rrbracket$ \end{turn}
&\begin{turn}{-90} $\llbracket20;5\rrbracket$ \end{turn}
&\begin{turn}{-90} $\llbracket22;5\rrbracket$ \end{turn}
&\begin{turn}{-90} $\llbracket24;5\rrbracket$ \end{turn}
&\begin{turn}{-90} $\llbracket26;5\rrbracket$ \end{turn}
%&\begin{turn}{-90}\# $\llbracket28;5\rrbracket \ $ \end{turn}
\\\hline \hline
10 &\textcolor{red}{0} & & & &
& & & & &
& & & &
 \\\hline
11 &1 &\textcolor{red}{0} & & &
& & & & &
& & & &  \\\hline
12 &0 &\textcolor{blue}{3} &\textcolor{red}{0} & &
& & & & &
& & & & \\\hline
13 &7 &6 &\textcolor{red}{4} &\textcolor{red}{0} &
& & & &
& & & & &
 \\\hline
14 &10 &26 &\textcolor{blue}{11} &\textcolor{red}{4} &\textcolor{red}{0}
& & & & &
& & & &
\\\hline
15 &37 &58 &\textcolor{blue}{46} &\textcolor{red}{12} &\textcolor{red}{4}
&\textcolor{red}{0} & & & &
& & & &
 \\\hline
16 &70 &163 &111 &\textcolor{blue}{54} &\textcolor{red}{12}
&\textcolor{red}{4} &\textcolor{red}{0} & & &
& & & &
 \\
\hline
\end{tabular}
%\end{longtable}
%\end{sidewaysfigure}

\vspace{0.7cm}

%\begin{sidewaysfigure}
\begin{tabular}{|c|| c|c|c|c|c| c|c|c|c|}
%\begin{longtable}{|c|| c|c|c|c|c| c|c|c|c|c| c|c|c|c|c| c|c|c|c|c| c|c|c|c|c| c|} % \small
 \hline
\begin{turn}{-90}$ \ \ap m^2\ $  \end{turn}
&\begin{turn}{-90} $\llbracket0;7\rrbracket$ \end{turn}
&\begin{turn}{-90} $\llbracket2;7\rrbracket$ \end{turn}
&\begin{turn}{-90} $\llbracket4;7\rrbracket$ \end{turn}
&\begin{turn}{-90} $\llbracket6;7\rrbracket$ \end{turn}
&\begin{turn}{-90} $\llbracket8;7\rrbracket$ \end{turn}
&\begin{turn}{-90} $\llbracket10;7\rrbracket$ \end{turn}
&\begin{turn}{-90} $\llbracket12;7\rrbracket$ \end{turn}
&\begin{turn}{-90} $\llbracket14;7\rrbracket \ $ \end{turn}
\\\hline \hline 
18 &\textcolor{red}{0} & & & &
& & &   \\\hline
19 &1 &\textcolor{red}{0} & & &
& & &   \\\hline
20 &0 &\textcolor{blue}{3} &\textcolor{red}{0} & &
& & &   \\\hline
21 &7 &6 &\textcolor{red}{4} &\textcolor{red}{0} &
& & &   \\\hline
22 &11 &27 &\textcolor{blue}{11} &\textcolor{red}{4} &\textcolor{red}{0}
& & &   \\\hline
23 &41 &63 &\textcolor{blue}{47} &\textcolor{red}{12} &\textcolor{red}{4}
&\textcolor{red}{0} & &   \\\hline
24 &78 &180 &120 &\textcolor{blue}{55} &\textcolor{red}{12}
&\textcolor{red}{4} &\textcolor{red}{0} &   \\\hline
25 &214 &402 &336 &\textcolor{blue}{150} &\textcolor{red}{56}
&\textcolor{red}{12} &\textcolor{red}{4} &\textcolor{red}{0}   \\\hline
\end{tabular}
%\end{longtable}
%\end{sidewaysfigure}
\end{center}
\end{scriptsize}
%\caption{${\cal N}_{4d}=1$ multiplets at $U(1)_R$ charge $Q=7$}
%\label{tab4d7}
%\end{table}

\vspace{0.8cm} 
%%%%%%%%%%%%%%%%%%%%%%%%%
%%%%%%%%%%%%%%%%%%%%%%%%%

\subsection{8 supercharges in six dimensions}
\label{app6}

The tables in this subsection are based on the ${\cal N}_{6d}=(1,0)$ partition function (\ref{NSandR6d1}), organized in terms of multiplicity generating functions $G_{n_1,n_2,p}(q)$, see (\ref{GGG6}).

%%% j,2,0

%\begin{table}
\begin{tiny}
\begin{center}
%\begin{sidewaysfigure}
\begin{tabular}{|c|| c|c|c|c|c| c|c|c|c|c| c|c|}
%\begin{longtable}{|c|| c|c|c|c|c| c|c|c|c|c| c|c|c|c|c| c|c|c|c|c| c|c|c|c|c| c|} % \small
 \hline
\begin{turn}{-90}$\ \ap m^2 \ $  \end{turn}
&\begin{turn}{-90} $\llbracket0,2;0\rrbracket$ \end{turn}
&\begin{turn}{-90} $\llbracket1,2;0\rrbracket$ \end{turn}
&\begin{turn}{-90} $\llbracket2,2;0\rrbracket$ \end{turn}
&\begin{turn}{-90} $\llbracket3,2;0\rrbracket$ \end{turn}
&\begin{turn}{-90} $\llbracket4,2;0\rrbracket$ \end{turn}
&\begin{turn}{-90} $\llbracket5,2;0\rrbracket$ \end{turn}
&\begin{turn}{-90} $\llbracket6,2;0\rrbracket$ \end{turn}
&\begin{turn}{-90} $\llbracket7,2;0\rrbracket$ \end{turn}
&\begin{turn}{-90} $\llbracket8,2;0\rrbracket$ \end{turn}
&\begin{turn}{-90} $\llbracket9,2;0\rrbracket$ \end{turn}
&\begin{turn}{-90} $\llbracket10,2;0\rrbracket\ $ \end{turn}
&\begin{turn}{-90} $\llbracket11,2;0\rrbracket$ \end{turn}
%&\begin{turn}{-90}\# $\llbracket12,2;0\rrbracket$ \end{turn}
%&\begin{turn}{-90}\# $\llbracket13,2;0\rrbracket$ \end{turn}
%
\\\hline \hline
1 &\textcolor{red}{0} & & & &
& & & & &
& &   \\\hline
2 &\textcolor{red}{1} &\textcolor{red}{0} & & &
& & & & &
& &   \\\hline
3 &\textcolor{blue}{1} &\textcolor{red}{1} &\textcolor{red}{0} & &
& & & & &
& &   \\\hline
4 &\textcolor{blue}{4} &\textcolor{red}{2} &\textcolor{red}{1} &\textcolor{red}{0} &
& & & & & & &   \\\hline
5 &6 &\textcolor{blue}{7} &\textcolor{red}{2} &\textcolor{red}{1} &\textcolor{red}{0}
& & & & &
& &  \\\hline
6 &19 &\textcolor{blue}{13} &\textcolor{red}{8} &\textcolor{red}{2} &\textcolor{red}{1}
&\textcolor{red}{0} & & & &
& &   \\\hline
7 &34 &\textcolor{blue}{38} &\textcolor{blue}{16} &\textcolor{red}{8} &\textcolor{red}{2}
&\textcolor{red}{1} &\textcolor{red}{0} & & &
& &   \\
\hline
\end{tabular}
%\end{longtable}
%\end{sidewaysfigure}
\end{center}
\end{tiny}
%\caption{${\cal N}_{6d}=(1,0)$ multiplets with $SO(5)$ quantum numbers $[j,2]$ and $SU(2)_R$ spin 0}
%\label{tab6d20}
%\end{table}

%%%% j,4,0

\vspace{0.8cm}

%\vfill 
%\begin{table}
\begin{tiny}
\begin{center}
%\begin{sidewaysfigure}
\begin{tabular}{|c|| c|c|c|c|c| c|c|c|c|c| c|c|c|}
%\begin{longtable}{|c|| c|c|c|c|c| c|c|c|c|c| c|c|c|c|c| c|c|c|c|c| c|c|c|c|c| c|} % \small
 \hline
\begin{turn}{-90}$\ \ap m^2 \ $  \end{turn}
&\begin{turn}{-90} $\llbracket0,4;0\rrbracket$ \end{turn}
&\begin{turn}{-90} $\llbracket1,4;0\rrbracket$ \end{turn}
&\begin{turn}{-90} $\llbracket2,4;0\rrbracket$ \end{turn}
&\begin{turn}{-90} $\llbracket3,4;0\rrbracket$ \end{turn}
&\begin{turn}{-90} $\llbracket4,4;0\rrbracket$ \end{turn}
&\begin{turn}{-90} $\llbracket5,4;0\rrbracket$ \end{turn}
&\begin{turn}{-90} $\llbracket6,4;0\rrbracket$ \end{turn}
&\begin{turn}{-90} $\llbracket7,4;0\rrbracket$ \end{turn}
&\begin{turn}{-90} $\llbracket8,4;0\rrbracket$ \end{turn}
&\begin{turn}{-90} $\llbracket9,4;0\rrbracket$ \end{turn}
&\begin{turn}{-90} $\llbracket10,4;0\rrbracket\ $ \end{turn}
&\begin{turn}{-90} $\llbracket11,4;0\rrbracket$ \end{turn}
&\begin{turn}{-90} $\llbracket12,4;0\rrbracket$ \end{turn}
%&\begin{turn}{-90}\# $\llbracket13,4;0\rrbracket$ \end{turn}
%
\\\hline \hline
4 &\textcolor{red}{0} & & & &
& & & & &
& & &  \\\hline
5 &\textcolor{red}{1} &\textcolor{red}{0} & & &
& & & & &
& & &  \\\hline
6 &\textcolor{blue}{4} &\textcolor{red}{1} &\textcolor{red}{0} & &
& & & &
& & & &  \\\hline
7 &\textcolor{blue}{9} &\textcolor{red}{5} &\textcolor{red}{1} &\textcolor{red}{0} &
& & & & &
 & & &  \\\hline
8 &\textcolor{blue}{25} &\textcolor{blue}{13} &\textcolor{red}{5} &\textcolor{red}{1} &\textcolor{red}{0}
& & & & &
& & &  \\\hline
9 &61 &\textcolor{blue}{38} &\textcolor{red}{14} &\textcolor{red}{5} &\textcolor{red}{1}
&\textcolor{red}{0} & & & &
& &  &  \\\hline
10 &142 &\textcolor{blue}{95} &\textcolor{blue}{42} &\textcolor{red}{14} &\textcolor{red}{5}
&\textcolor{red}{1} &\textcolor{red}{0} & & &
& & &  \\
\hline
\end{tabular}
%\end{longtable}
%\end{sidewaysfigure}
\end{center}
\end{tiny}
%\caption{${\cal N}_{6d}=(1,0)$ multiplets with $SO(5)$ quantum numbers $[j,4]$ and $SU(2)_R$ spin 0}
%\label{tab6d40}
%\end{table}
%
%%%% j,6,0

\vspace{0.8cm}

%\begin{table}
\begin{tiny}
\begin{center}
%\begin{sidewaysfigure}
\begin{tabular}{|c|| c|c|c|c|c| c|c|c|c|c| c|c|c|c|}
%\begin{longtable}{|c|| c|c|c|c|c| c|c|c|c|c| c|c|c|c|c| c|c|c|c|c| c|c|c|c|c| c|} % \small
 \hline
\begin{turn}{-90}$\ \ap m^2 \ $  \end{turn}
&\begin{turn}{-90} $\llbracket0,6;0\rrbracket$ \end{turn}
&\begin{turn}{-90} $\llbracket1,6;0\rrbracket$ \end{turn}
&\begin{turn}{-90} $\llbracket2,6;0\rrbracket$ \end{turn}
&\begin{turn}{-90} $\llbracket3,6;0\rrbracket$ \end{turn}
&\begin{turn}{-90} $\llbracket4,6;0\rrbracket$ \end{turn}
&\begin{turn}{-90} $\llbracket5,6;0\rrbracket$ \end{turn}
&\begin{turn}{-90} $\llbracket6,6;0\rrbracket$ \end{turn}
&\begin{turn}{-90} $\llbracket7,6;0\rrbracket$ \end{turn}
&\begin{turn}{-90} $\llbracket8,6;0\rrbracket$ \end{turn}
&\begin{turn}{-90} $\llbracket9,6;0\rrbracket$ \end{turn}
&\begin{turn}{-90} $\llbracket10,6;0\rrbracket\ $ \end{turn}
&\begin{turn}{-90} $\llbracket11,6;0\rrbracket$ \end{turn}
&\begin{turn}{-90} $\llbracket12,6;0\rrbracket$ \end{turn}
&\begin{turn}{-90} $\llbracket13,6;0\rrbracket$ \end{turn}
\\\hline \hline
7 &\textcolor{red}{0} & & & &
& & & & &
& & & &  \\\hline
8 &\textcolor{red}{1} &\textcolor{red}{0} & & &
& & & & &
& & & &  \\\hline
9 &\textcolor{blue}{4} &\textcolor{red}{1} &\textcolor{red}{0} & &
& & & & &
& & & &  \\\hline
10 &\textcolor{blue}{13} &\textcolor{red}{5} &\textcolor{red}{1} &\textcolor{red}{0} &
& & & & &
& & & &  \\\hline
11 &\textcolor{blue}{35} &\textcolor{blue}{17} &\textcolor{red}{5} &\textcolor{red}{1}
&\textcolor{red}{0} & & & & &
& & & &  \\\hline
12 &\textcolor{blue}{101} &\textcolor{blue}{48} &\textcolor{red}{18} &\textcolor{red}{5} &\textcolor{red}{1}
&\textcolor{red}{0} & & & &
& & & &  \\\hline
13 &238 &\textcolor{blue}{140} &\textcolor{blue}{52} &\textcolor{red}{18} &\textcolor{red}{5}
&\textcolor{red}{1} &\textcolor{red}{0} & & &
& & & &  \\
\hline
\end{tabular}
%\end{longtable}
%\end{sidewaysfigure}
\end{center}
\end{tiny}
%\caption{${\cal N}_{6d}=(1,0)$ multiplets with $SO(5)$ quantum numbers $[j,6]$ and $SU(2)_R$ spin 0}
%\label{tab6d60}
%\end{table}

%%%% j,0,2

%\begin{table}
\begin{tiny}
\begin{center}
%\begin{sidewaysfigure}
\begin{tabular}{|c|| c|c|c|c|c| c|c|c|c|c| c|c|}
%\begin{longtable}{|c|| c|c|c|c|c| c|c|c|c|c| c|c|c|c|c| c|c|c|c|c| c|c|c|c|c| c|} % \small
 \hline
\begin{turn}{-90}$ \ \ap m^2 \ $  \end{turn}
&\begin{turn}{-90} $\llbracket0,0;2\rrbracket$ \end{turn}
&\begin{turn}{-90} $\llbracket1,0;2\rrbracket$ \end{turn}
&\begin{turn}{-90} $\llbracket2,0;2\rrbracket$ \end{turn}
&\begin{turn}{-90} $\llbracket3,0;2\rrbracket$ \end{turn}
&\begin{turn}{-90} $\llbracket4,0;2\rrbracket$ \end{turn}
&\begin{turn}{-90} $\llbracket5,0;2\rrbracket$ \end{turn}
&\begin{turn}{-90} $\llbracket6,0;2\rrbracket$ \end{turn}
&\begin{turn}{-90} $\llbracket7,0;2\rrbracket$ \end{turn}
&\begin{turn}{-90} $\llbracket8,0;2\rrbracket$ \end{turn}
&\begin{turn}{-90} $\llbracket9,0;2\rrbracket$ \end{turn}
&\begin{turn}{-90} $\llbracket10,0;2\rrbracket\ $ \end{turn}
&\begin{turn}{-90} $\llbracket11,0;2\rrbracket$ \end{turn}
%&\begin{turn}{-90}\# $\llbracket12,0;2\rrbracket$ \end{turn}
%&\begin{turn}{-90}\# $\llbracket13,0;2\rrbracket$ \end{turn}
%
\\\hline \hline
2 &\textcolor{red}{0} & & & &
& & & & &
& &   \\\hline
3 &1 &\textcolor{red}{0} & & &
& & & & &
& &  \\\hline
4 &0 &\textcolor{blue}{2} &\textcolor{red}{0} &
& & & & & &
& & \\\hline
5 &3 &3 &\textcolor{red}{3} &\textcolor{red}{0} &
& & & & &
& &   \\\hline
6 &4 &9 &\textcolor{blue}{4} &\textcolor{red}{3} &\textcolor{red}{0}
& & & & &
& &   \\\hline
7 &13 &20 &\textcolor{blue}{17} &\textcolor{red}{5} &\textcolor{red}{3}
&\textcolor{red}{0} & & & &
& &  \\\hline
8 &20 &50 &34 &\textcolor{blue}{19} &\textcolor{red}{5}
&\textcolor{red}{3} &\textcolor{red}{0} & & &
& &  \\
\hline
\end{tabular}
%\end{longtable}
%\end{sidewaysfigure}
\end{center}
\end{tiny}
%\caption{${\cal N}_{6d}=(1,0)$ multiplets with $SO(5)$ quantum numbers $[j,0]$ and $SU(2)_R$ spin 1}
%\label{tab6d02}
%\end{table}

%%%% j,2,2

\vspace{0.8cm}

%\begin{table}
\begin{tiny}
\begin{center}
%\begin{sidewaysfigure}
\begin{tabular}{|c|| c|c|c|c|c| c|c|c|c|c| c|c|}
%\begin{longtable}{|c|| c|c|c|c|c| c|c|c|c|c| c|c|c|c|c| c|c|c|c|c| c|c|c|c|c| c|} % \small
 \hline
\begin{turn}{-90}$\ \ap m^2 \ $  \end{turn}
&\begin{turn}{-90} $\llbracket0,2;2\rrbracket$ \end{turn}
&\begin{turn}{-90} $\llbracket1,2;2\rrbracket$ \end{turn}
&\begin{turn}{-90} $\llbracket2,2;2\rrbracket$ \end{turn}
&\begin{turn}{-90} $\llbracket3,2;2\rrbracket$ \end{turn}
&\begin{turn}{-90} $\llbracket4,2;2\rrbracket$ \end{turn}
&\begin{turn}{-90} $\llbracket5,2;2\rrbracket$ \end{turn}
&\begin{turn}{-90} $\llbracket6,2;2\rrbracket$ \end{turn}
&\begin{turn}{-90} $\llbracket7,2;2\rrbracket$ \end{turn}
&\begin{turn}{-90} $\llbracket8,2;2\rrbracket$ \end{turn}
&\begin{turn}{-90} $\llbracket9,2;2\rrbracket$ \end{turn}
&\begin{turn}{-90} $\llbracket10,2;2\rrbracket\ $ \end{turn}
&\begin{turn}{-90} $\llbracket11,2;2\rrbracket$ \end{turn}
%&\begin{turn}{-90}\# $\llbracket12,2;2\rrbracket$ \end{turn}
%&\begin{turn}{-90}\# $\llbracket13,2;2\rrbracket$ \end{turn}
%
\\\hline \hline
3 &\textcolor{red}{0} & & & &
& & & & &
& & \\\hline
4 &\textcolor{red}{1} &\textcolor{red}{0} & & &
& & & & & & &  \\\hline
5 &\textcolor{blue}{3} &\textcolor{red}{1} &\textcolor{red}{0} & &
& & & & &
& &  \\\hline
6 &9 &\textcolor{red}{6} &\textcolor{red}{1} &\textcolor{red}{0} &
& & & & &
& &  \\\hline
7 &22 &\textcolor{blue}{16} &\textcolor{red}{6} &\textcolor{red}{1} &\textcolor{red}{0}
& & & & &
 & &  \\\hline
8 &54 &\textcolor{blue}{47} &\textcolor{red}{19} &\textcolor{red}{6} &\textcolor{red}{1}
&\textcolor{red}{0} & & & &
 & &  \\\hline
9 &122 &114 &\textcolor{blue}{57} &\textcolor{red}{19} &\textcolor{red}{6}
&\textcolor{red}{1} &\textcolor{red}{0} & & &
& & \\
\hline
\end{tabular}
%\end{longtable}
%\end{sidewaysfigure}
\end{center}
\end{tiny}
%\caption{${\cal N}_{6d}=(1,0)$ multiplets with $SO(5)$ quantum numbers $[j,2]$ and $SU(2)_R$ spin 1}
%\label{tab6d22}
%\end{table}
%
%%%% j,4,2

\vspace{0.8cm}

%\begin{table}
\begin{tiny}
\begin{center}
%\begin{sidewaysfigure}
\begin{tabular}{|c|| c|c|c|c|c| c|c|c|c|c| c|c|c|}
%\begin{longtable}{|c|| c|c|c|c|c| c|c|c|c|c| c|c|c|c|c| c|c|c|c|c| c|c|c|c|c| c|} % \small
 \hline
\begin{turn}{-90}$\ \ap m^2 \ $  \end{turn}
&\begin{turn}{-90} $\llbracket0,4;2\rrbracket$ \end{turn}
&\begin{turn}{-90} $\llbracket1,4;2\rrbracket$ \end{turn}
&\begin{turn}{-90} $\llbracket2,4;2\rrbracket$ \end{turn}
&\begin{turn}{-90} $\llbracket3,4;2\rrbracket$ \end{turn}
&\begin{turn}{-90} $\llbracket4,4;2\rrbracket$ \end{turn}
&\begin{turn}{-90} $\llbracket5,4;2\rrbracket$ \end{turn}
&\begin{turn}{-90} $\llbracket6,4;2\rrbracket$ \end{turn}
&\begin{turn}{-90} $\llbracket7,4;2\rrbracket$ \end{turn}
&\begin{turn}{-90} $\llbracket8,4;2\rrbracket$ \end{turn}
&\begin{turn}{-90} $\llbracket9,4;2\rrbracket$ \end{turn}
&\begin{turn}{-90} $\llbracket10,4;2\rrbracket\ $ \end{turn}
&\begin{turn}{-90} $\llbracket11,4;2\rrbracket$ \end{turn}
&\begin{turn}{-90} $\llbracket12,4;2\rrbracket$ \end{turn}
%&\begin{turn}{-90}\# $\llbracket13,4;2\rrbracket$ \end{turn}
%
\\\hline \hline
6 &\textcolor{red}{0} & & & &
& & & & &
& & &  \\\hline
7 &\textcolor{red}{3} &\textcolor{red}{0} & & &
& & & & &
& & &  \\\hline
8 &\textcolor{blue}{9} &\textcolor{red}{3} &\textcolor{red}{0} & &
& & & & &
& & &  \\\hline
9 &\textcolor{blue}{33} &\textcolor{red}{12} &\textcolor{red}{3} &\textcolor{red}{0} &
& & & & &
& & &  \\\hline
10 &81 &\textcolor{blue}{45} &\textcolor{red}{12} &\textcolor{red}{3} &\textcolor{red}{0}
& & & & &
& & &  \\\hline
11 &218 &\textcolor{blue}{126} &\textcolor{red}{48} &\textcolor{red}{12} &\textcolor{red}{3}
&\textcolor{red}{0} & & & &
& & &  \\\hline
12 &504 &\textcolor{blue}{345} &\textcolor{blue}{138} &\textcolor{red}{48} &\textcolor{red}{12}
&\textcolor{red}{3} &\textcolor{red}{0} & & &
& & &  \\
\hline
\end{tabular}
%\end{longtable}
%\end{sidewaysfigure}
\end{center}
\end{tiny}
%\caption{${\cal N}_{6d}=(1,0)$ multiplets with $SO(5)$ quantum numbers $[j,4]$ and $SU(2)_R$ spin 1}
%\label{tab6d42}
%\end{table}
%
%%%% j,0,4

%\begin{table}
\begin{scriptsize}
\begin{center}
%\begin{sidewaysfigure}
\begin{tabular}{|c|| c|c|c|c|c| c|c|c|c|c| c|c|c|c|}
%\begin{longtable}{|c|| c|c|c|c|c| c|c|c|c|c| c|c|c|c|c| c|c|c|c|c| c|c|c|c|c| c|} % \small
 \hline
\begin{turn}{-90}$\ \ap m^2 \ $  \end{turn}
&\begin{turn}{-90} $\llbracket0,0;4\rrbracket$ \end{turn}
&\begin{turn}{-90} $\llbracket1,0;4\rrbracket$ \end{turn}
&\begin{turn}{-90} $\llbracket2,0;4\rrbracket$ \end{turn}
&\begin{turn}{-90} $\llbracket3,0;4\rrbracket$ \end{turn}
&\begin{turn}{-90} $\llbracket4,0;4\rrbracket$ \end{turn}
&\begin{turn}{-90} $\llbracket5,0;4\rrbracket$ \end{turn}
&\begin{turn}{-90} $\llbracket6,0;4\rrbracket$ \end{turn}
&\begin{turn}{-90} $\llbracket7,0;4\rrbracket$ \end{turn}
&\begin{turn}{-90} $\llbracket8,0;4\rrbracket$ \end{turn}
&\begin{turn}{-90} $\llbracket9,0;4\rrbracket$ \end{turn}
&\begin{turn}{-90} $\llbracket10,0;4\rrbracket\ $ \end{turn}
&\begin{turn}{-90} $\llbracket11,0;4\rrbracket$ \end{turn}
&\begin{turn}{-90} $\llbracket12,0;4\rrbracket$ \end{turn}
&\begin{turn}{-90} $\llbracket13,0;4\rrbracket$ \end{turn}
\\\hline \hline
6 &0 &\textcolor{red}{0} & & &
& & & & &
& & & &  \\\hline
7 &1 &\textcolor{red}{1} &\textcolor{red}{0} & &
& & & & &
& & & &  \\\hline
8 &2 &2 &\textcolor{red}{1} &\textcolor{red}{0} &
& & & & &
& & & &  \\\hline
9 &5 &10 &\textcolor{red}{4} &\textcolor{red}{1} &\textcolor{red}{0}
& & & & &
& & & &  \\\hline
10 &12 &20 &\textcolor{blue}{15} &\textcolor{red}{4} &\textcolor{red}{1} &\textcolor{red}{0} & & & &
& & & &  \\\hline
11 &30 &58 &38 &\textcolor{red}{18} &\textcolor{red}{4}
&\textcolor{red}{1} &\textcolor{red}{0} & & &
& & & &  \\\hline
12 &61 &125 &104 &\textcolor{blue}{44} &\textcolor{red}{18}
&\textcolor{red}{4} &\textcolor{red}{1} &\textcolor{red}{0} & &
& & & &  \\
\hline
\end{tabular}
%\end{longtable}
%\end{sidewaysfigure}
\end{center}
\end{scriptsize}
%\caption{${\cal N}_{6d}=(1,0)$ multiplets with $SO(5)$ quantum numbers $[j,0]$ and $SU(2)_R$ spin 2}
%\label{tab6d04}
%\end{table}
%
%%%% j,2,4

\vspace{-0.3cm}

%\begin{table}
\begin{scriptsize}
\begin{center}
%\begin{sidewaysfigure}
\begin{tabular}{|c|| c|c|c|c|c| c|c|c|c|c| c|c|c|}
%\begin{longtable}{|c|| c|c|c|c|c| c|c|c|c|c| c|c|c|c|c| c|c|c|c|c| c|c|c|c|c| c|} % \small
 \hline
\begin{turn}{-90}$\ \ap m^2 \ $  \end{turn}
&\begin{turn}{-90} $\llbracket0,2;4\rrbracket$ \end{turn}
&\begin{turn}{-90} $\llbracket1,2;4\rrbracket$ \end{turn}
&\begin{turn}{-90} $\llbracket2,2;4\rrbracket$ \end{turn}
&\begin{turn}{-90} $\llbracket3,2;4\rrbracket$ \end{turn}
&\begin{turn}{-90} $\llbracket4,2;4\rrbracket$ \end{turn}
&\begin{turn}{-90} $\llbracket5,2;4\rrbracket$ \end{turn}
&\begin{turn}{-90} $\llbracket6,2;4\rrbracket$ \end{turn}
&\begin{turn}{-90} $\llbracket7,2;4\rrbracket$ \end{turn}
&\begin{turn}{-90} $\llbracket8,2;4\rrbracket$ \end{turn}
&\begin{turn}{-90} $\llbracket9,2;4\rrbracket$ \end{turn}
&\begin{turn}{-90} $\llbracket10,2;4\rrbracket\ $ \end{turn}
&\begin{turn}{-90} $\llbracket11,2;4\rrbracket$ \end{turn}
&\begin{turn}{-90} $\llbracket12,2;4\rrbracket$ \end{turn}
%&\begin{turn}{-90}\# $\llbracket13,2;4\rrbracket$ \end{turn}
%
\\\hline \hline
7 &\textcolor{red}{0} & & & &
& & & & &
& & &  \\\hline
8 &\textcolor{blue}{3} &\textcolor{red}{0} & & &
& & & & &
& & &  \\\hline
9 &6 &\textcolor{red}{4} &\textcolor{red}{0} & &
& & & & &
& & &  \\\hline
10 &25 &\textcolor{blue}{13} &\textcolor{red}{4} &\textcolor{red}{0} &
& & & & &
& & &  \\\hline
11 &57 &\textcolor{blue}{47} &\textcolor{red}{14} &\textcolor{red}{4} &\textcolor{red}{0}
& & & & &
& & &  \\\hline
12 &152 &128 &\textcolor{blue}{57} &\textcolor{red}{14} &\textcolor{red}{4}
&\textcolor{red}{0} & & & &
& & &  \\\hline
13 &338 &338 &\textcolor{blue}{159} &\textcolor{red}{58} &\textcolor{red}{14}
&\textcolor{red}{4} &\textcolor{red}{0} & & &
& & &  \\
\hline
\end{tabular}
%\end{longtable}
%\end{sidewaysfigure}
\end{center}
\end{scriptsize}
%\caption{${\cal N}_{6d}=(1,0)$ multiplets with $SO(5)$ quantum numbers $[j,2]$ and $SU(2)_R$ spin 2}
%\label{tab6d24}
%\end{table}
%%
%%%% j,0,6
%

\vspace{-0.3cm}

%\begin{table}
\begin{scriptsize}
\begin{center}
%\begin{sidewaysfigure}
\begin{tabular}{|c|| c|c|c|c|c| c|c|c|c|c| c|c|c|c|}
%\begin{longtable}{|c|| c|c|c|c|c| c|c|c|c|c| c|c|c|c|c| c|c|c|c|c| c|c|c|c|c| c|} % \small
 \hline
\begin{turn}{-90}$\ \ap m^2 \ $  \end{turn}
&\begin{turn}{-90} $\llbracket0,0;6\rrbracket$ \end{turn}
&\begin{turn}{-90} $\llbracket1,0;6\rrbracket$ \end{turn}
&\begin{turn}{-90} $\llbracket2,0;6\rrbracket$ \end{turn}
&\begin{turn}{-90} $\llbracket3,0;6\rrbracket$ \end{turn}
&\begin{turn}{-90} $\llbracket4,0;6\rrbracket$ \end{turn}
&\begin{turn}{-90} $\llbracket5,0;6\rrbracket$ \end{turn}
&\begin{turn}{-90} $\llbracket6,0;6\rrbracket$ \end{turn}
&\begin{turn}{-90} $\llbracket7,0;6\rrbracket$ \end{turn}
&\begin{turn}{-90} $\llbracket8,0;6\rrbracket$ \end{turn}
&\begin{turn}{-90} $\llbracket9,0;6\rrbracket$ \end{turn}
&\begin{turn}{-90} $\llbracket10,0;6\rrbracket\ $ \end{turn}
&\begin{turn}{-90} $\llbracket11,0;6\rrbracket$ \end{turn}
&\begin{turn}{-90} $\llbracket12,0;6\rrbracket$ \end{turn}
&\begin{turn}{-90} $\llbracket13,0;6\rrbracket$ \end{turn}
\\\hline \hline
10 &\textcolor{red}{0} & & & &
& & & & &
& & & &  \\\hline
11 &1 &\textcolor{red}{0} & & &
& & & & &
& & & &  \\\hline
12 &0 &\textcolor{blue}{2} &\textcolor{red}{0} & &
& & & & &
& & & &  \\\hline
13 &5 &5 &\textcolor{red}{3} &\textcolor{red}{0} &
& & & & &
& & & &  \\\hline
14 &8 &16 &\textcolor{blue}{7} &\textcolor{red}{3} &\textcolor{red}{0}
& & & & &
& & & &  \\\hline
15 &27 &42 &\textcolor{blue}{30} &\textcolor{red}{8} &\textcolor{red}{3}
&\textcolor{red}{0} & & & &
& & & &  \\\hline
16 &50 &110 &74 &\textcolor{blue}{34} &\textcolor{red}{8}
&\textcolor{red}{3} &\textcolor{red}{0} & & &
& & & &  \\
\hline
\end{tabular}
%\end{longtable}
%\end{sidewaysfigure}
\end{center}
\end{scriptsize}
%\caption{${\cal N}_{6d}=(1,0)$ multiplets with $SO(5)$ quantum numbers $[j,0]$ and $SU(2)_R$ spin 3}
%\label{tab6d06}
%\end{table}

%%%% j,1,1
%
%\begin{table}
\begin{tiny}
\begin{center}
%\begin{sidewaysfigure}
\begin{tabular}{|c|| c|c|c|c|c| c|c|c|c|c| c|c|}
%\begin{longtable}{|c|| c|c|c|c|c| c|c|c|c|c| c|c|c|c|c| c|c|c|c|c| c|c|c|c|c| c|} % \small
 \hline
\begin{turn}{-90}$\ \ap m^2 \ $  \end{turn}
&\begin{turn}{-90} $\llbracket0,1;1\rrbracket$ \end{turn}
&\begin{turn}{-90} $\llbracket1,1;1\rrbracket$ \end{turn}
&\begin{turn}{-90} $\llbracket2,1;1\rrbracket$ \end{turn}
&\begin{turn}{-90} $\llbracket3,1;1\rrbracket$ \end{turn}
&\begin{turn}{-90} $\llbracket4,1;1\rrbracket$ \end{turn}
&\begin{turn}{-90} $\llbracket5,1;1\rrbracket$ \end{turn}
&\begin{turn}{-90} $\llbracket6,1;1\rrbracket$ \end{turn}
&\begin{turn}{-90} $\llbracket7,1;1\rrbracket$ \end{turn}
&\begin{turn}{-90} $\llbracket8,1;1\rrbracket$ \end{turn}
&\begin{turn}{-90} $\llbracket9,1;1\rrbracket$ \end{turn}
&\begin{turn}{-90} $\llbracket10,1;1\rrbracket\ $ \end{turn}
&\begin{turn}{-90} $\llbracket11,1;1\rrbracket$ \end{turn}
%&\begin{turn}{-90}\# $\llbracket12,1;1\rrbracket$ \end{turn}
\\\hline \hline
1 &\textcolor{red}{0} & & & &
& & & & &
& &    \\\hline
2 &\textcolor{blue}{1} &\textcolor{red}{0} & & &
& & & & &
& &  \\\hline
3 &1 &\textcolor{red}{2} &\textcolor{red}{0} & &
& & & & &
& &   \\\hline
4 &4 &\textcolor{blue}{3} &\textcolor{red}{2} &\textcolor{red}{0} &
& & & & &
& &    \\\hline
5 &8 &\textcolor{blue}{9} &\textcolor{red}{4} &\textcolor{red}{2} &\textcolor{red}{0}
& & & & &
& &   \\\hline
6 &18 &23 &\textcolor{blue}{12} &\textcolor{red}{4} &\textcolor{red}{2}
&\textcolor{red}{0} & & & &
& &   \\\hline
7 &39 &51 &\textcolor{blue}{31} &\textcolor{red}{13} &\textcolor{red}{4}
&\textcolor{red}{2} &\textcolor{red}{0} & & &
& &  \\
\hline
\end{tabular}
%\end{longtable}
%\end{sidewaysfigure}
\end{center}
\end{tiny}
%\caption{${\cal N}_{6d}=(1,0)$ multiplets with $SO(5)$ quantum numbers $[j,1]$ and $SU(2)_R$ spin 1/2}
%\label{tab6d11}
%\end{table}

\vspace{0.7cm}

%%%% j,3,1
%
%\begin{table}
\begin{tiny}
\begin{center}
%\begin{sidewaysfigure}
\begin{tabular}{|c|| c|c|c|c|c| c|c|c|c|c| c|c|}
%\begin{longtable}{|c|| c|c|c|c|c| c|c|c|c|c| c|c|c|c|c| c|c|c|c|c| c|c|c|c|c| c|} % \small
 \hline
\begin{turn}{-90}$\ \ap m^2 \ $  \end{turn}
&\begin{turn}{-90} $\llbracket0,3;1\rrbracket$ \end{turn}
&\begin{turn}{-90} $\llbracket1,3;1\rrbracket$ \end{turn}
&\begin{turn}{-90} $\llbracket2,3;1\rrbracket$ \end{turn}
&\begin{turn}{-90} $\llbracket3,3;1\rrbracket$ \end{turn}
&\begin{turn}{-90} $\llbracket4,3;1\rrbracket$ \end{turn}
&\begin{turn}{-90} $\llbracket5,3;1\rrbracket$ \end{turn}
&\begin{turn}{-90} $\llbracket6,3;1\rrbracket$ \end{turn}
&\begin{turn}{-90} $\llbracket7,3;1\rrbracket$ \end{turn}
&\begin{turn}{-90} $\llbracket8,3;1\rrbracket$ \end{turn}
&\begin{turn}{-90} $\llbracket9,3;1\rrbracket$ \end{turn}
&\begin{turn}{-90} $\llbracket10,3;1\rrbracket\ $ \end{turn}
&\begin{turn}{-90} $\llbracket11,3;1\rrbracket$ \end{turn}
%&\begin{turn}{-90}\# $\llbracket12,3;1\rrbracket$ \end{turn}
%&\begin{turn}{-90}\# $\llbracket13,3;1\rrbracket$ \end{turn}
%
\\\hline \hline
3 &\textcolor{red}{0} & & & &
& & & & &
& &\\\hline
4 &\textcolor{red}{1} &\textcolor{red}{0} & & &
& & & & &
& &\\\hline
5 &\textcolor{blue}{4} &\textcolor{red}{1} &\textcolor{red}{0} & &
& & & & &
& & \\\hline
6 &\textcolor{blue}{9} &\textcolor{red}{5} &\textcolor{red}{1} &\textcolor{red}{0} &
& & & & &
& & \\\hline
7 &26 &\textcolor{blue}{15} &\textcolor{red}{5} &\textcolor{red}{1} &\textcolor{red}{0}
& & & & &
& & \\\hline
8 &61 &\textcolor{blue}{42} &\textcolor{red}{16} &\textcolor{red}{5} &\textcolor{red}{1}
&\textcolor{red}{0} & & & &
& &  \\\hline
9 &140 &\textcolor{blue}{109} &\textcolor{blue}{48} &\textcolor{red}{16} &\textcolor{red}{5}
&\textcolor{red}{1} &\textcolor{red}{0} & & &
& &  \\
\hline
\end{tabular}
%\end{longtable}
%\end{sidewaysfigure}
\end{center}
\end{tiny}
%\caption{${\cal N}_{6d}=(1,0)$ multiplets with $SO(5)$ quantum numbers $[j,3]$ and $SU(2)_R$ spin 1/2}
%\label{tab6d31}
%\end{table}

\vspace{0.7cm}

%%%% j,5,1
%
%\begin{table}
\begin{tiny}
\begin{center}
%\begin{sidewaysfigure}
\begin{tabular}{|c|| c|c|c|c|c| c|c|c|c|c| c|c|c|}
%\begin{longtable}{|c|| c|c|c|c|c| c|c|c|c|c| c|c|c|c|c| c|c|c|c|c| c|c|c|c|c| c|} % \small
 \hline
\begin{turn}{-90}$\ \ap m^2 \ $  \end{turn}
&\begin{turn}{-90} $\llbracket0,5;1\rrbracket$ \end{turn}
&\begin{turn}{-90} $\llbracket1,5;1\rrbracket$ \end{turn}
&\begin{turn}{-90} $\llbracket2,5;1\rrbracket$ \end{turn}
&\begin{turn}{-90} $\llbracket3,5;1\rrbracket$ \end{turn}
&\begin{turn}{-90} $\llbracket4,5;1\rrbracket$ \end{turn}
&\begin{turn}{-90} $\llbracket5,5;1\rrbracket$ \end{turn}
&\begin{turn}{-90} $\llbracket6,5;1\rrbracket$ \end{turn}
&\begin{turn}{-90} $\llbracket7,5;1\rrbracket$ \end{turn}
&\begin{turn}{-90} $\llbracket8,5;1\rrbracket$ \end{turn}
&\begin{turn}{-90} $\llbracket9,5;1\rrbracket$ \end{turn}
&\begin{turn}{-90} $\llbracket10,5;1\rrbracket\ $ \end{turn}
&\begin{turn}{-90} $\llbracket11,5;1\rrbracket$ \end{turn}
&\begin{turn}{-90} $\llbracket12,5;1\rrbracket$ \end{turn}
%&\begin{turn}{-90} $\llbracket13,5;1\rrbracket$ \end{turn}
%
\\\hline \hline
6 &\textcolor{red}{0} & & & &
& & & & &
& & &  \\\hline
7 &\textcolor{red}{1} &\textcolor{red}{0} & & &
& & & & &
& & &  \\\hline
8 &\textcolor{blue}{6} &\textcolor{red}{1} &\textcolor{red}{0} & &
& & & & &
& & &  \\\hline
9 &\textcolor{blue}{17} &\textcolor{red}{7} &\textcolor{red}{1} &\textcolor{red}{0} &
& & & & &
& & &  \\\hline
10 &\textcolor{blue}{54} &\textcolor{blue}{23} &\textcolor{red}{7} &\textcolor{red}{1} &\textcolor{red}{0} & & & &
& & & &  \\\hline
11 &138 &\textcolor{blue}{73} &\textcolor{red}{24} &\textcolor{red}{7} &\textcolor{red}{1}
&\textcolor{red}{0} & & & &
& & & \\\hline
12 &341 &\textcolor{blue}{202} &\textcolor{blue}{79} &\textcolor{red}{24} &\textcolor{red}{7}
&\textcolor{red}{1} &\textcolor{red}{0} & & &
& & &  \\
\hline
\end{tabular}
%\end{longtable}
%\end{sidewaysfigure}
\end{center}
\end{tiny}
%\caption{${\cal N}_{6d}=(1,0)$ multiplets with $SO(5)$ quantum numbers $[j,5]$ and $SU(2)_R$ spin 1/2}
%\label{tab6d51}
%\end{table}
%

%%%% j,1,3
%
%\begin{table}
\begin{tiny}
\begin{center}
%\begin{sidewaysfigure}
\begin{tabular}{|c|| c|c|c|c|c| c|c|c|c|c| c|c|}
%\begin{longtable}{|c|| c|c|c|c|c| c|c|c|c|c| c|c|c|c|c| c|c|c|c|c| c|c|c|c|c| c|} % \small
 \hline
\begin{turn}{-90}$\ \ap m^2 \ $  \end{turn}
&\begin{turn}{-90} $\llbracket0,1;3\rrbracket$ \end{turn}
&\begin{turn}{-90} $\llbracket1,1;3\rrbracket$ \end{turn}
&\begin{turn}{-90} $\llbracket2,1;3\rrbracket$ \end{turn}
&\begin{turn}{-90} $\llbracket3,1;3\rrbracket$ \end{turn}
&\begin{turn}{-90} $\llbracket4,1;3\rrbracket$ \end{turn}
&\begin{turn}{-90} $\llbracket5,1;3\rrbracket$ \end{turn}
&\begin{turn}{-90} $\llbracket6,1;3\rrbracket$ \end{turn}
&\begin{turn}{-90} $\llbracket7,1;3\rrbracket$ \end{turn}
&\begin{turn}{-90} $\llbracket8,1;3\rrbracket$ \end{turn}
&\begin{turn}{-90} $\llbracket9,1;3\rrbracket$ \end{turn}
&\begin{turn}{-90} $\llbracket10,1;3\rrbracket\ $ \end{turn}
&\begin{turn}{-90} $\llbracket11,1;3\rrbracket$ \end{turn}
%&\begin{turn}{-90}\# $\llbracket12,1;3\rrbracket$ \end{turn}
%&\begin{turn}{-90}\# $\llbracket13,1;3\rrbracket$ \end{turn}
%
\\\hline \hline
4 &\textcolor{red}{0} & & & &
& & & & &
& &  \\\hline
5 &1 &\textcolor{red}{0} & & &
& & & & &
& &  \\\hline
6 &3 &\textcolor{red}{2} &\textcolor{red}{0} & &
& & & & &
& &   \\\hline
7 &7 &\textcolor{blue}{7} &\textcolor{red}{2} &\textcolor{red}{0} &
& & & & &
& &  \\\hline
8 &19 &20 &\textcolor{red}{9} &\textcolor{red}{2} &\textcolor{red}{0}
& & & & &
& & \\\hline
9 &44 &53 &\textcolor{blue}{27} &\textcolor{red}{9} &\textcolor{red}{2}
&\textcolor{red}{0} & & & &
& & \\\hline
10 &100 &130 &\textcolor{blue}{76} &\textcolor{red}{29} &\textcolor{red}{9}
&\textcolor{red}{2} &\textcolor{red}{0} & & &
& & \\
\hline
\end{tabular}
%\end{longtable}
%\end{sidewaysfigure}
\end{center}
\end{tiny}
%\caption{${\cal N}_{6d}=(1,0)$ multiplets with $SO(5)$ quantum numbers $[j,1]$ and $SU(2)_R$ spin 3/2}
%\label{tab6d13}
%\end{table}

%%%% j,3,3

%\begin{table}
\begin{tiny}
\begin{center}
%\begin{sidewaysfigure}
\begin{tabular}{|c|| c|c|c|c|c| c|c|c|c|c| c|c|}
%\begin{longtable}{|c|| c|c|c|c|c| c|c|c|c|c| c|c|c|c|c| c|c|c|c|c| c|c|c|c|c| c|} % \small
 \hline
\begin{turn}{-90}$ \ \ap m^2 \ $  \end{turn}
&\begin{turn}{-90} $\llbracket0,3;3\rrbracket$ \end{turn}
&\begin{turn}{-90} $\llbracket1,3;3\rrbracket$ \end{turn}
&\begin{turn}{-90} $\llbracket2,3;3\rrbracket$ \end{turn}
&\begin{turn}{-90} $\llbracket3,3;3\rrbracket$ \end{turn}
&\begin{turn}{-90} $\llbracket4,3;3\rrbracket$ \end{turn}
&\begin{turn}{-90} $\llbracket5,3;3\rrbracket$ \end{turn}
&\begin{turn}{-90} $\llbracket6,3;3\rrbracket$ \end{turn}
&\begin{turn}{-90} $\llbracket7,3;3\rrbracket$ \end{turn}
&\begin{turn}{-90} $\llbracket8,3;3\rrbracket$ \end{turn}
&\begin{turn}{-90} $\llbracket9,3;3\rrbracket$ \end{turn}
&\begin{turn}{-90} $\llbracket10,3;3\rrbracket\ $ \end{turn}
&\begin{turn}{-90} $\llbracket11,3;3\rrbracket$ \end{turn}
%&\begin{turn}{-90}\# $\llbracket12,3;3\rrbracket$ \end{turn}
%&\begin{turn}{-90}\# $\llbracket13,3;3\rrbracket$ \end{turn}
%
\\\hline \hline
6 &\textcolor{red}{0} & & & &
& & & & & & &  \\\hline
7 &\textcolor{red}{2} &\textcolor{red}{0} & & &
& & & & &
 & &  \\\hline
8 &\textcolor{blue}{7} &\textcolor{red}{2} &\textcolor{red}{0} & &
& & & & &
 & &  \\\hline
9 &24 &\textcolor{red}{10} &\textcolor{red}{2} &\textcolor{red}{0} &
 & & & & & & &  \\\hline
10 &63 &\textcolor{blue}{38} &\textcolor{red}{10} &\textcolor{red}{2} &\textcolor{red}{0}
& & & & &
& &  \\\hline
11 &163 &\textcolor{blue}{109} &\textcolor{red}{41} &\textcolor{red}{10} &\textcolor{red}{2}
&\textcolor{red}{0} & & & &
& &  \\\hline
12 &385 &295 &\textcolor{blue}{124} &\textcolor{red}{41} &\textcolor{red}{10}
&\textcolor{red}{2} &\textcolor{red}{0} & & &
& &   \\
\hline
\end{tabular}
%\end{longtable}
%\end{sidewaysfigure}
\end{center}
\end{tiny}
%\caption{${\cal N}_{6d}=(1,0)$ multiplets with $SO(5)$ quantum numbers $[j,3]$ and $SU(2)_R$ spin 3/2}
%\label{tab6d33}
%\end{table}

%%%% j,1,5
%
%\begin{table}
\begin{tiny}
\begin{center}
%\begin{sidewaysfigure}
\begin{tabular}{|c|| c|c|c|c|c| c|c|c|c|c| c|c|c|c|}
%\begin{longtable}{|c|| c|c|c|c|c| c|c|c|c|c| c|c|c|c|c| c|c|c|c|c| c|c|c|c|c| c|} % \small
 \hline
\begin{turn}{-90}$\ \ap m^2 \ $  \end{turn}
&\begin{turn}{-90} $\llbracket0,1;5\rrbracket$ \end{turn}
&\begin{turn}{-90} $\llbracket1,1;5\rrbracket$ \end{turn}
&\begin{turn}{-90} $\llbracket2,1;5\rrbracket$ \end{turn}
&\begin{turn}{-90} $\llbracket3,1;5\rrbracket$ \end{turn}
&\begin{turn}{-90} $\llbracket4,1;5\rrbracket$ \end{turn}
&\begin{turn}{-90} $\llbracket5,1;5\rrbracket$ \end{turn}
&\begin{turn}{-90} $\llbracket6,1;5\rrbracket$ \end{turn}
&\begin{turn}{-90} $\llbracket7,1;5\rrbracket$ \end{turn}
&\begin{turn}{-90} $\llbracket8,1;5\rrbracket$ \end{turn}
&\begin{turn}{-90} $\llbracket9,1;5\rrbracket$ \end{turn}
&\begin{turn}{-90} $\llbracket10,1;5\rrbracket\ $ \end{turn}
&\begin{turn}{-90} $\llbracket11,1;5\rrbracket$ \end{turn}
&\begin{turn}{-90} $\llbracket12,1;5\rrbracket$ \end{turn}
&\begin{turn}{-90} $\llbracket13,1;5\rrbracket$ \end{turn}
\\\hline \hline
8 &\textcolor{red}{0} & & & &
& & & & &
& & & &  \\\hline
9 &1 &\textcolor{red}{0} & & &
& & & & &
& & & &  \\\hline
10 &3 &\textcolor{red}{2} &\textcolor{red}{0} & &
& & & & &
& & & &  \\\hline
11 &9 &\textcolor{blue}{8} &\textcolor{red}{2} &\textcolor{red}{0} &
& & & & &
& & & &  \\\hline
12 &26 &25 &\textcolor{red}{10} &\textcolor{red}{2} &\textcolor{red}{0}
& & & & &
& & & &  \\\hline
13 &62 &73 &\textcolor{blue}{34} &\textcolor{red}{10} &\textcolor{red}{2}
&\textcolor{red}{0} & & & &
& & & &  \\\hline
14 &148 &188 &\textcolor{blue}{105} &\textcolor{red}{36} &\textcolor{red}{10}
&\textcolor{red}{2} &\textcolor{red}{0} & & &
& & & &  \\
\hline
\end{tabular}
%\end{longtable}
%\end{sidewaysfigure}
\end{center}
\end{tiny}
%\caption{${\cal N}_{6d}=(1,0)$ multiplets with $SO(5)$ quantum numbers $[j,1]$ and $SU(2)_R$ spin 5/2}
%\label{tab6d15}
%\end{table}

%%%
%%% 
%%%
%%%

\subsection{16 supercharges in ten dimensions}
\label{app10}

The tables in this subsection are based on the ${\cal N}_{10d}=1$ partition function (\ref{NSandR10d}), organized in terms of multiplicity generating functions $G_{n_1,n_2,n_3,n_4}(q)$, see (\ref{GGG10}).

%%% Now [j,1,0,0]

%\begin{table}
\begin{tiny}
\begin{center}
%\begin{sidewaysfigure}
\begin{tabular}{|c|| c|c|c|c|c| c|c|c|c|c| c|c|c|c|c|}
%\begin{longtable}{|c|| c|c|c|c|c| c|c|c|c|c| c|c|c|c|c| c|c|c|c|c| c|c|c|c|c| c|} % \small
 \hline
\begin{turn}{-90}$ \ \ap m^2 \ $  \end{turn}
&\begin{turn}{-90} $\llbracket0,1,0,0\rrbracket$ \end{turn}
&\begin{turn}{-90} $\llbracket1,1,0,0\rrbracket$ \end{turn}
&\begin{turn}{-90} $\llbracket2,1,0,0\rrbracket$ \end{turn}
&\begin{turn}{-90} $\llbracket3,1,0,0\rrbracket$ \end{turn}
&\begin{turn}{-90} $\llbracket4,1,0,0\rrbracket$ \end{turn}
&\begin{turn}{-90} $\llbracket5,1,0,0\rrbracket$ \end{turn}
&\begin{turn}{-90} $\llbracket6,1,0,0\rrbracket$ \end{turn}
&\begin{turn}{-90} $\llbracket7,1,0,0\rrbracket$ \end{turn}
&\begin{turn}{-90} $\llbracket8,1,0,0\rrbracket$ \end{turn}
&\begin{turn}{-90} $\llbracket9,1,0,0\rrbracket$ \end{turn}
&\begin{turn}{-90} $\llbracket10,1,0,0\rrbracket\ $ \end{turn}
&\begin{turn}{-90} $\llbracket11,1,0,0\rrbracket$ \end{turn}
&\begin{turn}{-90} $\llbracket12,1,0,0\rrbracket$ \end{turn}
&\begin{turn}{-90} $\llbracket13,1,0,0\rrbracket$ \end{turn}
&\begin{turn}{-90} $\llbracket14,1,0,0\rrbracket$ \end{turn}
%&\begin{turn}{-90}\# $\llbracket15,0,0,0\rrbracket$ \end{turn}
%
\\\hline \hline
3 &\textcolor{red}{0} & & & &
& & & & &
& & & & & \\\hline
4 &\textcolor{red}{1} &\textcolor{red}{0} & & &
& & & & &
& & & & & \\\hline
5 &\textcolor{blue}{1} &\textcolor{red}{1} &\textcolor{red}{0} & &
& & & & &
& & & & & \\\hline
6 &\textcolor{blue}{1} &\textcolor{red}{2} &\textcolor{red}{1} &\textcolor{red}{0} &
& & & & &
& & & & & \\\hline
7 &\textcolor{blue}{2} &\textcolor{blue}{2} &\textcolor{red}{2} &\textcolor{red}{1} &\textcolor{red}{0}
& & & & &
& & & & & \\\hline
8 &5 &\textcolor{blue}{5} &\textcolor{red}{3} &\textcolor{red}{2} &\textcolor{red}{1}
&\textcolor{red}{0} & & & &
& & & & & \\\hline
9 &7 &\textcolor{blue}{9} &\textcolor{blue}{6} &\textcolor{red}{3} &\textcolor{red}{2}
&\textcolor{red}{1} &\textcolor{red}{0} & & &
& & & & & \\
\hline
\end{tabular}
%\end{longtable}
%\end{sidewaysfigure}
\end{center}
\end{tiny}
%\caption{${\cal N}_{10d}=(1,0)$ multiplets with $SO(9)$ quantum numbers $[j,X,Y,Z]$}
%\label{tab10d100}
%\end{table}

\vspace{0.5cm}

%%% Now [j,2,0,0]

%\begin{table}
\begin{tiny}
\begin{center}
%\begin{sidewaysfigure}
\begin{tabular}{|c|| c|c|c|c|c| c|c|c|c|c| c|c|c|c|c|}
%\begin{longtable}{|c|| c|c|c|c|c| c|c|c|c|c| c|c|c|c|c| c|c|c|c|c| c|c|c|c|c| c|} % \small
 \hline
\begin{turn}{-90}$\ \ap m^2 \ $  \end{turn}
&\begin{turn}{-90} $\llbracket0,2,0,0\rrbracket$ \end{turn}
&\begin{turn}{-90} $\llbracket1,2,0,0\rrbracket$ \end{turn}
&\begin{turn}{-90} $\llbracket2,2,0,0\rrbracket$ \end{turn}
&\begin{turn}{-90} $\llbracket3,2,0,0\rrbracket$ \end{turn}
&\begin{turn}{-90} $\llbracket4,2,0,0\rrbracket$ \end{turn}
&\begin{turn}{-90} $\llbracket5,2,0,0\rrbracket$ \end{turn}
&\begin{turn}{-90} $\llbracket6,2,0,0\rrbracket$ \end{turn}
&\begin{turn}{-90} $\llbracket7,2,0,0\rrbracket$ \end{turn}
&\begin{turn}{-90} $\llbracket8,2,0,0\rrbracket$ \end{turn}
&\begin{turn}{-90} $\llbracket9,2,0,0\rrbracket$ \end{turn}
&\begin{turn}{-90} $\llbracket10,2,0,0\rrbracket\ $ \end{turn}
&\begin{turn}{-90} $\llbracket11,2,0,0\rrbracket$ \end{turn}
&\begin{turn}{-90} $\llbracket12,2,0,0\rrbracket$ \end{turn}
&\begin{turn}{-90} $\llbracket13,2,0,0\rrbracket$ \end{turn}
&\begin{turn}{-90} $\llbracket14,2,0,0\rrbracket$ \end{turn}
%&\begin{turn}{-90}\# $\llbracket15,0,0,0\rrbracket$ \end{turn}
%
\\\hline \hline
6 &\textcolor{red}{0} & & & &
& & & & &
& & & & & \\\hline
7 &\textcolor{red}{1} &\textcolor{red}{0} & & &
& & & & &
& & & & & \\\hline
8 &\textcolor{blue}{1} &\textcolor{red}{1} &\textcolor{red}{0} & &
& & & & &
& & & & & \\\hline
9 &\textcolor{blue}{4} &\textcolor{red}{2} &\textcolor{red}{1} &\textcolor{red}{0} &
& & & & &
& & & & & \\\hline
10 &\textcolor{blue}{5} &\textcolor{blue}{5} &\textcolor{red}{2} &\textcolor{red}{1} &\textcolor{red}{0}
& & & & &
& & & & & \\\hline
11 &\textcolor{blue}{13} &\textcolor{blue}{9} &\textcolor{red}{6} &\textcolor{red}{2} &\textcolor{red}{1}
&\textcolor{red}{0} & & & &
& & & & & \\\hline
12 &21 &\textcolor{blue}{21} &\textcolor{blue}{10} &\textcolor{red}{6} &\textcolor{red}{2}
&\textcolor{red}{1} &\textcolor{red}{0} & & &
& & & & & \\
\hline
\end{tabular}
%\end{longtable}
%\end{sidewaysfigure}
\end{center}
\end{tiny}
%\caption{${\cal N}_{10d}=(1,0)$ multiplets with $SO(9)$ quantum numbers $[j,X,Y,Z]$}
%\label{tab10d200}
%\end{table}

%%% Now [j,3,0,0]

\vspace{0.5cm}

%\begin{table}
\begin{scriptsize}
\begin{center}
%\begin{sidewaysfigure}
\begin{tabular}{|c|| c|c|c|c|c| c|c|c|c|c| c|c|c|c|c|}
%\begin{longtable}{|c|| c|c|c|c|c| c|c|c|c|c| c|c|c|c|c| c|c|c|c|c| c|c|c|c|c| c|} % \small
 \hline
\begin{turn}{-90}$\ \ap m^2 \ $  \end{turn}
&\begin{turn}{-90} $\llbracket0,3,0,0\rrbracket$ \end{turn}
&\begin{turn}{-90} $\llbracket1,3,0,0\rrbracket$ \end{turn}
&\begin{turn}{-90} $\llbracket2,3,0,0\rrbracket$ \end{turn}
&\begin{turn}{-90} $\llbracket3,3,0,0\rrbracket$ \end{turn}
&\begin{turn}{-90} $\llbracket4,3,0,0\rrbracket$ \end{turn}
&\begin{turn}{-90} $\llbracket5,3,0,0\rrbracket$ \end{turn}
&\begin{turn}{-90} $\llbracket6,3,0,0\rrbracket$ \end{turn}
&\begin{turn}{-90} $\llbracket7,3,0,0\rrbracket$ \end{turn}
&\begin{turn}{-90} $\llbracket8,3,0,0\rrbracket$ \end{turn}
&\begin{turn}{-90} $\llbracket9,3,0,0\rrbracket$ \end{turn}
&\begin{turn}{-90} $\llbracket10,3,0,0\rrbracket\ $ \end{turn}
&\begin{turn}{-90} $\llbracket11,3,0,0\rrbracket$ \end{turn}
&\begin{turn}{-90} $\llbracket12,3,0,0\rrbracket$ \end{turn}
&\begin{turn}{-90} $\llbracket13,3,0,0\rrbracket$ \end{turn}
&\begin{turn}{-90} $\llbracket14,3,0,0\rrbracket$ \end{turn}
%&\begin{turn}{-90}\# $\llbracket15,0,0,0\rrbracket$ \end{turn}
%
\\\hline \hline
9 &\textcolor{red}{0} & & & &
& & & & &
& & & & & \\\hline
10 &\textcolor{red}{1} &\textcolor{red}{0} & & &
& & & & &
& & & & & \\\hline
11 &\textcolor{blue}{1} &\textcolor{red}{1} &\textcolor{red}{0} & &
& & & & &
& & & & & \\\hline
12 &\textcolor{blue}{4} &\textcolor{red}{2} &\textcolor{red}{1} &\textcolor{red}{0} &
& & & & &
& & & & & \\\hline
13 &\textcolor{blue}{8} &\textcolor{blue}{5} &\textcolor{red}{2} &\textcolor{red}{1} &\textcolor{red}{0}
& & & & &
& & & & & \\\hline
14 &\textcolor{blue}{18} &\textcolor{blue}{12} &\textcolor{red}{6} &\textcolor{red}{2} &\textcolor{red}{1}
&\textcolor{red}{0} & & & &
& & & & & \\\hline
15 &\textcolor{blue}{34} &\textcolor{blue}{26} &\textcolor{blue}{13} &\textcolor{red}{6} &\textcolor{red}{2}
&\textcolor{red}{1} &\textcolor{red}{0} & & &
& & & & & \\
\hline
\end{tabular}%
%\end{longtable}
%\end{sidewaysfigure}
\end{center}
\end{scriptsize}
%\caption{${\cal N}_{10d}=(1,0)$ multiplets with $SO(9)$ quantum numbers $[j,X,Y,Z]$}
%\label{tab10d003}
%\end{table}

%%%%%%%%%%%%%%%%%%%%%%%
%%%%%%%%%%%%%%%%%%%%%%
%%%%%%%%%%%%%%%%%%%%%

\section{Large spin asymptotics of super Poincar\'e multiplicities}
\label{app:data}

This appendix contains some more data on the large spin asymptotics of ${\cal N}_{6d}=(1,0)$ and ${\cal N}_{10d}=1$ spectra. The leading and subleading Regge trajectories $\tau^{k,p}_{\ell}(q)$ and $\tau^{x,y,z}_{\ell}(q)$ are defined through the expansion (\ref{6dimtau}) and (\ref{10dimtau}) of super Poincar\'e multiplicity generating functions in terms of $q^n$ powers (with $n$ denoting the first $SO(d-1)$ Dynkin label). They have been computed on the basis of the $\ap m^2 \leq 25$ data tabulated in appendix \ref{app6} and \ref{app10}, respectively.

\subsection[${\cal N}_{6d}=(1,0)$ multiplets at $SO(5)$ Dynkin labels $n \rightarrow \infty, k$]{$\bm{{\cal N}_{6d}=(1,0)}$ multiplets at $\bm{SO(5)}$ Dynkin labels $\bm{[n \rightarrow \infty, k]}$}
\label{app:data6}

For the universal ${\cal N}_{6d}=(1,0)$ multiplets $\llbracket n \rightarrow \infty, k;p \rrbracket$ we display some $\tau_{\ell \leq 5}^{k,p}(q)$ associated with super Poincar\'e quantum numbers $(k,p)=(2,2), (4,2), (2,4), (3,1), (1,3), (5,1), (3,3), (1,5)$ here. The former three multiplets are bosonic and characterized by the following asymptotic behaviour:
\begin{itemize}
\item $SO(5)$ Dynkin labels $[n \rightarrow \infty, 2]$ and $SU(2)_R$ representation $[2]$
\begin{align}
\tau_1^{2,2}(q) \eq& q^{4} \,(1 + 6 q + 19 q^2 + 60 q^3 + 160 q^4 + 421 q^5 + 1015 q^6 + 2400 q^7 + 
 5398 q^8 \notag \\
 & \ \ + 11900 q^9 + 25371 q^{10} + 53107 q^{11} + 108500 q^{12} + 
 218074 q^{13} \notag \\
 & \ \ + 430116 q^{14} + 836194 q^{15} + 1600889 q^{16}+\ldots)
 \notag \\
 \tau_2^{2,2}(q) \eq& q^{5} \,(3 + 13 q + 49 q^2 + 151 q^3 + 439 q^4 + 1166 q^5 + 2956 q^6 + 
 7119 q^7 \notag \\
 & \ \ + 16566 q^8 + 37224 q^9 + 81414 q^{10} + 173493 q^{11}+\ldots)
 \notag \\
 \tau_3^{2,2}(q) \eq& q^{6} \,(3 + 12 q + 53 q^2 + 171 q^3 + 537 q^4 + 1486 q^5 + 3960 q^6 + 9876 q^7+\ldots)
 \notag \\
 \tau_4^{2,2}(q) \eq& q^{7} \,(1 + 8 q + 35 q^2 + 134 q^3 + 434 q^4+\ldots)
 \notag \\
 \tau_5^{2,2}(q) \eq& q^{9} \,(4+\ldots)
\end{align}
\item $SO(5)$ Dynkin labels $[n \rightarrow \infty, 4]$ and $SU(2)_R$ representation $[2]$
\begin{align}
\tau_1^{4,2}(q) \eq& q^{7} \,(3 + 12 q + 48 q^2 + 141 q^3 + 408 q^4 + 1052 q^5 + 2632 q^6 + 
 6194 q^7 + 14200 q^8 \notag \\
 &   + 31309 q^9 + 67467 q^{10} + 141443 q^{11} + 
 290805 q^{12} + 585447 q^{13} + 1159182 q^{14}+\ldots) \notag \\
 \tau_2^{4,2}(q) \eq& q^{8} \,(3 + 15 q + 63 q^2 + 206 q^3 + 623 q^4 + 1714 q^5 + 4464 q^6 + 
 11006 q^7 + 26108 q^8 \notag \\
 &   + 59679 q^9 + 132452 q^{10}+\ldots) \notag \\
 \tau_3^{4,2}(q) \eq& q^{10} \,(3 + 16 q + 76 q^2 + 262 q^3 + 847 q^4 + 2427 q^5 + 6599 q^6+\ldots) \notag \\
 \tau_4^{4,2}(q) \eq& q^{12} \,(1 + 11 q + 52 q^2+\ldots) 
  \end{align}
\item $SO(5)$ Dynkin labels $[n \rightarrow \infty, 2]$ and $SU(2)_R$ representation $[4]$
\begin{align}
\tau_1^{2,4}(q) \eq& q^{8} \, (4 + 14 q + 58 q^2 + 170 q^3 + 492 q^4 + 1264 q^5 + 3165 q^6 + 
 7432 q^7 + 17012 q^8 \notag \\
 & \ \ + 37428 q^9 + 80496 q^{10} + 168377 q^{11} + 
 345433 q^{12}+\ldots) \notag \\
 \tau_2^{2,4}(q) \eq& q^{8} \, (1 + 11 q + 45 q^2 + 169 q^3 + 523 q^4 + 1505 q^5 + 3992 q^6 + 
 10086 q^7 + 24241 q^8+\ldots) \notag \\
 \tau_3^{2,4}(q) \eq& q^{9} \, (3 + 15 q + 70 q^2 + 241 q^3 + 781 q^4+\ldots) \notag \\
 \tau_4^{2,4}(q) \eq& q^{10} \, (3 + 15 q+\ldots) 
 \end{align}
\end{itemize}
In addition, let us display some $\tau_{\ell}^{k,p}(q)$ associated with fermionic supermultiplets:
\begin{itemize}
\item $SO(5)$ Dynkin labels $[n \rightarrow \infty, 3]$ and $SU(2)_R$ representation $[1]$
\begin{align}
\tau_1^{3,1}(q) \eq& q^{4} \, (1 + 5 q + 16 q^2 + 49 q^3 + 134 q^4 + 343 q^5 + 840 q^6 + 1971 q^7 + 
 4460 q^8 \notag \\
 & \ \ + 9810 q^9 + 21006 q^{10} + 43952 q^{11} + 90078 q^{12} + 
 181178 q^{13} + 358196 q^{14} \notag \\
 & \ \  + 697195 q^{15} + 1337468 q^{16}+\ldots) \notag \\
 \tau_2^{3,1}(q) \eq& q^{5} \, (1 + 7 q + 25 q^2 + 84 q^3 + 247 q^4 + 674 q^5 + 1733 q^6 + 4252 q^7 + 
 10005 q^8 \notag \\
 & \ \ + 22774 q^9 + 50306 q^{10} + 108276 q^{11}+\ldots) \notag \\
 \tau_3^{3,1}(q) \eq& q^{7} \, (2 + 11 q + 46 q^2 + 158 q^3 + 486 q^4 + 1369 q^5 + 3622 q^6+\ldots) \notag \\
 \tau_4^{3,1}(q) \eq& q^{9} \, (2 + 13 q + 57 q^2+\ldots)
  \end{align}
%%%%%%%%
\item $SO(5)$ Dynkin labels $[n \rightarrow \infty, 1]$ and $SU(2)_R$ representation $[3]$
\begin{align}
\tau_1^{1,3}(q) \eq& q^{5} \,( 2 + 9 q + 29 q^2 + 86 q^3 + 233 q^4 + 591 q^5 + 1426 q^6 + 
 3308 q^7 + 7408 q^8 + 16117 q^9 \notag \\
 & \! \! \! + 34176 q^{10}+ 70842 q^{11} + 
 143887 q^{12} + 286959 q^{13} + 562767 q^{14} + 1086923 q^{15}+\ldots) \notag \\
 \tau_2^{1,3}(q) \eq& q^{5} \,( 2 + 10 q + 39 q^2 + 125 q^3 + 366 q^4 + 990 q^5 + 2530 q^6 + 
 6157 q^7 + 14414 q^8  \notag \\
 & \! \! \! + 32604 q^9 + 71640 q^{10} + 153380 q^{11}+\ldots) \notag \\
 \tau_3^{1,3}(q) \eq& q^{5} \,( 1 + 6 q + 24 q^2 + 87 q^3 + 275 q^4 + 799 q^5 + 2168 q^6 + 5570 q^7 + 
 13669 q^8 +\ldots) \notag \\
 \tau_4^{1,3}(q) \eq& q^{6} \,( 2 + 9 q + 38 q^2 + 135 q^3 + 428 q^4+\ldots) \notag \\
 \tau_5^{1,3}(q) \eq& q^{7} \,( 2 + 11 q+\ldots) 
\end{align}
%%%%%%%%%
\item $SO(5)$ Dynkin labels $[n \rightarrow \infty, 5]$ and $SU(2)_R$ representation $[1]$
\begin{align}
\tau_1^{5,1}(q) \eq& q^{7} \, (1 + 7 q + 24 q^2 + 80 q^3 + 228 q^4 + 610 q^5 + 1533 q^6 + 3691 q^7 + 
 8520 q^8 \notag \\
 & + 19063 q^9 + 41409 q^{10} + 87751 q^{11} + 181781 q^{12} +  369134 q^{13} + 735899 q^{14}+\ldots) \notag \\
 \tau_2^{5,1}(q) \eq& q^{8} \, (1 + 7 q + 26 q^2 + 92 q^3 + 281 q^4 + 791 q^5 + 2090 q^6 + 5251 q^7 + 
 12618 q^8 \notag \\
 & + 29264 q^9 + 65731 q^{10}+\ldots) \notag \\
 \tau_3^{5,1}(q) \eq& q^{11} \, (2 + 12 q + 55 q^2 + 196 q^3 + 625 q^4 + 1808 q^5+\ldots) \notag \\
 \tau_4^{5,1}(q) \eq& q^{14} \, (2 + 15 q+\ldots)
\end{align}
%%%%%%%%
\item $SO(5)$ Dynkin labels $[n \rightarrow \infty, 3]$ and $SU(2)_R$ representation $[3]$
\begin{align}
\tau_1^{3,3}(q) \eq& q^{7} \,(2 + 10 q + 41 q^2 + 127 q^3 + 369 q^4 + 977 q^5 + 2453 q^6 + 
 5856 q^7 + 13474 q^8 \notag \\
 & + 29947 q^9 + 64743 q^{10} + 136433 q^{11} + 
 281245 q^{12} +  568184 q^{13} + 1127435 q^{14}+\ldots)
 \notag \\
 \tau_2^{3,3}(q) \eq& q^{8} \,(3 + 18 q + 75 q^2 + 252 q^3 + 762 q^4 + 2111 q^5 + 5496 q^6 + 
 13580 q^7 + 32188 q^8 \notag \\
 & + 73580 q^9 + 163122 q^{10}+\ldots)
 \notag \\
 \tau_3^{3,3}(q) \eq& q^{9} \,(1 + 11 q + 49 q^2 + 189 q^3 + 617 q^4 + 1841 q^5 + 5079 q^6+\ldots)
 \notag \\
 \tau_4^{3,3}(q) \eq& q^{11} \,(3 + 19 q + 84 q^2+\ldots)
\end{align}
\item $SO(5)$ Dynkin labels $[n \rightarrow \infty, 1]$ and $SU(2)_R$ representation $[5]$
\begin{align}
\tau_1^{1,5}(q) \eq& q^{9} \, (2 + 10 q + 36 q^2 + 118 q^3 + 335 q^4 + 893 q^5 + 2237 q^6 + 
 5356 q^7 + 12311 q^8 \notag \\
 & \ \ + 27406 q^9 + 59236 q^{10} + 124892 q^{11}+\ldots) \notag \\
 \tau_2^{1,5}(q) \eq& q^{9} \, (2 + 13 q + 54 q^2 + 186 q^3 + 573 q^4 + 1609 q^5 + 4237 q^6 + 
 10575 q^7+\ldots) \notag \\
 \tau_3^{1,5}(q) \eq& q^{9} \, (2 + 10 q + 45 q^2 + 161 q^3 + 518 q^4+\ldots) \notag \\
 \tau_4^{1,5}(q) \eq& q^{9} \, (1 + 6 q + 26 q^2+\ldots) 
\end{align}
\end{itemize}
The results listed in this appendix confirm the trend observed in subsection \ref{emp6dim}: The $\tau_{\ell}^{k,p}(q)$ expansion (\ref{6dimtau}) of $G_{n,k,p}(q)$ converges more rapidly at large vales of the second Dynkin label $k$ and small values $SU(2)_R$ spin $p$.

%%%%%%%%%%%%%%%%%%%%%%%%%%%%%%%

\subsection[${\cal N}_{10d}=1$ multiplets at $SO(9)$ Dynkin labels $n \rightarrow \infty, x,y,z$]{$\bm{{\cal N}_{10d}=1}$ multiplets at $\bm{SO(9)}$ Dynkin labels $\bm{[n \rightarrow \infty, x,y,z]}$}
\label{app:data10}

Also for ${\cal N}_{10d}=1$ multiplets $\llbracket n \rightarrow \infty, x,y,z\rrbracket$ we would like to list some more $\tau_{\ell \leq 5}^{x,y,z}(q)$ beyond those of subsection \ref{emp10dim}, specifically for Dynkin labels $(x,y,z)=(1,1,0), (1,0,2), (2,1,0), (1,0,1), (0,1,1), $ $(2,0,1), (0,0,3), (1,1,1)$. We focus on three bosonic families
\begin{itemize}
\item $SO(9)$ Dynkin labels $[n \rightarrow \infty,1,1,0]$
\begin{align}
\tau_1^{1,1,0} (q)\eq& q^{8} \, (1 + 2 q + 8 q^2 + 19 q^3 + 45 q^4 + 100 q^5 + 217 q^6 + 446 q^7 + 
 905 q^8 + 1779 q^9 \notag \\
 & \ \ + 3440 q^{10} + 6521 q^{11} + 12181 q^{12} + 22396 q^{13}+\ldots)
\notag \\
\tau_2^{1,1,0}(q) \eq& q^{9} \, (1 + 2 q + 9 q^2 + 23 q^3 + 61 q^4 + 143 q^5 + 330 q^6 + 715 q^7 + 
 1524 q^8 + 3128 q^9+\ldots)
\notag \\
\tau_3^{1,1,0}(q) \eq& q^{12} \, (1 + 4 q^1 + 16 q^2 + 46 q^3 + 125 q^4+\ldots)
\end{align}
\item $SO(9)$ Dynkin labels $[n \rightarrow \infty,1,0,2]$
\begin{align}
\tau_1^{1,0,2}(q) \eq& q^{9} \, (1 + 4 q + 12 q^2 + 31 q^3 + 75 q^4 + 172 q^5 + 375 q^6 + 791 q^7 + 
 1615 q^8 \notag \\
 & \ \ + 3225 q^9 + 6287 q^{10} + 12044 q^{11} + 22652 q^{12}+\ldots) \notag \\
 \tau_2^{1,0,2}(q) \eq& q^{10} \, (1 + 4 q + 14 q^2 + 39 q^3 + 104 q^4 + 252 q^5 + 587 q^6 + 1300 q^7 + 
 2794 q^8+\ldots) \notag \\
 \tau_3^{1,0,2}(q) \eq& q^{13} \, (2 + 8 q + 30 q^2 + 87 q^3+\ldots) 
\end{align}
\item $SO(9)$ Dynkin labels $[n \rightarrow \infty,2,1,0]$
\begin{align}
\tau_1^{2,1,0}(q) \eq& q^{11} \, (1 + 2 q + 9 q^2 + 22 q^3 + 59 q^4 + 132 q^5 + 306 q^6 + 646 q^7 + 
 1369 q^8 + 2756 q^9 \notag \\
 & \ \ + 5514 q^{10} + 10682 q^{11}+\ldots) \notag \\
\tau_2^{2,1,0}(q) \eq& q^{12} \, (1 + 2 q + 9 q^2 + 23 q^3 + 63 q^4 + 150 q^5 + 357 q^6 + 791 q^7 + 
 1728 q^8+\ldots) \notag \\
 \tau_3^{2,1,0}(q) \eq& q^{16} \, (1 + 4 q + 18 q^2 + 51 q^3+\ldots)
  \end{align}
\end{itemize}
and five fermionic families of supermultiplets:
\begin{itemize}
\item $SO(9)$ Dynkin labels $[n \rightarrow \infty,1,0,1]$
\begin{align}
\tau_1^{1,0,1}(q) \eq& q^{6} \,(1 + 3 q + 7 q^2 + 17 q^3 + 37 q^4 + 79 q^5 + 162 q^6 + 325 q^7 + 
 635 q^8 + 1220 q^9 \notag \\
 & \ \ + 2298 q^{10} + 4266 q^{11} + 7807 q^{12} + 
 14110 q^{13} + 25197 q^{14} + 44530 q^{15}+\ldots)
 \notag \\
\tau_2^{1,0,1}(q) \eq& q^{7} \,(1 + 3 q + 9 q^2 + 24 q^3 + 57 q^4 + 131 q^5 + 288 q^6 + 610 q^7 + 
 1256 q^8 + 2523 q^9 \notag \\
 & \ \ + 4957 q^{10} + 9557 q^{11}+\ldots)
 \notag \\
 \tau_3^{1,0,1}(q) \eq& q^{10} \,(2 + 7 q + 22 q^2 + 61 q^3 + 155 q^4 + 367 q^5 + 835 q^6+\ldots)
 \notag \\
 \tau_4^{1,0,1}(q) \eq& q^{13} \,(2 + 9 q + 31 q^2+\ldots)
\end{align}
\item $SO(9)$ Dynkin labels $[n \rightarrow \infty,0,1,1]$
\begin{align}
\tau_1^{0,1,1}(q) \eq& q^{8} \, (1 + 4 q + 10 q^2 + 27 q^3 + 63 q^4 + 141 q^5 + 302 q^6 + 628 q^7 + 
 1264 q^8 \notag \\
 & \ \ + 2494 q^9 + 4811 q^{10} + 9119 q^{11} + 17005 q^{12} + 31260 q^{13}+\ldots) \notag \\
 \tau_2^{0,1,1}(q) \eq& q^{9} \, (1 + 5 q + 16 q^2 + 44 q^3 + 113 q^4 + 269 q^5 + 610 q^6 + 1330 q^7 \notag \\
 & \ \ + 
 2804 q^8 + 5748 q^9+\ldots) \notag \\
 \tau_3^{0,1,1}(q) \eq& q^{11} \, (1 + 6 q + 19 q^2 + 59 q^3 + 160 q^4 + 404 q^5+\ldots) \notag \\
 \tau_4^{0,1,1}(q) \eq& q^{14} \, (2 + 9 q+\ldots)
\end{align}
\item $SO(9)$ Dynkin labels $[n \rightarrow \infty,2,0,1]$
\begin{align}
\tau_1^{2,0,1}(q) \eq& q^{9} \, (1 + 3 q + 9 q^2 + 23 q^3 + 55 q^4 + 123 q^5 + 267 q^6 + 556 q^7 + 
 1132 q^8 \notag \\
 & \ \  + 2244 q^9 + 4362 q^{10} + 8318 q^{11} + 15616 q^{12} + 28873 q^{13}+\ldots) \notag \\
 \tau_2^{2,0,1}(q) \eq& q^{10} \, (1 + 3 q + 9 q^2 + 25 q^3 + 63 q^4 + 150 q^5 + 342 q^6 + 749 q^7 + 
 1591 q^8 \notag \\
 & \ \ + 3289 q^9 + 6640 q^{10}+\ldots) \notag \\
 \tau_3^{2,0,1}(q) \eq& q^{14} \, (2 + 8 q + 27 q^2 + 77 q^3 + 204 q^4+\ldots) 
 \end{align}
\item $SO(9)$ Dynkin labels $[n \rightarrow \infty,0,0,3]$
\begin{align}
\tau_1^{0,0,3}(q) \eq& q^{10} \, (2 + 5 q + 15 q^2 + 38 q^3 + 90 q^4 + 201 q^5 + 437 q^6 + 905 q^7 + 
 1838 q^8 \notag \\
 & \ \ + 3633 q^9 + 7038 q^{10} + 13374 q^{11}+\ldots) \notag \\
\tau_2^{0,0,3}(q) \eq& q^{11} \, (2 + 8 q + 25 q^2 + 69 q^3 + 176 q^4 + 418 q^5 + 949 q^6 + 2069 q^7+\ldots) \notag \\
 \tau_3^{0,0,3}(q) \eq& q^{13} \, (3 + 11 q + 38 q^2 + 109 q^3+\ldots) \notag \\
 \tau_4^{0,0,3}(q) \eq& q^{15} \, (1+\ldots)
\end{align}
\item $SO(9)$ Dynkin labels $[n \rightarrow \infty,1,1,1]$
\begin{align}
\tau_1^{1,1,1}(q) \eq& q^{11} \, (1 + 5 q + 16 q^2 + 45 q^3 + 116 q^4 + 276 q^5 + 624 q^6 + 1358 q^7 + 
 2852 q^8 \notag \\
 & \ \  + 5825 q^9 + 11616 q^{10} + 22669 q^{11}+\ldots)
 \notag \\
 \tau_2^{1,1,1}(q) \eq& q^{12} \, (1 + 5 q + 17 q^2 + 52 q^3 + 142 q^4 + 358 q^5 + 855 q^6 + 1950 q^7 + 
 4279 q^8+\ldots)
 \notag \\
 \tau_3^{1,1,1}(q) \eq& q^{15} \, (1 + 7 q + 26 q^2 + 84 q^3 + 243 q^4+\ldots)
  \end{align}
  \end{itemize}
These results confirm the trend observed in subsection \ref{emp10dim}: The $\tau_{\ell}^{x,y,z}(q)$ expansion (\ref{10dimtau}) of multiplicity generating functions $G_{n,x,y,z}(q)$ converges more quickly at higher values of the Dynkin labels $x,y,z$.

%%%%%%%%%%%%%%%%%%%%
%%%%%%%%%%%%%%%%%%%%%
%%%%%%%%%%%%%%%%%%%%%%

\section[Deriving the asymptotic formulae for ${\cal N}_{4d}=1$ multiplicity generating functions]{Deriving the asymptotic formulae for $\bm{{\cal N}_{4d}=1}$ multiplicity generating functions}
\label{deriv}

In this appendix, we derive the asymptotic results on multiplicity generating function $G_{n,Q}(q)$ in the limit $n \rightarrow \infty$ presented in subsection \ref{asympt4dim}.

In what follows, we will exploit the $n \rightarrow \infty$ behaviour of objects $T_p(m,k) := {m \choose k} - {m \choose k-p}$,
\bea
T_{2n+2} (2m+1, m+n+1-k) &\sim {2m+1 \choose m+n+1-k}~, \nn \\
T_{2n+2} (2m, m+n-k) &\sim {2m \choose m+n-k}~. \label{approxTbybin}
\eea
assuming that $m, k \geq 0$

\subsection[Warm-up: Multiplicities of $\lb 2n+1,0 \rb$ and $\lb 2n,1 \rb$ as $n \rightarrow \infty$]{Warm-up: Multiplicities of $\bm{\lb 2n+1,0 \rb}$ and $\bm{\lb 2n,1 \rb}$ as $\bm{n \rightarrow \infty}$}

In order to get familiar with the asymptotic methods in the ${\cal N}_{4d}=1$ context, we shall first of all discuss the large spin regime of supermultiplets with $U(1)_R$ neutral Clifford vacuum.

The multiplicity generating function for the representation $\lb2n+1,0 \rb$ can be written as
\bea
G_{2n+1,0}(q)= \sum_{k= 0}^\infty \sum_{m= 0}^\infty \sum_{p=0}^{\infty}  \frM_{\lb 2n+1,0 \rb} (m, -p-1, k;q) + \sum_{k= 0}^\infty \sum_{p=0}^{\infty} \frM_{\lb 2n+1,0 \rb} (p, p, k;q)~, \label{mulsumodd0}
\eea
where the function $\mathfrak{M}_{\lb 2n+1,2Q \rb}$ and $\mathfrak{M}_{\lb 2n+1,2Q \rb}$ are defined in (\ref{shorthd1}) and (\ref{shorthd2}) and, as $n \rightarrow \infty$, 
\bea
&\mathfrak{M}_{\lb 2n+1,0 \rb} (m,p,k; q) \sim (-1)^{-m-p} \Bigg[ F^{\NS}_{k,p} (q) {m-p \choose 2m+1} {2m+1 \choose m+n+1-k}  \nn \\
& \hspace{4cm} + F^{\Ra}_{k,p} (q) {m-p\choose 2m} {2m \choose m+n-k} \Bigg]~.
\eea
Note that the binomial coefficient ${\alpha \choose \beta}$ increases as $\beta$ increases from $0$ to $\lfloor \alpha/2 \rfloor$ and then decreases as $\beta$ increases from $\lfloor \alpha/2 \rfloor+1$ to $\alpha$.  

Observe that $\frM_{\lb 2n+1,0 \rb} (m, -p-1, k;q)$ is sharply peaked near $(m,p,k)=(0,0,n)$ for $n$ large.  Therefore, the dominant contribution to the first set of summations in \eref{mulsumodd0} comes from 
\bea \label{firstsetsum}
&   \sum_{m= 0}^\infty \sum_{p=0}^{\infty} \sum_{k= 0}^\infty  \frM_{\lb 2n+1,0 \rb} (m, -p-1, k;q) \nn \\
& \sim \sum_{m= 0}^{\lceil \epsilon_1 \rceil} \sum_{p=0}^{\lceil \epsilon_2 \rceil} \sum_{k= \lfloor n(1- \epsilon_3) \rfloor}^{\lceil n(1+\epsilon_3) \rceil} \frM_{\lb 2n+1,0 \rb} (m, -p-1, k;q) \qquad \text{any $\epsilon_1, \epsilon_2, \epsilon_3 >0$}, \quad n \rightarrow \infty \nn \\
 & \sim \sum_{m =0}^\infty \sum_{p =0}^\infty \sum_{\delta =-\infty}^\infty \frM_{\lb 2n+1,0 \rb} (m, -p-1, n+\delta;q), \qquad n \rightarrow \infty~.
\eea
In the limit of large $k$, we can use asymptotic formulae \eref{asympFNSkp} and \eref{asympFRkp} for $F^{\NS}_{k,p} (q)$ and $F^{\Ra}_{k,p} (q)$. The summation over $\delta$ from $-\infty$ to $\infty$ can be readily computed using the fact that 
\bea
\sum_{\delta=-\infty}^\infty q^\delta {2 m \choose m-\delta} &= \sum_{\delta=-m}^m q^\delta {2 m \choose m-\delta}  =  q^{-m} (1+q)^{2m}~, \nn \\
\sum_{\delta=-\infty}^\infty q^\delta {2 m+1 \choose m-\delta+1} &= \sum_{\delta=-(m+1)}^{m+1} q^\delta {2 m+1 \choose m-\delta+1} = q^{-m} (1+q)^{2m+1}~.
\eea
Next, the summation over $m$ from $0$ to $\infty$ can be computed using the following identities:
\bea
\sum_{m=0}^\infty (-q)^{-m} (1+q)^{2m} {1+m+p \choose  2m} &= (-q)^{-p-1} \frac{1-q^{2p+3}}{1-q}~, \nn \\
\sum_{m=0}^\infty (-q)^{-m} (1+q)^{2m+1} {1+m+p \choose 1+ 2m} &= (-q)^{-p} \frac{1-q^{2p+2}}{1-q}~.
\eea
Thus, from \eref{firstsetsum}, we find that
\bea
\sum_{m= 0}^\infty \sum_{p=0}^{\infty} \sum_{k= 0}^\infty  &\frM_{\lb 2n+1,0 \rb} (m, -p-1, k;q) = \frac{(1-q)^2 q^{n-\frac{1}{2}}}{2(q,q)_{\infty}^6} \nn \\
&\times \  \Big\{ u_1(\sqrt{q}) \vartheta_2 (1, q)^2 - \left[ u_2(\sqrt{q}) \vartheta_3(1, q)^2-u_2(-\sqrt{q}) \vartheta_4(1,q)^2 \right] \Big\}~,
\eea
where the functions $u_1(q)$ and $u_2(q)$ are defined as follows:
\bea
u_1(q) &= \sum_{p=0}^\infty q^{2\left(p+\frac{3}{2} \right)^2} \frac{1-q^{4p+6}}{(1+q^{2p+2})(1+q^{2p+4})}~, \nn \\
u_2(q) &=  \sum_{p=0}^\infty q^{2\left(p+1 \right)^2} \frac{1-q^{4p+4}}{(1+q^{2p+1})(1+q^{2p+3})}~.
\eea
It remains unclear whether $u_1(q)$ and $u_2(q)$ can be written in terms of known functions (if this is useful at all).  In practice, it is easy to compute the power series $u_1(q)$ and $u_2(q)$ up to a high order in $q$.  Moreover, their asymptotic formulae can be easily derived in the limit $q \rightarrow 0$.  We shall come back to this point later.

Let us now examine the second set of summations in \eref{mulsumodd0}.  The function $\frM_{\lb 2n+1,0 \rb} (p, p, k;q)$ is sharply peaked near $(p,k)=(0,n)$ for large $n$.  Thus,
\bea
\sum_{k= 0}^\infty \sum_{p=0}^{\infty} \frM_{\lb 2n+1,0 \rb} (p, p, k;q) &\sim \frM_{\lb 2n+1,0 \rb} (0, 0, n;q)~, \qquad n \rightarrow \infty \nn \\
&= \frac{1}{4 (q;q)_\infty^6}  \frac{(1-q)^3}{1+q} q^{n-\frac{1}{4}} \vartheta_2 (1,q)^2~. \label{secondsetsum}
\eea

From \eref{mulsumodd0}, we simply add \eref{firstsetsum} and \eref{secondsetsum} together and obtain the expression (\ref{simpleeven}) for $Q_{2n+1,0}$, in agreement with the stable pattern in table \ref{tab4d0}.

From recurrence relation \eref{recrel2} for $G_{n,Q}$, the asymptotic behaviour of multiplicity generating functions $U(1)_R$ charge $Q=1$ is given by
\bea
G_{2n,1}(q) = \frac{1}{2} \big[ F^{\NS}_{n,0} (q) - G_{2n-1 , 0}(q) - G_{2n+1 , 0 } (q) \big] \ .
\eea
Using the asymptotics $G_{2n-1,Q} \sim q^{-1} G_{2n+1,Q}$ as well as (\ref{simpleeven}) for $G_{2n+1,Q}$ and (\ref{asympFNSkp}) for $F^{\NS}_{n,0} $, we arrive at (\ref{simpleodd}). This also agrees with the stable pattern tabulated in appendix \ref{app4}.

\subsection[Multiplicities of $\lb 2n+1,2Q \rb$ and $\lb 2n,2Q+1 \rb$ with $Q=O(1)$ as $n \rightarrow \infty$]{Multiplicities of $\bm{\lb 2n+1,2Q \rb}$ and $\bm{\lb 2n,2Q+1 \rb}$ as $\bm{n \rightarrow \infty, \ Q = {\cal O}(1)}$}

This subsection generalizes the asymptotic results from the $Q=0$ (or $Q=1$) sector to generic $U(1)_R$ charges. The multiplicity generating function for $\lb 2n+1,2Q \rb$ can be written as
\bea 
G_{2n+1 , 2Q }(q)= &\sum_{k= 0}^\infty \sum_{m= 0}^\infty  \Bigg[ \sum_{p=0}^{\infty} \Big\{ \frM_{\lb 2n+1,2Q \rb} (m, -p-1, k;q) +\frM_{\lb 2n+1,2Q \rb} (m+p, p, k;q) \Big \} \nn \\
& \hspace{1.5cm} + \sum_{p=0}^{Q-1}  \frM_{\lb 2n+1,2Q \rb} (m, m+p+1, k;q) \Bigg]~.
\eea
where the $\mathfrak{M}_{\lb 2n+1,2Q \rb}$ function follows the following $n \rightarrow \infty$ behaviour:
\bea
&\mathfrak{M}_{\lb 2n+1,2Q \rb} (m,p,k; q) = (-1)^{Q-m-p} \Bigg[ F^{\NS}_{k,p} (q) {Q+m-p \choose 2m+1} {2m+1 \choose m+n+1-k}  \nn \\
& \hspace{5.5cm} + F^{\Ra}_{k,p} (q) {Q+m-p\choose 2m} {2m \choose m+n-k} \Bigg]
\eea
The dominant contribution to $G_{2n+1 , 2Q}(q)$  comes from
\bea \label{mulsumoddevenasymp}
G_{2n+1,2Q}(q) &\sim  \sum_{m= 0}^\infty  \sum_{p=0}^{\lceil \epsilon_2 \rceil}  \sum_{k= \lfloor n(1-\epsilon_1) \rfloor}^{\lceil n(1+\epsilon_1) \rceil}   \Big [ \frM_{\lb 2n+1,2Q \rb} (m, -p-1, k;q) +\frM_{\lb 2n+1,2Q \rb} (m+p, p, k;q) \Big ] \nn \\
& \qquad +   \sum_{m= 0}^\infty   \sum_{p=0}^{Q-1} \sum_{k= \lfloor n(1-\epsilon_1) \rfloor}^{\lceil n(1+\epsilon_1) \rceil} \frM_{\lb 2n+1,2Q \rb} (m,m+ p+1, k ;q) ~, \qquad \epsilon_1, \epsilon_2 >0~, n \rightarrow \infty \nn \\
&\sim  \sum_{m= 0}^\infty   \sum_{p=0}^{\infty} \sum_{\delta= -\infty}^{\infty}   \Big [ \frM_{\lb 2n+1,2Q \rb} (m, -p-1, n+\delta;q) +\frM_{\lb 2n+1,2Q \rb} (m+p, p, n+\delta;q) \Big]  \nn \\
& \qquad +  \sum_{m= 0}^\infty   \sum_{p=0}^{Q-1} \sum_{\delta= -\infty}^{\infty}   \frM_{\lb 2n+1,2Q \rb} (m, m+p+1, n+\delta ;q) ~, \qquad n \rightarrow \infty~.
\eea
The first set of summations can be evaluated as follows:
\bea
& \sum_{m= 0}^\infty   \sum_{p=0}^{\infty} \sum_{\delta= -\infty}^{\infty}  \frM_{\lb 2n+1,2Q \rb} (m, -p-1, n+\delta;q) = \frac{(1-q)^2 q^{n-Q-\frac{1}{2}}}{2 (q;q)_\infty^6} \nn \\
& \ \ \ \ \ \times  \Big\{ u_1(\sqrt{q},Q) \vartheta_2 (1, q)^2  - \left[ u_2(\sqrt{q},Q) \vartheta_3(1, q)^2-u_2(-\sqrt{q},Q) \vartheta_4(1,q)^2 \right] \Big\}~,
\eea
where
\bea
u_1(q,Q) &= \sum_{p=0}^\infty q^{2\left(p+\frac{3}{2} \right)^2} \frac{1-q^{4p+4Q+6}}{(1+q^{2p+2})(1+q^{2p+4})}~, \nn \\
u_2(q,Q) &=  \sum_{p=0}^\infty q^{2\left(p+1 \right)^2} \frac{1-q^{4p+4Q+4}}{(1+q^{2p+1})(1+q^{2p+3})}~.
\eea
The next set of summations in \eref{mulsumoddeven} can be evaluated as follows:
\bea
 \sum_{m= 0}^\infty   \sum_{p=0}^{\infty}& \sum_{\delta= -\infty}^{\infty} \frM_{\lb 2n+1,2Q \rb} (m+p, p, n+ \delta;q) = \frac{(-1)^Q (1-q)^3 q^{n-\frac{3}{2}}}{2 (q;q)_\infty^6}\nn \\
&\times \ \Big\{ v_1(\sqrt{q},Q) \vartheta_2 (1, q)^2   + \left[ v_2(\sqrt{q},Q) \vartheta_3(1, q)^2-v_2(-\sqrt{q},Q) \vartheta_4(1,q)^2 \right] \Big\}~,
\eea
where\footnote{Upon obtaining the hypergeometric functions, we make use of the following identities for $p\geq0$:
\bea
\sum_{m=0}^Q (-1)^m q^{-m} (1+q)^{2m} {Q+m \choose 2p+2 m } &= {Q \choose 2p} {}_3F_2\left[{1,\ Q+1,\ 2p-Q \atop p+1/2,\ p+1} ; \ \frac{(1+q)^2}{4q}\right]~, \nn \\
\sum_{m=0}^Q (-1)^m q^{-m} (1+q)^{2m+1} {Q+m \choose 1+2p+2 m } &= {Q \choose 2 p+1} {}_3F_2\left[{1,\ Q+1,\ 2p+1-Q \atop p+1,\ p+3/2};\ \frac{(1+q)^2}{4q} \right]~.
\eea}
\bea
%v_1(q,Q) &= \sum_{p=0}^{\lfloor Q/2 \rfloor} \frac{q^{2(p-\frac{1}{2})^2}(1+q^2)^{2p}}{(1+q^{2p-2})(1+q^{2p})} \sigma_1(Q,p) ~, \nn \\
%v_2(q,Q) &= \sum_{p=0}^{\lfloor Q/2 \rfloor} \frac{q^{2 p^2}(1+q^2)^{2p}}{(1+q^{2p-1})(1+q^{2p+1})} \sigma_2(Q,p) ~, \nn\\
%\sigma_1(Q,p) &= {Q \choose 2 p} {}_3F_2 \left(1, Q+1, 2p-Q; p+\frac{1}{2}, p+1; \frac{(1+q)^2}{4q} \right)~, \nn \\
%\sigma_2(Q,p) &= (1+q) {Q \choose 2 p+1} {}_3F_2 \left(1, Q+1, 2p-Q+1; p+1, p+\frac{3}{2}; \frac{(1+q)^2}{4q} \right)~. \nn \\
%
%v_1(q,Q) &= \sum_{p=0}^{\lfloor Q/2 \rfloor} \frac{q^{2(p-\frac{1}{2})^2}(1+q^2)^{2p}}{(1+q^{2p-2})(1+q^{2p})} {Q \choose 2 p} {}_3F_2 \left(1, Q+1, 2p-Q; p+\frac{1}{2}, p+1; \frac{(1+q)^2}{4q} \right) ~,  \\
%v_2(q,Q) &= \sum_{p=0}^{\lfloor Q/2 \rfloor} \frac{ (1+q) q^{2 p^2}(1+q^2)^{2p}}{(1+q^{2p-1})(1+q^{2p+1})}  {Q \choose 2 p+1} {}_3F_2 \left(1, Q+1, 2p-Q+1; p+1, p+\frac{3}{2}; \frac{(1+q)^2}{4q} \right) ~, \nn 
%\\
%
v_1(q,Q) &= \sum_{p=0}^{\lfloor Q/2 \rfloor} \frac{q^{2(p-\frac{1}{2})^2}(1+q^2)^{2p}}{(1+q^{2p-2})(1+q^{2p})} {Q \choose 2 p} {}_3F_2\left[{1,\ Q+1,\ 2p-Q \atop p+1/2,\ p+1} ; \ \frac{(1+q)^2}{4q}\right] ~,  \nn \\
v_2(q,Q) &= \sum_{p=0}^{\lfloor Q/2 \rfloor} \frac{ (1+q) q^{2 p^2}(1+q^2)^{2p}}{(1+q^{2p-1})(1+q^{2p+1})}  {Q \choose 2 p+1} {}_3F_2\left[{1,\ Q+1,\ 2p+1-Q \atop p+1,\ p+3/2};\ \frac{(1+q)^2}{4q} \right] ~, 
\eea
The last set of summations in \eref{mulsumoddeven} can be evaluated as follows:
\bea
 \sum_{m= 0}^\infty   \sum_{p=0}^{Q-1} &\sum_{\delta= -\infty}^{\infty} \frM_{\lb 2n+1,2Q \rb} (m, m+p+1, n+\delta ;q)  = \frac{(-1)^Q (1-q)^3 q^{n-\frac{7}{4}}}{2 (q;q)_\infty^6}  \nn \\
&\times \ \Big\{ w_1(\sqrt{q},Q) \vartheta_2 (1, q)^2    + q^{\frac{9}{4}}\left[ w_2(\sqrt{q},Q) \vartheta_3(1, q)^2-w_2(-\sqrt{q},Q) \vartheta_4(1,q)^2 \right] \Big\}~,
\eea
where
\bea
w_1 (q,Q) &= \sum_{m= 0}^\infty   \sum_{p=0}^{Q-1} \frac{ (-1)^{p+1}q^{1+2(1+m+p)^2-2m}\left(1+q^2\right)^{2m} {Q-1-p \choose 2 m}}{ \left(1+q^{2(m+p)}\right) \left(1+q^{2(1+m+p)}\right) }~, \nn \\
w_2 (q,Q) &= q^{-\frac{9}{2}} \sum_{m= 0}^\infty   \sum_{p=0}^{Q-1} \frac{(-1)^{p+1}q^{2\left(m+p+\frac{3}{2}\right)^2-2m}\left(1+q^2\right)^{2m+1} {Q-1-p \choose 1+2 m}}{\left(1+q^{1+2m+2p}\right) \left(1+q^{3+2m+2p}\right)}~.
\eea
Combining the three sets of summations into \eref{mulsumoddeven}, we have
\bea
G_{2n+1 , 2Q }(q)
&= \frac{(1-q)^2 q^{n} }{2 q^{ \frac{3}{2}}(q;q)_\infty^6} \Bigg \{ \vartheta_2(1,q)^2 \Big[ q^{1-Q} u_1(\sqrt{q},Q)+(-1)^Q (1-q)  (v_1(\sqrt{q},Q)+q^{-1/4} w_1(\sqrt{q},Q)) \Big] \nn \\
& \quad +\vartheta_3(1,q)^2 \Big[ -q^{1-Q} u_2(\sqrt{q},Q)+(-1)^Q (1-q)  (v_2(\sqrt{q},Q)+ q^2 w_2(\sqrt{q},Q)) \Big] \nn \\
& \quad +\vartheta_4(1,q)^2 \Big [ q^{1-Q} u_2(-\sqrt{q},Q)-(-1)^Q (1-q)  (v_2(-\sqrt{q},Q)+q^2 w_2(-\sqrt{q},Q)) \Big] \Bigg\}
\eea
which exactly (\ref{asympmuloddeven}) with the definition (\ref{defF}) for the function $\CF(q,Q)$ in the curly brackets. Note that this formula reproduces \eref{simpleeven} when $Q=0$.

This allows to quickly infer asymptotic $\lb 2n,2Q+1 \rb$ multiplicities through the recursion \eref{recrel5} and the asymptotic relations $G_{2n+2,2Q+1} (q) \sim q G_{2n,2Q+1} (q)$ as $ n \rightarrow \infty$:
\bea
G_{2n,2Q+1}(q) &\sim \frac{1}{1+q} \left[ F^\Ra_{n,Q+1} (q) - G_{2n+1,2Q}(q) - G_{2n+1,2Q+2} (q)\right] 
\eea
The asymptotic formula \eref{asympFRkp} for $F^\Ra_{n,Q+1} (q)$ and the definition (\ref{defF}) for the function $\CF(q,Q)$ then leads to (\ref{mainferm}). 

\bibliographystyle{ytphys}
\bibliography{ref}

\providecommand{\href}[2]{#2}\begingroup\raggedright\begin{thebibliography}{10}

\bibitem{Narain:1986am}
K.~Narain, M.~Sarmadi, and E.~Witten, ``{A Note on Toroidal Compactification of
  Heterotic String Theory},''
\href{http://dx.doi.org/10.1016/0550-3213(87)90001-0}{{\em Nucl.Phys.}
  {\bfseries B279} (1987) 369}.
%%CITATION = NUPHA,B279,369;%%.

\bibitem{Anchordoqui:2012fq}
L.~A. Anchordoqui, I.~Antoniadis, H.~Goldberg, X.~Huang, D.~L{\"u}st, {\em et
  al.}, ``{Vacuum Stability of Standard Model$^{++}$},''
\href{http://arxiv.org/abs/1208.2821}{{\ttfamily arXiv:1208.2821 [hep-ph]}}.
%%CITATION = ARXIV:1208.2821;%%.

\bibitem{Feng:2010yx}
W.-Z. Feng, D.~L{\"u}st, O.~Schlotterer, S.~Stieberger, and T.~R. Taylor,
  ``{Direct Production of Lightest Regge Resonances},''
  \href{http://dx.doi.org/10.1016/j.nuclphysb.2010.10.013}{{\em Nucl.Phys.}
  {\bfseries B843} (2011) 570--601},
\href{http://arxiv.org/abs/1007.5254}{{\ttfamily arXiv:1007.5254 [hep-th]}}.
%%CITATION = ARXIV:1007.5254;%%.

\bibitem{Curtright:1986di}
T.~L. Curtright and C.~B. Thorn, ``{Symmetry Patterns in the Mass Spectra of
  Dual String Models},''
\href{http://dx.doi.org/10.1016/0550-3213(86)90525-0}{{\em Nucl. Phys.}
  {\bfseries B274} (1986) 520}.
%%CITATION = NUPHA,B274,520;%%.

\bibitem{Hanany:2010da}
A.~Hanany, D.~Forcella, and J.~Troost, ``{The covariant perturbative string
  spectrum},'' \href{http://dx.doi.org/10.1016/j.nuclphysb.2011.01.002}{{\em
  Nucl. Phys.} {\bfseries B846} (2011) 212--225},
\href{http://arxiv.org/abs/1007.2622}{{\ttfamily arXiv:1007.2622 [hep-th]}}.
%%CITATION = 1007.2622;%%.

\bibitem{Feng:2007ur}
B.~Feng, A.~Hanany, and Y.-H. He, ``{Counting Gauge Invariants: the Plethystic
  Program},'' \href{http://dx.doi.org/10.1088/1126-6708/2007/03/090}{{\em JHEP}
  {\bfseries 03} (2007) 090},
\href{http://arxiv.org/abs/hep-th/0701063}{{\ttfamily arXiv:hep-th/0701063}}.
%%CITATION = HEP-TH/0701063;%%.

\bibitem{Benvenuti:2006qr}
S.~Benvenuti, B.~Feng, A.~Hanany, and Y.-H. He, ``{Counting BPS operators in
  gauge theories: Quivers, syzygies and plethystics},''
  \href{http://dx.doi.org/10.1088/1126-6708/2007/11/050}{{\em JHEP} {\bfseries
  11} (2007) 050},
\href{http://arxiv.org/abs/hep-th/0608050}{{\ttfamily arXiv:hep-th/0608050}}.
%%CITATION = HEP-TH/0608050;%%.

\bibitem{Forcella:2008bb}
D.~Forcella, A.~Hanany, Y.-H. He, and A.~Zaffaroni, ``{The Master Space of N=1
  Gauge Theories},''
  \href{http://dx.doi.org/10.1088/1126-6708/2008/08/012}{{\em JHEP} {\bfseries
  08} (2008) 012},
\href{http://arxiv.org/abs/0801.1585}{{\ttfamily arXiv:0801.1585 [hep-th]}}.
%%CITATION = 0801.1585;%%.

\bibitem{Gray:2008yu}
J.~Gray, A.~Hanany, Y.-H. He, V.~Jejjala, and N.~Mekareeya, ``{SQCD: A
  Geometric Apercu},''
  \href{http://dx.doi.org/10.1088/1126-6708/2008/05/099}{{\em JHEP} {\bfseries
  0805} (2008) 099},
\href{http://arxiv.org/abs/0803.4257}{{\ttfamily arXiv:0803.4257 [hep-th]}}.
%%CITATION = ARXIV:0803.4257;%%.

\bibitem{Hanany:2008kn}
A.~Hanany and N.~Mekareeya, ``{Counting Gauge Invariant Operators in SQCD with
  Classical Gauge Groups},''
  \href{http://dx.doi.org/10.1088/1126-6708/2008/10/012}{{\em JHEP} {\bfseries
  0810} (2008) 012},
\href{http://arxiv.org/abs/0805.3728}{{\ttfamily arXiv:0805.3728 [hep-th]}}.
%%CITATION = ARXIV:0805.3728;%%.

\bibitem{Hanany:2008qc}
A.~Hanany, N.~Mekareeya, and A.~Zaffaroni, ``{Partition Functions for Membrane
  Theories},'' \href{http://dx.doi.org/10.1088/1126-6708/2008/09/090}{{\em
  JHEP} {\bfseries 0809} (2008) 090},
\href{http://arxiv.org/abs/0806.4212}{{\ttfamily arXiv:0806.4212 [hep-th]}}.
%%CITATION = ARXIV:0806.4212;%%.

\bibitem{Benvenuti:2010pq}
S.~Benvenuti, A.~Hanany, and N.~Mekareeya, ``{The Hilbert Series of the One
  Instanton Moduli Space},''
  \href{http://dx.doi.org/10.1007/JHEP06(2010)100}{{\em JHEP} {\bfseries 06}
  (2010) 100},
\href{http://arxiv.org/abs/1005.3026}{{\ttfamily arXiv:1005.3026 [hep-th]}}.
%%CITATION = 1005.3026;%%.

\bibitem{Hanany:2010qu}
A.~Hanany and N.~Mekareeya, ``{Tri-vertices and SU(2)'s},''
  \href{http://dx.doi.org/10.1007/JHEP02(2011)069}{{\em JHEP} {\bfseries 02}
  (2011) 069},
\href{http://arxiv.org/abs/1012.2119}{{\ttfamily arXiv:1012.2119 [hep-th]}}.
%%CITATION = 1012.2119;%%.

\bibitem{Hanany:2012dm}
A.~Hanany, N.~Mekareeya, and S.~S. Razamat, ``{Hilbert Series for Moduli Spaces
  of Two Instantons},''
\href{http://arxiv.org/abs/1205.4741}{{\ttfamily arXiv:1205.4741 [hep-th]}}.
%%CITATION = ARXIV:1205.4741;%%.

\bibitem{Cheng:2010pq}
M.~C. Cheng, ``{K3 Surfaces, N=4 Dyons, and the Mathieu Group M24},'' {\em
  Commun.Num.Theor.Phys.} {\bfseries 4} (2010) 623--658,
\href{http://arxiv.org/abs/1005.5415}{{\ttfamily arXiv:1005.5415 [hep-th]}}.
%%CITATION = ARXIV:1005.5415;%%.

\bibitem{Gaberdiel:2010ch}
M.~R. Gaberdiel, S.~Hohenegger, and R.~Volpato, ``{Mathieu twining characters
  for K3},'' \href{http://dx.doi.org/10.1007/JHEP09(2010)058}{{\em JHEP}
  {\bfseries 1009} (2010) 058},
\href{http://arxiv.org/abs/1006.0221}{{\ttfamily arXiv:1006.0221 [hep-th]}}.
%%CITATION = ARXIV:1006.0221;%%.

\bibitem{Gaberdiel:2010ca}
M.~R. Gaberdiel, S.~Hohenegger, and R.~Volpato, ``{Mathieu Moonshine in the
  elliptic genus of K3},''
  \href{http://dx.doi.org/10.1007/JHEP10(2010)062}{{\em JHEP} {\bfseries 1010}
  (2010) 062},
\href{http://arxiv.org/abs/1008.3778}{{\ttfamily arXiv:1008.3778 [hep-th]}}.
%%CITATION = ARXIV:1008.3778;%%.

\bibitem{Eguchi:2010fg}
T.~Eguchi and K.~Hikami, ``{Note on Twisted Elliptic Genus of K3 Surface},''
  \href{http://dx.doi.org/10.1016/j.physletb.2010.10.017}{{\em Phys.Lett.}
  {\bfseries B694} (2011) 446--455},
\href{http://arxiv.org/abs/1008.4924}{{\ttfamily arXiv:1008.4924 [hep-th]}}.
%%CITATION = ARXIV:1008.4924;%%.

\bibitem{Govindarajan:2011em}
S.~Govindarajan, ``{Unravelling Mathieu Moonshine},''
\href{http://arxiv.org/abs/1106.5715}{{\ttfamily arXiv:1106.5715 [hep-th]}}.
%%CITATION = ARXIV:1106.5715;%%.

\bibitem{Cheng:2012tq}
M.~C. Cheng, J.~F. Duncan, and J.~A. Harvey, ``{Umbral Moonshine},''
\href{http://arxiv.org/abs/1204.2779}{{\ttfamily arXiv:1204.2779 [math.RT]}}.
%%CITATION = ARXIV:1204.2779;%%.

\bibitem{Green:1987sp}
M.~B. Green, J.~Schwarz, and E.~Witten, {\em {Superstring Theory, Volume 1:
  Introduction}}.
\newblock
1987.
\newblock
%%CITATION = INSPIRE-250488;%%.

\bibitem{Green:1987mn}
M.~B. Green, J.~Schwarz, and E.~Witten, {\em {Superstring Theory, Volume 2:
  Loop Amplitudes, Anomalies and Phenomenology}}.
\newblock
1987.
\newblock
%%CITATION = INSPIRE-252419;%%.

\bibitem{Berkooz:1996km}
M.~Berkooz, M.~R. Douglas, and R.~G. Leigh, ``{Branes intersecting at
  angles},'' \href{http://dx.doi.org/10.1016/S0550-3213(96)00452-X}{{\em
  Nucl.Phys.} {\bfseries B480} (1996) 265--278},
\href{http://arxiv.org/abs/hep-th/9606139}{{\ttfamily arXiv:hep-th/9606139
  [hep-th]}}.
%%CITATION = HEP-TH/9606139;%%.

\bibitem{BenderOrszag}
C.~Bender and S.~Orszag, {\em Advanced mathematical methods for scientists and
  engineers: Asymptotic methods and perturbation theory}.
\newblock Advanced Mathematical Methods for Scientists and Engineers. Springer,
  1978.

\bibitem{HardyWright}
G.~Hardy and E.~Wright, {\em An introduction to the theory of numbers}.
\newblock Oxford science publications. Clarendon Press, 1979.

\bibitem{Banks:1987cy}
T.~Banks, L.~J. Dixon, D.~Friedan, and E.~J. Martinec, ``{Phenomenology and
  Conformal Field Theory Or Can String Theory Predict the Weak Mixing
  Angle?},''
\href{http://dx.doi.org/10.1016/0550-3213(88)90551-2}{{\em Nucl.Phys.}
  {\bfseries B299} (1988) 613--626}.
%%CITATION = NUPHA,B299,613;%%.

\bibitem{Banks:1988yz}
T.~Banks and L.~J. Dixon, ``{Constraints on String Vacua with Space-Time
  Supersymmetry},''
\href{http://dx.doi.org/10.1016/0550-3213(88)90523-8}{{\em Nucl.Phys.}
  {\bfseries B307} (1988) 93--108}.
%%CITATION = NUPHA,B307,93;%%.

\bibitem{Ferrara:1989ud}
S.~Ferrara, D.~L{\"u}st, and S.~Theisen, ``{World sheet versus Spectrum
  Symmetries in Heterotic and Type II Superstrings},''
\href{http://dx.doi.org/10.1016/0550-3213(89)90464-1}{{\em Nucl.Phys.}
  {\bfseries B325} (1989) 501}.
%%CITATION = NUPHA,B325,501;%%.

\bibitem{Feng:2012bb}
W.-Z. Feng, D.~L{\"u}st, and O.~Schlotterer, ``{Massive Supermultiplets in
  Four-Dimensional Superstring Theory},''
  \href{http://dx.doi.org/10.1016/j.nuclphysb.2012.03.010}{{\em Nucl. Phys. B}
  {\bfseries 861} (2012) 175--235},
\href{http://arxiv.org/abs/1202.4466}{{\ttfamily arXiv:1202.4466 [hep-th]}}.
%%CITATION = ARXIV:1202.4466;%%.

\bibitem{Odake:1989dm}
S.~Odake, ``{Character Formulas of an Extended Superconformal Algebra Relevant
  to String Compactification},''
\href{http://dx.doi.org/10.1142/S0217751X90000428}{{\em Int. J. Mod. Phys.}
  {\bfseries A5} (1990) 897}.
%%CITATION = IMPAE,A5,897;%%.

\bibitem{Eguchi:1988af}
T.~Eguchi and A.~Taormina, ``{On the Unitary Representations of N=2 and N=4
  Superconformal Algebras},''
\href{http://dx.doi.org/10.1016/0370-2693(88)90360-7}{{\em Phys. Lett.}
  {\bfseries B210} (1988) 125}.
%%CITATION = PHLTA,B210,125;%%.

\bibitem{Eguchi:2008gc}
T.~Eguchi and K.~Hikami, ``{Superconformal Algebras and Mock Theta
  Functions},'' \href{http://dx.doi.org/10.1088/1751-8113/42/30/304010}{{\em
  J.Phys.A} {\bfseries A42} (2009) 304010},
\href{http://arxiv.org/abs/0812.1151}{{\ttfamily arXiv:0812.1151 [math-ph]}}.
%%CITATION = ARXIV:0812.1151;%%.

\bibitem{Eguchi:2009cq}
T.~Eguchi and K.~Hikami, ``{Superconformal Algebras and Mock Theta Functions 2.
  Rademacher Expansion for K3 Surface},''
\href{http://arxiv.org/abs/0904.0911}{{\ttfamily arXiv:0904.0911 [math-ph]}}.
%%CITATION = ARXIV:0904.0911;%%.

\bibitem{Boels:2011zz}
R.~H. Boels and C.~Schwinn, ``{On-shell supersymmetry for massive
  multiplets},'' \href{http://dx.doi.org/10.1103/PhysRevD.84.065006}{{\em
  Phys.Rev.} {\bfseries D84} (2011) 065006},
\href{http://arxiv.org/abs/1104.2280}{{\ttfamily arXiv:1104.2280 [hep-th]}}.
%%CITATION = ARXIV:1104.2280;%%.

\bibitem{Berkovits:1997zd}
N.~Berkovits and M.~M. Leite, ``{First massive state of the superstring in
  superspace},'' \href{http://dx.doi.org/10.1016/S0370-2693(97)01269-0}{{\em
  Phys.Lett.} {\bfseries B415} (1997) 144--148},
\href{http://arxiv.org/abs/hep-th/9709148}{{\ttfamily arXiv:hep-th/9709148
  [hep-th]}}.
%%CITATION = HEP-TH/9709148;%%.

\bibitem{Berkovits:1998ua}
N.~Berkovits and M.~M. Leite, ``{Superspace action for the first massive states
  of the superstring},''
  \href{http://dx.doi.org/10.1016/S0370-2693(99)00334-2}{{\em Phys.Lett.}
  {\bfseries B454} (1999) 38--42},
\href{http://arxiv.org/abs/hep-th/9812153}{{\ttfamily arXiv:hep-th/9812153
  [hep-th]}}.
%%CITATION = HEP-TH/9812153;%%.

\bibitem{Zwegers2002}
S.~Zwegers, ``{Mock Theta Functions (Ph.D. thesis, Utrecht University,
  2002)},'' \href{http://arxiv.org/abs/0807.4834}{{\ttfamily arXiv:0807.4834
  [math-NT]}}.

\bibitem{Boels:2012ie}
R.~H. Boels and D.~O'Connell, ``{Simple superamplitudes in higher
  dimensions},''
\href{http://arxiv.org/abs/1201.2653}{{\ttfamily arXiv:1201.2653 [hep-th]}}.
%%CITATION = ARXIV:1201.2653;%%.

\bibitem{Boels:2012if}
R.~H. Boels, ``{Three particle superstring amplitudes with massive legs},''
  \href{http://dx.doi.org/10.1007/JHEP06(2012)026}{{\em JHEP} {\bfseries 1206}
  (2012) 026},
\href{http://arxiv.org/abs/1201.2655}{{\ttfamily arXiv:1201.2655 [hep-th]}}.
%%CITATION = ARXIV:1201.2655;%%.

\bibitem{Boels:progress}
R.~H. Boels and O.~Schlotterer, ``{work in progress},''
  \href{http://arxiv.org/abs/12XX.YYYY}{{\ttfamily arXiv:12XX.YYYY [hep-th]}}.

\bibitem{deWit:1999ir}
B.~de~Wit and D.~L{\"u}st, ``{BPS amplitudes, helicity supertraces and
  membranes in M theory},''
  \href{http://dx.doi.org/10.1016/S0370-2693(00)00137-4}{{\em Phys.Lett.}
  {\bfseries B477} (2000) 299--308},
\href{http://arxiv.org/abs/hep-th/9912225}{{\ttfamily arXiv:hep-th/9912225
  [hep-th]}}.
%%CITATION = HEP-TH/9912225;%%.

\bibitem{Maldacena:1997re}
J.~M. Maldacena, ``{The Large N limit of superconformal field theories and
  supergravity},'' {\em Adv.Theor.Math.Phys.} {\bfseries 2} (1998) 231--252,
\href{http://arxiv.org/abs/hep-th/9711200}{{\ttfamily arXiv:hep-th/9711200
  [hep-th]}}.
%%CITATION = HEP-TH/9711200;%%.

\bibitem{Aharony:1999ti}
O.~Aharony, S.~S. Gubser, J.~M. Maldacena, H.~Ooguri, and Y.~Oz, ``{Large N
  field theories, string theory and gravity},''
  \href{http://dx.doi.org/10.1016/S0370-1573(99)00083-6}{{\em Phys.Rept.}
  {\bfseries 323} (2000) 183--386},
\href{http://arxiv.org/abs/hep-th/9905111}{{\ttfamily arXiv:hep-th/9905111
  [hep-th]}}.
%%CITATION = HEP-TH/9905111;%%.

\bibitem{Gerigk:2012cq}
S.~Gerigk, ``{String States on $AdS_3 x S^3$ from the Supergroup},''
\href{http://arxiv.org/abs/1208.0345}{{\ttfamily arXiv:1208.0345 [hep-th]}}.
%%CITATION = ARXIV:1208.0345;%%.

\bibitem{Mazzucato:2011jt}
L.~Mazzucato, ``{Superstrings in AdS},''
\href{http://arxiv.org/abs/1104.2604}{{\ttfamily arXiv:1104.2604 [hep-th]}}.
%%CITATION = ARXIV:1104.2604;%%.

\bibitem{Bianchi:2003wx}
M.~Bianchi, J.~F. Morales, and H.~Samtleben, ``{On stringy $AdS_5 x S^5$ and
  higher spin holography},'' {\em JHEP} {\bfseries 0307} (2003) 062,
\href{http://arxiv.org/abs/hep-th/0305052}{{\ttfamily arXiv:hep-th/0305052
  [hep-th]}}.
%%CITATION = HEP-TH/0305052;%%.

\bibitem{Gaberdiel:2010ar}
M.~R. Gaberdiel, R.~Gopakumar, and A.~Saha, ``{Quantum $W$-symmetry in
  $AdS_3$},'' \href{http://dx.doi.org/10.1007/JHEP02(2011)004}{{\em JHEP}
  {\bfseries 1102} (2011) 004},
\href{http://arxiv.org/abs/1009.6087}{{\ttfamily arXiv:1009.6087 [hep-th]}}.
%%CITATION = ARXIV:1009.6087;%%.

\bibitem{Gaberdiel:2010pz}
M.~R. Gaberdiel and R.~Gopakumar, ``{An $AdS_3$ Dual for Minimal Model CFTs},''
  \href{http://dx.doi.org/10.1103/PhysRevD.83.066007}{{\em Phys.Rev.}
  {\bfseries D83} (2011) 066007},
\href{http://arxiv.org/abs/1011.2986}{{\ttfamily arXiv:1011.2986 [hep-th]}}.
%%CITATION = ARXIV:1011.2986;%%.

\bibitem{Gaberdiel:2011zw}
M.~R. Gaberdiel, R.~Gopakumar, T.~Hartman, and S.~Raju, ``{Partition Functions
  of Holographic Minimal Models},''
  \href{http://dx.doi.org/10.1007/JHEP08(2011)077}{{\em JHEP} {\bfseries 1108}
  (2011) 077},
\href{http://arxiv.org/abs/1106.1897}{{\ttfamily arXiv:1106.1897 [hep-th]}}.
%%CITATION = ARXIV:1106.1897;%%.

\bibitem{Kiritsis:1997hj}
E.~Kiritsis, ``{Introduction to superstring theory},''
\href{http://arxiv.org/abs/hep-th/9709062}{{\ttfamily arXiv:hep-th/9709062
  [hep-th]}}.
%%CITATION = HEP-TH/9709062;%%.

\end{thebibliography}\endgroup


\providecommand{\href}[2]{#2}\begingroup\raggedright\endgroup

\end{document}